\documentclass[twocolumn,floatfix,linenubers]{aastex63}

\usepackage{xcolor}
\definecolor{blue}{rgb}{0,0,1}
\definecolor{BLUE}{rgb}{0,0,1}
\definecolor{green}{rgb}{0,1,0}
\definecolor{GREEN}{rgb}{0,1,0}

\definecolor{red}{cmyk}{0,0,0,1}
\definecolor{RED}{cmyk}{0,0,0,1}

\received{}
\revised{}
\accepted{}
\submitjournal{ApJ}

\begin{document}

\title{On the importance of 
 wave  planet interactions
 for the migration of two super-Earths embedded 
in a protoplanetary disk 
}

\correspondingauthor{Ewa Szuszkiewicz}
\email{ewa.szuszkiewicz@usz.edu.pl}

\author{Zijia Cui}
\affiliation{ Institute of Physics and CASA*, University of Szczecin, \\
Wielkopolska 15, 70-451 Szczecin, Poland}

\author{John C. B. Papaloizou}
\affiliation{DAMTP, University of Cambridge, Wilberforce Road, Cambridge, CB3 0WA, UK}

\author[0000-0002-7881-2805]{Ewa Szuszkiewicz}
\affiliation{ Institute of Physics and CASA*, University of Szczecin, \\
Wielkopolska 15, 70-451 Szczecin, Poland}











\begin{abstract}
\textcolor{red} {
We investigate a repulsion mechanism between two low-mass planets
migrating in a  protoplanetary disk, for which the relative 
migration switches from convergent to  divergent. This mechanism invokes 
density waves emitted by one planet transferring angular
momentum to the coorbital region of the other and then directly to it 
through the horseshoe drag. 
We formulate simple analytical estimates, which indicate
when the repulsion  mechanism is  effective.
One condition for a planet to be  repelled
is that it forms a partial gap in the disk and
another is that this should contain enough material
to support  angular momentum exchange with it.
Using two-dimensional 
hydrodynamical simulations we obtain divergent migration of
two super-Earths embedded in a protoplanetary disk because  of
repulsion between them and verify these conditions.
To investigate the importance of resonant interaction 
we study the  migration of  planet pairs  near first-order commensurabilities. 
It appears that proximity  to resonance is significant  but 
not essential. In this context we find repulsion  still occurs when the 
gravitational interaction between the planets is removed sugesting  the 
importance  of  angular momentum transfer through  waves excited 
by another planet. This may occur through the scattering of 
coorbital material (the horseshoe drag), or  material orbiting close by.
Our results indicate that if  conditions  favor the
repulsion between two planets described above, 
we expect to observe planet pairs with their period ratios  greater,
often only slightly greater, than  resonant values or possibly rarity 
of commensurability.}
\end{abstract}



\section{Introduction} 
\label{sec:intro}
The early stages of the evolution of multi-planet systems that 
occurs after their  formation, when  gaseous protoplanetary disks 
are still present, have been studied in many works 
\citep[e.g.][]{Nelson2002, Kley2004, Papa2005, KleyNelson2012}.
One of the important outcomes of this evolutionary phase is an
inevitable process of planetary migration taking place in a
gaseous environment.
It is expected that when there is convergent migration of two
planets they  can be trapped in  mean-motion resonances (MMRs). 
The frequency of the occurrence of MMRs is therefore a robust 
feature to be tested against predictions based on the results 
of numerical simulations.
The observational data indicates that planet pairs with period 
ratios close to commensurability are not as common as simulations, 
that adopt  simple prescriptions for migration rates, indicate.
Moreover, the multiple systems observed by the Kepler mission 
are such  that the planet pairs near resonances tend to have 
period ratios slightly larger than those required for strict 
commensurability, especially in the case of the 3:2 and 2:1
resonances
\citep[e.g][]{Lissauer2011, Fabrycky2014, Steffen2015}.
This has been found for both pairs of giant planets and systems 
composed of two low-mass planets.

One  example of these is the Kepler-59 system, which contains 
two planets with masses of $5M_{\oplus}$ and $4.6M_{\oplus}$ 
being the inner and outer planet respectively. The orbital period 
ratio is  1.5141 \citep[]{Olivera2020}.
Another example is the Kepler-128 system in which the inner 
and outer planets have masses equal to $0.77M_{\oplus}$ and 
$0.9M_{\oplus}$, respectively and their period ratio is 
1.5112 \citep[]{Hadden2016}. In both planetary systems, the 
orbital period ratio of two planets is slightly larger than 
$3/2.$

Furthermore, several two-planet systems with planet masses  
in the super-Earth  range possess period ratios close to other 
first-order resonances.
The two low-mass planets in the Kepler-177 system are near the
4:3 resonance \citep[]{Hadden2017,Vissapragada2020}. In the  
Kepler-307 system, two super-Earths are close to 5:4 resonance 
\citep{Jontof2016} and in  Kepler-36 the period ratio is close
to 7:6 \citep{Carter2012, Vissapragada2020}. 
\textcolor{red} {It is worth mentioning that in the case of
the Kepler-59 and Kepler-307 systems, the outer planet is less
massive than the inner one.}
Detailed information 
about the planetary systems mentioned above is given in 
Table~\ref{tab:nearto}.

\begin{deluxetable*}{llllccl}
\tablecaption{Two-planet systems with the period ratios near
the first-order resonances \label{tab:nearto}}
\tablecolumns{6}
\tablewidth{0pt}
\tablehead{
\colhead{System} & \colhead{Planet pair}
& \colhead{Period (days)} & \colhead{Mass ($M_{\star}$)} & \colhead{Resonance } & \colhead{Deviation of period ratio } & \colhead{Reference} \\
& & & &$p+1:p$ & from $(p+1)/p$ }
\startdata
Kepler-59 & Kepler-59b & 11.8715 & $1.10 \times 10^{-5}$ & 3:2 & 0.0141  & \citet{Olivera2020} \\
          & Kepler-59c & 17.9742 & $1.02 \times 10^{-5}$ &     &         &                    \\
\hline
Kepler-128 & Kepler-128b & 15.090 & $2.12 \times 10^{-6}$ & 3:2 & 0.0112  & \citet{Hadden2016} \\
           & Kepler-128c & 22.804 & $2.48 \times 10^{-6}$ &     &         &                    \\
\hline
Kepler-177 & Kepler-177b & 35.860 & $1.90 \times 10^{-5}$ & 4:3 & 0.0445 & \citet{Vissapragada2020} \\
           & Kepler-177c & 49.409 & $4.79 \times 10^{-5}$ &     &         &                    \\
\hline
Kepler-307 & Kepler-307b & 10.4208 & $2.46 \times 10^{-5}$ & 5:4 & 0.0045  & \citet{Jontof2016} \\
           & Kepler-307c & 13.0729 & $1.20 \times 10^{-5}$ &     &         &                    \\
\hline
Kepler-36  & Kepler-36b & 13.8683& $1.11 \times 10^{-5}$ & 7:6 & 0.0028  & \citet{Vissapragada2020} \\
           & Kepler-36c & 16.2187 & $2.07 \times 10^{-5}$ &     &         &                    \\
\hline
\enddata
\end{deluxetable*}

A better understanding of the distribution of the observed period 
ratios is a key ingredient in any model of the formation of compact
planetary systems. Many mechanisms have been put forward to explain 
the reasons for the departure from strict mean-motion resonance 
that would be expected as a result of migration in the protoplanetary 
disk.  In the case of planets orbiting close to their parent stars, 
a mechanism driven by dissipation induced by tides raised on the 
planets by the central  star can be particularly effective 
\citep[]{JohnCaroline2010, John2011, Lithwick2012, Batygin2013, 
Lee2013, Delisle2014}.

A mechanism to account for the departure from resonance or more 
generally for the forestallment of the attainment of commensurability, 
regardless of the distance from which a planet pair orbits its parent 
star, has been identified. This considers the effects of the density 
waves excited by one of the planets on the other one. In particular 
the transfer of the angular momentum carried by the waves to the other  
planet can lead to divergent migration, thus preventing the planets
from being closely locked into any MMR.

In \citet{Podlewska2012}, these effects were investigated in a system 
containing an inner giant planet (a source of the outward propagating
density waves) and an outer super-Earth, using both global 
two-dimensional hydrodynamical calculations and local shearing box 
simulations. They showed that the inward migration of the super-Earth
could be stopped or even reversed due to the angular momentum transfer
by the outgoing density waves.

Similar interactions between the planets and the disk were studied for 
planet pairs consisting of two Saturn-like planets as well as two 
less massive Uranus-like ones  in \citet{Baruteau2013}.
They found that the disk-driven repulsion,  described above, that leads 
to the divergent migration of the planets, is particularly efficient 
when at least one of the planets opens a partial gap in the disk.

The migration of planets, which are able to form a partial gap has 
attracted a lot of attention  recently \citep[e.g.][]{duffell2014,
durmann2015,durmann2017,kanagawa2018}. The way in which such planets
migrate is crucial for answering  questions about several aspects
of planetary system formation and early evolution, and in particular 
about the possibility of forming mean-motion resonances. This issue  
has been recently investigated by \citet{kanagawa2020}.

The aim of this paper is to extend  previous studies to  pairs of
low-mass planets with masses in the super-Earth range embedded in 
a disk in which they are  capable of opening a partial gap in 
which they orbit.

\textcolor{red}{The main focus is put on} the  repulsion mechanism 
responsible for the 
divergent migration that  prevents capture into a strict 
commensurability.
We check that the angular momentum flow between the planets
is consistent with theoretical expectation for the density waves 
they produce and that the angular momentum flow resulting from the 
density waves emitted by one into the horseshoe region of the other 
can account for the torque required to produce the switch from 
convergent  to divergent migration.
\textcolor{red} {As a result, we are able to identify the nature of the
repulsion between low-mass planets migrating in the protoplanetary
disk and demonstrate that it can occur
for super-Earths with masses below 10 Earth masses. Finally,
we provide simple analytical estimates, which
indicate when this repulsion is effective.
The observational consequences of the repulsion mechanism found in this 
paper  might be important but to come up with more \textcolor{red} {quantitative}
picture requires further studies.
}

The plan of this paper is as follows. In Section~\ref{sec:setup} we 
describe the disk and planet parameters adopted in our investigations.
In Section~\ref{sec:results} we describe two-dimensional hydrodynamical 
simulations of two super-Earths that after an initial period of 
convergent migration towards the vicinity of the 3:2 MMR ultimately 
undergo divergent migration.

In order to separate genuine single planet evolution from the
effects caused by the presence of a second planet,
in  Section~\ref{sec:single} we analyse the migration of a
single super-Earth corresponding to the outer component
of the system considered in Section~\ref{sec:results}.
Because this planet opens a partial gap in the disk, standard 
type I migration does not apply.
Accordingly we provide simple fits to our numerical data for
different surface density profiles, using these to quantify 
deviations seen in the two planet simulations in 
Section~\ref{sec:twoplanetcase}.
In Section~\ref{Coorbgaps} we describe a simple model for how 
density waves emitted by one planet being absorbed in the horseshoe 
region of the other can lead to planet-planet repulsion, giving 
criteria for the mechanism to be effective.
We go on to verify these for systems having a range of mass ratios
in Section~\ref{sec:innermass} and study the dependence of the 
initial rate of convergent migration on the disk surface density 
and planet  mass ratios in Section~\ref{sec:repulsion}.

The consistency of the angular momentum flow associated with
density waves with  the torque necessary to provide  the repulsion 
between planets is investigated  in Section~\ref{sec:horseshoe}.
We then establish the robustness of our results to changes of the 
surface density profile, the adopted equation of state,  and the 
effect of neglecting the disk self-gravity in 
Sections~ \ref{sec:surfacedensity} - \ref{sec:selfg}.
{\textcolor{red}{ In order to consider  protoplanetary disks
with significantly smaller surface densities that 
may occur during the dispersal process possibly
produced by photoevaporation,}
\textcolor{red}{we examine the repulsion  \textcolor{red}{mechanism} in  disks with significantly 
reduced surface densities \textcolor{red}{ in Section~\ref{sec:lowdensity} and 
finally} we demonstrate the effectiveness of the \textcolor{red}{repulsion} mechanism
for the super-Earths migrating in a viscous
disk \textcolor{red}{ in which the temperature is determined by the balance
between local heating and cooling
and which supports a} constant angular momentum flux 
in Section~\ref{sec:viscousdisk}.
\textcolor{red} This situation may arise when the inner disk is
disrupted by the magnetosphere of the               
central star \citep[eg.][]{Clarke1996}. }
Finally, we summarise our results and conclude in 
Section~\ref{sec:discussion}.


\section{Disk model and numerical setup} 
\label{sec:setup}
We consider a system of two planets with masses $m_{1}$ (inner planet)
and $m_{2}$ (outer planet), 
orbiting a central star of  mass  $M_{\star}$,  while being embedded 
in a protoplanetary disk. We will find it convenient to make use of 
the planet-to-star mass ratios  $q_i = m_i/M_{\star}$ with $i=1$ 
denoting the inner planet and $i=2$ denoting the outer one. 
A two-dimensional disk model is adopted together with a cylindrical 
polar coordinate system ($r, \phi$, z) with  origin located at the 
central star which is regarded as a point mass. 
\textcolor{red}{ 
We start our investigations 
from the simple disk
model described by the} 
continuity equation and the equation of motion 
\textcolor{red} {in}
the form:

\begin{equation}
    \frac{\partial \Sigma}{\partial t}=-\nabla \cdot (\Sigma \bf{v})
	\label{eq:continuity}
\end{equation}
\noindent and
\begin{equation}
    \frac{\partial\bf{v}}{\partial t} + {\bf{v}} \cdot \nabla {\bf{v}} = 
- \frac{1}{\Sigma} \nabla P - \nabla \Phi + {\bf{f}}_{\nu}
	\label{eq:motion}
\end{equation}

\noindent where $\Sigma$ and ${\bf v} $ denote the surface density of 
the disk and the velocity, while $\Phi$ is the gravitational potential 
and ${\bf f}_{\nu}$ is the viscous force per unit mass. We adopt a 
locally isothermal equation of state, where the vertically integrated 
pressure can be expressed as $P=\Sigma c_{s}^{2}$, with the sound speed 
related to the pressure scale height of the disk, $H$, through
$c_{s}=(H/r)(GM_{\star}/r)^{1/2}$. Here $G$ is the gravitational 
constant. The aspect ratio of the disk, $h=H/r$, is assumed to be 
constant in the simulation, which leads to $c_{s}^{2} \propto r^{-\beta}$ 
with $\beta=1.$ Thus, the power-law index of the radial temperature 
profile is equal to $-1$. 
\textcolor{red}{We go on to perform a calculation with an adiabatic
equation of state  where $P$ is given by $P=(\gamma -1) e\Sigma$. 
Here $\gamma$ is the adiabatic index and $e$ is the specific internal
energy \textcolor{red}{ in Section \ref{sec:adiab}.} }The self-gravity of the disk is neglected in our simulations. 
The effect of this assumption has been investigated and discussed in 
Section~\ref{sec:selfg1}. 
\textcolor{red}{Finally, we adopt a viscous disk model,  
in which the temperature is determined by the balance
between local heating and cooling and which supports
a constant angular momentum flux. The formulation of this model
is given in 
Section~\ref{sec:viscousdisk}.}

The system of units adopted in this paper is as follows: The unit of 
mass is the mass of the central star $M_{\star}$. The unit of length 
is the initial orbital radius of the inner planet $r_{1}$. The time unit 
is the initial orbital period of the inner planet $P_{1}$. \textcolor{red}{To fix on particular parameters
we can think of
initiating the inner planet in a circular orbit at $r_{1}=1 \rm ~au$ 
and take $M_{\star}$ to be the mass of the Sun. Then, the time unit is 
$1 \rm ~yr$. However, we note that the results can be scaled to apply
to other values of  $r_1$ and $M_{\star}.$}

In our simulations the initial orbital eccentricities of both planets 
are set to be zero. The initial orbital radii of the two planets in 
the disk are $r_{1}=1$ for the inner planet and $r_{2}=1.48$ for the 
outer one.  The masses of the inner and outer planets are in the 
super-Earth mass range. 

The initial surface density profile $\Sigma(r)$, adopted in the majority
of our simulations \textcolor{red}{that incorporate a central disk cavity}  is given by
\begin{eqnarray}
\label{disk}
\Sigma(r) & = 1.25\Sigma_0 r 
- 0.25\Sigma_{0} \hspace{5.5mm} {\rm for} & r_{\rm min} < r < 1, \nonumber \\
\Sigma(r) & = 
\Sigma_0r^{-\alpha} \hspace{20mm}  \ \ {\rm for } & 1 \leq r < r_{\rm max},
\end{eqnarray}

\noindent where $\Sigma_{0}$ is a scaling parameter 
while $r_{\rm min}$ and $r_{\rm max}$ are the inner and outer boundaries 
of the computational domain, respectively. In this 
work, if not stated differently, we take $\Sigma_{0}=6 \times 10^{-5}$ 
in units of $M_{\star}/r_{1}^{2}$ and $\alpha=1/2$.
\textcolor{red}{ We remark that $\pi\Sigma_0$ is approximately the disk mass in units of $M_{\star}$ within 
$r_1.$ In this case for $M_{\star}= 1M_{\odot},$ this is    $0.19$ Jupiter masses.
For $r_1 = 1$ au this is about five times smaller than expected for the minimum mass solar nebula. } 
 This particular 
surface density profile is found to guarantee the convergent migration 
of super-Earths in the early stages of their evolution. The surface density 
of the inner part of the disk decreases sharply on moving inwards resulting 
in the formation of a trap  where the migration of an incoming low mass 
planet is halted \citep[]{Masset2006b}. In this way, the migration of the 
inner planet may be stopped while the outer planet being external to the 
trap continues  to migrate inwards. Thus ensures that in the early stages
of the simulations the planets will undergo convergent migration. 

The governing hydrodynamical equations are solved by using the numerical 
code FARGO3D \citep[][]{BenitezM2016}.
A constant kinematic viscosity, $\nu$, is adopted to model the transport
presumed to result from turbulence.  
The computational domain in the radial direction extends from 
$r_{\rm min}=0.2$ to $r_{\rm max}=2.6$ and covers the whole $2\pi$ 
domain in azimuth. The resolution in the calculations is 900 equal cells 
forming a staggered grid in the radial direction and 1800 equal cells in 
the azimuthal direction. 
We found it advantageous to adopt a  rotating frame that corotates with 
the Keplerian angular velocity at the initial location of the inner planet. 
The standard outflow boundary conditions are applied at the  disk 
boundaries and  wave killing-zones \citep[]{Val2006} operate in the domains
$[0.2,0.32]$ and $[2.36,2.6]$ which are connected to the inner and outer 
boundaries of the computational domain, respectively.

The gravitational potential at any point, $\Phi,$ is the sum of the 
potential due to the central star and the potential due to the planets, 
$\Phi_{i}, i =1,2.$  When working out the force per unit mass on any planet, 
there is in addition to the contribution due to direct gravitational 
interaction, an indirect term that arises on account of  the acceleration 
of the origin of the coordinate system that is constrained to be centred 
on the star. The softened potential due to a planet with mass $m_{i}$ has 
the form:
\begin{equation}
\Phi_{i}=-\frac{Gm_{i}}{\sqrt{r^{2}-2rr_{i}\cos(\phi-\phi_{i})
+r_{i}^{2}+b^{2}r_{i}^{2}}}
	\label{eq:potential}
\end{equation}  

\noindent where $m_{i}$ is the mass of the planet, $(r_{i}, \phi_{i})$ 
are the cylindrical coordinates of the planet, and  $b$ is the softening 
parameter. The latter is adopted to take account of the fact that the 
disk material is distributed over the vertical extent of the disk in an 
approximate manner.  The value of $b$ is taken to be equal to $0.6h$. 
 
In all simulations \textcolor{red} {in which the equation of state
is} \textcolor{red}{locally isothermal}
the disk aspect ratio is adopted to be $h=0.02$ and 
the viscosity is taken as $\nu=1.2 \times 10^{-6}$ in units of 
$r_{1}^{2}(GM_{\star}/r_{1}^{3})^{1/2}$.
These correspond to the expected  aspect ratio at $0.1 \rm ~au$  given 
$h =0.05$ at $5\rm ~au$ and the scaling $h\propto r^{1/4}$ corresponding 
to the source of heating being radiation from the central star,
and a \citet{Sha1973} $\alpha$-viscosity parameter equal to $0.003.$

For the disk parameters taken in our study, the two super-Earths 
are such that the parameters $q_i^{1/3}/h,$ and $q_i/(40\nu)$ which 
respectively measure the degree of nonlinearity and the ratio of tidal 
to viscous torques  are respectively $\sim 1$ and $\sim 0.25.$ 
Thus the planets are expected to open a partial gap in the disk around 
their orbits \citep[see][]{Lin1993,Kory1996, Crida2006}.

\begin{figure*}[htb!]
\centerline{
\vbox{
\hbox{
\includegraphics[width=0.5\columnwidth]{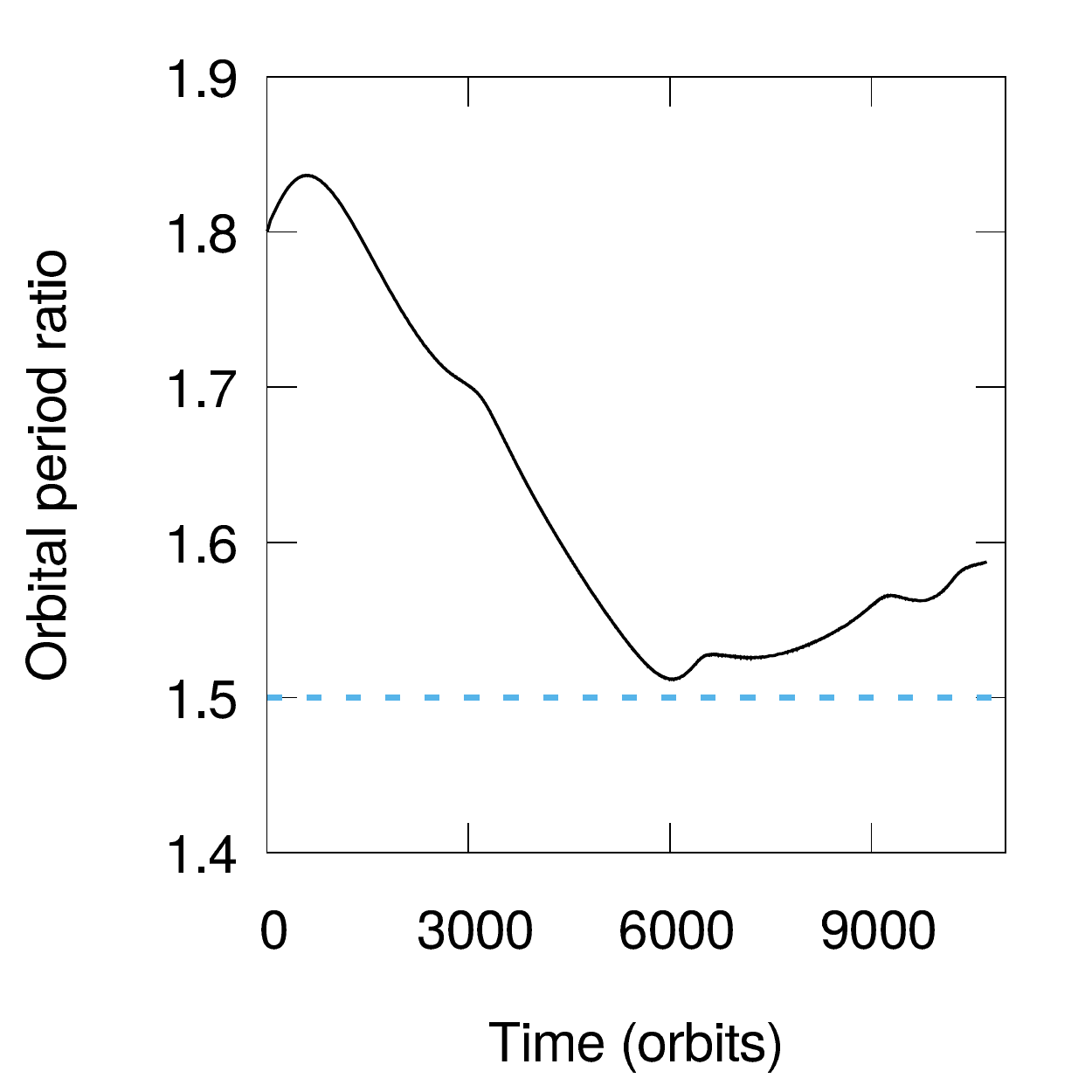}
\includegraphics[width=0.5\columnwidth]{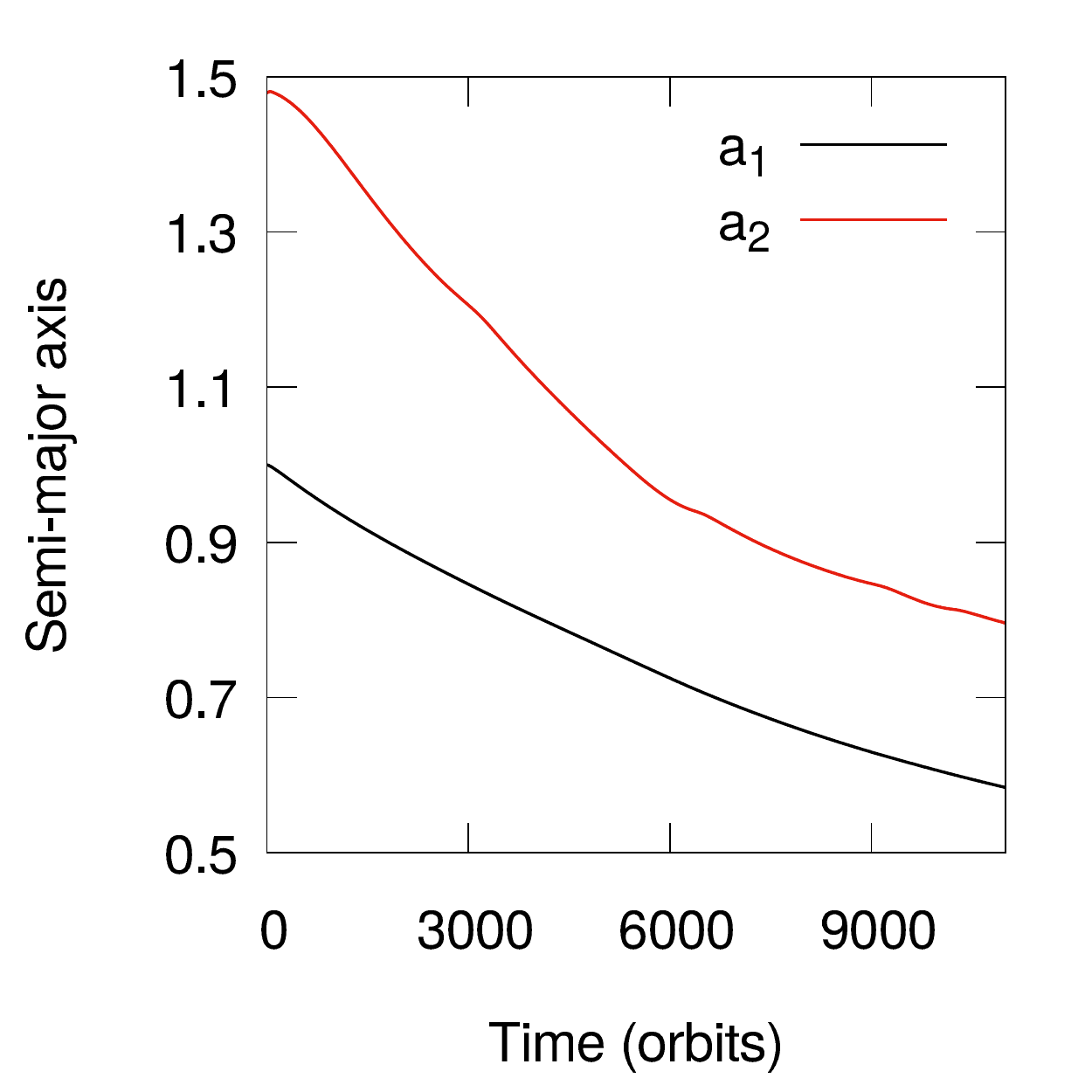}
\includegraphics[width=0.5\columnwidth]{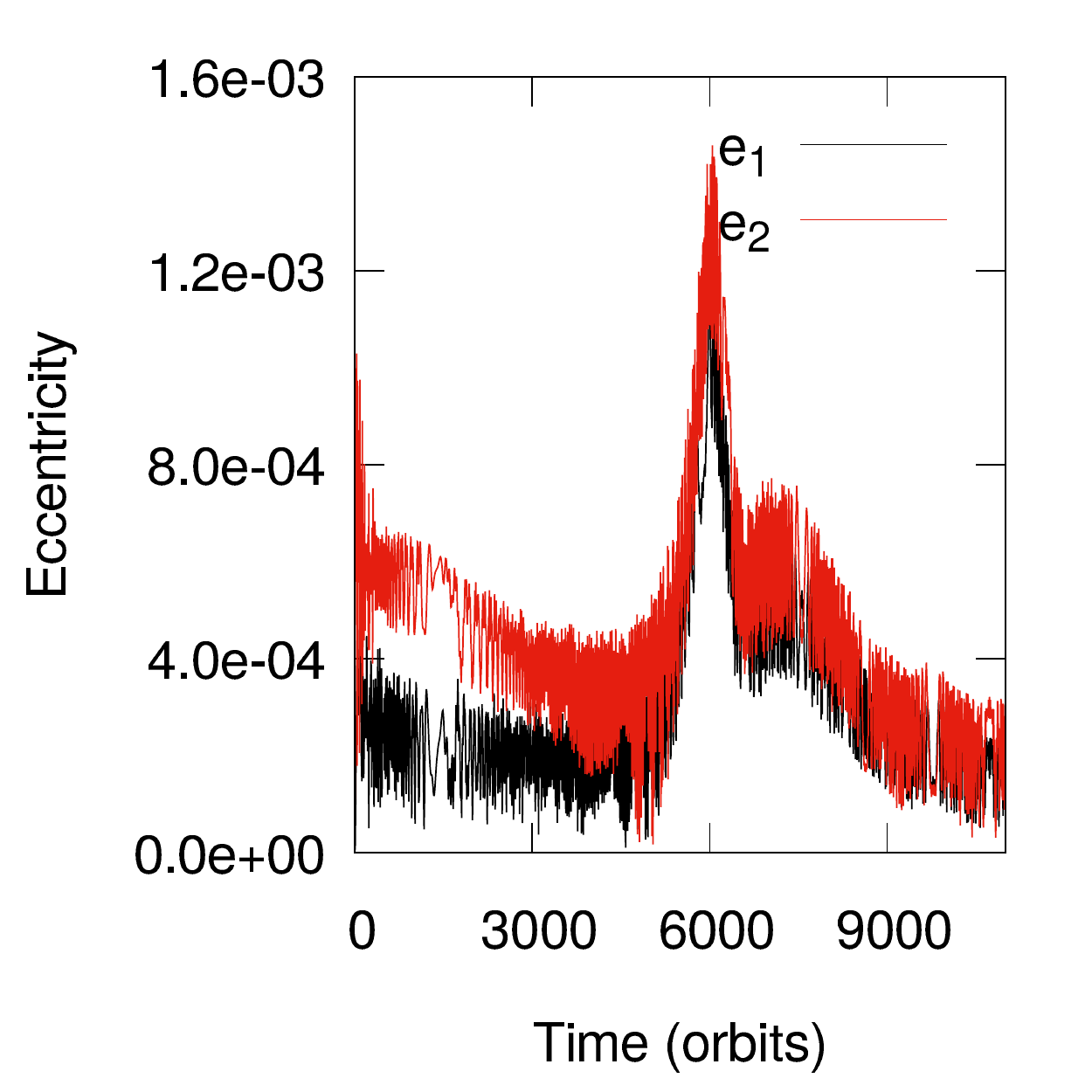}
\includegraphics[width=0.5\columnwidth]{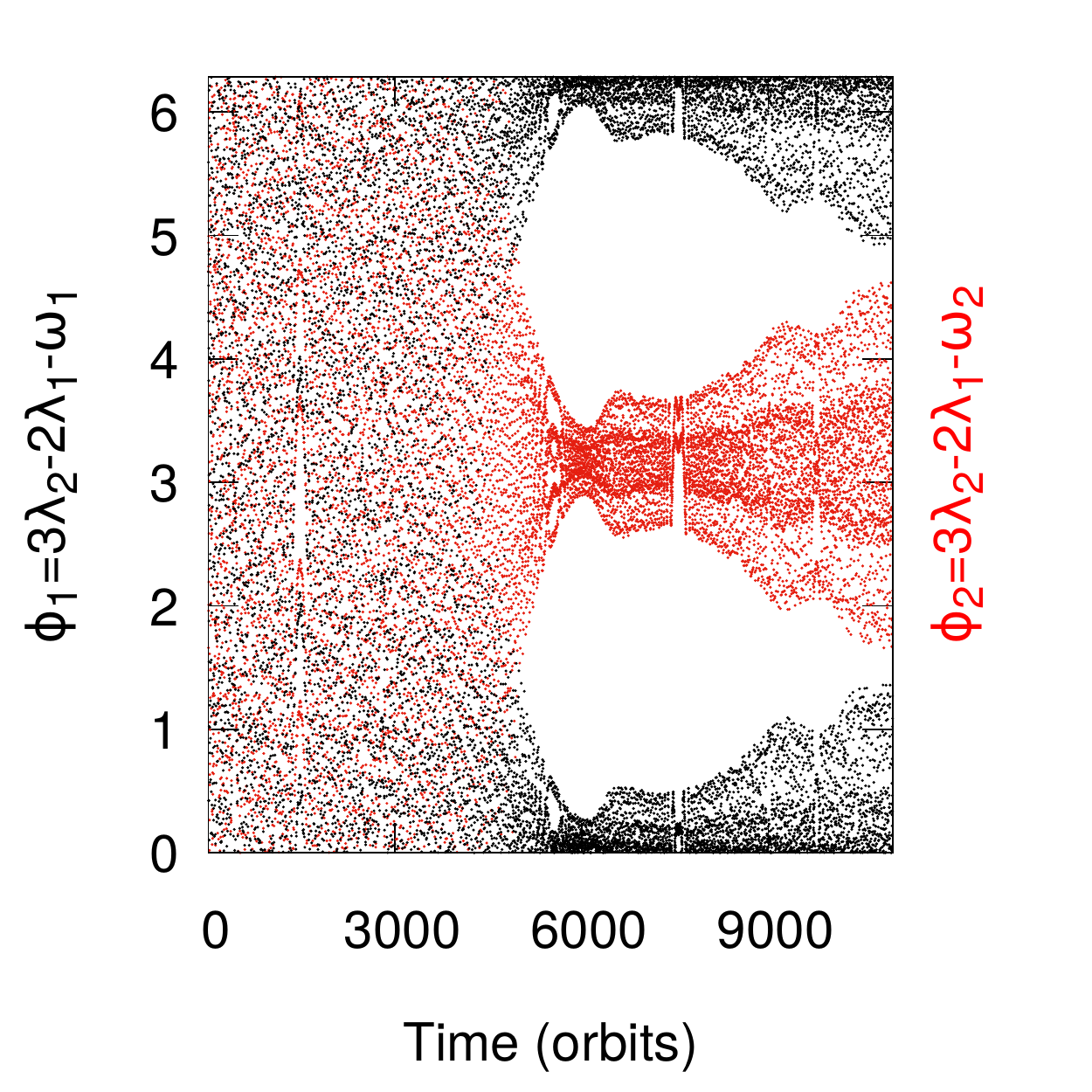}
}
}
}
\caption{The results of the hydrodynamical simulation of two super-Earths
with $q_{1}=1.3 \times 10^{-5}$ and $q_{2}=1.185 \times 10^{-5}$
migrating in a protoplanetary disk with
$\Sigma=8\times 10^{-5} r^{-1/2}$, $h=0.02$ and $\nu = 1.2
\times 10^{-6}$.
The evolution of the  orbital period ratio, the semi-major axes, 
the eccentricities and the resonance angles are shown in the panels 
from left to right. The horizontal dashed blue line in the first panel 
indicates the position of the 3:2 commensurability.
}
\label{fig:mig_scaling}
\end{figure*}

\begin{figure*}[htb!]
\centerline{
\vbox{
\hbox{
\includegraphics[width=1.8\columnwidth]{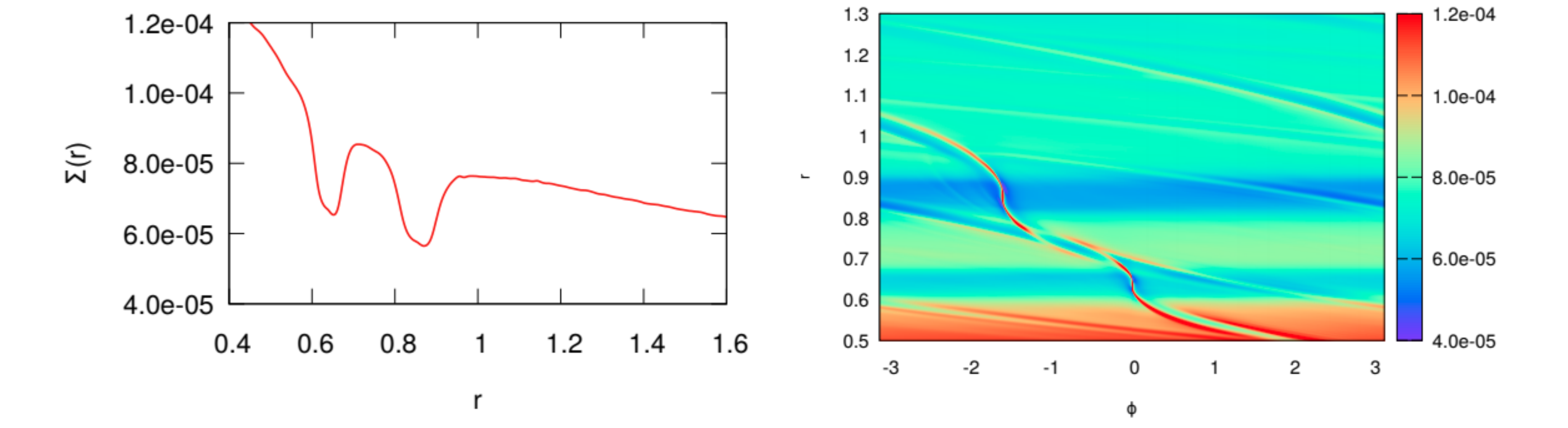}
}
}
}
\caption{The results of the hydrodynamical simulation  of two super-Earths
with $q_{1}=1.3 \times 10^{-5}$ and $q_{2}=1.185 \times 10^{-5}$
migrating in a protoplanetary disk with $\Sigma=8\times 10^{-5} r^{-1/2}$, 
$h=0.02$ and $\nu = 1.2 \times 10^{-6}$.
The azimuthally averaged surface density profile and a contour plot of the 
surface density of the disk in the vicinity of two planets at $t \sim 8560$ 
orbits are shown.
}
\label{fig:figure2}
\end{figure*}


\section{The divergent migration  of two super-Earths in a protoplanetary 
disk following a period of convergent migration} 
\label{sec:results}
In this Section we describe the evolution of two super-Earths evolving 
dynamically in a gaseous protoplanetary disk in the vicinity of the 3:2 
mean-motion resonance using full two-dimensional hydrodynamic simulations. 
The aim of the calculations is to verify if the system of two super-Earths  
having undergone a period of convergent migration  can see this reversed 
and subsequently undergo  divergent migration. This phenomenon is similar 
to that  observed in  systems containing a  Jupiter mass planet and a
super-Earth \citep{Podlewska2012}, two gaseous giants with the masses 
similar to  that of Saturn, and two ice giants with  masses similar to 
that of Uranus \citep{Baruteau2013}. 

The initiation of the migrating planets close to a first-order 
commensurability gives us an opportunity to investigate the role, if any, 
of the mean-motion resonance in determining the orbital evolution of two 
super-Earths.
We consider two low-mass planets with $q_{1}=1.3\times 10^{-5}$ and
$q_{2}=1.185\times 10^{-5}$. Both planets will open a partial gap in
the disk with  aspect ratio $h=0.02$ and  viscosity $\nu = 1.2
\times 10^{-6}$  that we consider here. This is our flagship case and it 
will serve as a reference for numerous simulations presented in the 
present paper.

At the beginning of the simulation the planets are located in circular 
orbits with $r_{2}/r_{1}=1.48.$ The initial orbital period ratio 
being $1.8$ exceeds 3:2 (see Figure~\ref{fig:mig_scaling}).
For this simulation the initial surface density is taken to be 
$\Sigma=\Sigma_0 r^{-\alpha}$, where $\Sigma_0=8\times 10^{-5}$ and 
$\alpha =0.5$. In this case we did not adopt the surface density profile
given by Equation~(\ref{disk}) as it is not required to establish  
initially  convergent migration of the planets.

\begin{figure*}[htb!]
\centerline{
\vbox{
\hbox{
\includegraphics[width=0.5\columnwidth]{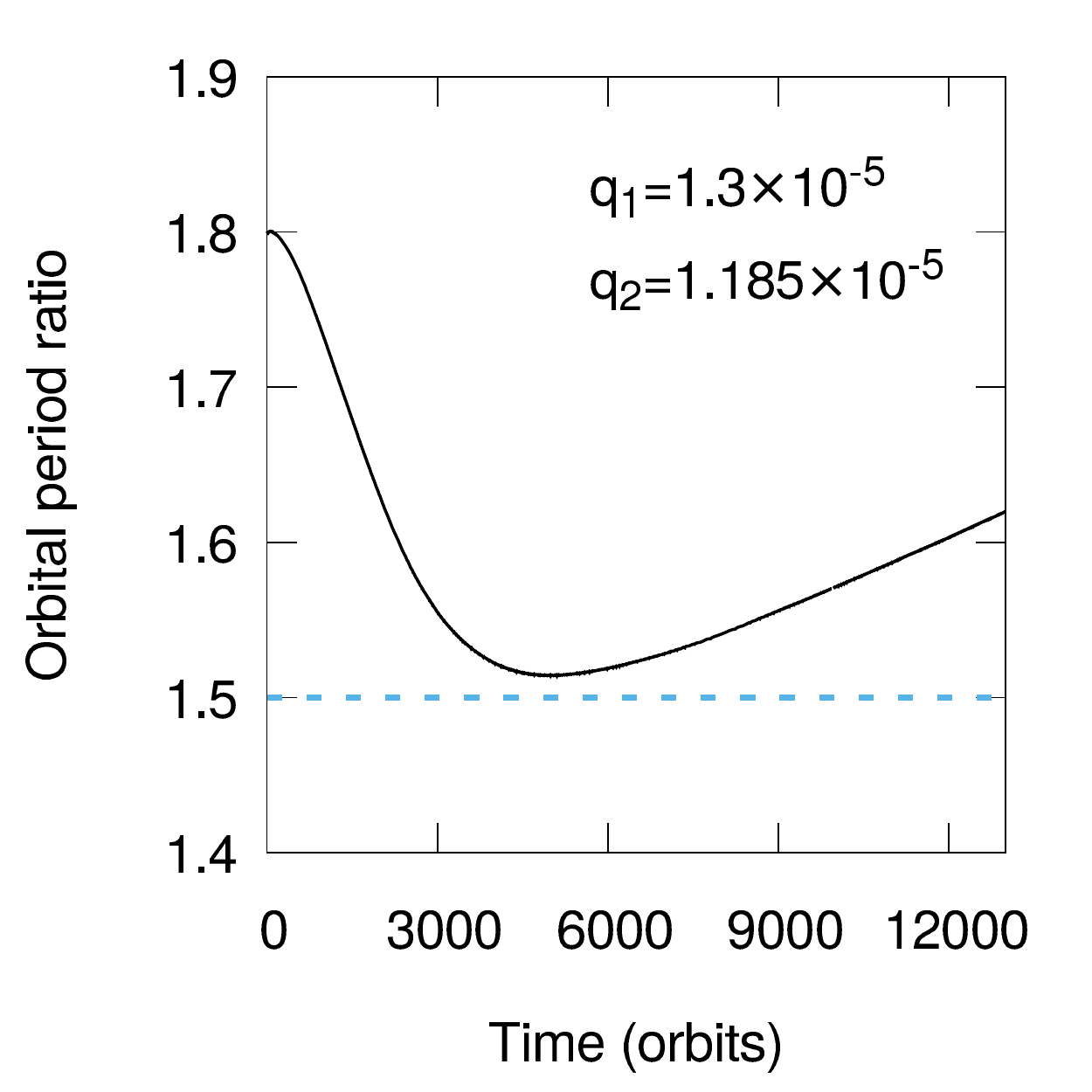}
\includegraphics[width=0.5\columnwidth]{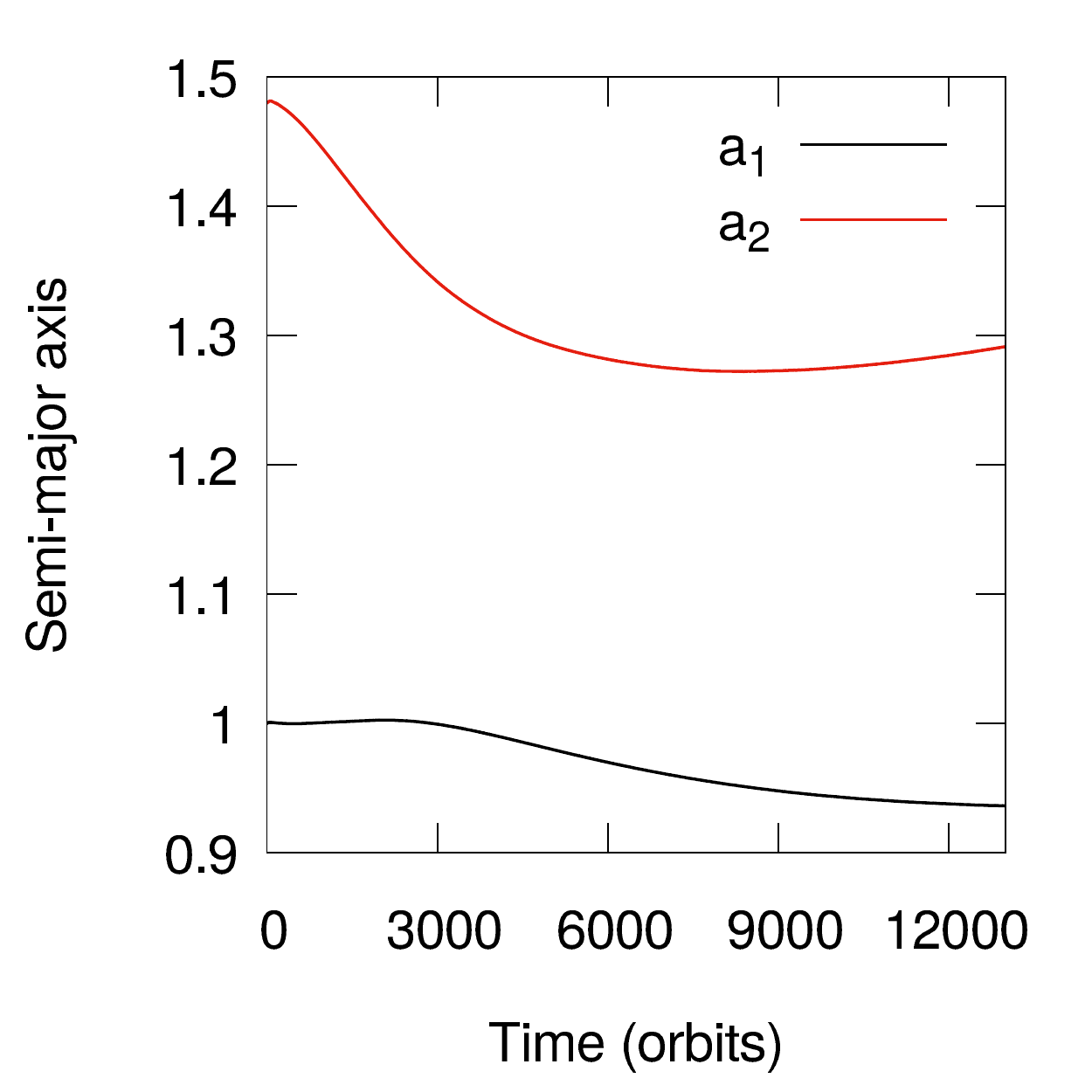}
\includegraphics[width=0.5\columnwidth]{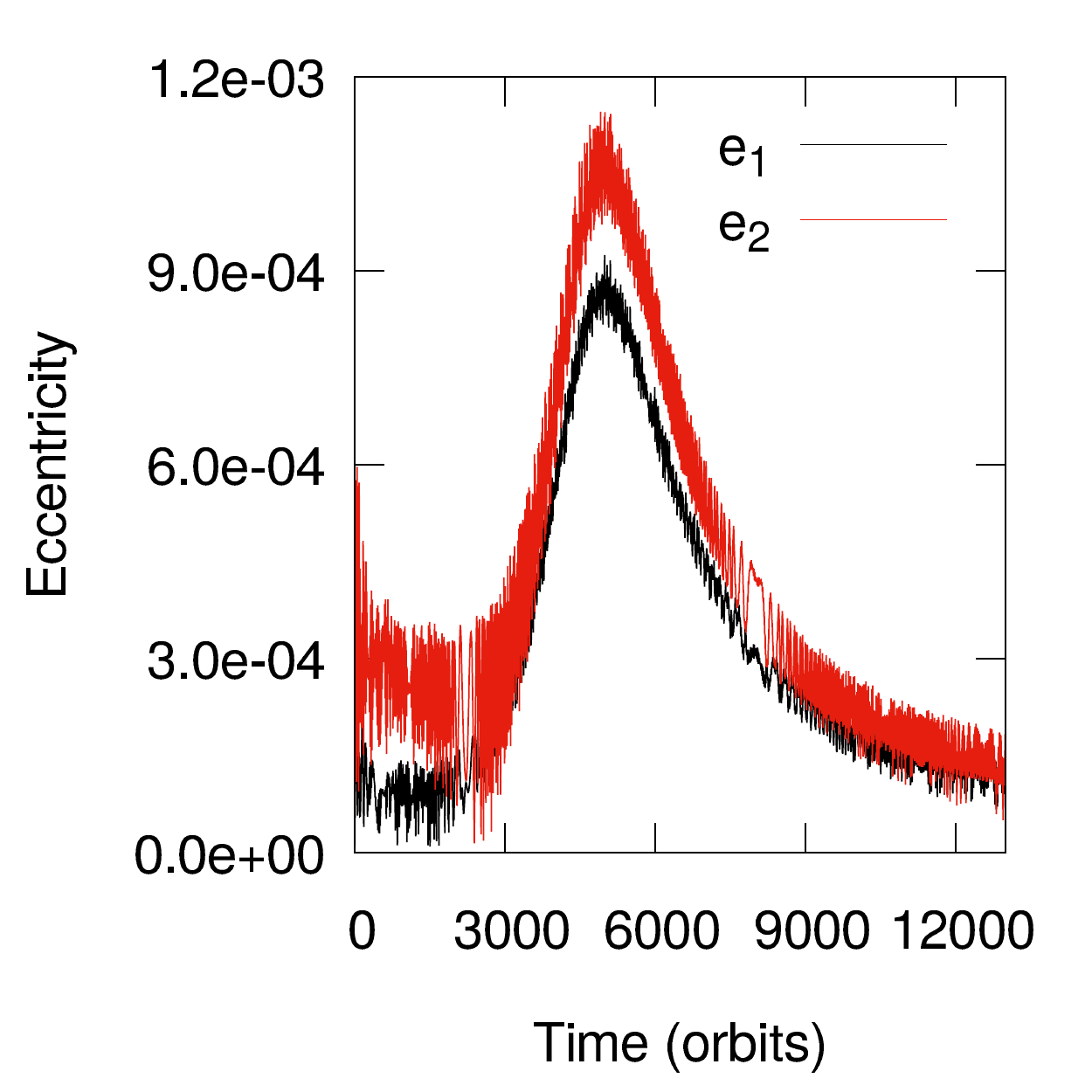}
\includegraphics[width=0.5\columnwidth]{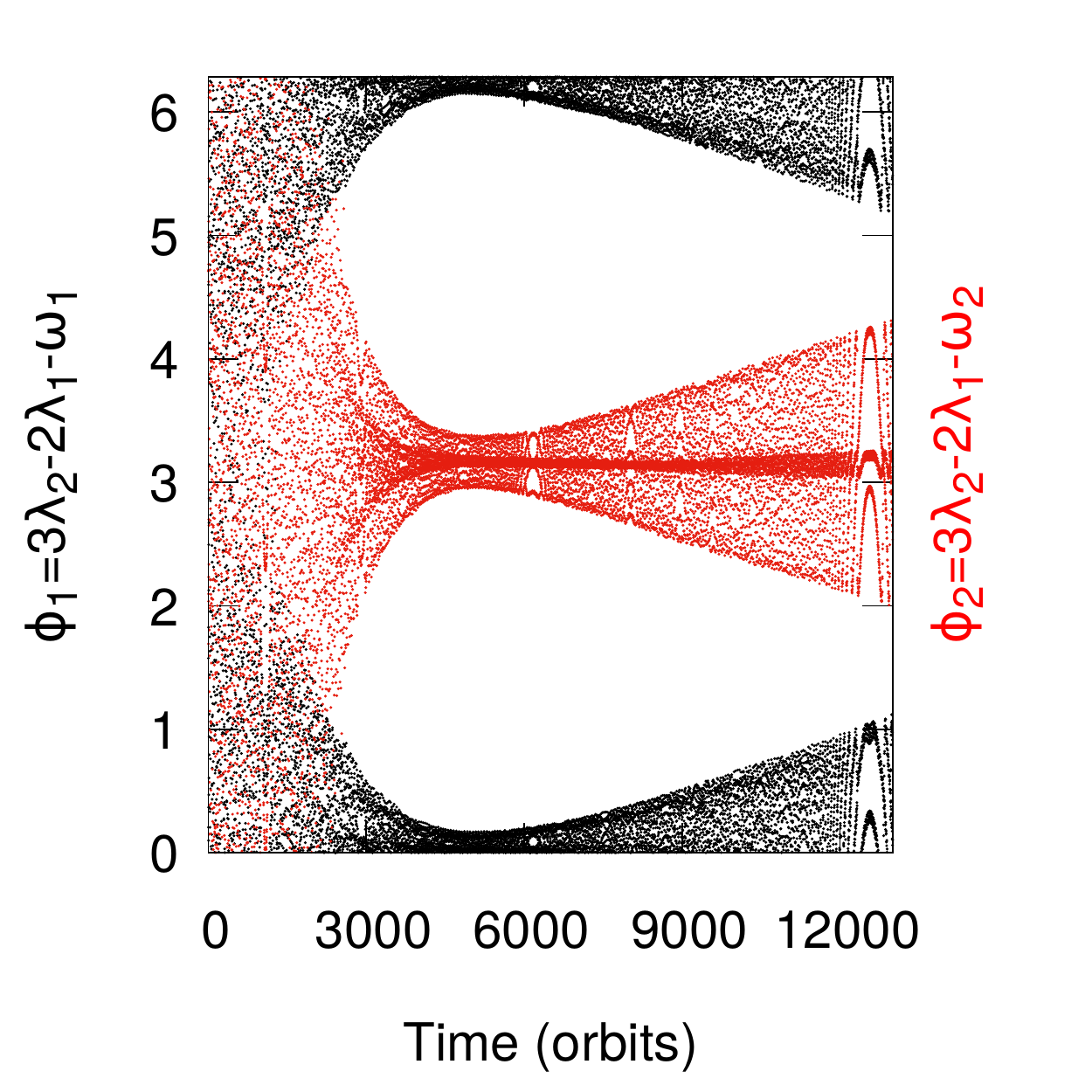}
}
}
}
\caption{The results of the hydrodynamical simulation of the orbital 
evolution of the two super-Earths with $q_{1}=1.3 \times 10^{-5}$ and 
$q_{2}=1.185 \times 10^{-5}$ migrating in a protoplanetary disk with 
the initial surface density given by Equation~(\ref{disk}),
$\Sigma_0=6\times 10^{-5}$, $h=0.02$ and $\nu = 1.2
\times 10^{-6}$. The evolution of the planets' orbital period ratio, 
the semi-major axes, the eccentricities and the resonance angles
are shown in the panels from left to right. The horizontal dashed blue
line in the first panel indicates the position of the 3:2
commensurability.}
\label{fig:fix_scaling}
\end{figure*}

The results of the simulation for the evolution of the orbital 
period ratio, the semi-major axes, the eccentricities $e_{1}$, $e_2$ 
and the resonance angles associated with the 3:2 mean-motion resonance,  
are shown in Figure~\ref{fig:mig_scaling}.
As can be seen from this figure, both planets migrate inwards for the 
duration of the  calculation. The migration rate of the outer planet 
slows down  noticeably at times $t \sim 3000$  and $t \sim 6000$ orbital 
periods measured at the initial location of the inner planet,
hereafter simply denoted as orbits. 

The first slowdown in the migration rate of the outer planet at around 
$t \sim 3000$ orbits can be identified  with the passage of the two 
planets through the 5:3 mean-motion resonance, while the second at 
$t \sim 6000$ orbits with the vicinity of the 3:2 commensurability. 
The characteristic rise in both planet eccentricities close to
the 3:2 resonance is clearly visible.
The eccentricities are excited by a factor of ${\sim}$ three and 
that of the inner planet reaches a value of \textcolor{red} {0.0011}. 
The temporary 
capture into the 3:2 resonance is also illustrated by the evolution 
of the  resonance angles, defined in Figure~\ref{fig:mig_scaling},  
which are seen to enter into libration. 

Beyond $t \sim 6000$ orbits, the migration has changed from convergent 
to divergent. The orbital period ratio is increasing and the 
eccentricities are decreasing.  At $t \sim 6500$ orbits  the
migration reverts to being  weakly convergent  and the eccentricities 
are increasing again. This behaviour does not last long and at
$t \sim 7500$ orbits the relative migration changes back to being 
divergent.

Another short episode of convergent migration takes place at around 
$t \sim 9000$ orbits, but apart from these two brief periods of 
convergent migration, the overall migration is obviously divergent 
from $t \sim 6000$ orbits till the end of the calculation.

In Figure~\ref{fig:figure2} we illustrate the azimuthally averaged
surface density profile at $t \sim 8560$ orbits. A contour plot of 
the disk surface density in the vicinity of the two planets is shown 
in the right panel. Both planets develop a partial gap  in the disk. 
As illustrated in the left panel these  gaps are well separated.

\subsection{Migration in a disk with an inner protoplanet trap}
\label{innertrap}
The detailed analysis of the mechanism, responsible for the
observed evolution, necessitates studies of a range of masses of 
the inner planet. For this purpose, the particular surface 
density profile given by Equation~(\ref{disk}) with 
$\Sigma_0 =6 \times 10^{-5}$ is adopted to guarantee the convergent 
migration of the super-Earths in the early stages of the evolution.
For consistency with these calculations we have rerun the simulation 
already described above starting with this surface density
profile. The results are shown in  Figure~\ref{fig:fix_scaling}. 
They are qualitatively very similar to those shown in 
Figure~\ref{fig:mig_scaling}. We illustrate this fact in 
Figure~\ref{fig:fix_xm1_1} where  the evolution of the orbital 
period ratio for both simulations are shown for comparison in the 
same panel.
  
\begin{figure}[htb!]
\centerline{
\vbox{
\hbox{
\includegraphics[width=0.8\columnwidth]{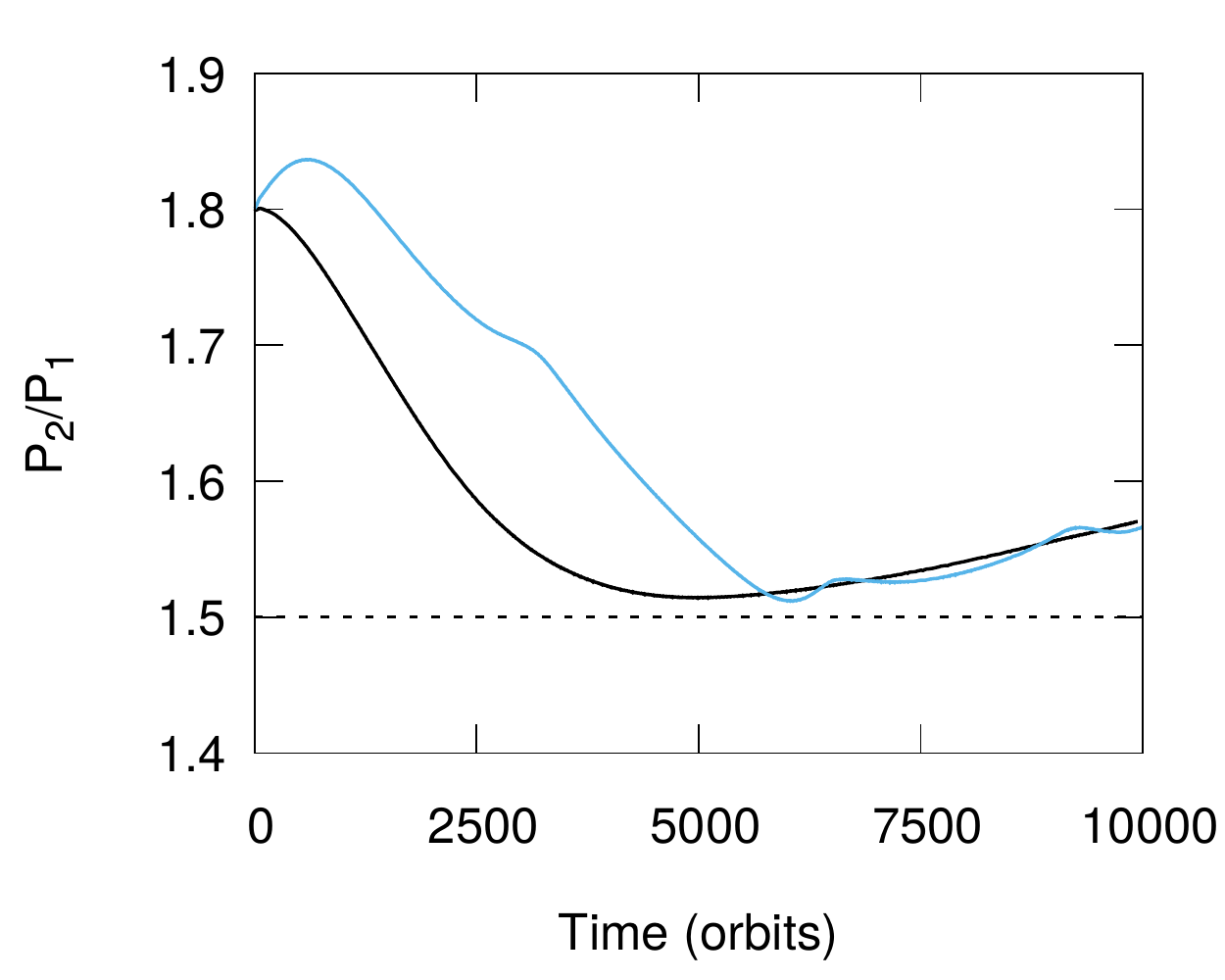}
}
}
}
\caption{The comparison between the period ratio evolution for two
super-Earths with $q_1=1.3 \times 10^{-5}$ and 
$q_{2}=1.185 \times 10^{-5}$ migrating in protoplanetary disks with 
initial surface density given by Equation~(\ref{disk}) with
$\Sigma_0=6\times 10^{-5}$ (black line), and with initial surface 
density  profile given by $\Sigma=8\times 10^{-5} r^{-1/2}$ 
(light blue). 
The disk aspect ratio $h=0.02$ and viscosity $\nu = 1.2
\times 10^{-6}$ are the same in both runs.
}
\label{fig:fix_xm1_1}
\end{figure}

The planets undergo the convergent migration at the beginning of 
the simulation. The outer planet migrates inwards  and the inner one 
almost does not migrate as a result of the initial surface density 
profile adopted in this simulation decreasing rapidly inwards. 
A consequence of this is that the relative migration is initially  
faster in this case and there is no sign of passage through the 5:3 
resonance on the way to the 3:2 commensurability.

At $t \sim 4000$ orbits, the system is in the vicinity of the 3:2 
resonance. The eccentricities of two planets are increasing while 
both resonance angles associated with the 3:2 MMR begin to librate. 
As the resonance is entered, the inner planet is released from the 
trap and begins to migrate slowly inwards. 
The two planets then migrate inwards together briefly maintaining 
an approximately fixed period ratio under the effect of the 
mean-motion resonance. 

At $t \sim 5000$ orbits, the planets are very close to the 3:2 MMR. 
The eccentricities have attained maximum values 
$e_{1} \sim 8.5 \times 10^{-4}$ and $e_{2} \sim 1.0 \times 10^{-3}$. 
Subsequently, the convergent migration of two planets changes to 
being divergent. The orbital period ratio of the system increases  
and the planets separate from the 3:2 MMR. The eccentricities 
subsequently decrease and the libration amplitudes of the resonance 
angles increase. The inward migration rate of the outer planet slows 
down and when $t \sim 8000$ orbits is reached the migration of the 
outer planet reverses and it starts to migrate outwards. The migration 
rate of the inner planet also decreases becoming almost zero at the 
end of the calculation.

In order to obtain an understanding of the mechanisms responsible 
for the results of the above simulations, in particular the switch 
from convergent to divergent migration and the increasing separation 
from resonance, we study in detail the way an isolated super-Earth 
migrates in disks with varying background surface density profiles
in which it creates a partial gap.
This enables us to separate features that can be understood in terms
of isolated planets from those that are related to influences of 
one planet on another.


\section{The migration of an isolated super-Earth that is capable 
of forming a partial gap}
\label{sec:single}
In this Section we provide a close look at the evolution of the
outer super-Earth ($q=1.185 \times 10^{-5}$) without the presence 
of the inner planet in the disk. We consider a range of background 
surface density profiles.

\subsection{Setup of the simulations}
\label{simsetup} 
In these calculations the outer super-Earth is placed on the same 
initial circular orbit in the disk as in the two-planet case.
We adopt the same initial disk properties as in the case of two planet 
simulations, starting with the surface density profile given by 
Equation~(\ref{disk}).  Thus $h=0.02$, $\nu=1.2\times 10^{-6}$, 
$\Sigma_0=6\times 10^{-5}$ and for the first simulation $\alpha =0.5.$ 

In addition to the run with $\alpha=0.5$, we perform two additional 
runs, the first with $\alpha=-0.067$ corresponding to an almost flat
surface density profile, and the second with $\alpha = -0.3$ 
corresponding to a profile that increases outwards. 
All other disk parameters are unchanged. 
The choices of $\alpha$ are based entirely on the form of the evolving
surface density profile observed in the vicinity of the outer planet 
in the calculation performed with two planets.
In that calculation the slope of the surface density distribution 
at the outer planet location evolves in time, starting from the
initial slope with $\alpha =0.5$, passing through the flat profile
($\alpha =0.0$) and ending with the positive slope ($\alpha \approx
- 0.3$).

\subsection{Relevant length scales}
\label{lengthscales}
Before describing the migration of the single planet in the disk
we first discuss the  relevant length  scales in the calculations. 
The Hill radius,  $r_H= (q/3)^{1/3}r_p = 0.0158r_p$, where $r_{p}$ 
is the radius at which the planet is located, is comparable to the  
local thickness of the disk, which is, as we have already stated, 
$ 0.02r_p.$ The half-width of the horseshoe region, $x_s/r_p $,
can be determined from the topology of the gas flow in the vicinity 
of the planetary orbit.
It has been determined by \cite{Paardekooper2009} to be 
$x_s \sim  r_{p}\sqrt{q/h}.$
This value can be understood in the following way. The streamline 
which passes through the location of the planet (possible as the 
potential is softened) starts at large azimuthal distance with
radial separation $x_s.$ The speed relative to the planet is 
$3\Omega_p x_s/2,$ where $\Omega_p$ is the Keplerian angular velocity 
at the planet's location. In reaching the planet the kinetic energy  
per unit mass of material on this streamline is changed by 
$\sim \Omega_p^2 x_s^2.$  This must be comparable to the change in 
planet potential $ GqM_{\star}/(r_pb).$ Note the use of the 
softening length here as we anticipate this will exceed $x_s$ for small
$q.$ Hence we deduce $x_s \sim r_p\sqrt{q/b}.$ Note that $b$ is 
comparable to $h$ here and \cite{Paardekooper2009} found that in the 
limit $b\rightarrow 0,$ the effect of back pressure is to limit the 
magnitude of the potential in a way that effectively replaces $b$ 
by $h.$ Hence  $x_s \sim  r_{p}\sqrt{q/h}.$
 
We show a contour plot of the surface density of the disk for the 
calculation with $\alpha=0.5$ in the vicinity of the planet together 
with the gas flow velocity field at $t=1000$ orbits in 
Figure~\ref{fig:single-streamline}.  

\begin{figure*}[htb!]
\centerline{
\vbox{
\hbox{
\includegraphics[width=1.9\columnwidth]{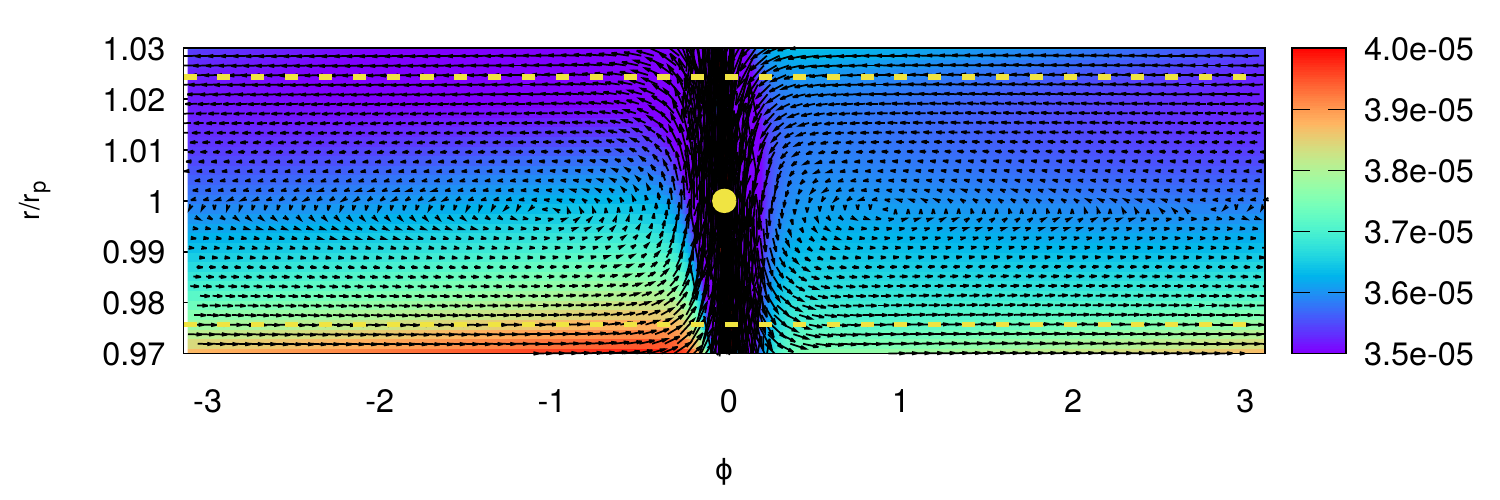}
}
}
}
\caption{A snapshot of the surface density and the  gas velocity field
in the vicinity of the planet at $t=1000$ orbits in the simulation 
with $\alpha=0.5$. The yellow solid circle indicates the position of the 
planet and the dashed horizontal yellow lines show the position 
of a putative separatrix at $r=r_{p} \pm r_{p}\sqrt{q/h}$.}
\label{fig:single-streamline}
\end{figure*}

\begin{figure}[htb!]
\centerline{
\vbox{
\hbox{
\includegraphics[width=0.9\columnwidth]{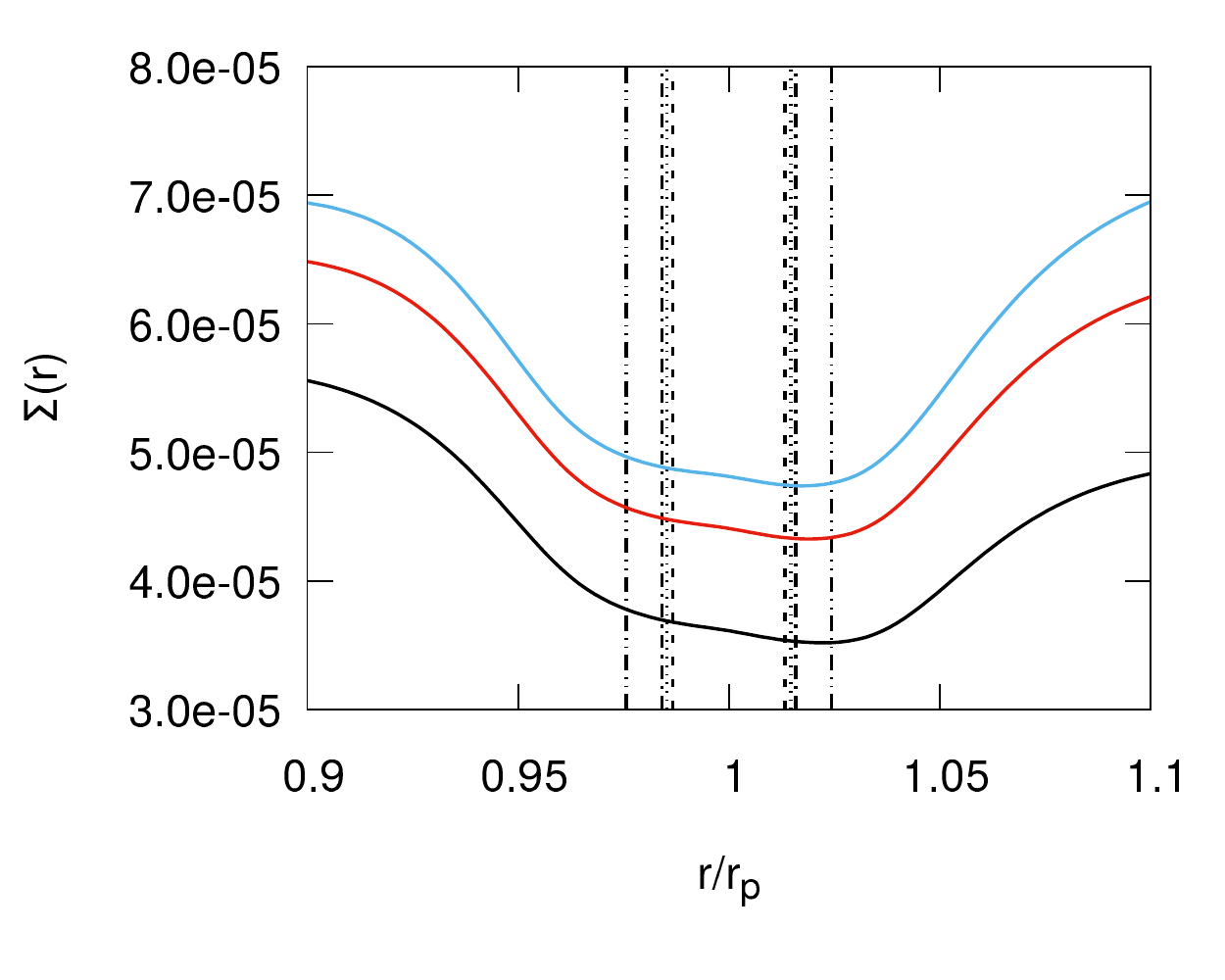}
}
}
}
\caption{The distances from the orbit of the planet of the radial 
locations determining the characteristic scales in the problem, listed 
starting from the closest to the planet and proceeding to the most 
distant, ($(2/3)H$ - dashed, $l_{sh}$ - dotted, $r_H$ - dash-dotted
and $x_s$ - dash double-dotted lines).
Also plotted are the surface density profiles of the disk in the vicinity 
of the planet after $t=1000$ orbits when it had reached $r_p \sim1.413$ 
in the simulations with different values of $\alpha$ (0.5 - black,
$-0.067$ - red  and $-0.3$ - blue lines).
}
\label{fig:scales-gap}
\end{figure}

Following the trend indicated by arrows directed along the gas velocity 
vectors in the disk, it can be concluded that the horseshoe region extends 
approximately 
from $r = r_{p}-r_{p}\sqrt{q/h}$ to $r_{p}+r_{p}\sqrt{q/h}$, 
as indicated by the dashed lines in the figure. It can be inferred that 
a streamline that is  close to one of these lines will pass through the 
location of the planet, hence the half-width of the region $\sim x_s$ is 
such that

 \begin{equation}
\label{eq:hswidth}
\frac{x_s}{r_p} = \sqrt{\frac{q}{h}}=0.0243.
\end{equation}
This is in a good agreement with  previous numerical studies as for example 
in \cite{Paardekooper2009}. In our other two simulations the width of the 
horseshoe region is very similar.

Another potentially relevant scale is the non-linear shocking length
$l_{sh}.$ This is the distance from the planet at which the density wave
produced by the planet becomes nonlinear and shocks, $q,$ being assumed 
to be small enough that the wave is linear close to the planet. It is
defined through \citep{Dong2011} 

\begin{equation}
\frac{l_{sh}}{r_p} \approx 0.8 \left( \frac{\gamma +1 }{12/5}
\frac{m_p}{M_{th}}\right) ^{-2/5}h
\label{eq:shock}
\end{equation}
with $M_{th}= c_s^3/(\Omega_p G)$, $m_p$ and $\gamma$ being 
mass of the planet and the adiabatic index, respectively. 
Here, for our set of parameters, its value is equal to $0.0147r_p.$
This is smaller than both $x_s$ and $2H/3$ indicating that
density  waves are nonlinear when launched. This is not unexpected as 
$q$ is large enough to be in the partial gap forming regime and it means 
that Equation~(\ref{eq:shock}) is inapplicable.
  
All mentioned characteristic scales are shown in Figure~\ref{fig:scales-gap}
together with the profiles of the partial gaps formed by the planet in the 
disk.  The shape of the gaps is  represented in a more global context in 
Figure~\ref{fig:single-sdensity}, where the surface density of the disk 
in a larger neighbourhood of  the planet at $t=1000$ orbits in the three 
simulations is  shown. 
The width of the gap measured between the points at which the surface 
density corresponds to  $50\%$ of  the maximum gap depth is
larger than the disk aspect ratio and the value of the surface density
at the planet position is about 68\% of its unperturbed value.

\begin{figure}[htb!]
\centerline{
\vbox{
\hbox{
\includegraphics[width=0.9\columnwidth]{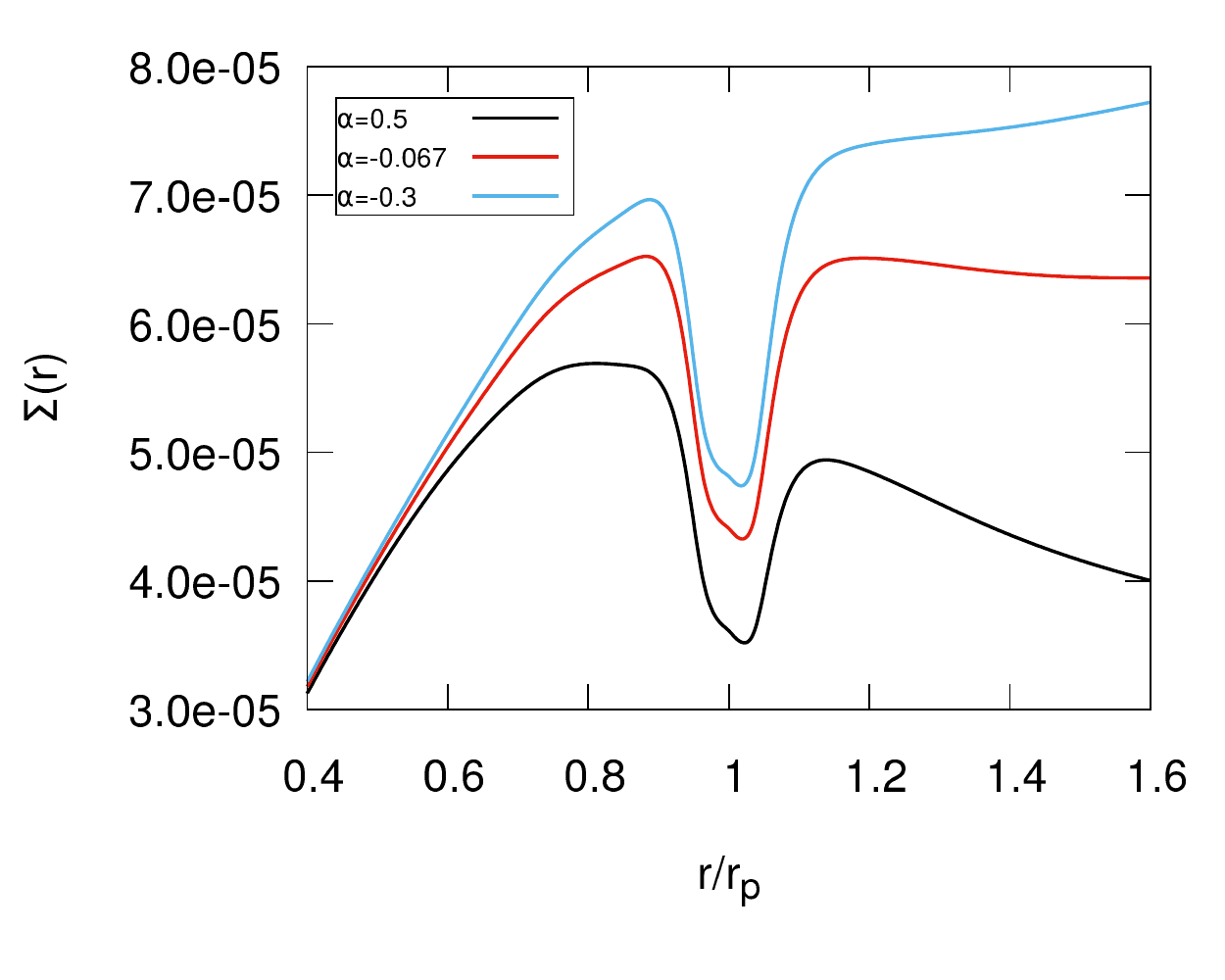}
}
}
}
\caption{The surface density of the disk in the neighbourhood of the 
planet at $t=1000$ orbits in the simulations with different values of 
$\alpha$. }
\label{fig:single-sdensity}
\end{figure}

\subsection{Characterisation of the torque exerted on the disk by a planet 
with a partial gap}
\label{Chpg}						
Now we are ready to go on to analyse the torques exerted on the disk by 
the planet that result in its  migration.  In Figure~\ref{fig:single-torque} 
we plot the total torque $\Gamma_{\rm total}$ exerted by the disk on the 
planet obtained from the three hydrodynamical simulations with  different 
initial surface density slopes mentioned above with black lines.
The first 600 orbits of the evolution is dominated by the process of  
partial gap formation in the disk. Later on, the gap profile does not 
change significantly in time and can be considered as quasi-stationary.

The total torque, in general, can be expressed as a sum of the differential 
Lindblad torque $\Gamma_{L}$ and the corotation torque $\Gamma_{C}$.
However, when the horseshoe region extends further than (2/3)$H$ from the 
orbit linear theory cannot be used in its complete form.  In that theory 
the Lindblad torque is obtained  by summing contributions for different 
azimuthal mode number, $m,$ with significant contributions up to the torque 
cut off corresponding to $m \sim 3r/(2H ),$ and with Lindblad resonance 
separated by  approximately $2H/3$ from the planet. 
It is clear from the Figure~\ref{fig:scales-gap} that in our calculations 
the horseshoe region invades the zone in which most of the linear Lindblad 
torque is produced.
In order to limit the linear theory to its domain of applicability, we require 
the interaction to take place at a  distance exceeding $x_s$ from the planet.
To see how this works we remark that \citet{Lin1979} and \citet{Papa1984} 
performed a local scattering calculation applicable to circulating fluid 
elements and argued that only contributions from those originating at a 
distance greater than $x_s$ should be considered. As the one sided torque 
is inversely proportional to the cube of this distance we then  expect this 
to be reduced by a factor $\sim (2H/(3x_s))^3.$

\subsection{Comparison with type I migration}
First, we  compare  the results of our calculations with classical type I 
migration in which the low-mass planet is not able to form a partial gap 
in the disk and the Lindblad torque is obtained from linear theory.
In such a situation, the total torque acting on the planet in a locally 
isothermal limit consists of the linear Lindblad torque and the corotation 
torque, which can be expressed in the form \citep{Paardekooper2010}: 

\begin{equation}
\Gamma_{L}/\Gamma_{0}=-3.15+0.075\alpha_{p}\label{LinLindblad}
\end{equation}
\begin{equation}
\Gamma_{C}/\Gamma_{0}=1.905-0.73\alpha_{p}
\end{equation}
\noindent with
\begin{equation}
\Gamma_{0}=(q/h)^{2}\Sigma_{p, un}r_{p}^{4}\Omega_{p}^{2}
\end{equation}
\noindent where $\alpha_{p}$ is the slope of the surface density 
fitted in the vinicity of the planet, $\Sigma_{p, un}$ is the 
unperturbed surface density at the position of the planet and 
$\Omega_{p}$ is the angular velocity of the planet.
The total torque than reads

\begin{eqnarray}
\label{eq:paardekooper0}
\frac{\Gamma_{\rm total}}{\Gamma_0}= 
\frac{\Gamma_{L}}{\Gamma_0} +
\frac{\Gamma_{C}}{\Gamma_0}=-1.245-0.655\alpha_{p} 
\end{eqnarray}
The total torque calculated from Equation~(\ref{eq:paardekooper0}) is 
indicated by the orange asterisks in Figure~\ref{fig:single-torque}. 
It is important to note that $\alpha_p$  we consider here, applies to  
the partial gap and not the background profile. It was determined by 
matching a power law fit of the form $\Sigma \propto r^{-\alpha_p}$ to 
the azimuthally averaged surface density in the partial gap region 
$r_p-x_s < r < r_p+x_s.$ 
Some justification for this approach can be obtained by noting that 
the Lindblad torque is insensitive to the surface density profile but 
the corotation torque, which is determined in the horseshoe region 
that is located in the partial gap, is not. Accordingly the form of 
the surface density profile there is what is significant.
The values of $\alpha_p$ determined for  the different cases at various 
times are listed in Table \ref{table2}.

It is interesting to note that the torques indicated by the orange
asterisks agree quite well with the results obtained from the
numerical simulations (black curves) for the case  for which the   
initial surface density profile has a 
\textcolor{red}{ negative slope ($\alpha >0$)}.
The agreement is not as good for the flat initial profile 
($\alpha \sim 0),$  and the largest differences can be seen for the 
initial profile with 
\textcolor{red}{ positive surface density slope ($\alpha <0).$} 
The latter discrepancy most likely arises on account of the depth 
of the  partial gap affecting the corotation torque as well as the 
truncation of the linear Lindblad torque.  We attempt to correct 
for this in the next Section.
 
\begin{table}[htbp]
\centering
\caption{Numerical parameters for the torque calculation in the 
isolated super-Earth cases}
\label{tab:single-se-parameter} 
\begin{tabular}{c|cccc} 
\hline
\hline
     \multicolumn{5}{c}{$\alpha = 0.5$} \\
\hline
Time ({\it orbits}) & $\alpha_{p}$ & $\Sigma_{p, un}$ ($\cdot 10^{-5}$) & $\Sigma_{p, min}$ ($\cdot 10^{-5}$) & $r_{p}$  \\ 
\hline 
159 &  1.098  & 5.00  &  4.25  &  1.4729 \\
\hline
318 &  1.258  & 5.08  &  3.96  &  1.4630 \\
\hline
478 &  1.386  & 5.39  &  3.82  &  1.4519 \\
\hline
637 &  1.437  & 5.38  &  3.73  &  1.4403 \\
\hline
796 &  1.477  & 5.37  &  3.67  &  1.4284 \\
\hline
955 &  1.471  & 5.35  &  3.63  &  1.4165 \\
\hline
      \multicolumn{5}{c}{$\alpha = -0.067$} \\
      \hline
Time ({\it orbits}) & $\alpha_{p}$ & $\Sigma_{p, un}$ ($\cdot 10^{-5}$) & $\Sigma_{p, min}$ ($\cdot 10^{-5}$) & $r_{p}$  \\ 
\hline 
159 &  0.783  & 6.21  &  5.32  &  1.4776 \\
\hline
318 &  0.986  & 6.27  &  4.92  &  1.4701 \\
\hline
478 &  1.047  & 6.33  &  4.72  &  1.4613 \\
\hline
637 &  1.108  & 6.59  &  4.59  &  1.4514 \\
\hline
796 &  1.113  & 6.56  &  4.50  &  1.4412 \\
\hline
955 &  1.155  & 6.54  &  4.43  &  1.4308 \\
\hline
      \multicolumn{5}{c}{$\alpha = -0.3$} \\
      \hline
Time ({\it orbits}) & $\alpha_{p}$ & $\Sigma_{p, un}$ ($\cdot 10^{-5}$) & $\Sigma_{p, min}$ ($\cdot 10^{-5}$) & $r_{p}$  \\ 
\hline 
159 &  0.564  & 6.75  &  5.82  &  1.4801 \\
\hline
318 &  0.809  & 6.93  &  5.38  &  1.4747 \\
\hline
478 &  0.841  & 7.10  &  5.16  &  1.4675 \\
\hline
637 &  0.937  & 7.16  &  5.03  &  1.4595 \\
\hline
796 &  0.978  & 7.18  &  4.93  &  1.4509 \\
\hline
955 &  0.997  & 7.15  &  4.83  &  1.4421 \\
\hline
\hline
\end{tabular}\label{table2}
\end{table}

\subsubsection{Taking account of the surface density depression 
in the  partial gap}
A consequence of partial gap formation is that the surface density 
at the planet position changes in time and is lower than the 
unperturbed value. Moreover, the slope of the surface density 
profile in the vicinity of the planet varies in time and  differs 
from  the initial one (see Table \ref{table2}.) 
In order to obtain an improved fit we  firstly rescale the above 
expressions for the corotation torque and the Lindblad torque so 
that they respectively become
\begin{eqnarray}
\Gamma_{C}'/\Gamma_{0}&=&
(\Sigma_{p,min}/\Sigma_{p,un})(\Gamma_C/\Gamma_{0}) \hspace{3mm} {\rm and} \\
\Gamma_{L}'/\Gamma_{0}&=&(\Sigma_{p,min}/\Sigma_{p,un})
(\Gamma_L/\Gamma_{0}),
\end{eqnarray} 
where $\Sigma_{p, min}$ is the surface density at the position of 
the planet in the partial gap.
The total torque can be therefore evaluated as:
\begin{eqnarray}\label{eq:paardekooper}
\frac{\Gamma_{\rm total}'}{\Gamma_0}=
\left( \frac{\Gamma_{L}}{\Gamma_0} +
\frac{\Gamma_{C}}{\Gamma_0}\right)
\frac{\Sigma_{p,min}}{\Sigma_{p,un}}= \\ \nonumber
=(-1.245-0.655\alpha_{p}) 
\frac{\Sigma_{p,min}}{\Sigma_{p,un}}
\end{eqnarray}
We compare this modified formula with the results of the 
hydrodynamical simulations. The total torque calculated from 
Equation~(\ref{eq:paardekooper}) is indicated by red asterisks 
in Figure~\ref{fig:single-torque}.  
It is clear that such a rescaling does not produce a good fit  
to the numerical results either. Since the Lindblad torque 
originates from beyond the partial gap and  may not be well 
represented by linear theory,  it is not unexpected that
Equation~(\ref{eq:paardekooper}) fails to represent the torque 
exerted by the planet opening a partial gap during its migration.

In order to obtain formulae for the total torque which are 
consistent with  our numerical simulations, we perform a fit to 
obtain values of $\Gamma_{\rm total}$ at $t=1000$ orbits for 
each of the three numerical simulations that were performed with
different initial profiles of the surface density in the disk. 
We are quite confident that at greater times in the simulations
the partial gap formed in the disk is to a good approximation 
stationary.
The fitted formula has the same form as in 
Equation~(\ref{eq:paardekooper}), the only difference is the 
presence of two constant coefficients $C_1$ and $C_2$ that 
separately rescale the contributions of the corotation and 
Lindblad torques, and it reads:
\begin{eqnarray}\label{eq:numerical-fit}
\frac{\Gamma_{\rm total}''}{\Gamma_0}=\left( C_{1}\frac{\Gamma_{L}}
{\Gamma_0}+ 
C_{2}\frac{\Gamma_{C}}{\Gamma_0}\right) \frac{\Sigma_{p, min}}
{\Sigma_{p, un}}=
\\ \nonumber
=(1.129-2.924\alpha_{p}) 
\frac{\Sigma_{p,min}}{\Sigma_{p,un}}
\end{eqnarray}
It was found that the numerical results could be well 
represented by this formula with $C_{1}=$ 2.2003 and 
$C_{2}=$  4.2310. This fit reduces the contribution of the 
Lindblad torque relative to the corotation torque consistent 
with the idea that the former should be smaller than the 
linear type I estimate.
 
Using  Equation~(\ref{eq:numerical-fit}), we calculate the 
total torque and it is shown with the blue asterisks in 
Figure~\ref{fig:single-torque}. From this figure it is clear 
that for times exceeding $200$ orbits the torque from the 
fitted formula is consistent with the numerical results in 
all three cases. It is important to stress that our fitting
procedure is not general and can only  be used in the limited 
parameter range determined by the three simulations on which 
it has been based. 

\begin{figure*}[htb!]
\centerline{
\vbox{
\hbox{
\includegraphics[width=0.66\columnwidth]{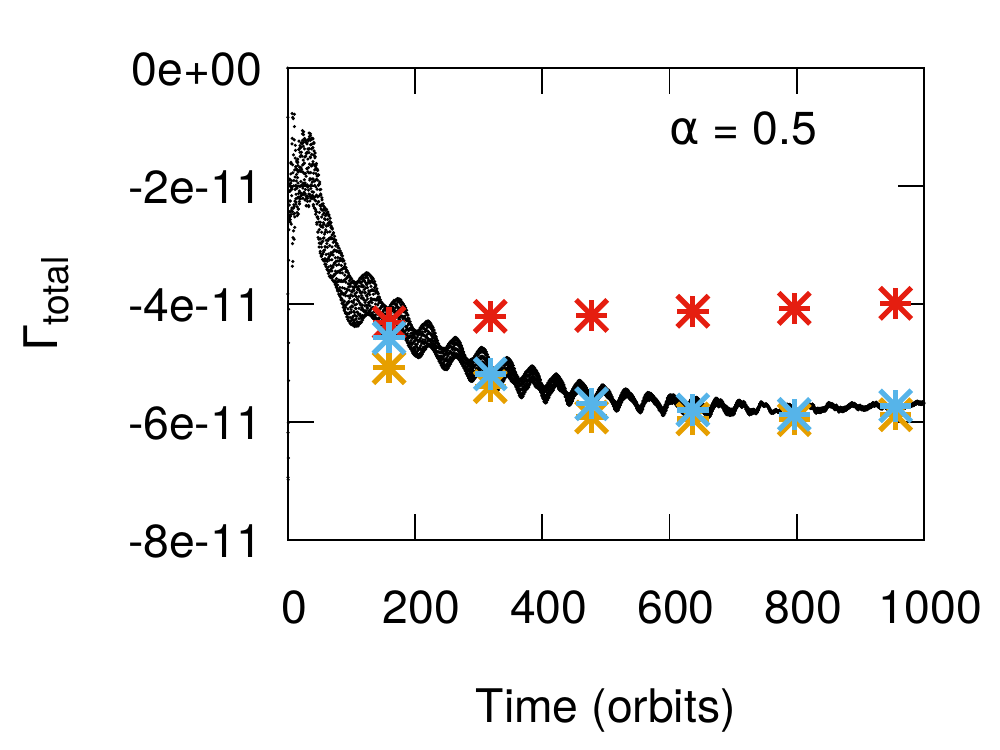}
\includegraphics[width=0.66\columnwidth]{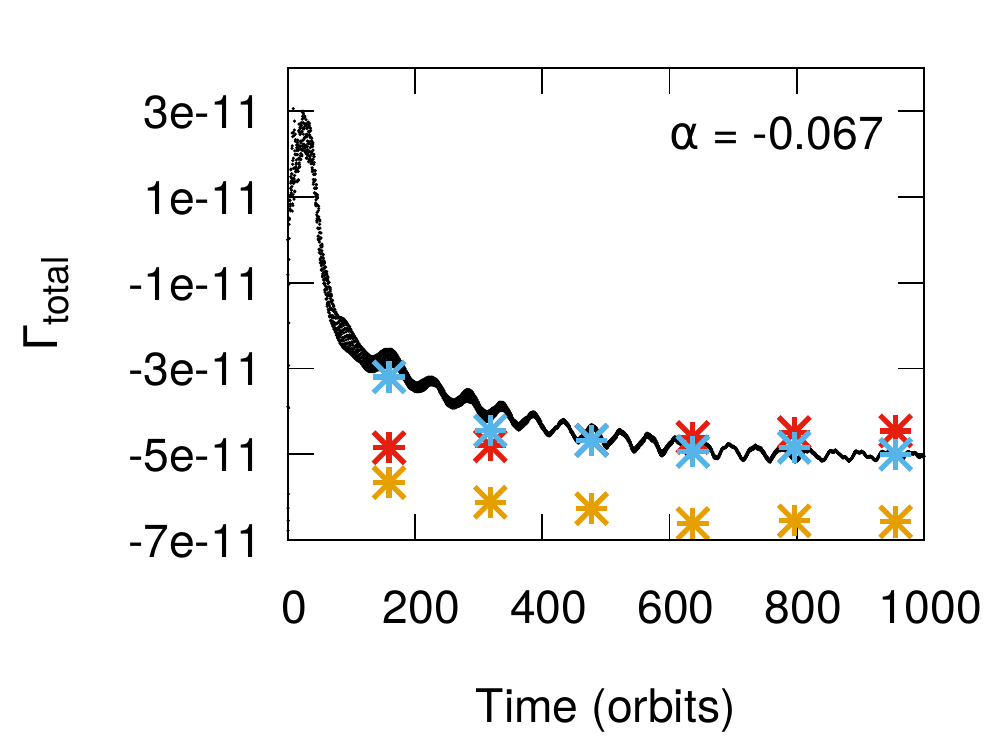}
\includegraphics[width=0.66\columnwidth]{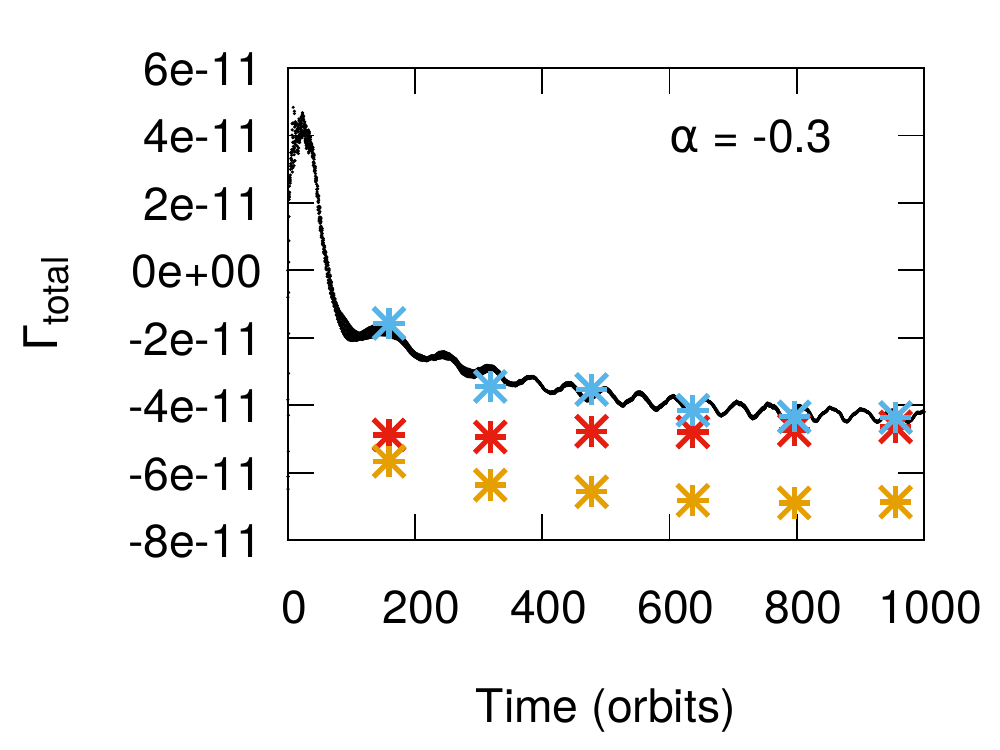}
}}}
\caption{The total torque from the disk acting on the planet 
in the simulations with different $\alpha$. The orange 
asterisks show the torque  expected from  type I migration of 
the planet embedded in the locally-isothermal disk. The red 
and blue asterisks respectively indicate the torque obtained 
from Equation~(\ref{eq:paardekooper}) and 
Equation~(\ref{eq:numerical-fit}).
}
\label{fig:single-torque}
\end{figure*}


\section{The migration of two super-Earths capable of forming 
partial gaps }
\label{sec:twoplanetcase}
In the previous Section we have derived a phenomenological 
formula for the torque exerted on an isolated migrating  
super-Earth by the disk. 
In this Section we use this to aid with the interpretation 
of the migration in the two-planet case.

\begin{figure}[htb!]
\centerline{
\vbox{
\hbox{
\includegraphics[width=0.9\columnwidth]{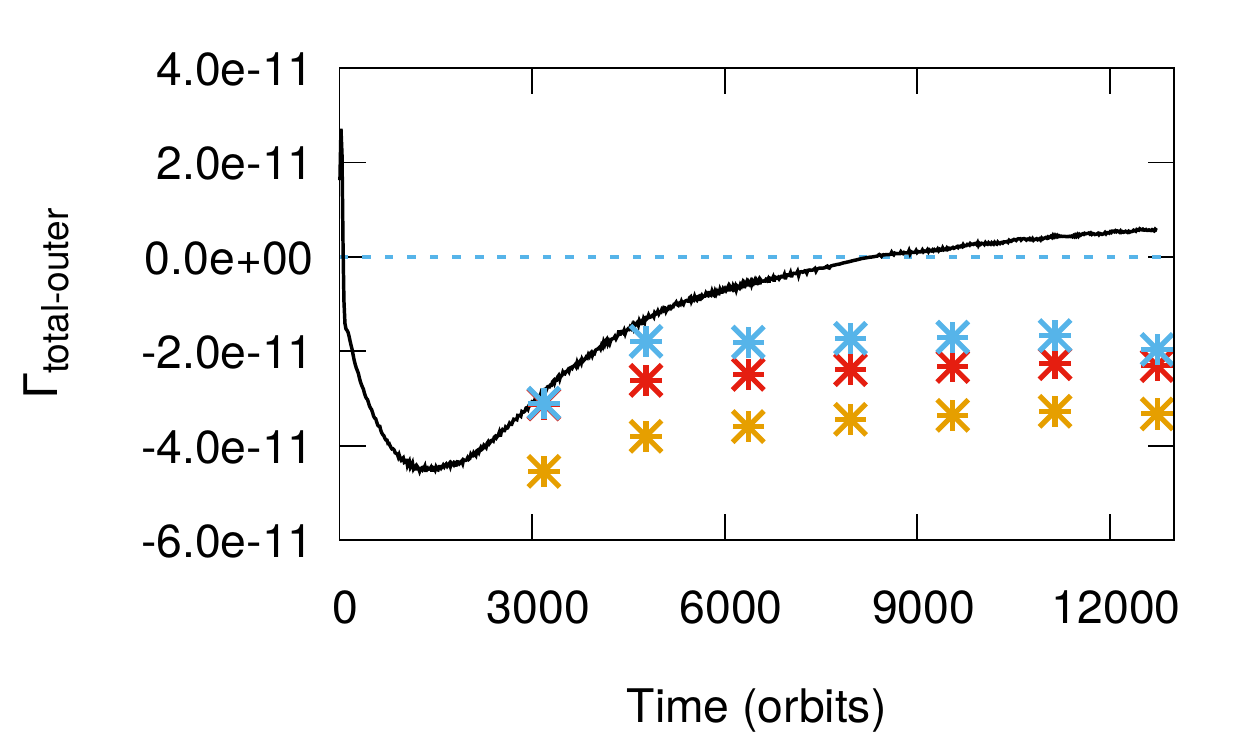}
}
}
}
\caption{The total torque from the disk acting on the outer planet 
in the simulation with two super-Earths. The orange asterisks show 
our expectation of the torque from assumed type I migration of the 
outer planet embedded in the locally-isothermal disk. The red and 
blue asterisks indicate the torque calculated from 
Equation~(\ref{eq:paardekooper}) and 
Equation~(\ref{eq:numerical-fit}) respectively.}
\label{fig:two-torque}
\end{figure}

In Figure~\ref{fig:two-torque}, we present the  torque 
$\Gamma_{\rm total-outer}$ from the disk acting on the outer planet 
in the numerical simulation with two super-Earths, migrating in a 
protoplanetary disk with the initial surface density given by 
Equation~(\ref{disk}) with $\alpha=0.5$, $\Sigma_0=6\times 10^{-5}$, 
$h=0.02$ and $\nu = 1.2 \times 10^{-6},$ in order to see  how the 
migration of the outer planet is affected by the presence of the 
second planet in the disk. In this figure  we also show the torques 
calculated from Equation~(\ref{eq:paardekooper0}), 
Equation~(\ref{eq:paardekooper}) and Equation~(\ref{eq:numerical-fit})
for the single migrating  planet case derived in the previous Section. 
 
We aim to compare the migration of the outer planet with and without 
the second planet in the disk. It is possible to do this only for 
times  exceeding 3000 orbits, because at earlier stages of the 
evolution, $\alpha_p,$ at the vicinity of the planet is outside 
the range covered by our single planet simulations.
From this comparison we expect that if there is no second planet 
in the disk, the torque acting on the planet should remain negative 
till the end of the simulation.
This is because the single planet always migrates inwards.  

From the comparison we can also see that the torques calculated for 
the single planet case are less (though of larger magnitude) than 
what is  seen in  the two planet case.  Moreover, at about 9000 
orbits the torque in the two planet case becomes positive and the 
outer planet starts to migrate outwards.  This means that in the  
two planets case, there is  an additional mechanism by which  
angular momentum is transferred between the planets, either by 
direct gravitational interaction between them, and/or to the 
region in the vicinity of the outer planet and then to the outer 
planet itself.
The first possibility is expected to be effective only very close 
to commensurability but we note that \citet{Baruteau2013} found 
that the mechanism still works when gravitational interaction 
between the planets was switched off in their simulations, 
indicating that something more is needed.
The nature of the mechanism mentioned as the second possibility 
is connected with the wave planet interaction and will be explored 
in the next Section. 


\section{Coorbital torques for pairs of migrating planets with partial
gaps}
\label{Coorbgaps}

In this Section we will derive the scaling relation for effective 
repulsion due to wave planet interactions. In order to do this we 
begin by  adopting an expression for the unmodified linear Lindblad 
torque induced by the unperturbed disk with background surface 
density, $\Sigma_{un},$ in the form
\begin{equation}
\label{eq:lindblad}
\Gamma_L = -3.075q^2\Sigma_{un} r^4 \Omega^2 h^{-2}.
\end{equation}
Noting that this is insensitive to the form of the background surface 
density profile we remark that it is obtained 
from Equation~(\ref{LinLindblad}) with the representative value 
$\alpha_p =1.$  Focusing on the inner planet, the net Lindblad torque 
is obtained by adding together the one sided Lindblad torques 
separately induced by the outer disk beyond the planet and the inner 
disk interior to it.
From \citet{Papa2007} we estimate that the outer one sided Lindblad 
torque is approximately given by  $\Gamma_{L(1s)} = \Gamma_L/(5.7h).$
Thus we adopt
 \begin{equation}
\label{eq:onesidedL}
\Gamma_{L(1s)} = - 0.54q^2\Sigma_{un} r^4 \Omega^2 h^{-3}.
\end{equation}
Although we have focused on the outer one sided Lindblad torque due 
to the inner planet a corresponding situation applies, with a sign 
change to the inner one sided Lindblad torque due to the  outer 
planet.
Note that the associated angular momentum flow so considered is 
directed away from the planet and towards the other planet in each 
case and so will repel it as it approaches provided that it can 
absorb some of the angular momentum flowing towards it.

Consider another planet approaching the inner one from larger radii 
(smaller radii could equally well be considered). Outward propagating 
density waves dissipate in its coorbital region. Angular momentum is 
transfered to the horseshoe region and then to the planet through 
horseshoe drag.  The effective angular momentum transfer rate  can be 
estimated as
\begin{equation}
\label{eq:transtohorseshoe}
|\Gamma_{L(1s)}| 2\lambda\sqrt{q_2/h} \sim 1.08\lambda\sqrt{q_2}q_1^2
\Sigma_{un} r^4 \Omega^2 h^{-7/2},
\end{equation}
where $q_1$ is the mass ratio of the emitting planet, here the inner 
one, $q_2$ is the mass ratio of the receiving planet, and we have 
assumed that a fraction $2\lambda x_s/r_p$ of the angular momentum 
flow is transferred to the horseshoe region of the receiving planet
with $\lambda$ being a dimensionless constant, which we expect to be 
of order unity, as well as expression (\ref{eq:hswidth}) for $x_s$. 
For this to be significant it should exceed the torques responsible 
for convergent migration.  We estimate these to be of magnitude 
$f|\Gamma_L|$ for the emitting planet with $f$  being a dimensionless 
constant that can be  of order but   usually less than unity.
Accordingly we require 
$|\Gamma_{L(1s)}|2\lambda\sqrt{q_2/h} \gtrsim f|\Gamma_L|$,
or equivalently
\begin{equation}
\label{eq:criterion1}
\frac{q_2}{h^3}\gtrsim 8.1\left(\frac{ f}{\lambda}\right)^2
\end{equation}
As the right hand side of the inequality is characteristically around 
unity, indeed it becomes so for $f \sim \lambda/3, $  it has the same 
form as the thermal condition for nonlinearity and gap formation.
Therefore the first condition for the effective wave planet interactions is 
of the same form as  one required for the
formation of a partial gap.

In the above discussion we limited consideration of transport of 
angular momentum brought in by waves to the horseshoe region. However, 
material just exterior to that may also transfer angular momentum to 
the planet through scattering. This may be incorporated within our 
simplified discussion by adopting a larger value for $\lambda.$
This would make our criteria easier to satisfy thus they would 
give a valid indication in this case also.

However, an issue of concern is that we used linear Lindblad torques
whereas as indicated in Section \ref{Chpg} $x_s/r_p> 2h/3$ for our 
simulations so we expect the torques to have been over estimated.
As indicated in Section \ref{Chpg} we allow for this in an approximate 
manner by replacing $h$ by $3x_s/(2r_1)$ with $x_s/r_1 = \sqrt{q_1/h},$ 
and $r_1$ being the orbital radius of he inner planet when calculating 
$\Gamma_L$ and $\Gamma_{L(1s)}.$ Doing this (\ref{eq:criterion1}) 
becomes simply
\begin{equation}
\label{eq:criterion1a}
\frac{q_2}{q_1}\gtrsim 18.3\left( \frac{f}{\lambda}\right)^2
\end{equation}
which will be satisfied for sufficiently slow convergent migration.
This would appear to have no reference to partial gap formation.
However (\ref{eq:criterion1a}) was derived under  the condition
$x_s> 2H/3$ or $\sqrt{q_1/h^3} > 2/3.$ If this is satisfied together
with (\ref{eq:criterion1a}) we conclude that  we must have
\begin{equation}
\frac{q_2}{h^3}> 8.1\left(\frac{f}{\lambda}\right)^2
\label{eq:criterion1b}
\end{equation}
which is the condition implying a partial gap as before.

\subsubsection{The value of $\lambda$} 
The condition (\ref{eq:criterion1})  involves the quantity $\lambda.$
We recall that this was defined such that the fraction of the outward 
wave flux of angular momentum produced by the inner planet that is 
absorbed in the horseshoe region of the outer planet 
is $2\lambda\sqrt{q_2/h}.$ Thus if $\lambda=1,$ this corresponds to 
the flux being absorbed uniformly over a scale of the outer planet's 
orbital radius, $r_2,$ with the amount of absorption in any  radial 
interval being proportional to its extent. However, in reality the 
flux is likely to decrease more rapidly with distance $x$ from the 
exciting planet. \citet{Dong2011} suggest that the flux 
$\propto x^{-5/4}.$ This implies that the fraction absorbed in the 
horseshoe region is $2.5\sqrt{q_2/h}(r_2/x) (L/x)^{5/4},$ where $L$ 
is the distance from the exciting planet beyond which the power law 
drop off is valid. In this context we remark that $L$ cannot be 
obtained from  \citet{Dong2011} on account of nonlinearity near to 
the planet. Instead we shall assume $L=2x_s,$ roughly corresponding
to the gap  width.
Then the fraction absorbed is 
$ 5.2^{1/4}\sqrt{q_2/h} (r_2/x) (x_s/x)^{5/4} .$
In that case $\lambda ~=~5.2^{-3/4}(r_2/x) (x_s/x)^{5/4}.$
Inserting $x_s/r_1=0.0243,$ the characteristic value for our 
simulations, and taking a separation corresponding to the 3:2 
resonance, we obtain $\lambda \sim 0.56.$ Thus although there is 
considerable uncertainty $\lambda$ is plausibly of order unity for 
parameters of interest.

\subsection{Effectiveness of the horseshoe drag}
In addition to the above condition expressed by (\ref{eq:criterion1}) 
we require that the material in the horseshoe region now in the gap 
region can transport the angular momentum deposited by waves to the 
planet through  horeshoe drag.  It is possible that material slightly 
exterior to the horseshoe region also transfers angular momentum
brought in by waves through scattering and as we indicated above this 
may be incorporated by the choice of the value of $\lambda.$
In this situation less transport from the horseshoe region would be 
required than is assumed in this Section.
Accordingly if the criteria  for horseshoe regions to be effective 
developed below are satisfied we can be assured that they will be 
able to sustain the required transport. 

In considering this we note that the strong density perturbation in the  
gap means  that we need to think about this in a different way to that 
appropriate for the situation where there is no gap.
We can consider the situation when the horseshoe drag is most effective.
This is when it only works on one side of the planet. In the case of 
outward angular momentum transport towards the outer planet, which we 
focus on below, this will be that leading the planet. On the trailing 
side we suppose that material approaching the planet absorbs angular 
momentum from the waves causing the horseshoe turn to occur at 
significant distances from the planet. The one sided horseshoe drag
results in
\begin{equation}
\label{eq:onesidedhorse}
\Gamma_{hsl} = \frac{1}{2}\Sigma_{min}\Omega^2 r x_s^3 =
\frac{1}{2}\Sigma_{min} \Omega^2 r^4 (q_2/h)^{3/2},
\end{equation}
where $r$ and $\Omega$ are evaluated at the outer planet's position with 
$\Sigma_{min}$  being  the surface density at the gap minimum.
Effective transfer to the outer planet can take place if 
\begin{equation}
\label{eq:effetransfer}
\Gamma_{hsl} > |\Gamma_{L(1s)}| 2\lambda\sqrt{q_2/h}
\end{equation}
or, recalling  that $\Gamma_{L(1s)}$ is evaluated for the inner planet 
that equivalently 
\begin{equation}
\label{eq:criterion2}
\left(\frac{\Sigma_{min}}{\Sigma_{un}} \right)_2 > \left(\frac{2.16 
\lambda q^2_1}{q_2 h^2}\right) \left( \frac{(\Sigma_{un} r^4\Omega^2)_1}
{(\Sigma_{un} r^4 \Omega^2)_2}\right),
\end{equation}
where the subscripts $1$ and $2$ attached to brackets indicate the planet 
location for evaluation.
We remark that we have used the  linear Lindblad torque which,  being an 
overestimate,  will not disturb  conclusions  obtained from the above 
condition. 
In order to evaluate the second term enclosed in large brackets we 
assume that, as for our simulations, $ -0.5 < \alpha < 0.5,$ and that
the planets are close to 3:2 resonance in a Keplerian disk.  Then this 
factor lies between $0.87$ and  $0.67.$ 
Given that by definition $(\Sigma_{min}/\Sigma_{un} )_2 <1$, if the  
above condition is always satisfied we should have
\begin{equation}
\label{eq:criterion2a}
\frac{1.9 \lambda q^2_1}{q_2 h^2}< 1.
\end{equation}
The two conditions we obtained for effective planet repulsion are
(\ref{eq:criterion1}) and (\ref{eq:criterion2}). Here the wave generating 
planet has mass ratio $q_1$ and the receiving planet mass ratio $q_2$. 
Note that the roles may be reversed in which case we interchange $q_1$ and 
$q_2$ and the subscripts $1$ and $2$ in these conditions. While doing this 
we assume that the value of $\lambda$ remains the same.

In our calculations for $q_1=1.3\times 10^{-5}$ and 
$q_2=1.185\times 10^{-5}$ the requirement that 
$q_2/h^3=1.48 > 8.1(f/\lambda)^2$ is satisfied for $f\lesssim 0.43 \lambda$ 
and also the second criterion is fulfilled, because, it is enough that
$(\Sigma_{min}/\Sigma_{un} )_2 >0.052\lambda$, which is exactly the case if 
$\lambda$ does not exceed  a factor of $\sim 4.$
If we apply the two criteria to the inner planet playing the role of the 
receiving planet then we obtain $1.6 > 8.1(f/\lambda)^2$, which is satisfied 
for $f\lesssim 0.44\lambda$.
The second criterion in this case is
\begin{equation}
\left(\frac{\Sigma_{min}}{\Sigma_{un}} \right)_1 > \left(\frac{2.16 
\lambda q^2_2}{q_1 h^2}\right) \left( \frac{(\Sigma_{un} r^4\Omega^2)_2}
{(\Sigma_{un} r^4 \Omega^2)_1}\right),
\end{equation}
This leads to the requirement that 
$(\Sigma_{min}/\Sigma_{un})_1 >0.059\lambda$ to ensure it is satisfied for 
the range of $\alpha$ considered.
This criterion  is also satisfied for $\lambda$ not exceeding  a factor of 
$\sim 4$ as above.
 
This  means that in our calculations we should expect effective angular
momentum transfer between the planets resulting from wave planet 
interactions, both from inner to outer planet and outer to inner planet can 
readily occur, particularly for slow enough convergent migration.
We can also verify an additional consequence, namely reversing
the planet positions, placing the inner one at the position of the outer 
planet and vice versa will lead to the same conclusion. Thus for either 
configuration repulsion due to wave planet interaction should be efficient. 
In the next Sections we will verify our prediction.  


\section{How does the migration depend on the mass of the planets?}
\label{sec:innermass}
For our particular choices of initial surface density profile we can study  
planet pairs with a range of mass ratios. We consider the mass ratios 
already discussed in detail, namely  $q_{1}=1.3 \times 10^{-5}$ and  
$q_{2}=1.185\times 10^{-5}$ with the initial surface density distribution 
being given by $\Sigma =\Sigma_0 r^{-\alpha}$with  $\alpha = 0.5$  and 
$\Sigma_0 =  8\times 10^{-5}.$  For the same mass ratios we also ran a 
case with the initial surface density profile given by 
Equation~(\ref{disk}) with $\alpha=0.5$ and $\Sigma_0 = 6\times 10^{-5}.$  
In addition we perform simulations, starting with the latter initial
surface density distribution and  with the same value of  $q_{2}$ but with 
$q_{1}$ respectively $1.185 \times 10^{-5}$, $1.95 \times 10^{-5}$
and $2.6 \times 10^{-5} .$ This suite  is completed by performing  
simulations with $q_1=1.185\times 10^{-5}$, $q_2=1.3\times 10^{-5}$ 
starting with the latter initial surface density distribution for
two different values of $\Sigma_0$, namely
$\Sigma_0 = 6\times 10^{-5}$  and $\Sigma_0 = 4.5\times 10^{-5}$.

Results are shown in Figure~\ref{fig:crit1crit2} in which we plot the 
position of both planets in each pair in the 
$$\left[\frac{q_1}{h^3} , \frac{h^2q_1}{\pi q_2^2}\left(\frac{\Sigma_{min}}
{\Sigma_{un}}\right)_1 \right]  \hspace{0mm}
\hspace{1mm}{\rm and} \hspace{0mm} \left[\frac{q_2}{h^3} , 
\frac{h^2q_2}{\pi q_1^2}\left(\frac{\Sigma_{min}}{\Sigma_{un}}\right)_2 
\right] 
\hspace{0mm}  \rm{planes.}$$ 
\begin{figure*}[htb!]
\centerline{
\vbox{
\hbox{
\includegraphics[width=0.9\columnwidth]{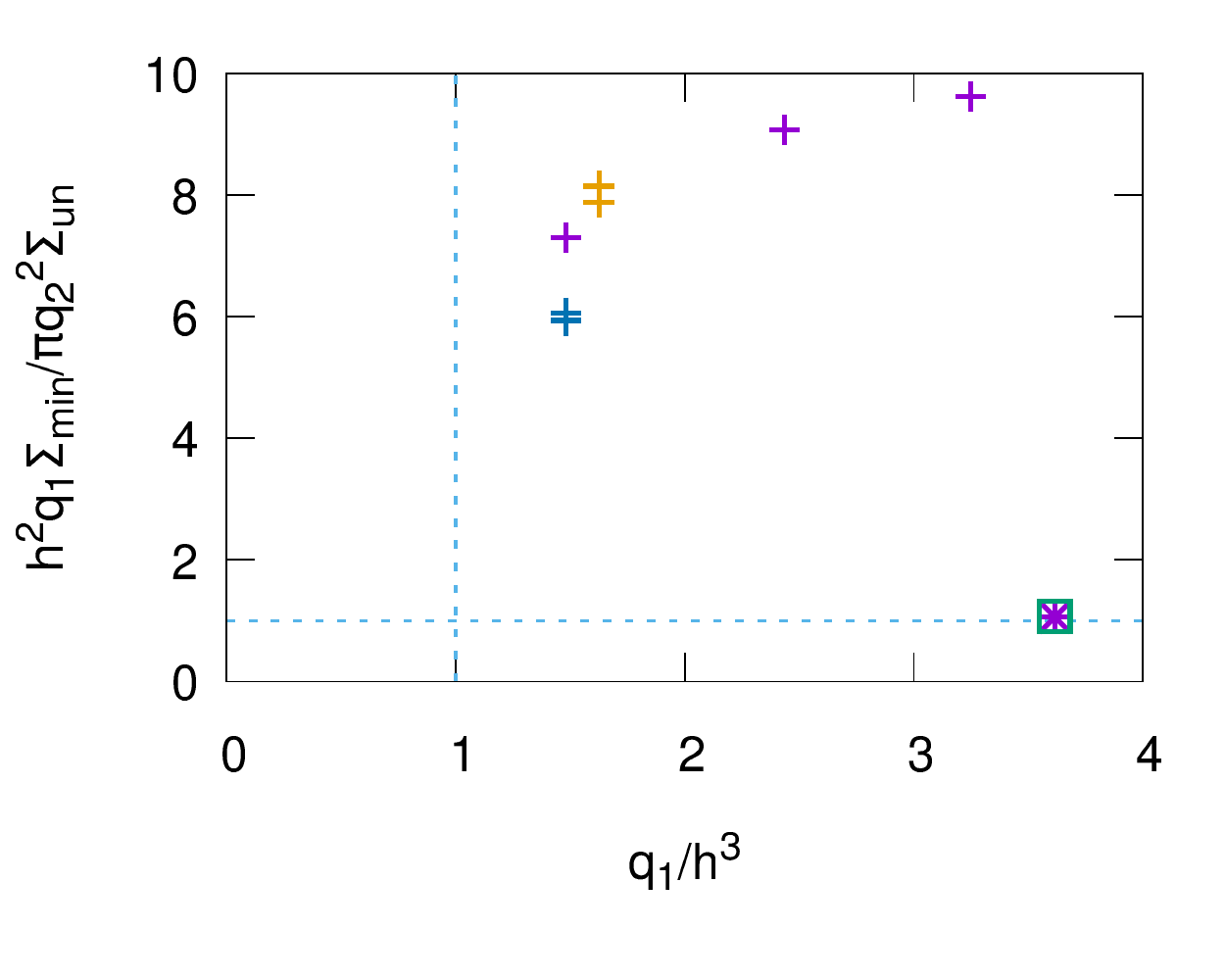}
\includegraphics[width=0.9\columnwidth]{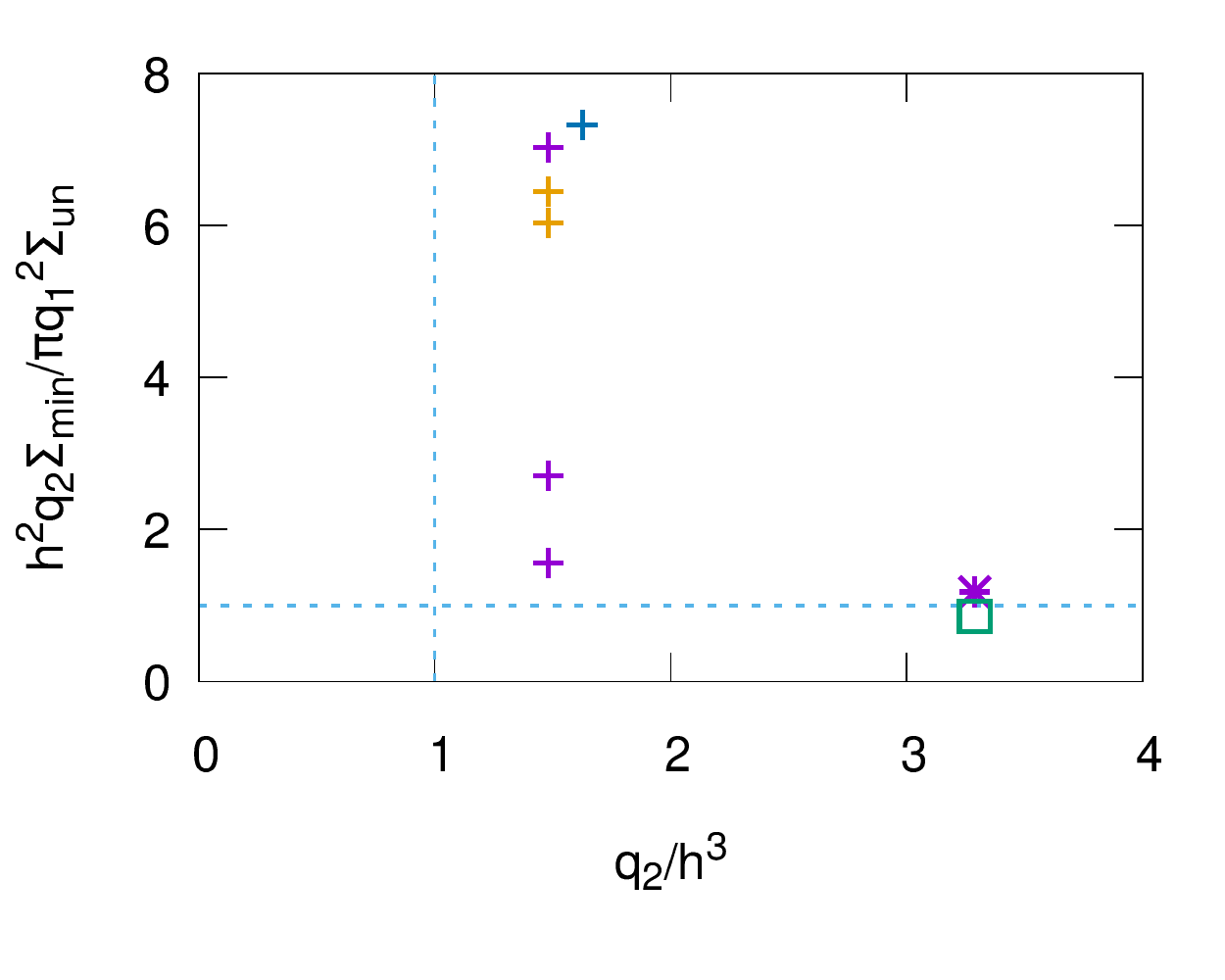}
}
}
}
\caption{The results of the hydrodynamical simulations of two super-Earths
migrating in a protoplanetary disk shown in the planes   indicating the 
effectiveness of the repulsion due to wave planet interactions (crosses). 
The asterisks and open squares  indicates the results obtained 
by recalculating the simulations with the parameters given in  
\citet{Baruteau2013} for Uranus-mass planets. The left panel gives 
information about  the case in which the inner planet with the mass ratio 
$q_1$ is viewed as the  receiver of waves emitted by the outer planet.  
The right panel gives information when the role of the planets in the pair 
is reversed, namely the outer planet is regarded  as a  receiver of waves 
emitted by the inner planet. See text for more details. 
}
\label{fig:crit1crit2}
\end{figure*}
The left panel of Figure~\ref{fig:crit1crit2} showing the former gives 
information about the inner  planet with the mass ratio $q_1$ as a 
receiver of waves emitted by the outer planet. The right panel showing 
the  latter gives information when the role of the planets in each pair 
is reversed, namely it gives information about the outer planet
regarded as a receiver of waves emitted by the inner planet.  

Regarding the left hand panel we remark that the counterpart of the  
criterion given by (\ref{eq:criterion1}) that is applicable 
to the inner planet will be satisfied for $f/\lambda < 0.35$ if  
${q_1}/{h^3} > 1$ indicating enough supply to it  as  receiver to 
enable the halting of convergent migration. Similarly the criterion 
$({h^2q_1})/({\pi q_2^2})\left({\Sigma_{min}}/{\Sigma_{un}}\right)_1 >
\lambda,$ recalling that we expect $\lambda$ to be of order unity,  
ensures that the condition (\ref{eq:criterion2a}) which indicates  
the effectiveness of the horseshoe drag for the inner planet
is satisfied  and the planets are near to the 3:2 resonance.
Similar remarks apply to the right hand panel with the role of the 
planets reversed and with   
${q_1}/{h^3} $ and $({h^2q_1})/({\pi q_2^2})\left({\Sigma_{min}}/
{\Sigma_{un}}\right)_1$ 
being respectively replaced by ${q_2}/{h^3} $ and $({h^2q_2})/
({\pi q_1^2})\left({\Sigma_{min}}/{\Sigma_{un}}\right)_2.$  
The regions delineated by the dashed lines  which extend to large 
distances from the origin in each panel are thus locations where 
wave planet interaction can be effective.
  
The two orange crosses in each panel indicate the results obtained 
from the two simulations performed with the mass ratios 
$q_1=1.3\times 10^{-5}$ and $q_2=1.185\times~10^{-5}.$
For each panel the upper orange cross indicates the simulation in 
which the two planets are placed in the disk with the initial 
surface density distribution with $\Sigma =\Sigma_0 r^{-\alpha}$ 
with $\alpha = 0.5$ and $\Sigma_0 =  8\times 10^{-5}$.  The lower 
one is the outcome of the simulation with the initial surface 
density profile given by  Equation~(\ref{disk}) with $\alpha=0.5$ 
and $\Sigma_0 = 6\times 10^{-5}$. The difference in the location 
of the crosses reflects the difference in the depths of the gaps.

The two turquoise crosses (in the left panel) and apparently one 
but in fact  there are actually two overlapping each other in the 
right panel illustrate the results of the simulations with 
$q_1=1.185\times 10^{-5}$, $q_2=1.3\times 10^{-5}$ and two 
different values of $\Sigma_0$, namely $\Sigma_0 = 6\times 10^{-5}$ 
(the lower turquoise cross in the left panel) and 
$\Sigma_0 = 4.5\times 10^{-5}$ (the upper turquoise cross in the
left panel). In both cases, the depth of the gap made by the planet 
with $q_2=1.3\times 10^{-5}$ is the same. That is why in the right 
panel one can see only one turquoise cross. 

The three violet crosses show the remaining simulations for which  
effective repulsion due to wave planet interactions takes place. 
These three simulations only differ in the choice of the mass 
ratio of the inner planet which takes values 
$q_1=1.185\times 10^{-5}$, $1.95\times 10^{-5}$
and $2.6\times 10^{-5}$ respectively. 
It is interesting that as the mass ratio of the inner planet 
increases the depth of the gap it produces becomes larger and the
depth of the gap the outer planet produces decreases, even though 
its mass ratio in all three cases is the same.

The position of the points in the planes discussed here is mostly 
determined by the mass ratios of the planets. In the left panel 
the location of the violet crosses increases from the left to the 
right following the increasing mass ratio of the inner planet. In 
the right panel, the violet crosses form a vertical line as $q_2$  
is the same for all of them. Those with larger $q_1$ are located 
below  those with  smaller $q_1.$
The above results indicate that as $q_1$ increases for fixed $q_2$ 
the horseshoe drag on the inner planet is increasingly able to 
sustain the wave planet interaction. On the other hand  the 
increasing Lindblad torques it produces make it harder for the 
outer planet to sustain the interaction at fixed $q_2.$ However, 
the reaction to this is likely to be that the disk planet 
interaction can be sustained with the planets being further apart.
 
According to our criteria, these  planet pairs  should eventually 
migrate divergently if they approach close enough to each other
and the convergent migration causing this is slow enough. 
This is expected for all our hydrodynamic simulations.

\subsection{The relationship to previous work}
In order to make a clear link with the previous studies by 
\citet{Baruteau2013} we have recalculated the migration of two 
planets presented in their paper with mass ratios 
$q_1=4.4\times 10^{-5}$ and $q_1=4\times 10^{-5}$.
The planets evolving in the disk with the initial surface density
given by $\Sigma =\Sigma_0 r^{-\alpha}$  with $\alpha =0.5$ and 
$\Sigma_0=8\times 10^{-5}$ enter into the 3:2 mean-motion resonance 
first and later on undergo divergent relative migration. The same 
planets evolving in the lower surface density disk with
$\Sigma_0=3\times 10^{-5}$ become locked in the 2:1 
commensurability and do not show signs of divergent migration.
The results corresponding to $\Sigma_0=8\times 10^{-5}$ (violet 
asterisk) and $\Sigma_0=3\times 10^{-5}$ (green empty square) are 
also plotted in Figure~\ref{fig:crit1crit2}.
The criterion expressed by (\ref{eq:criterion1}) and its 
counterpart corresponding to exchanging the planets are easily 
satisfied in both runs.  Both planets create a  partial gap. 
The second criterion corresponding to (\ref{eq:criterion2a}) with 
$\lambda=1$ and its counterpart corresponding to interchanging 
the planets are only just satisfied in the case eventually 
undergoing divergent migration with the criterion for the outer 
planet to be an effective receiver  marginally failing  when 
$ \lambda=1$ for the case that retained convergent migration. 
So the results of a  previous study  are fully consistent with 
the picture presented here.


\section{The effectiveness of  repulsion resulting from wave planet
interaction }\label{sec:repulsion}
We now go on to check the conditions under which we expect divergent 
migration of two super-Earths in the disk. In the previous Section 
we indicated that all the  cases considered there were likely to 
exhibit effective repulsion due to wave planet interaction on the 
basis of  our simple criteria based on the magnitude of wave fluxes 
and the potential of the horseshoe drag to communicate the angular 
momentum transported to the planet and we aim to verify this.

\subsection{Dependence on the rate of convergent migration}
We begin by comparing three particular cases. These form a sequence 
whereby the rate of  initial convergent migration is increasing and
either planet may have the largest mass ratio. Thus we are able to 
study how the location where the transition between convergent and 
divergent migration occurs depends on this. The cases we consider 
are one with the inner planet more massive than the outer one with
$q_{1}=1.3\times 10^{-5}$, $q_{2}= 1.185\times 10^{-5}$ (this case 
was already  discussed in Section~\ref{sec:results}), the equal mass 
case with $q_{1}=q_{2}= 1.185\times 10^{-5}$, and one for which
the inner planet is less massive than the outer one with
$q_{1}=1.185\times 10^{-5}$, $q_{2}= 1.3\times 10^{-5}.$

In all cases the initial surface density was given by 
Equation~(\ref{disk}) with $\alpha =0.5$ and
$ \Sigma_0 =6\times 10^{-5}.$
The period ratio of the two planets as a function of time for these  
simulations is illustrated in Figure~\ref{fig:inmass}.
\begin{figure}[htb!]
\centerline{
\vbox{
\hbox{
\includegraphics[width=0.9\columnwidth]{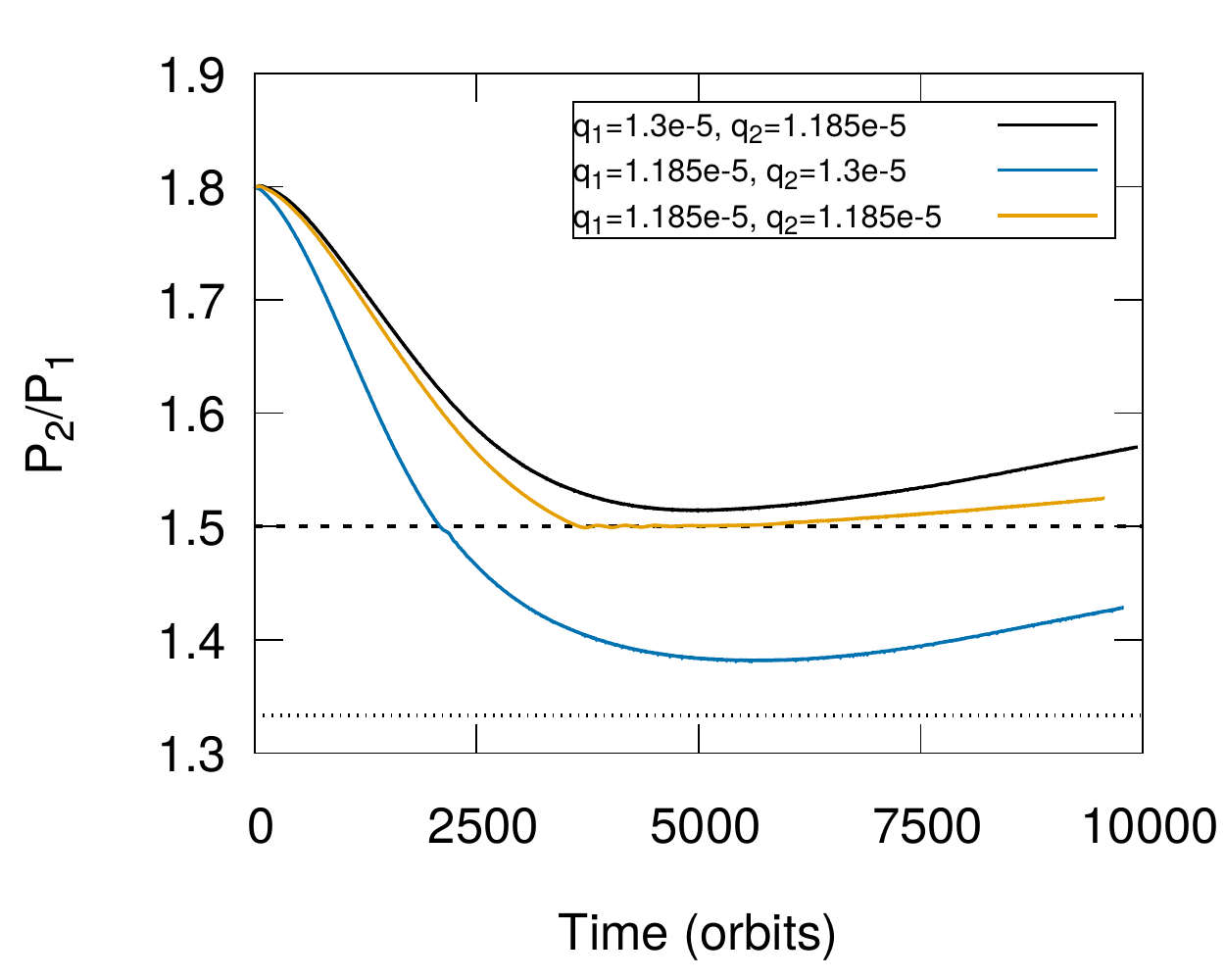}
}
}
}
\caption{The period ratio of the two planets as a function of time 
for the hydrodynamical simulations of two super-Earths with mass 
ratios of $q_{1}=q_2=1.185\times 10^{-5}$ (orange line), 
$q_1= 1.185\times 10^{-5}$, $q_2= 1.3 \times 10^{-5}$ (dark blue 
line) and $q_1=1.3 \times 10^{-5}$, $q_{2}=1.185 \times 10^{-5}$ (black line)
migrating in a protoplanetary disk. The horizontal dashed and 
dotted lines respectively indicate the locations of the 3:2 and 
4:3 resonances.}
\label{fig:inmass}
\end{figure}

We have already described the case with $q_1=1.3 \times 10^{-5}$ 
and $q_{2}=1.185 \times 10^{-5},$ represented by the black line 
in Figure~\ref{fig:inmass} in Section~\ref{sec:results}.
The system evolves towards the 3:2 resonance and at some point 
leaves the  vicinity of the commensurability and migrates 
divergently until  the end of the simulation.
The other two cases are qualitatively similar.

The case with $q_1=q_2$ represented by the orange line in 
Figure~\ref{fig:inmass}  arrives at the  3:2 resonance, stays 
there for a longer period of time as compared to the previous 
case and eventually leaves the resonance with a slower 
divergent migration rate.

In the simulation in which $q_2 > q_1$ represented by the dark 
blue line in Figure~\ref{fig:inmass}, the relative convergent 
migration at the beginning of the evolution is the fastest of 
the three. The relative migration rate is so fast that the 
system passes through the 3:2 commesurability, but later on,
the migration starts to be divergent. This shows that a close 
approach to strict commensurability is not necessary in order 
to induce  divergent migration, even though one might expect 
that the 4:3 resonance could have some effect here (see the 
bottom panel of Figure~\ref{fig:fix_xm1}).

\begin{figure*}[htb!]
\centerline{
\vbox{
\hbox{
\includegraphics[width=0.5\columnwidth]{pratio_fix_scaling.pdf}
\includegraphics[width=0.5\columnwidth]{semi_fix_scaling.pdf}
\includegraphics[width=0.5\columnwidth]{ecc_fix_scaling.pdf}
\includegraphics[width=0.5\columnwidth]{rangle_fix_scaling.pdf}
}
\hbox{
\includegraphics[width=0.5\columnwidth]{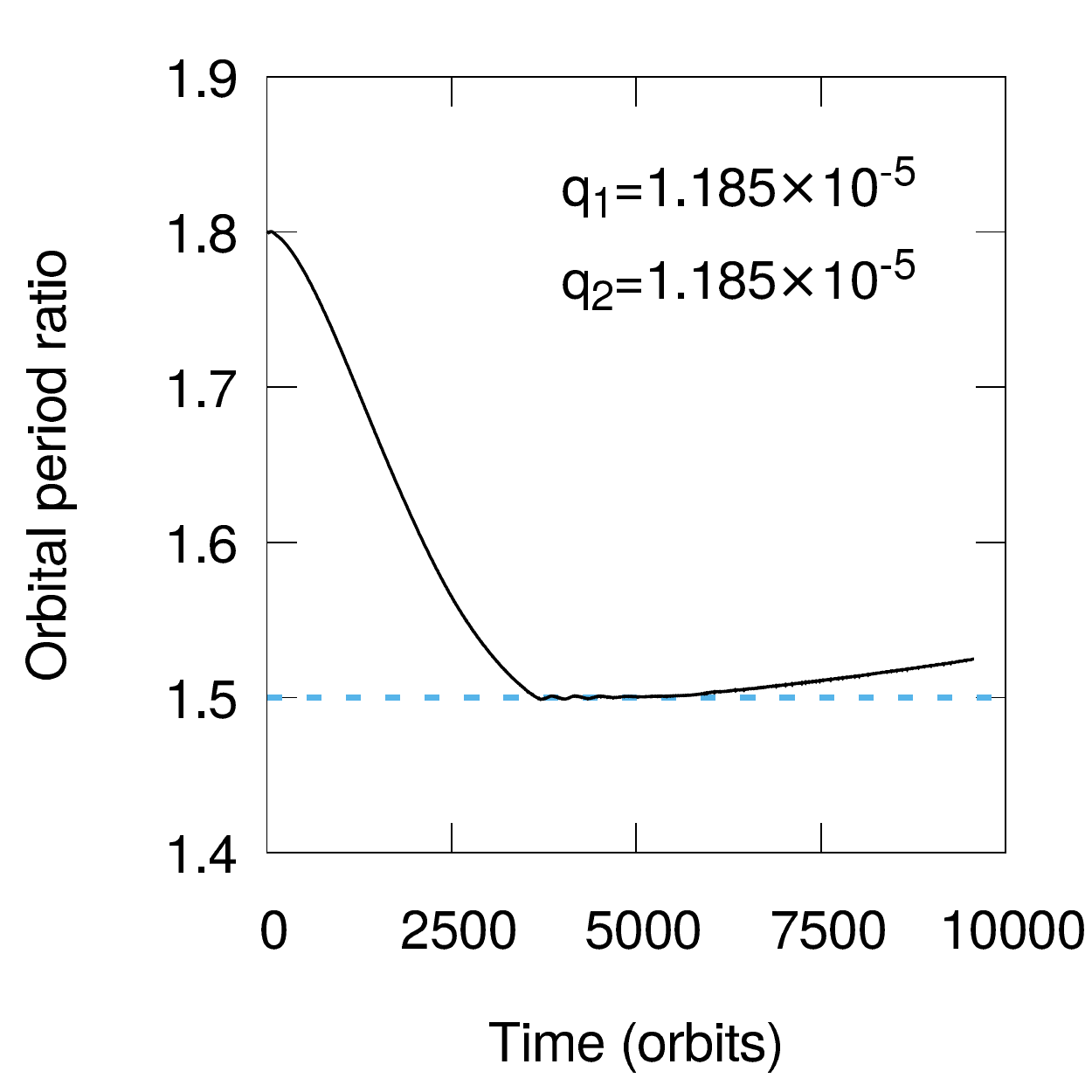}
\includegraphics[width=0.5\columnwidth]{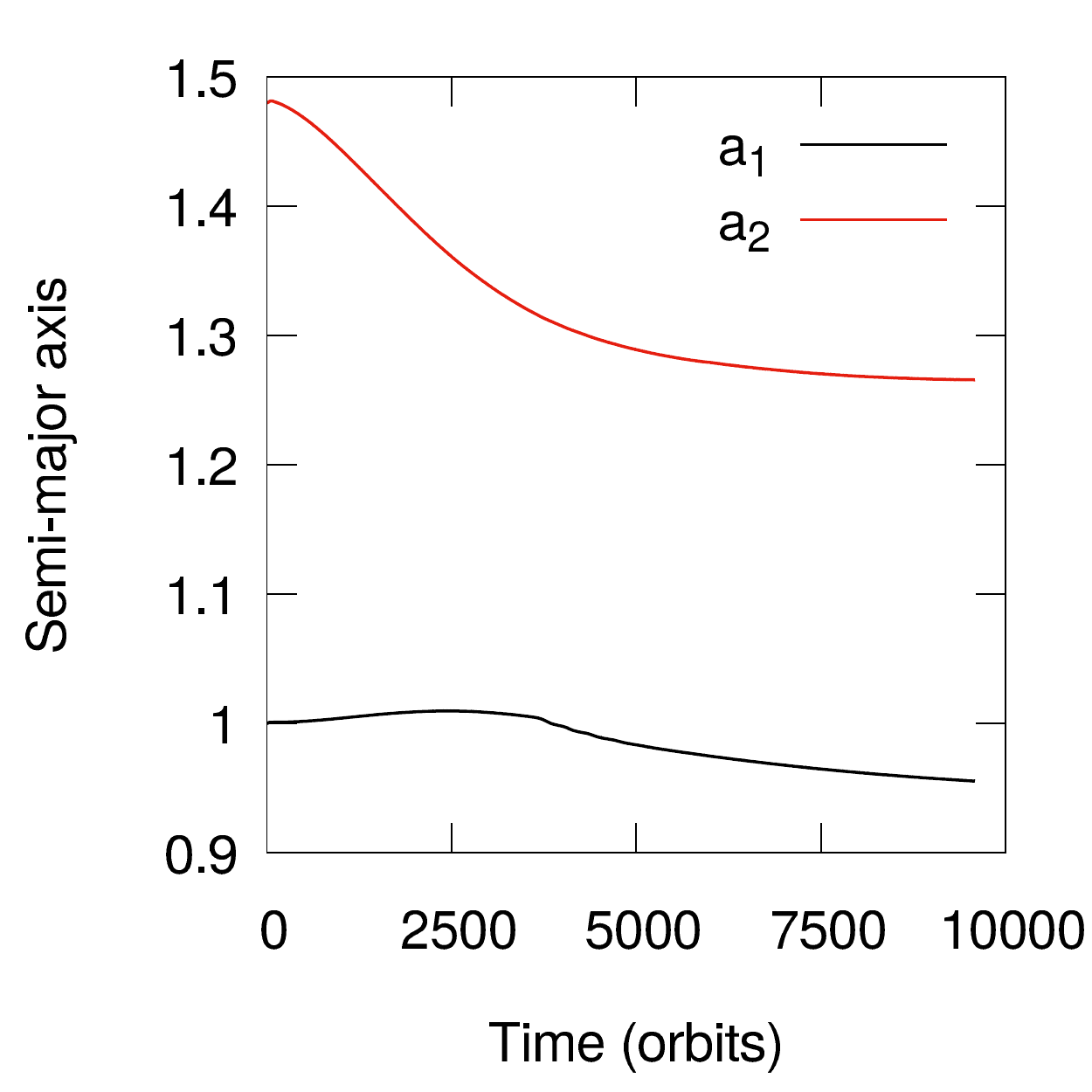}
\includegraphics[width=0.5\columnwidth]{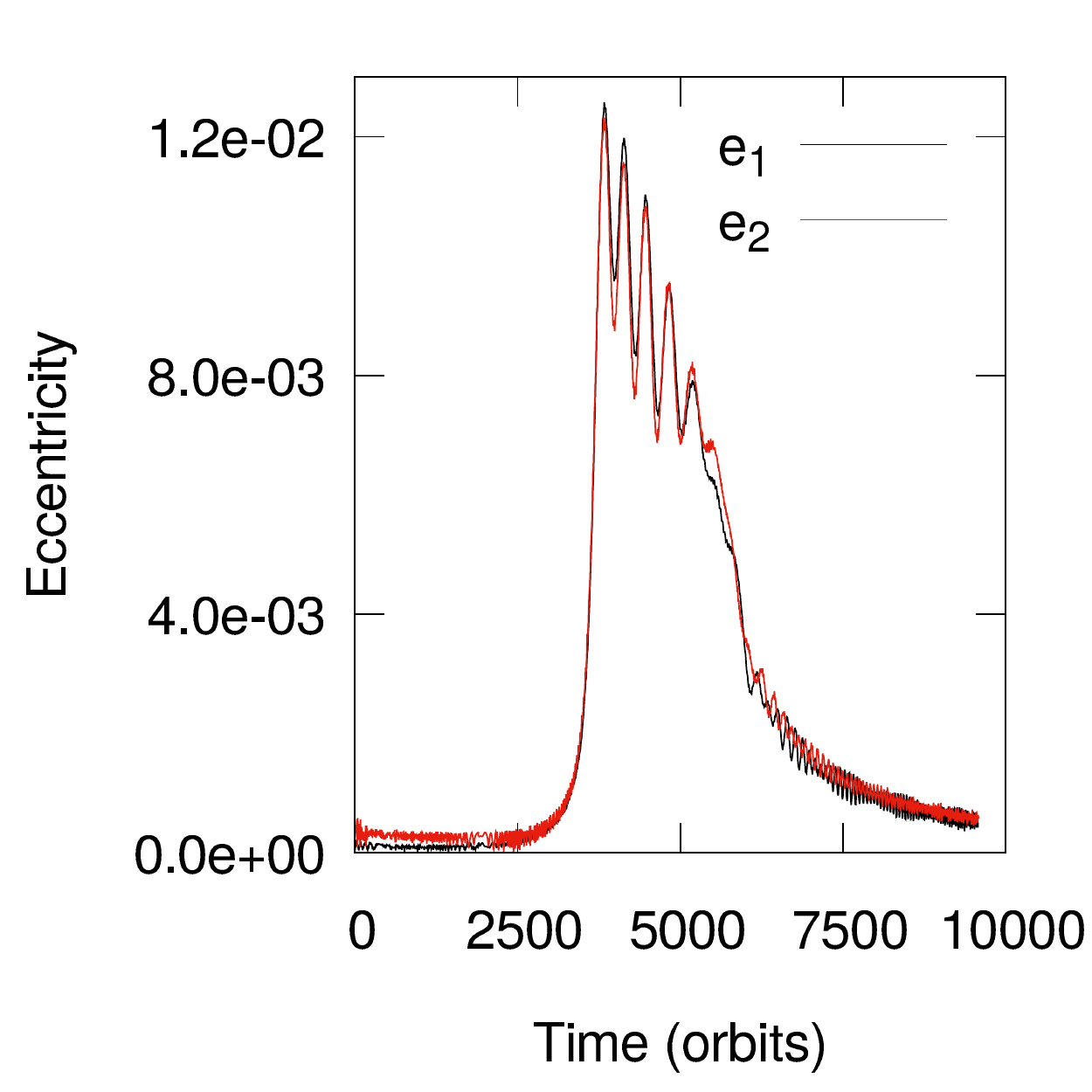}
\includegraphics[width=0.5\columnwidth]{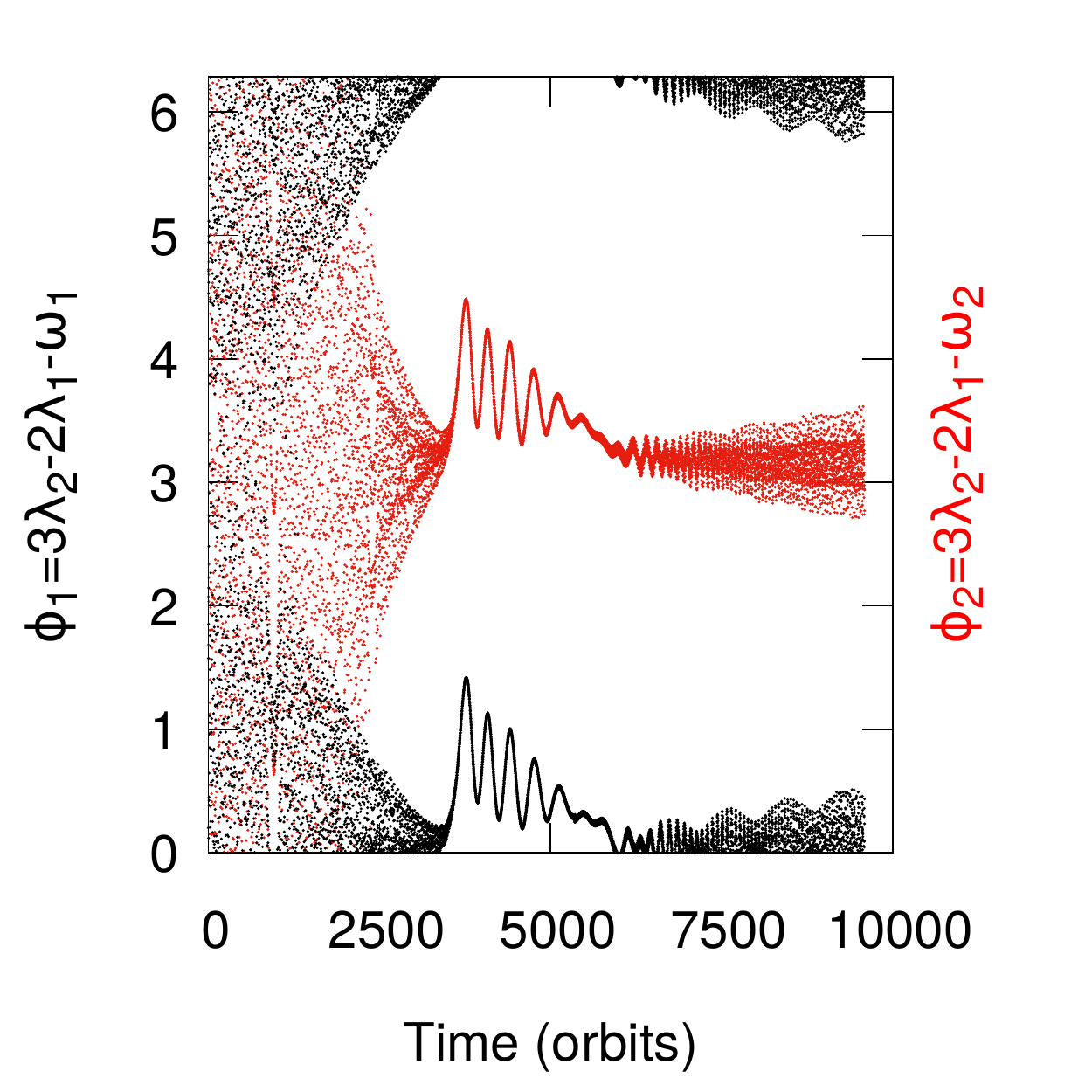}
}
\hbox{
\includegraphics[width=0.5\columnwidth]{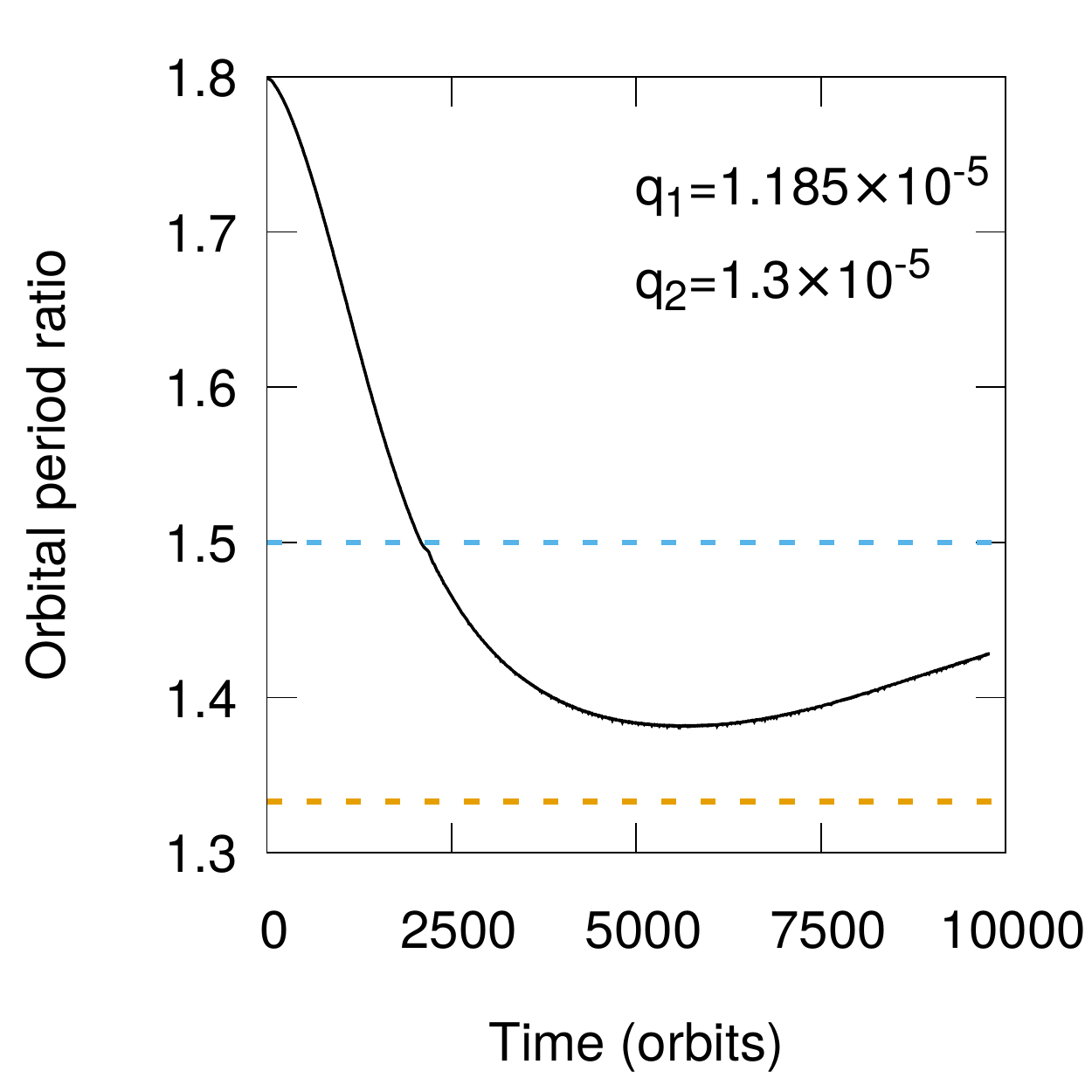}
\includegraphics[width=0.5\columnwidth]{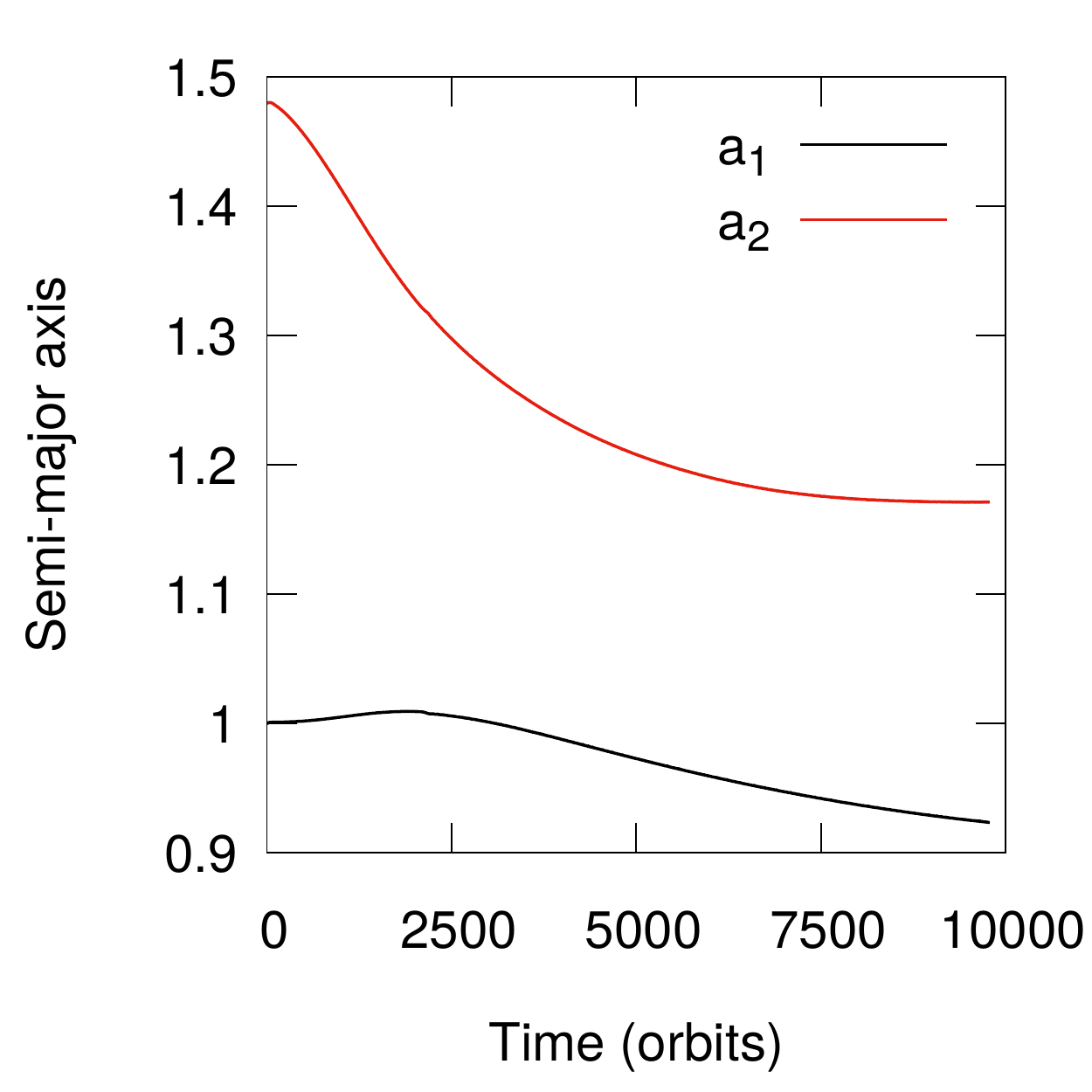}
\includegraphics[width=0.5\columnwidth]{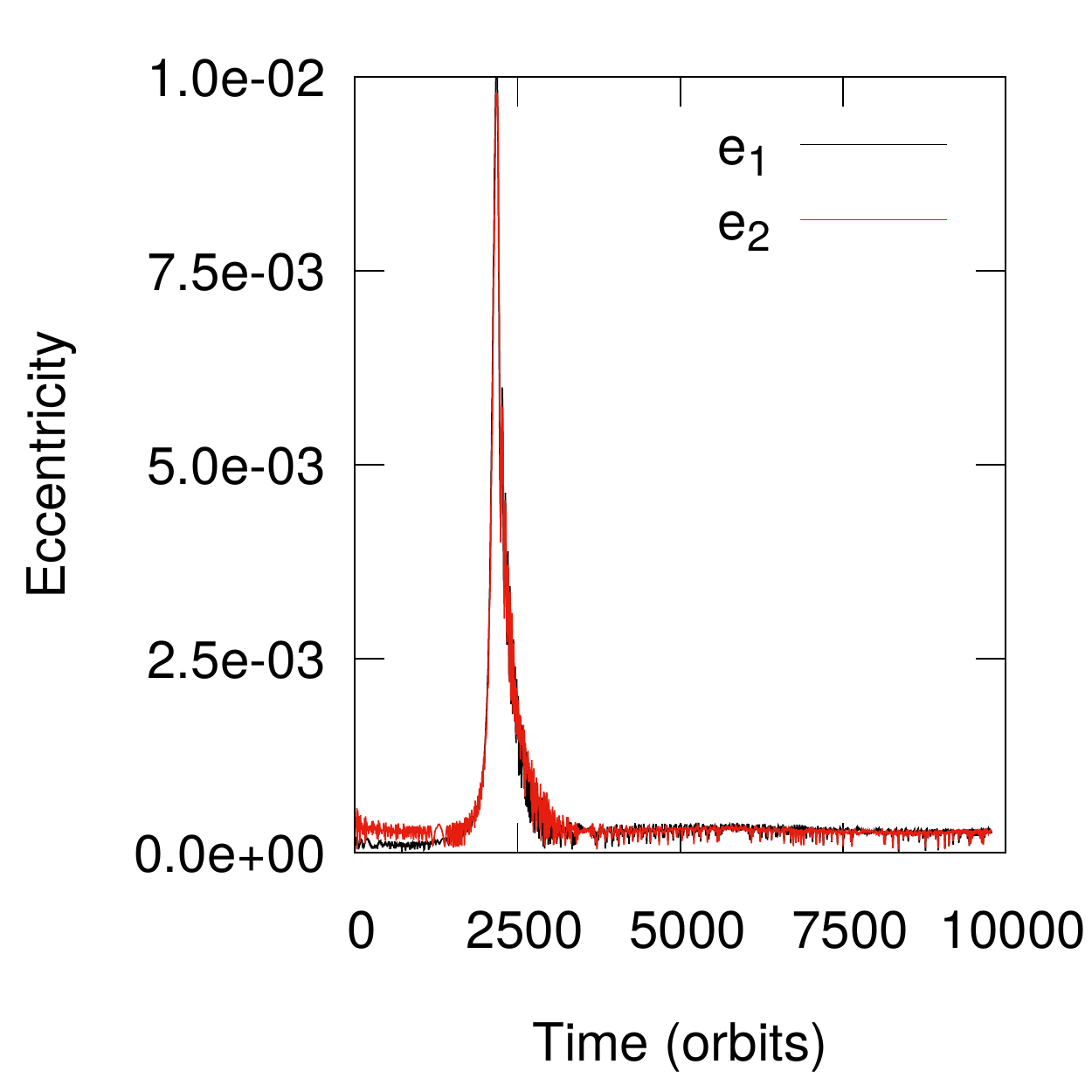}
\includegraphics[width=0.5\columnwidth]{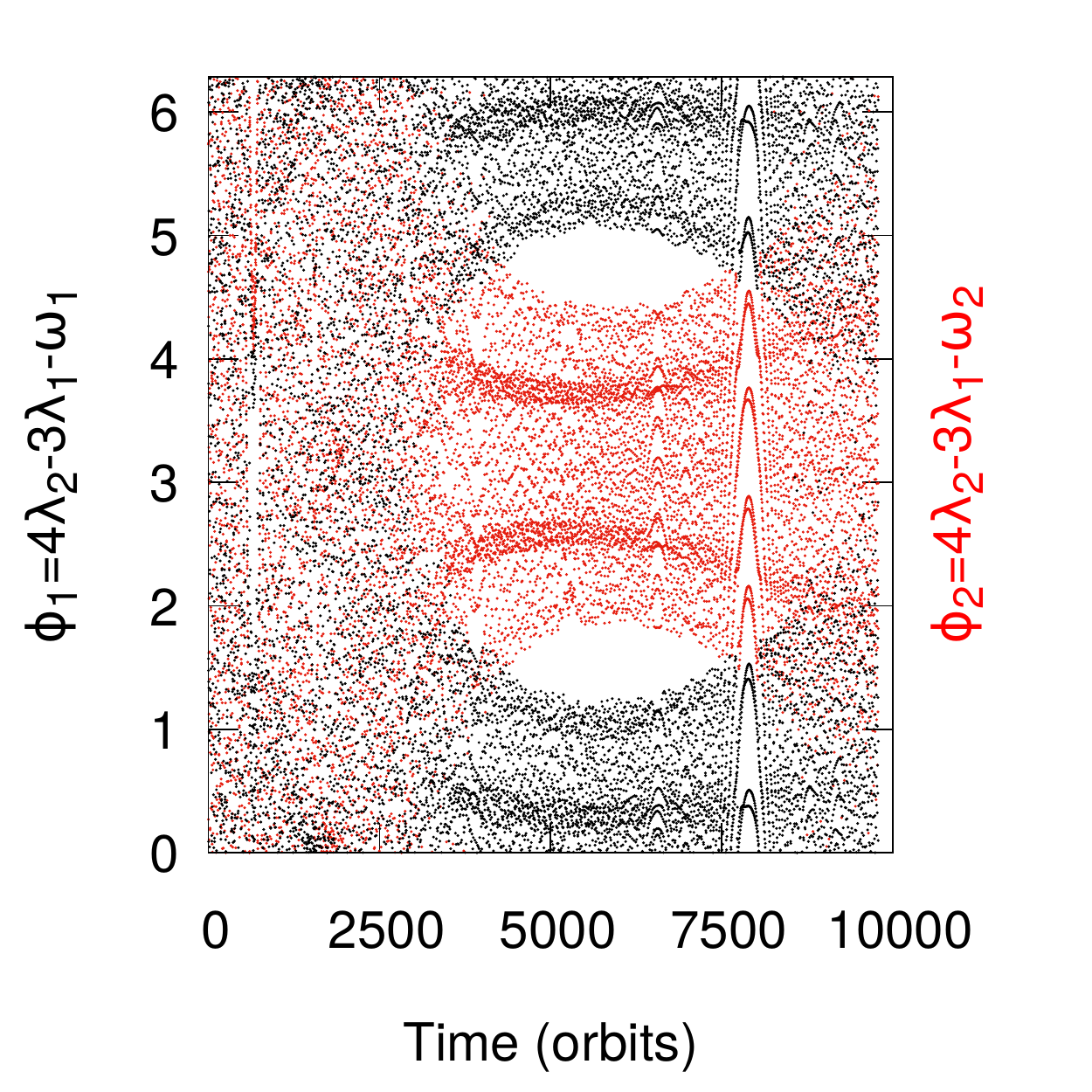}
}
}
}
\caption{Each row shows the results of the simulations done with 
the same disk parameters, but different mass ratios of the planets. 
In the first row the mass ratios are $q_1=1.3\times 10^{-5}$,
$q_{2}=1.185 \times 10^{-5}$, in the second $q_1=q_2=1.185 
\times 10^{-5}$, and in the third one $q_1=1.185\times 10^{-5}$,
$q_{2}=1.3 \times 10^{-5}$.
The panels in every row show the evolution of the planets' orbital 
period ratio, semi-major axes, eccentricities and the resonance 
angles, as indicated on the rightmost panel, from left to right 
in each simulation.
The horizontal dashed blue line in the first left panel in each
row indicates the position of the 3:2 commensurability. The 
horizontal dashed yellow line indicates the position of the 4:3 
resonance.}
\label{fig:fix_xm1}
\end{figure*}

More detailed information from the  three simulations is plotted  
in Figure~\ref{fig:fix_xm1}. This includes the individual planet 
migration rates, the evolution of eccentricities and the 
behaviour of the resonance angles for either the 3:2 or 4:3 mean 
motion resonances where relevant. 
For completeness and convenience  we  present the results already 
displayed  in Figure~\ref{fig:fix_scaling} in the uppermost row.

The second row shows the evolution of two equal mass planets with  
$q_{1}=q_{2}=1.185 \times 10^{-5}.$
From the figure we  see that before $t \sim 3500$ orbits
the inner planet migrates slowly outwards due to the particular 
surface density profile adopted, while the outer planet migrates 
inwards. The convergent migration brings planets closer  until
the system arrives to the 3:2 MMR.
The orbital period ratio of the two planets stays at around 1.5 
until $t \sim 6000$ orbits. During this period, both planets 
migrate inwards. The eccentricities are excited to $\sim 0.012$ 
at $t \sim 4000$ orbits and subsequently decrease.
One of the resonance angles associated with the 3:2 MMR is 
librating with mean value above $\pi$ that decreases with time, 
and the other with mean value in excess of zero that decreases 
with time.
After $t \sim 6000$ orbits, the orbital period ratio of the 
planets is increasing slowly while the eccentricities continue 
to decrease.  The resonance angles of the 3:2 MMR are librating  
with increasing amplitude. It is clear that the planets are 
leaving the 3:2 MMR.

The third row illustrates the case  when the mass ratios of the 
two planets in the simulation illustrated in the uppermost row
are interchanged. This has the effect of making the relative 
migration rate to be  significantly faster.
Thus it is not a surprise that in this case the relative 
migration is  too fast to allow  3:2 resonance capture.
Indeed, the planets passed through this resonance and at around 
6000 orbits the migration becomes divergent.
The 4:3 resonance may play a  significant role. Note the 
distance from  strict commensurability and the large amplitude 
librations of the resonance angles which occur around $6000$ 
orbits indicate a relatively weak effect though this could be 
enough to affect the evolution.

In summary, the three simulations illustrated in 
Figures~\ref{fig:inmass} and \ref{fig:fix_xm1} provide a sequence 
with increasing initial convergent migration rate.  An increased 
rate is found to enable the pair to approach each other more 
closely before the relative migration reverses.

\subsection{A slower initial convergent migration rate  and 
faster subsequent divergent migration rate obtained by increasing
$q_1$ at fixed $q_2.$}
We now investigate  how the evolution depends on the mass of the 
inner planet.  In addition to those already discussed we have run
another two simulations with $q_{1}=1.95 \times 10^{-5}$ and $2.6
\times 10^{-5}$ both of these having  $q_2=1.185 \times 10^{-5}.$
These enable us to consider  a sequence of four simulations, in 
which the outer planet has the fixed mass ratio 
$q_2=1.185 \times 10^{-5}$ 
with the inner planet respectively having 
the mass ratio $q_1=1.185 \times 10^{-5}$, $1.3 \times 10^{-5}$,
$1.95 \times 10^{-5}$ and $2.6 \times 10^{-5}.$
For all of these cases the initial surface density distribution 
was given by Equation~(\ref{disk}) with $\alpha =0.5$ and 
$\Sigma_0=6.0\times 10^{-5}.$

A comparison of the evolution of the orbital period ratio  for 
these simulations is given in Figure~\ref{fig:inmass2} and 
additional details are presented in Figure~\ref{fig:fix_inner}.

\begin{figure}[htb!]
\centerline{
\vbox{
\hbox{
\includegraphics[width=0.9\columnwidth]{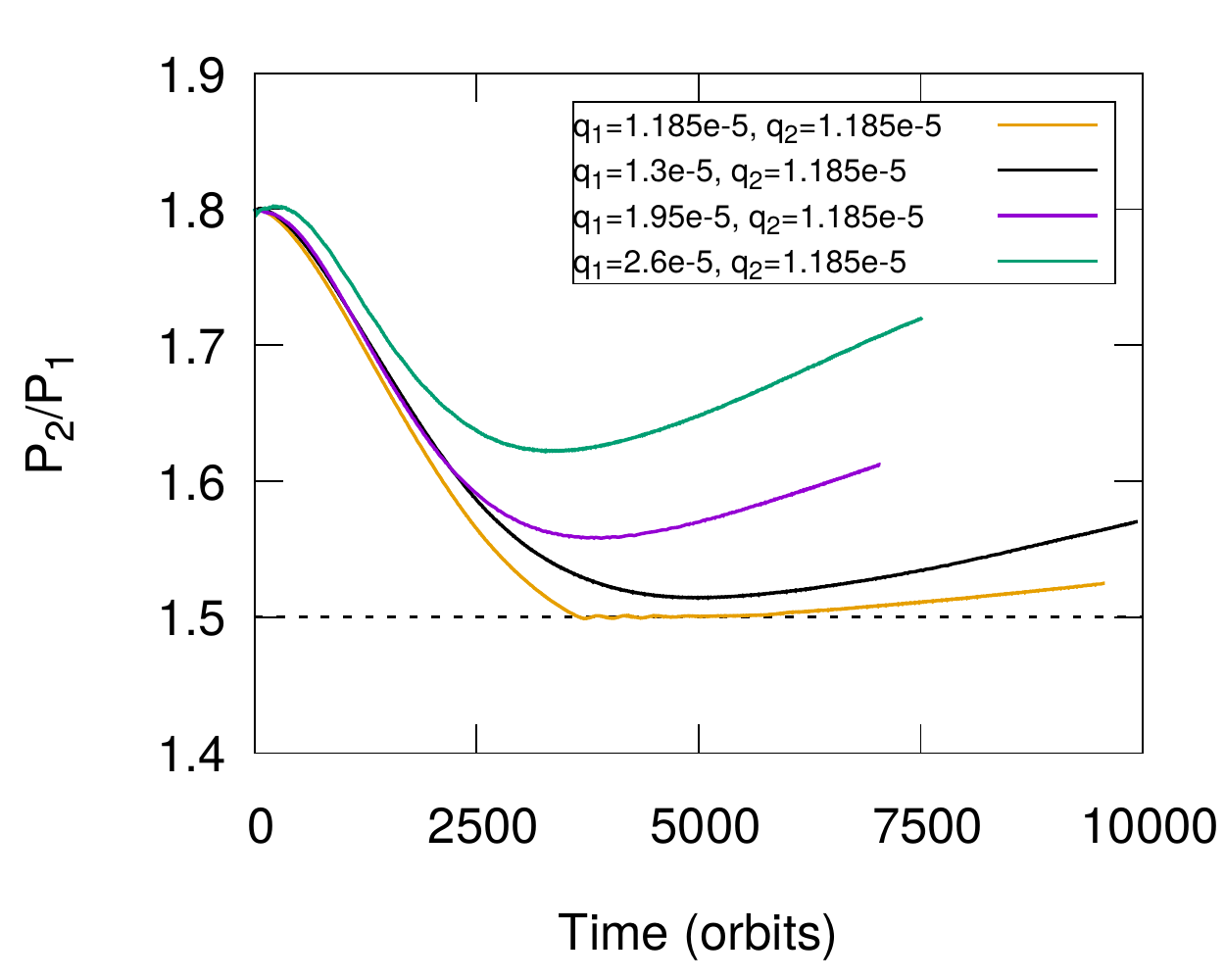}
}
}
}
\caption{The results of the hydrodynamical simulations of two 
super-Earths with mass ratios $q_{1}=q_2=1.185\times 10^{-5}$ 
(orange), $q_1= 1.3\times 10^{-5}$, $q_2= 1.185 \times 10^{-5}$ 
(black) and $q_1=1.95 \times 10^{-5}$, $q_{2}=1.185 \times 10^{-5}$ 
(violet) and $q_1=2.6 \times 10^{-5}$,
$q_{2}=1.185 \times 10^{-5}$ (green)
migrating in a protoplanetary disk.
}
\label{fig:inmass2}
\end{figure}

\begin{figure*}[htb!]
\centerline{
\vbox{
\hbox{
\includegraphics[width=0.5\columnwidth]{pratio_fix_equal.pdf}
\includegraphics[width=0.5\columnwidth]{semi_fix_equal.pdf}
\includegraphics[width=0.5\columnwidth]{ecc_fix_equal.pdf}
\includegraphics[width=0.5\columnwidth]{rangle_fix_equal.pdf}
}
\hbox{
\includegraphics[width=0.5\columnwidth]{pratio_fix_scaling.pdf}
\includegraphics[width=0.5\columnwidth]{semi_fix_scaling.pdf}
\includegraphics[width=0.5\columnwidth]{ecc_fix_scaling.pdf}
\includegraphics[width=0.5\columnwidth]{rangle_fix_scaling.pdf}
}
\hbox{
\includegraphics[width=0.5\columnwidth]{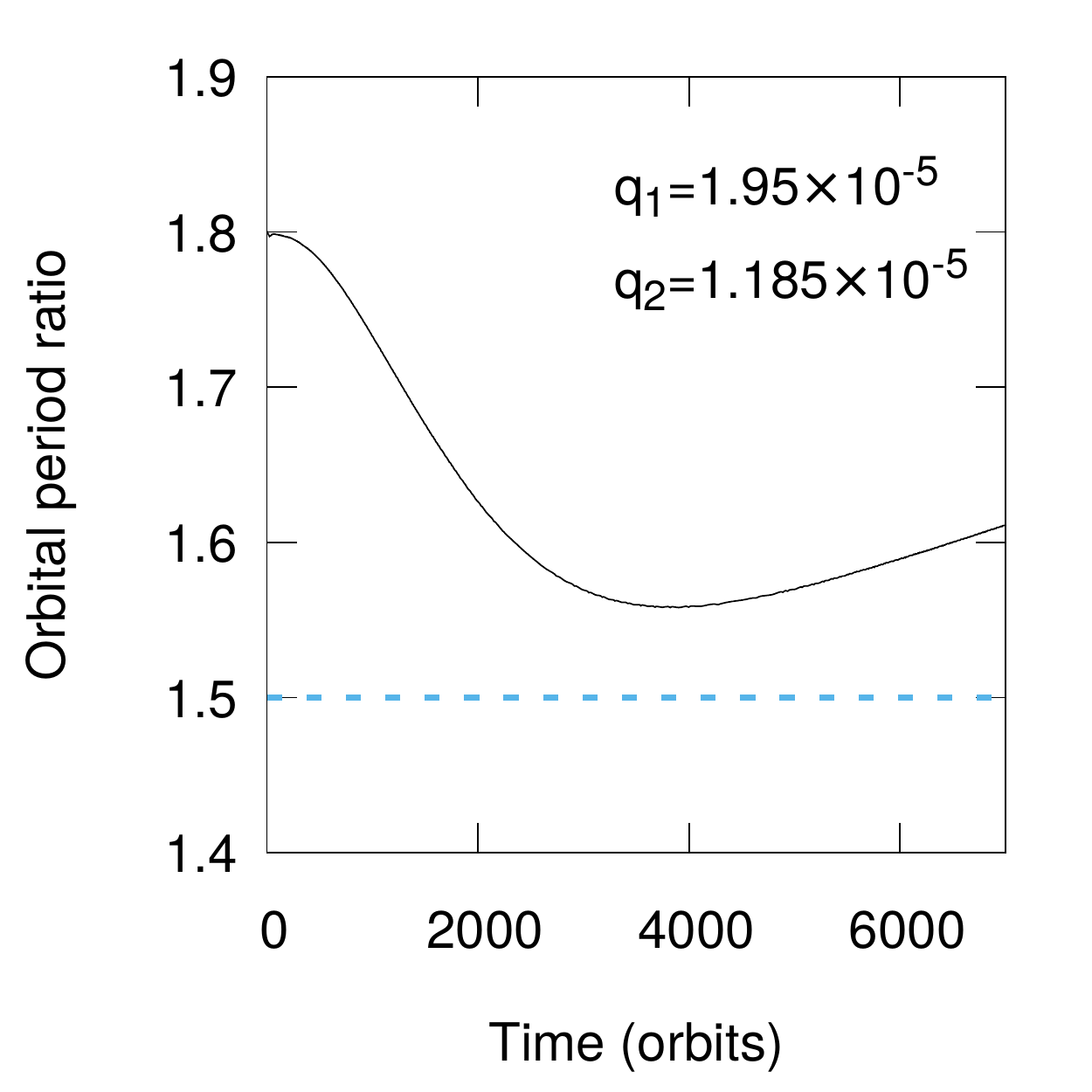}
\includegraphics[width=0.5\columnwidth]{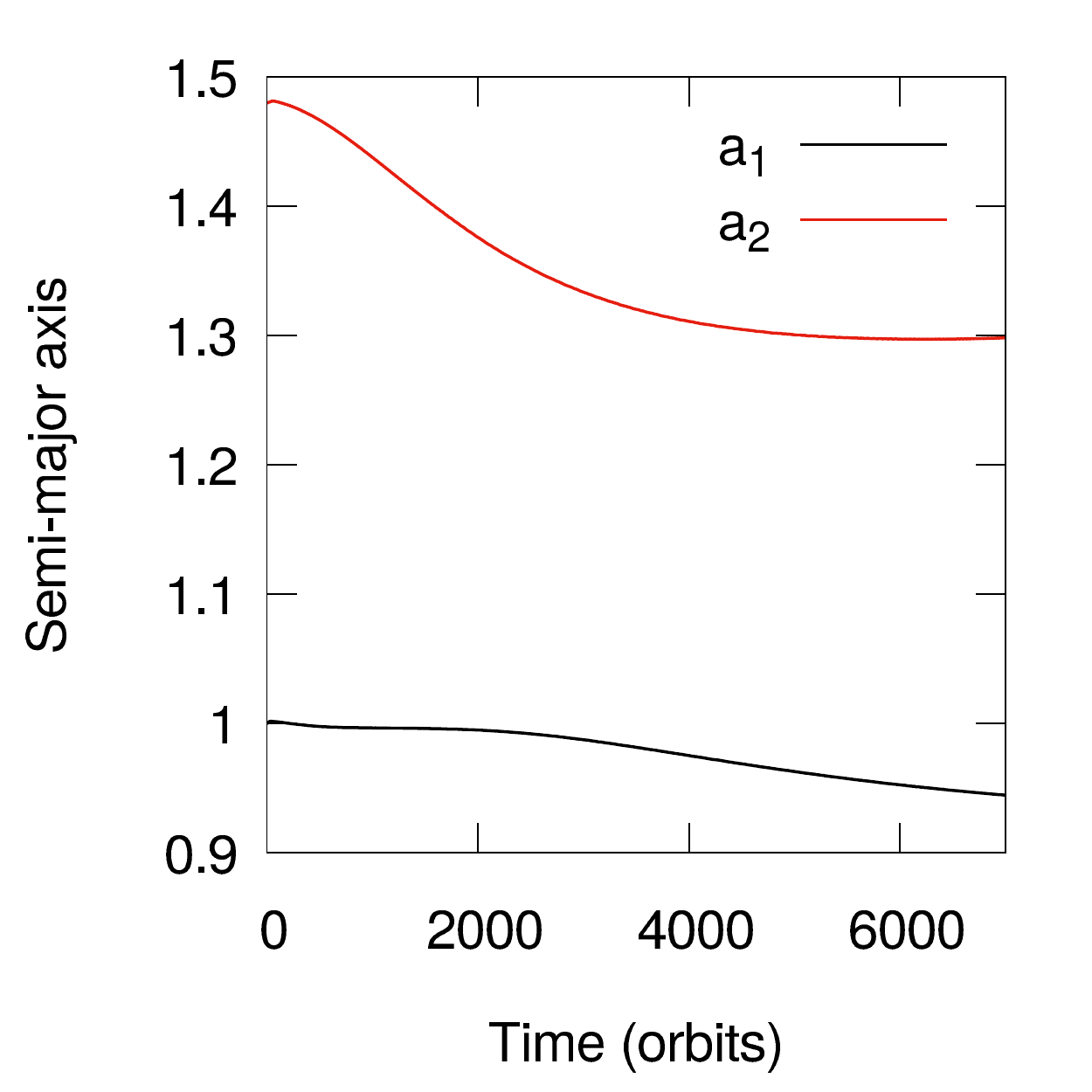}
\includegraphics[width=0.5\columnwidth]{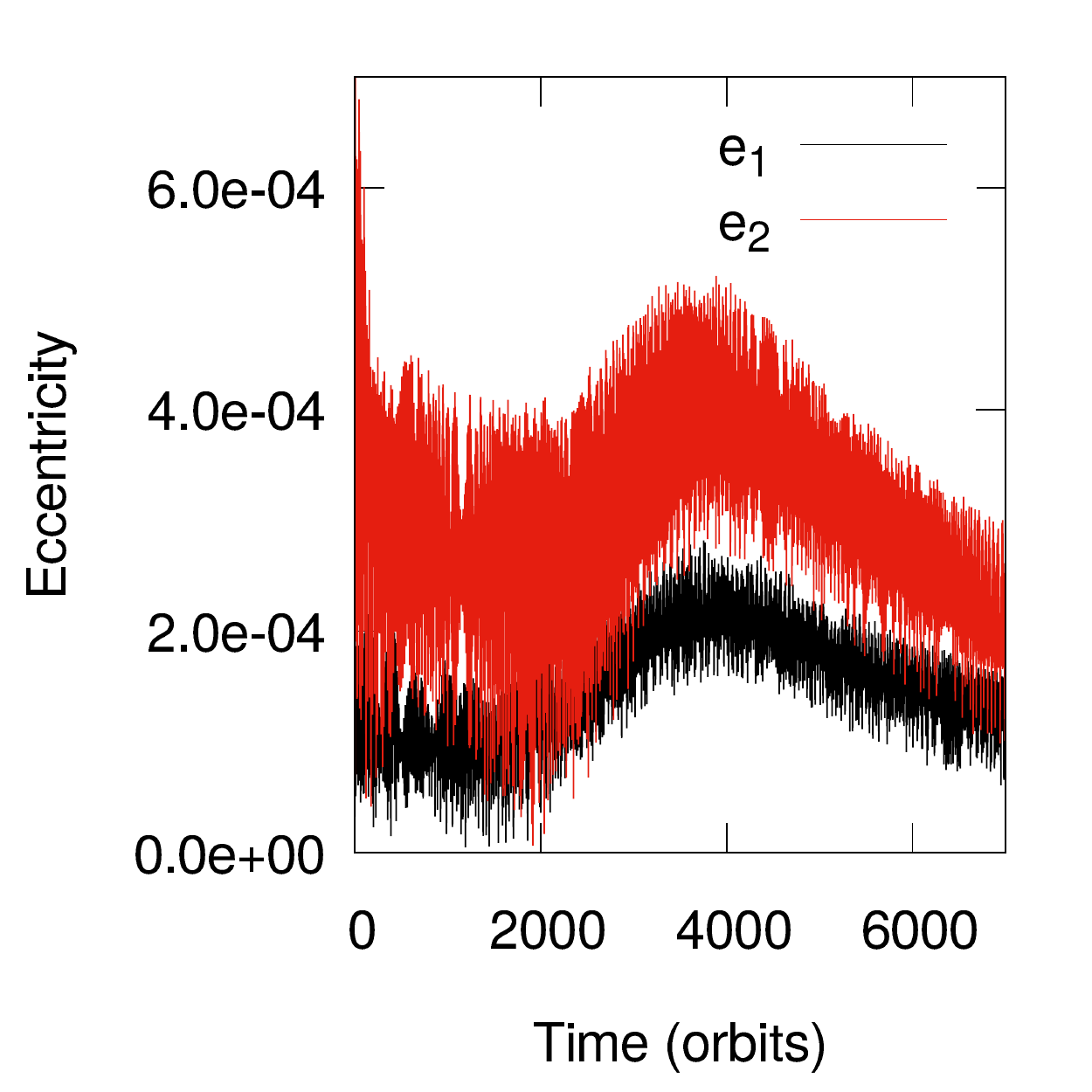}
\includegraphics[width=0.5\columnwidth]{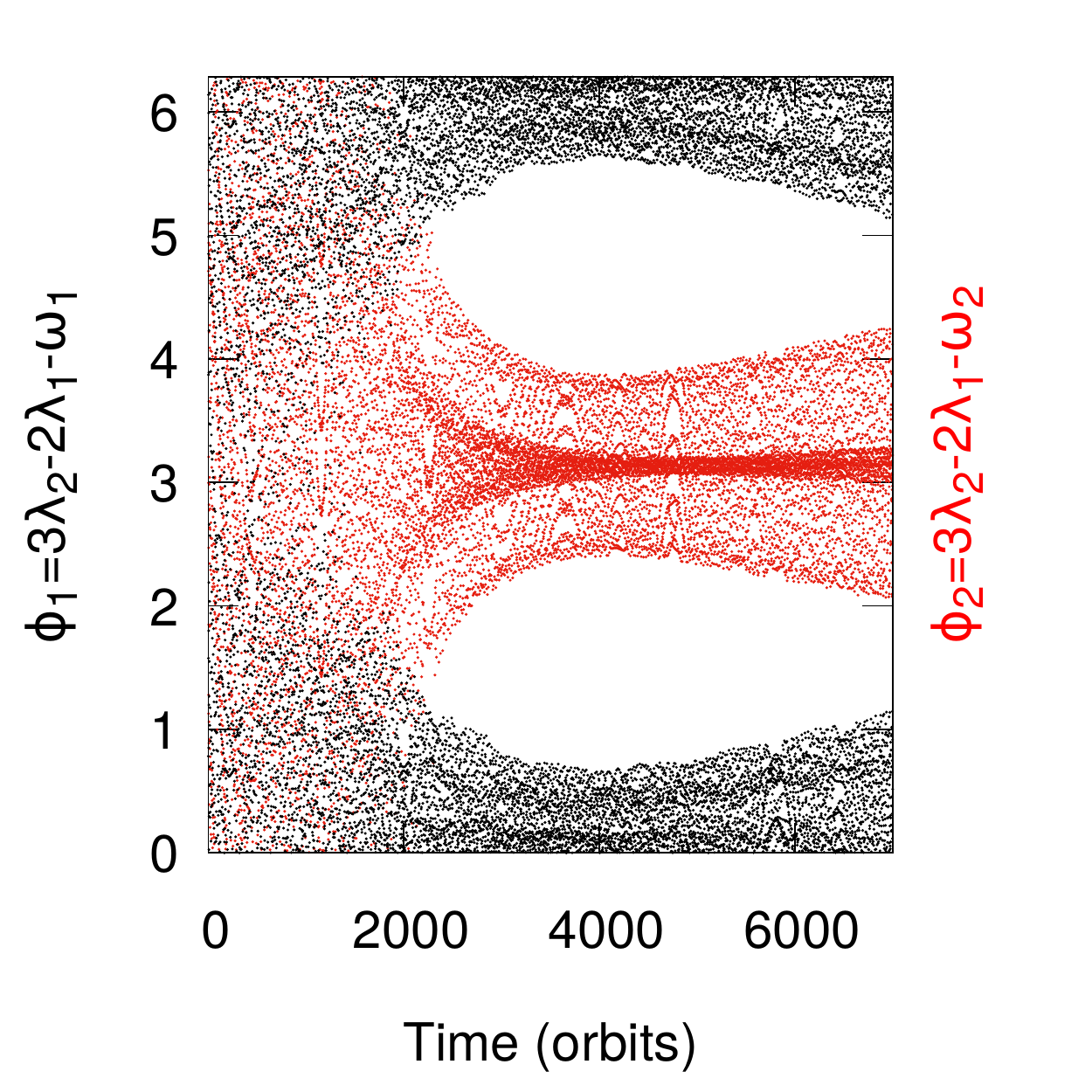}
}
\hbox{
\includegraphics[width=0.5\columnwidth]{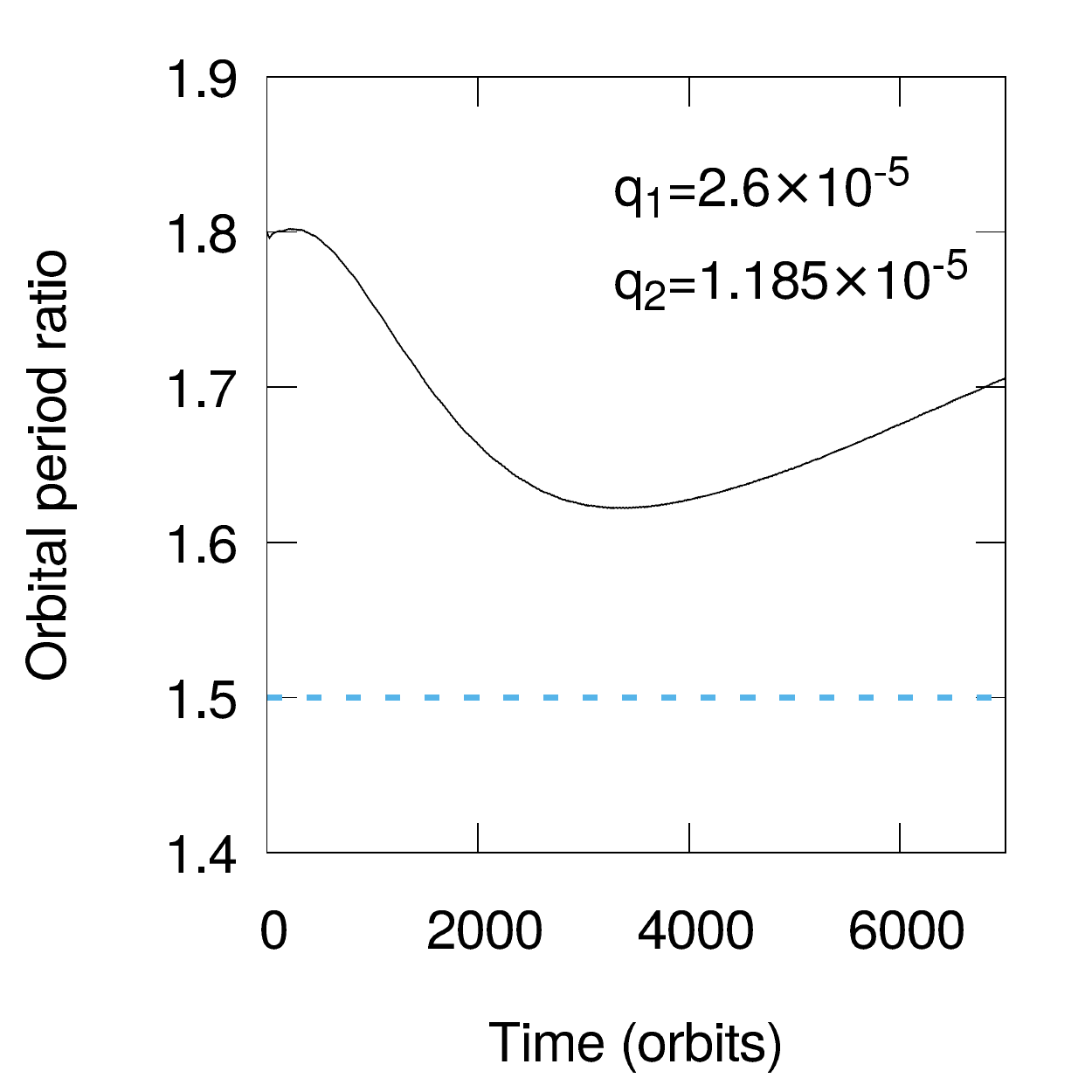}
\includegraphics[width=0.5\columnwidth]{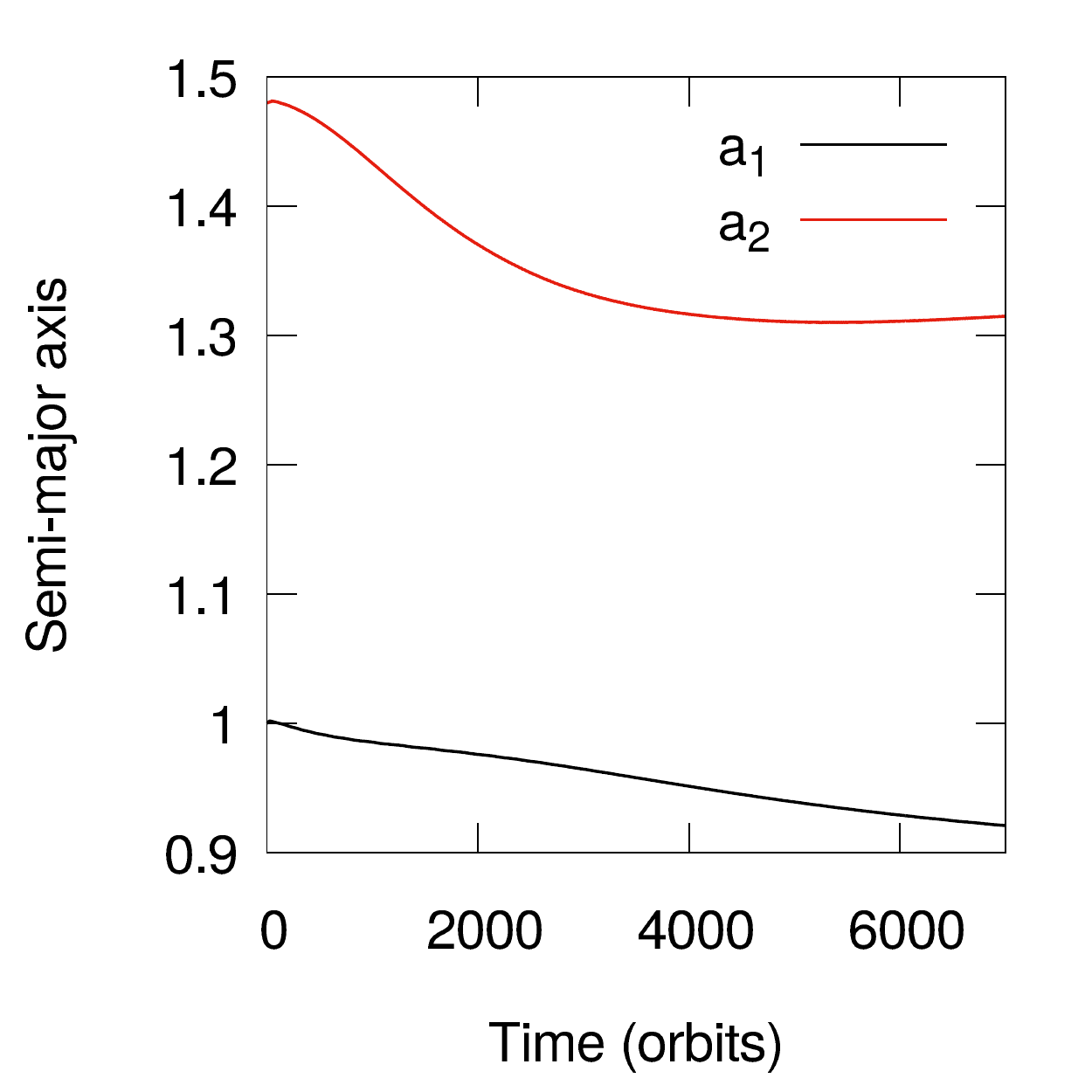}
\includegraphics[width=0.5\columnwidth]{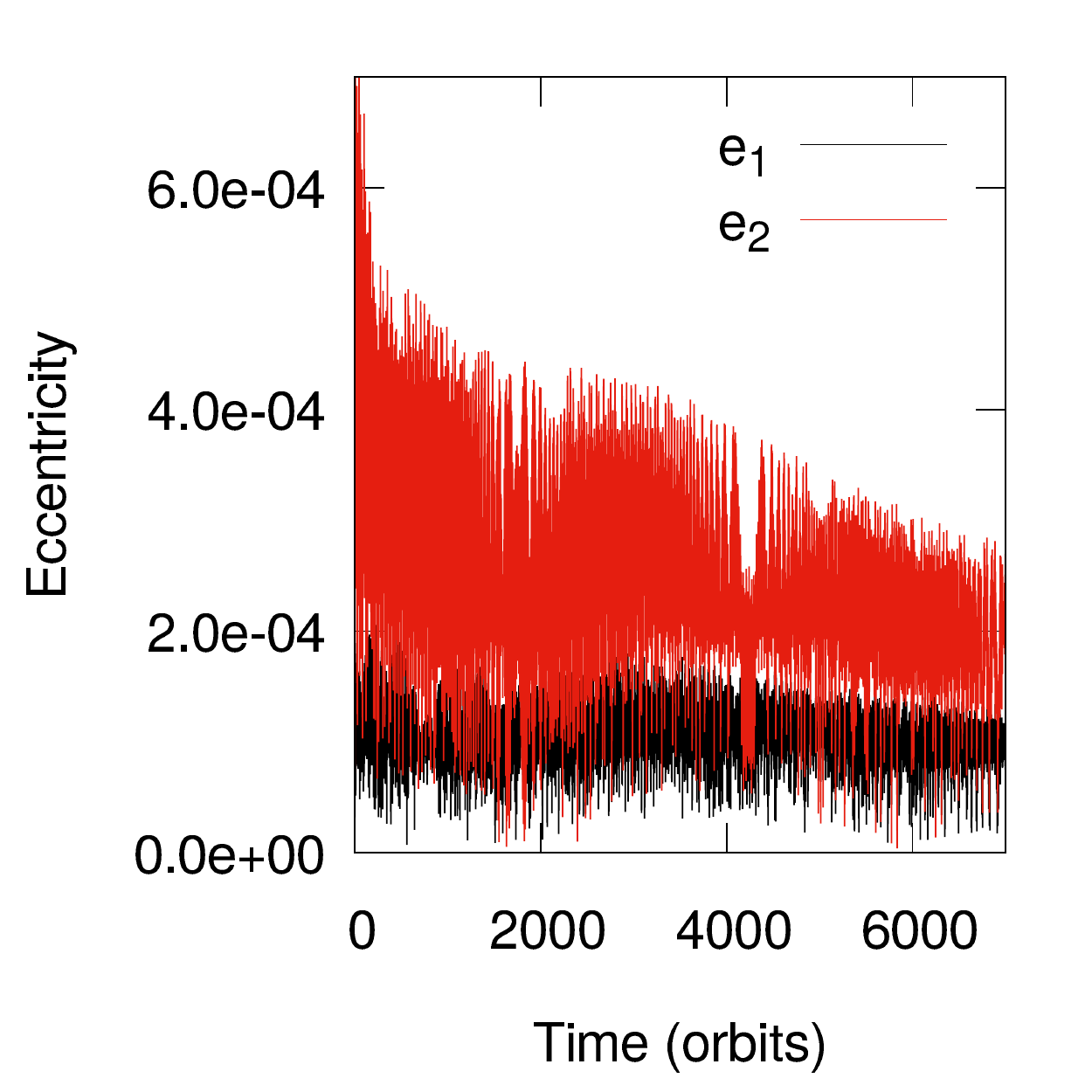}
\includegraphics[width=0.5\columnwidth]{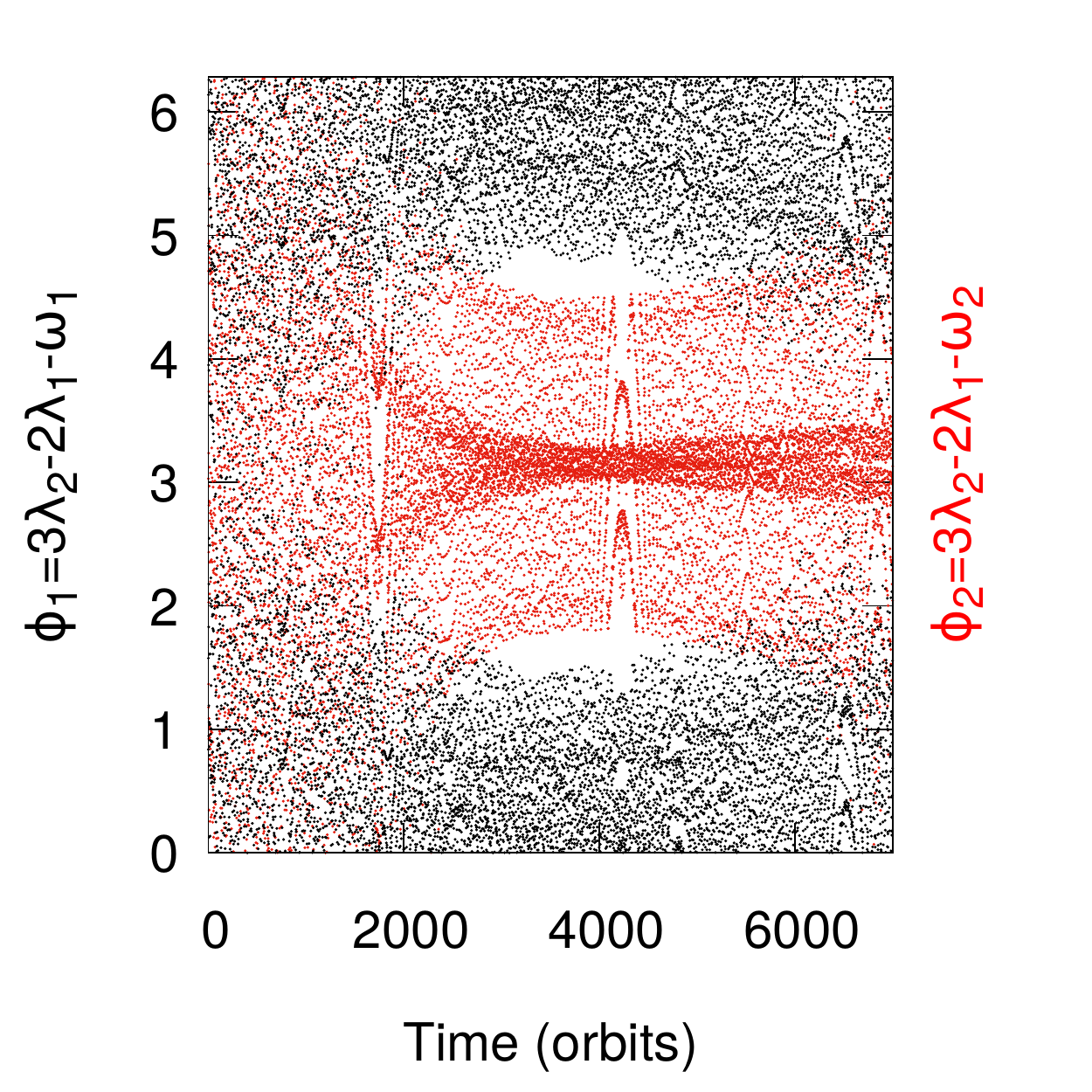}
}
}
}
\caption{Each row shows the results of the simulations done with 
the same disk parameters and the mass ratio of the outer planet 
$q_2=1.185 \times 10^{-5}$, but different mass ratios of the 
inner planet. In the first row the mass ratio of the inner planet 
is $q_{1}=1.185 \times 10^{-5}$, in the second 
$q_1=1.3 \times 10^{-5}$, in the third $q_1=1.95\times 10^{-5}$ 
and in the fourth $q_1=2.6 \times 10^{-5}$.
The panels in each row show the evolution of the planets' orbital 
period ratio, semi-major axes, eccentricities and the resonance 
angles from left to right for each simulation.
The horizontal dashed blue line in the leftmost panel in each
row indicates the position of the 3:2 commensurability.
}
\label{fig:fix_inner}
\end{figure*}

From Figure~\ref{fig:inmass2} we see that the initial relative 
migration rates in all four simulations are very similar but 
with a tendency to decrease as $q_1$ increases.
This  occurs because the surface density  decreases sharply in 
the inner region of the disk, which tends to halt inward migration 
exactly  where the inner planet is placed 
(see Equation~(\ref{disk})).

For the case where the inner planet has the lowest mass (orange 
line) the pair of super-Earths undergoes convergent migration 
and arrives at the position of the 3:2 resonance. However, stable 
resonance capture did not actually take place. This can be seen  
from the behaviour of the eccentricities of the planetary orbits 
displayed in the first row of Figure~\ref{fig:fix_inner}.  
When the resonance is first encountered the eccentricities are
excited  for a brief period before  starting  to decrease as the 
planets begin to undergo  divergent migration.

Increasing the mass of the inner planet results in slower initial 
convergent migration. This leads to divergent migration sooner at 
a higher period ratio. In fact, the planets do not have a close  
approach to  the 3:2 resonance.
The migration of the outer planet in all cases except the equal 
mass simulation with $q_{1}=q_{2}=1.185\times 10^{-5}$, at some 
point reverses  direction bringing the planet further away from 
its host star. It is likely that this would also  happen in the 
simulation with two equal planets if we had continued the 
simulation for a longer time.

We have  confirmation of the prediction that we should observe
divergent migration in all  configurations of the super-Earth 
pairs discussed here. It is of interest to perform a quantitative
analysis of the divergent migration obtained in our simulations
and its dependence  on the mass of the inner planet.
One of the characteristic properties of the  evolution is the
value of the semi-major axis ratio of two planets at the moment
of time at which the transition from  convergent to divergent 
migration takes place.
We denote this by $a_{\rm min}=(a_{2}/a_{1})_{\rm min}$.
Another relevant quantity is the relative migration rate during  
divergent migration
\begin{equation}
\dot{a}_{\rm div}= \left(\frac{d(a_2/a_1)}{dt}\right)_{\rm div}.
\end{equation}
The value of $\dot{a}_{\rm div}$ has been determined  as an 
average over the last 1500 orbits of  evolution. This and 
$a_{\rm min}$ are given in Table~\ref{tab:compare-xm1} for each 
simulation.
From this table  we can see that when $q_{1}$ is larger, 
$a_{\rm min}$ and $\dot{a}_{\rm div}$ are larger, which means
that a higher mass inner planet leads to a faster divergent 
migration rate even though the earlier convergent migration 
rate was slower. This is a natural expectation on account of 
the expected larger wave flux produced by a planet with larger 
$q_1.$
\begin{deluxetable}{ccc}
\tablecaption{The relative migration rate of two planets in simulations
with different $q_{1}$. \label{tab:compare-xm1}}
\tablecolumns{3}
\tablewidth{0pt}
\tablehead{
\colhead{$q_{1}$} & \colhead{$a_{\rm min}$} & \colhead{$\dot{a}_{\rm div}$}  \\
\colhead{($M_{\star}$)} & \colhead{} & \colhead{($ \rm yr^{-1}$)}
}
\startdata
$1.185 \times 10^{-5}$ & 1.310  & $3.954 \times 10^{-6}$  \\
\hline
$1.3 \times 10^{-5}$  & 1.319  & $9.263 \times 10^{-6}$  \\
\hline
$1.95 \times 10^{-5}$ & 1.344  & $1.235 \times 10^{-5}$  \\
\hline
$2.6 \times 10^{-5}$ & 1.381  & $1.627 \times 10^{-5}$   \\
\hline
\enddata
\end{deluxetable}

\subsection{Reducing the rate of convergent migration by
decreasing the surface density}
\label{Sigmascale}
We here explore the effect of reducing the initial convergent 
migration rate by reducing the surface density scale. To do 
this we consider an additional simulation for which the planets 
have mass ratios $q_1=1.185\times 10^{-5}$ and 
$q_2=1.3\times 10^{-5}$ but with the lower surface density 
scale with  $\Sigma_0 =4.5 \times 10^{-5}.$
Results for a simulation with the same mass ratios and initial 
surface density profile, but scaled with  
$\Sigma_0 =6 \times 10^{-5}$ was already presented in 
Figure~\ref{fig:inmass}. It will be seen from that figure that 
in that case the migration was fast enough for the planets
to be able to pass through the 3:2 resonance.

Our main motivation in performing the simulation with reduced 
surface density scale as we have already mentioned above, was 
to investigate the situation in which the outer planet is more 
massive than the inner one, and the relative  migration rate 
is not too fast for  3:2 resonance capture to take place.
We choose $\Sigma_0$ in such a way that the convergent 
migration rate of the planets is approximately the same as for 
the evolution of the planets with $q_1=1.3\times 10^{-5}$ and 
$q_2=1.185\times 10^{-5}$ illustrated in Figure~\ref{fig:inmass}.

In Figure~\ref{fig:fix_inmass2} the evolution of the period 
ratio is shown for the simulation with reduced surface density 
scale (red line) together with that for the simulation with  
$q_1=1.185\times 10^{-5}$ and $q_2=1.3\times 10^{-5}$ and 
$\Sigma_0=6\times 10^{-5}$ illustrated in 
Figure~\ref{fig:inmass} (dark blue line) and also the case 
with $q_1=1.3\times 10^{-5}$ and $q_2=1.185\times 10^{-5}$ 
and $\Sigma_0=6\times 10^{-5}$ (black line) can be seen in 
Figure~\ref{fig:fix_inmass2}. More details for the lower 
surface density case are shown in 
Figure~\ref{fig:fix_reverse}.

\begin{figure}[htb!]
\centerline{
\vbox{
\hbox{
\includegraphics[width=0.9\columnwidth]{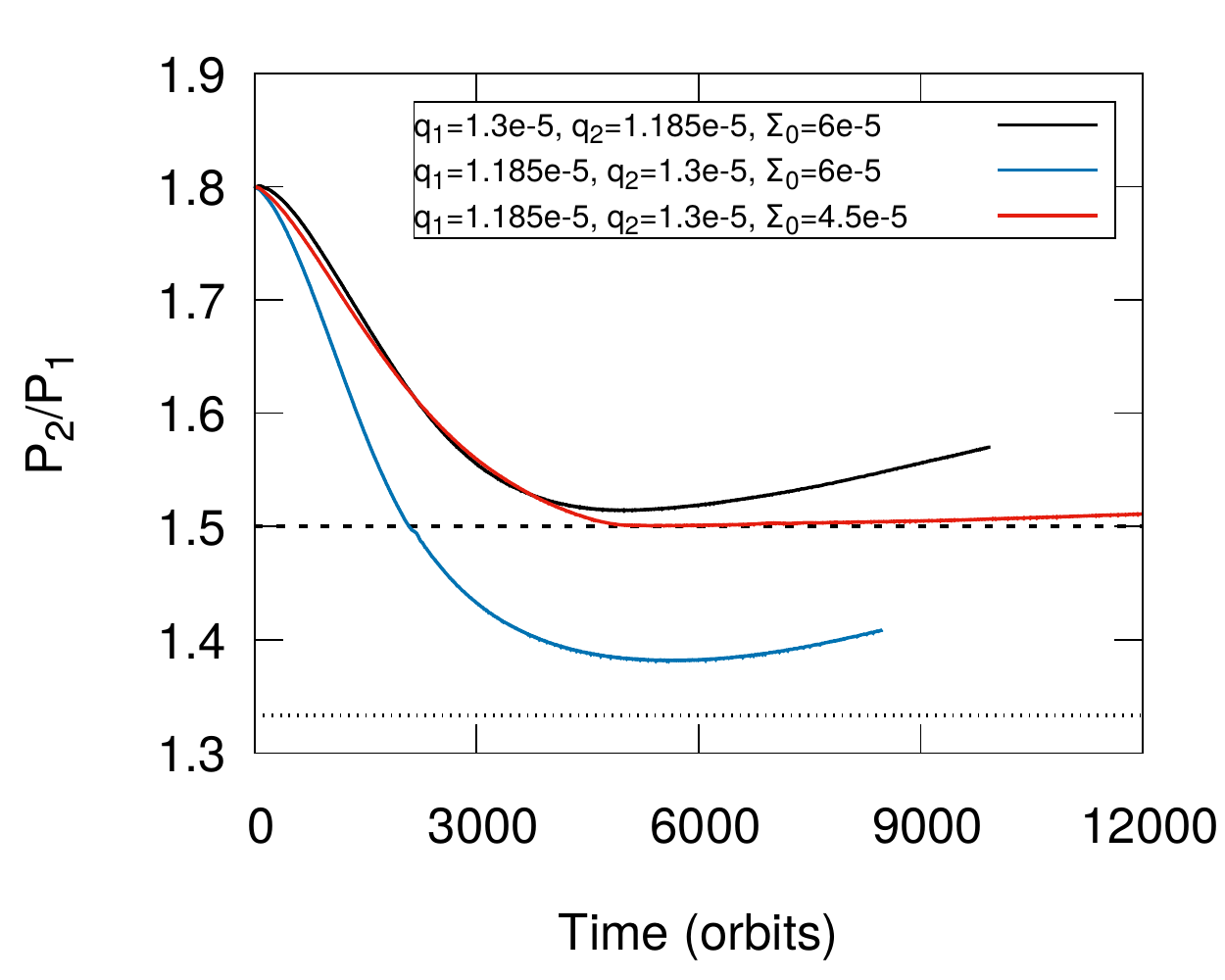}
}
}
}
\caption{The results of the hydrodynamical simulations of two 
super-Earths with mass ratios of $q_1= 1.185\times
10^{-5}$, $q_2= 1.3 \times 10^{-5}$ (dark blue) and 
$q_1=1.3 \times 10^{-5}$, $q_{2}=1.185 \times 10^{-5}$ (black)
migrating in a protoplanetary disk with 
$\Sigma_0 = 6\times 10^{-5}$ together with the results for two 
super-Earths with mass ratios of $q_1= 1.185\times
10^{-5}$, $q_2= 1.3 \times 10^{-5}$ migrating in the disk with
$\Sigma_0 = 4.5\times 10^{-5}$ (red).
}
\label{fig:fix_inmass2}
\end{figure}
\begin{figure*}[htb!]
\centerline{
\vbox{
\hbox{
\includegraphics[width=0.5\columnwidth]{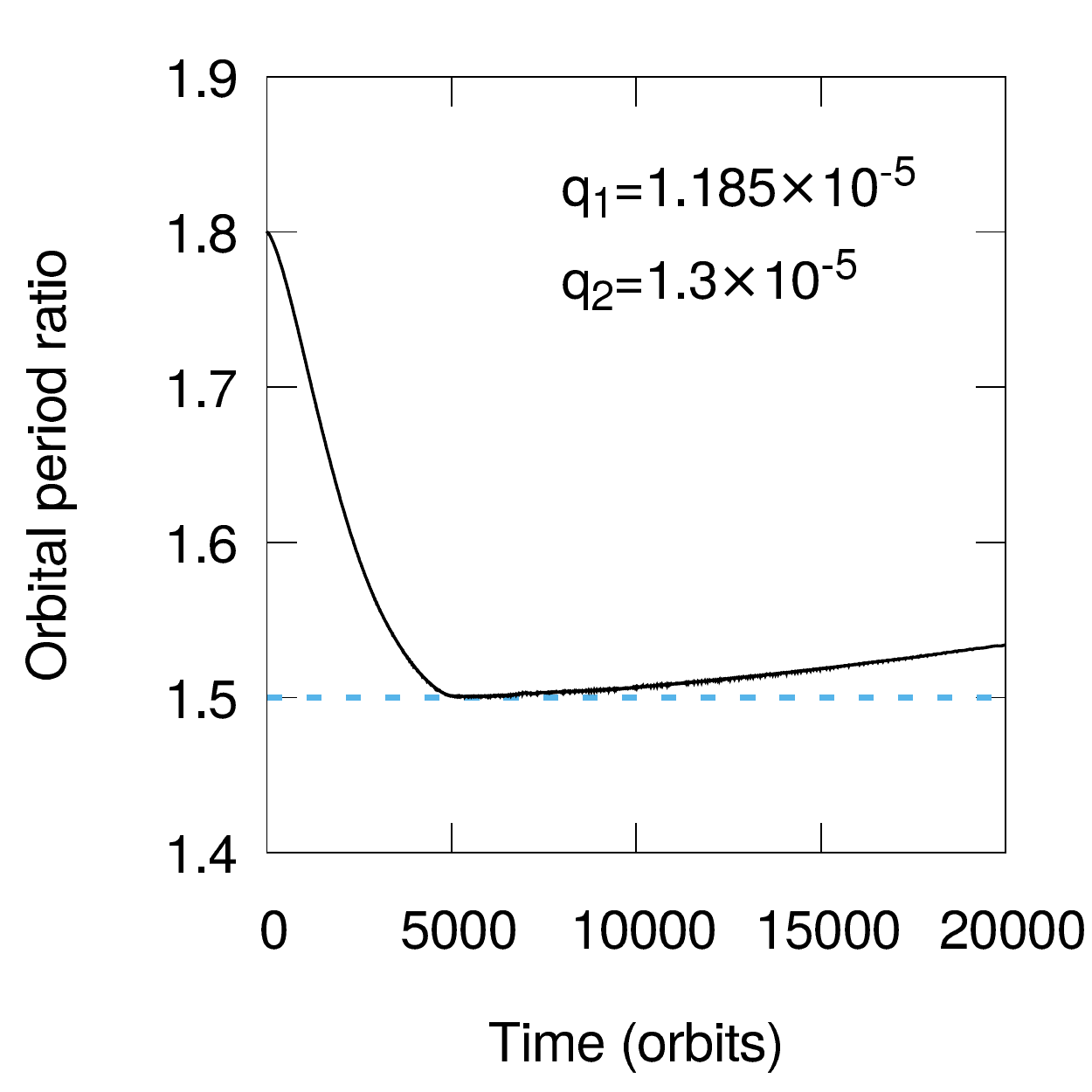}
\includegraphics[width=0.5\columnwidth]{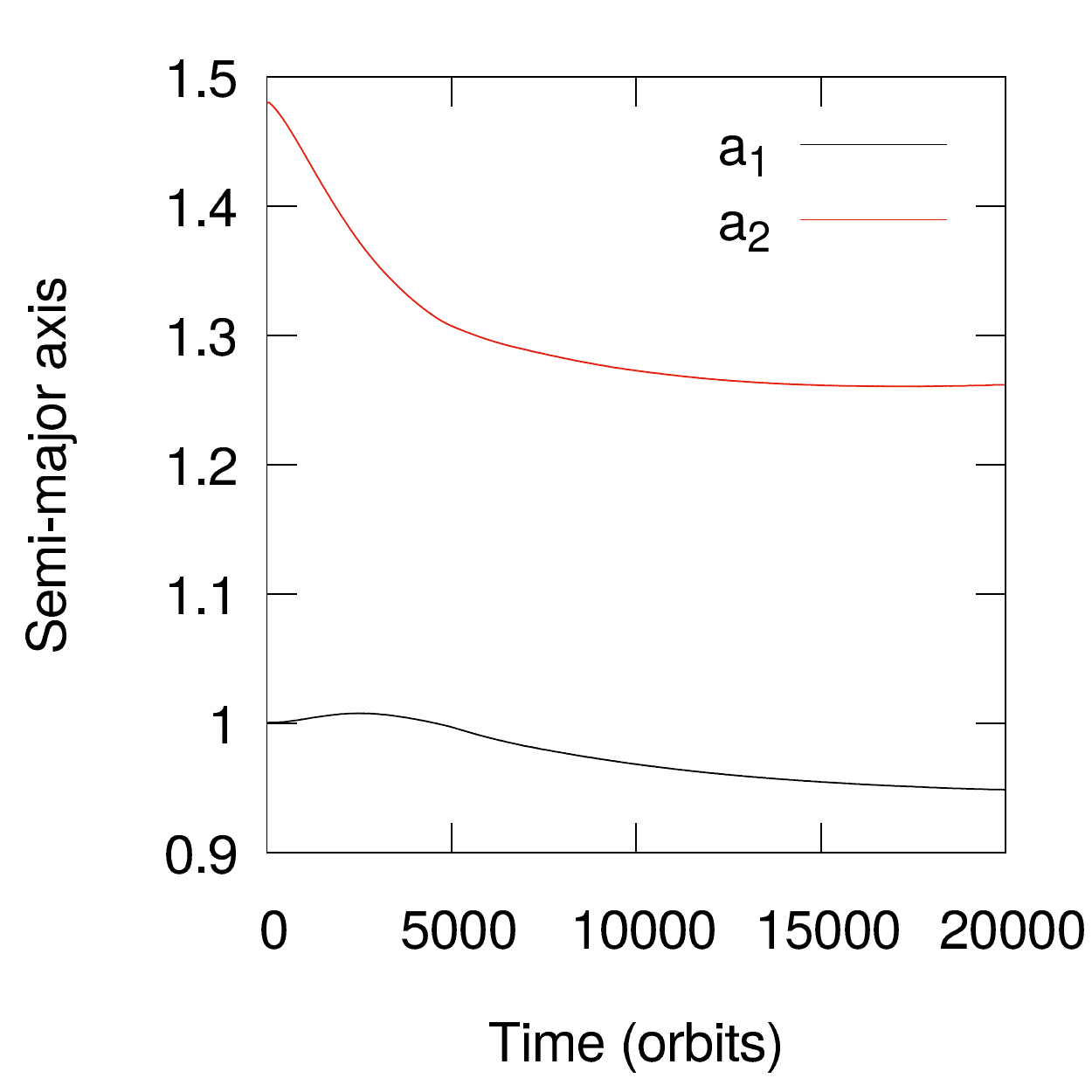}
\includegraphics[width=0.5\columnwidth]{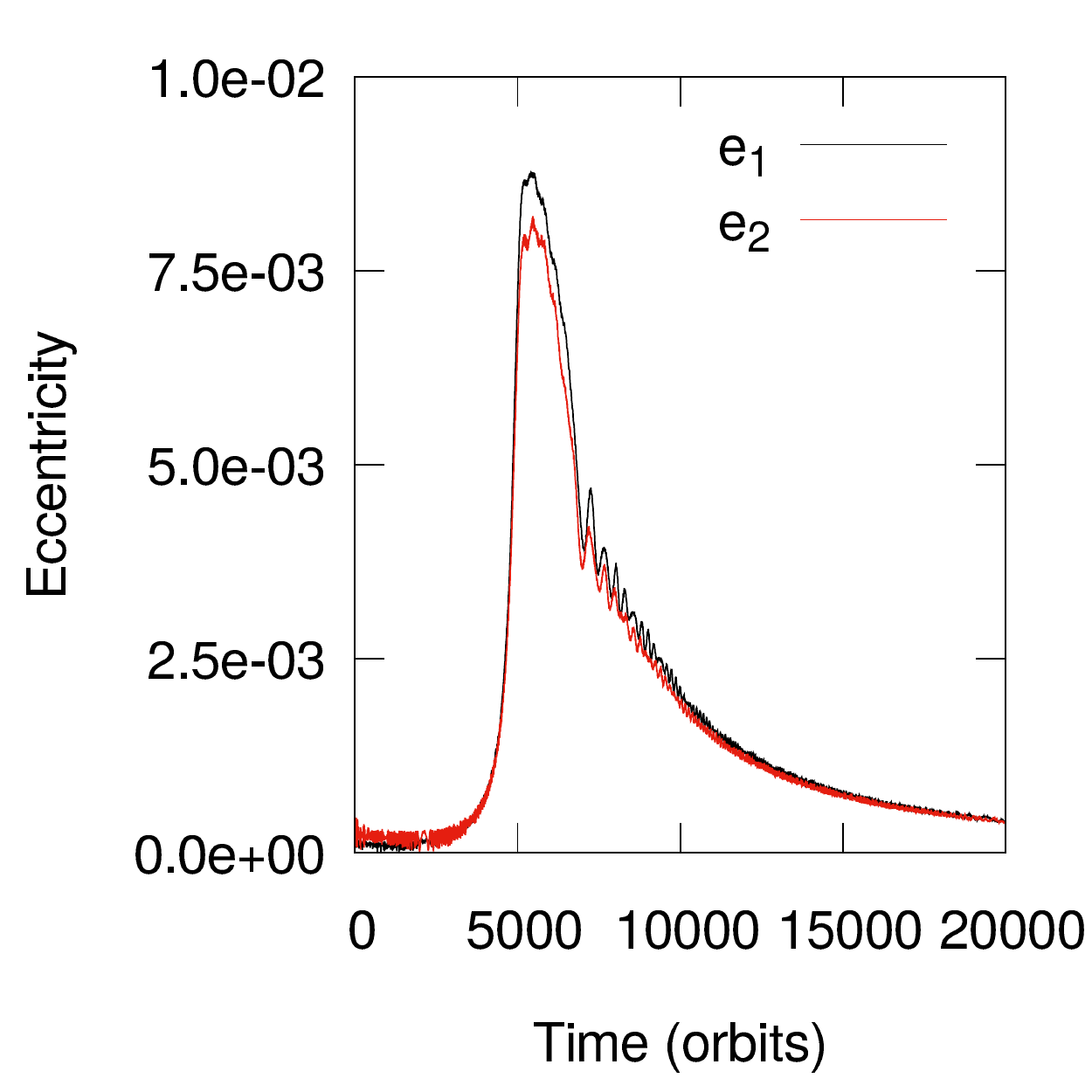}
\includegraphics[width=0.5\columnwidth]{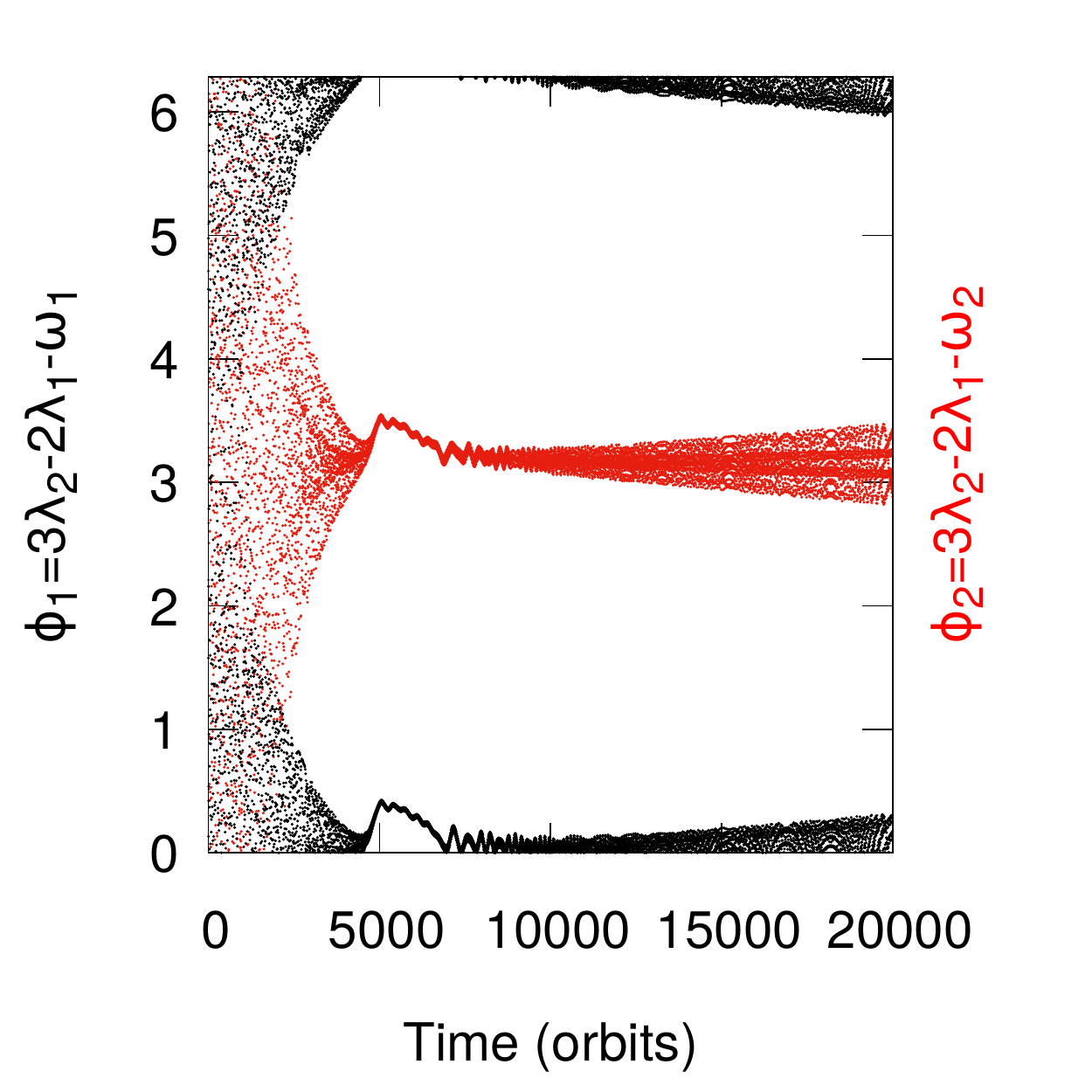}
}
}
}
\caption{Results of the hydrodynamical simulation of two 
super-Earths with masses of $q_{1}=1.185 \times 10^{-5}$ and 
$q_{2}=1.3 \times 10^{-5}$ migrating in a protoplanetary disk. 
The initial surface density scaling with 
$\Sigma_{0} = 4.5 \times 10^{-5}$ is adopted in this simulation. 
Evolution of the planets' orbital period ratio, semi-major axes, 
eccentricities and the resonant angles are shown in the panels 
from left to right. The horizontal dashed blue line in the 
leftmost panel indicates the position of the 3:2 commensurability.
}
\label{fig:fix_reverse}
\end{figure*}

As can be seen from Figure~\ref{fig:fix_reverse} at the 
beginning of this simulation the planets undergo convergent 
migration, at $t \sim 5000$ orbits, the orbital period ratio
has decreased to close to  1.5 while the eccentricities of 
the planets attain maximum values
$e_{1} = 8.0 \times 10^{-3}$ and $e_{2}=9.0 \times 10^{-3}$.
The 3:2 MMR resonance angles start to librate around values 
respectively slightly exceeding $0$ and $\pi$ with small 
amplitude. The period ratio stays close to 1.5 till 
$t \sim 7000$ orbits. At this time the eccentricities have 
decreased to $e_{1}=e_{2} \sim 4.0 \times 10^{-3}$ and 
the centres of libration of the 3:2 MMR resonance angles 
shift to $0$ and $\pi.$

At $t \sim 7000$ orbits, the migration rates of both planets
are decreasing, while the orbital period ratio is slowly  
increasing. The  eccentricities decrease continuously at a 
progressively slower rate as $t$ increases. In addition
the libration amplitudes of the resonance angles increases.
From $t \sim \textcolor{red}{16000}$ orbits until the end of 
the simulation the outer planet migrates outwards.

In summary we find  that in this case the planets enter 
into the 3:2 MMR, stay there for a couple of thousand of 
orbits and then slowly (slower than in the configuration 
with
$q_1=1.3 \times 10^{-5}$, and $q_{2}=1.185 \times 10^{-5}$
in a protoplanetary disk with $\Sigma_0 = 6\times 10^{-5}$),
moves away from the resonance.
This is consistent with the notion that while the system 
remains close to the 3:2 commensurability the convergent 
migration is halted by the resonant interaction with the 
wave planet interaction not being strong enough to separate 
the planets. However, this changes as the migration rates 
slow down at later times and separation from the resonance 
can take place.


\section{The mechanism at work and the importance of the horseshoe drag}
\label{sec:horseshoe}

In this Section we take closer look at the mechanism responsible for 
the repulsion between planets found in the hydrodynamical calculations.
We return to one of our flagship cases, presented at the beginning of 
the present paper, namely two super-Earths with $q_1=1.3\times 10^{-5}$ 
and $q_2=1.185 \times 10^{-5}$ evolving in the disk with $h=0.02$, 
$\nu = 1.2 \times 10^{-6},$  the initial surface density profile
determined by Equation~(\ref{disk}) with $\Sigma_0= 6 \times 10^{-5}$ 
and $\alpha = 0.5.$

\begin{figure}[htb!]
\vspace{0.8cm}
\centerline{
\vspace{-0.5cm}
\vbox{
\hbox{
\includegraphics[width=1.0\columnwidth]{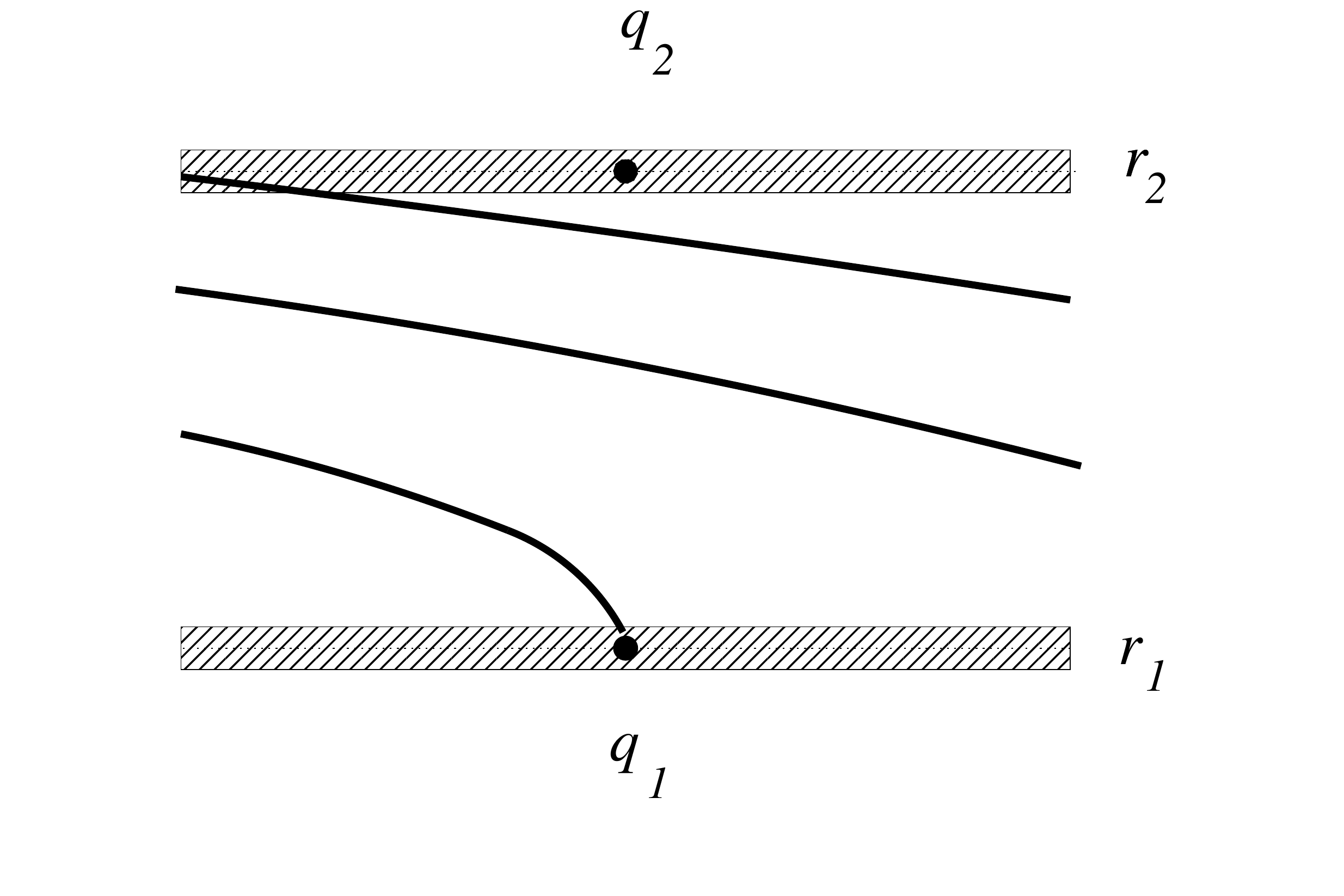}
}
}
}
\caption{The schematic view of the scenario considered here: Two 
planets (black circles) with the planet-to-star mass ratios $q_1$ 
and $q_2$ embedded in the disk at the distances $r_1$ ad $r_2$ 
from the central star.
The hatched regions along the orbits depict the horseshoe regions. 
The wave crests are associated with the density waves launched by 
the inner planet and are propagating outwards towards the horseshoe 
region of the outer planet where dissipation may occur.
}
\label{fig:repulsion}
\end{figure}

\subsection{The angular momentum flux associated with the 
planetary wakes}
We have postulated that  the repulsion between planets occurs 
due to the wave planet interaction. Outward propagating density 
waves excited by the inner planet dissipate in the coorbital 
region of the outer planet. The angular momentum carried by 
the waves is transfered to the horseshoe region and then to 
the planet through horseshoe drag (see Figure~\ref{fig:repulsion}).

To demonstrate this, first, we calculate the angular momentum 
flux carried by density waves.
The wave angular momentum flux as a function of $r$ has the form
\begin{equation}
F_{\rm wave} = r^{2} \bar \Sigma \int_{0}^{2\pi} (v_{\phi}
- \bar v_{\phi}) (v_{r} - \bar v_{r}) d\phi
\label{eq:amf-wave}
\end{equation}
where $\bar \Sigma$ is the azimuthally averaged surface density 
defined as
$\bar \Sigma = \frac{1}{2\pi} \int_{0}^{2\pi} \Sigma d \phi$ 
while $\bar v_{\phi}$ and $\bar v_{r}$ are the azimuthally 
averaged velocity in the azimuthal and radial direction, 
respectively.
This is correct to second order in perturbations around the 
background state and should be adequate for the small amplitude 
waves considered.  It is also important to note that this 
quantity, which we denote as a flux, represents the flow of 
angular momentum across a circle of radius, $r.$  
The results for the flagship case are  presented in the right 
panel of Figure~\ref{fig:amf-compare} (black line).

\begin{figure*}[htbp]
\centerline{
\vbox{
\hbox{
\includegraphics[width=0.9\columnwidth]{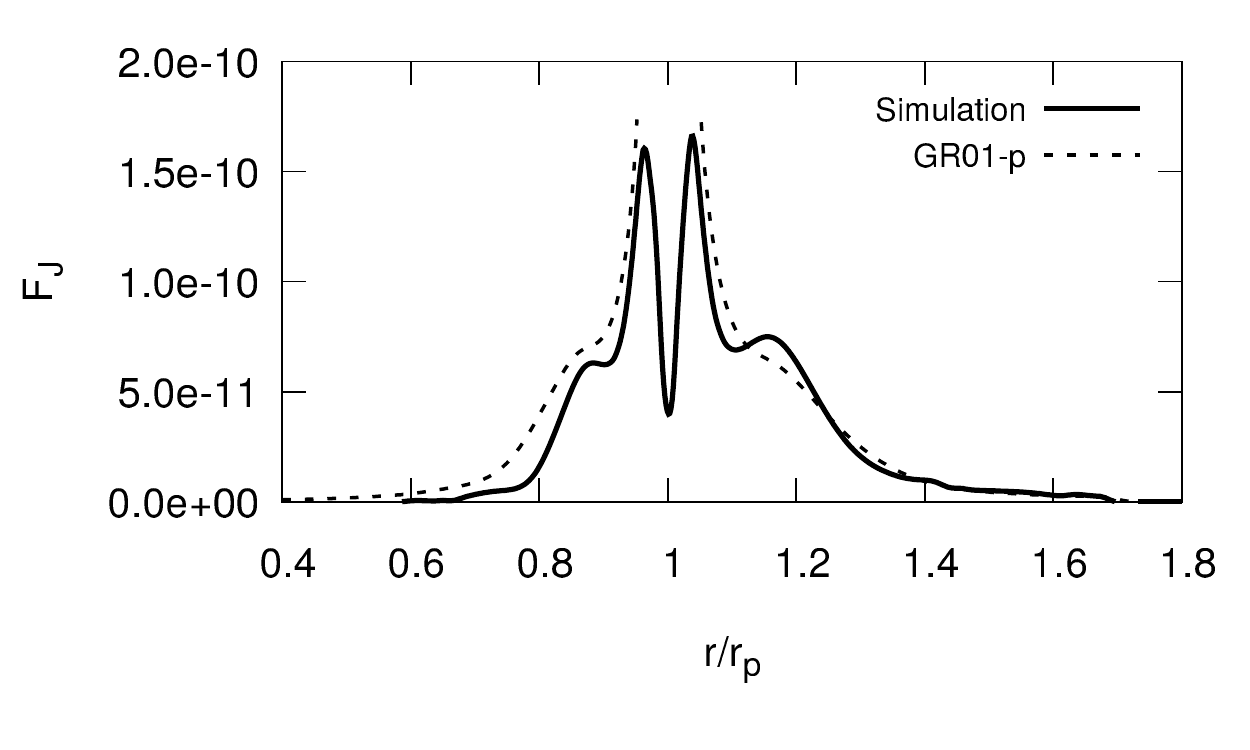}
\includegraphics[width=0.9\columnwidth]{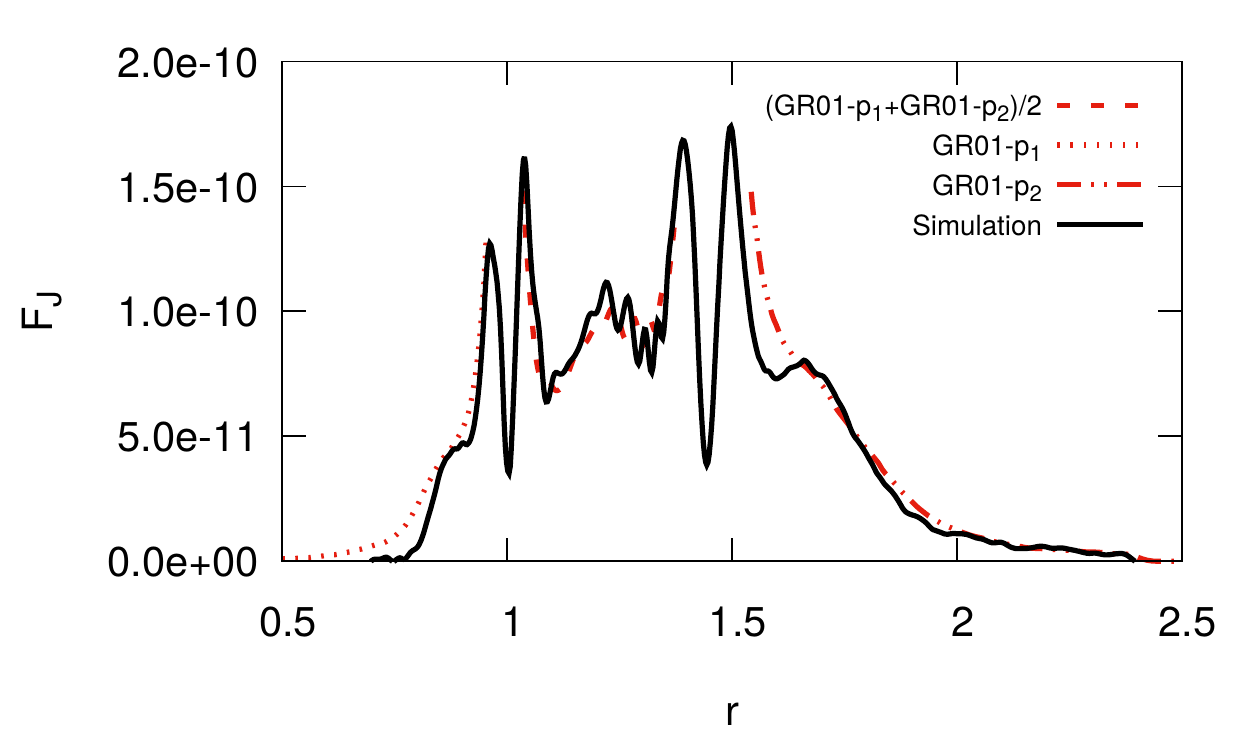}
}
}
}
\caption{A comparison of the angular momentum flux calculated by
Equation~(\ref{eq:amf-wave}) (full line), and Equation~(\ref{eq:amf-gr01}) 
(dashed line) in the single planet case at t = 1000 orbits is shown 
in the left panel. A similar comparison is made for the two planet case 
in the (right panel). See the text for a description of the curves.
}
\label{fig:amf-compare}
\end{figure*}

The wave flux $F_{\rm wave}$ shown in this figure has contributions 
from the wakes of two planets in the disk. Ideally, we would like 
to know what is the flux carried by the waves excited only by the 
inner planet, $F_{\rm wave-inner}$.

To estimate $F_{\rm wave-inner}$ we consider first a run with the 
same initial conditions as the flagship case but with the outer 
planet removed, we refer to this as the single planet case, and employ 
the expression given by \cite{Goodman}.
According to them the angular momentum flux carried by the density 
wave constituting the wake of a planet in a non-self-gravitating disk 
is given by
\begin{equation}
F_{\rm J}(r) = \frac{c_{s}^{3} r_{p}^{2}}{| x r d\Omega/dr|
\Sigma_{0}} \int_{-\pi }^{\pi } (\Sigma - \Sigma_{0})^{2} d \phi
\label{eq:amf-gr01}
\end{equation}
where $x$ is the distance to the planet in the radial direction 
defined as $x = r - r_{p}$ and $\Sigma_{0}$ denotes the unperturbed 
surface density. In the left panel of Figure~\ref{fig:amf-compare}, 
we present a comparison of the angular momentum flux calculated by 
Equation~(\ref{eq:amf-wave}) and Equation~(\ref{eq:amf-gr01}) for 
the simulation with a single planet in the disk. The agreement is 
very good. This shows that the wave flux generated by  a single 
planet on its own is well represented by Equation~(\ref{eq:amf-gr01}).

Next, we  calculate the angular momentum fluxes expected to be 
generated by the inner and outer planet separately and then combine 
them in a way that should represent the total angular momentum flux
in a radial interval between the planets when the wakes do not 
interact and then compare this with  Equation~(\ref{eq:amf-wave}).
The expression we use is given by
\begin{equation}
F_{\rm J}(r) =\left(\frac{r_1^2}{|x_1|}  + \frac{r_2^2}{|x_2|} \right)  \frac{c_{s}^{3}}{|2 r d\Omega/dr|
\Sigma_{0}} \int_{-\pi }^{\pi } (\Sigma - \Sigma_{0})^{2} d \phi,
\label{eq:amf-gr012}
\end{equation}
where $x_i = r-r_i.$ We remark that there is a contribution of both 
wakes to the surface density in Equation~(\ref{eq:amf-gr012}) hence the factor 
of two in the denominator. We note that at the intermediate location 
where $r_i^2/x_i$  is the same for $i=1,2,$ given the result of the 
comparison in the single planet case, apart from possible 
contributions from small regions where the wakes cross, we should 
obtain the correct total angular momentum flux. The same is true 
more generally if the surface density perturbations from the wakes 
are nearly equal as is found to be the case near the midpoint of 
the interval between the planets. However, use of Equation~(\ref{eq:amf-gr012}) 
is likely to give an underestimate as gap edges are approached.

The results of the comparison of results obtained from 
Equation~(\ref{eq:amf-wave}) (black curve), 
Equation~(\ref{eq:amf-gr012})
in the region between the planets (dashed red curve) and 
Equation~(\ref{eq:amf-gr01}) in the regions interior to the inner 
planet  (dotted red curve) and exterior to the outer planet 
(double dot dashed red curve) for the two planet case are shown 
in the right panel of Figure~\ref{fig:amf-compare}.

We  note that as expected, in the inner part of the disk, $F_{\rm wave}$
calculated by Equation~(\ref{eq:amf-wave}) and the  values obtained  
by applying the form of Equation~(\ref{eq:amf-gr01}) applicable to the 
inner planet, agree with each other. This is expected because in this 
region, the disk is mainly perturbed by the interior wake of the inner 
planet. Correspondingly, the outer part of the disk is mostly disturbed 
by the exterior wake of the outer planet and $F_{\rm wave}$ also agrees  
nicely with the values obtained from the form of 
Equation~(\ref{eq:amf-gr01}) applicable to the outer planet.

In the central part of the region, between two planets, the wakes of 
both planets perturb the disk with similar strength. Accordingly we 
found that $F_{\rm wave}$ is consistent with the contribution from the 
two planets expressed by Equation~(\ref{eq:amf-gr012}). Based on this
comparison, we confirm that the angular momentum fluxes calculated from
Equations~(\ref{eq:amf-gr01}) and (\ref{eq:amf-gr012}) as applicable,
which are based on the theoretical consideration of the effects 
arising from the planetary wakes, agree with the values of the angular 
momentum flux seen in our simulations.

\subsection{The horseshoe region}
Following the discussion in Section~\ref{Coorbgaps}, the transfer of 
the angular momentum to the horseshoe region was expressed in terms 
of the angular momentum  flow induced by a planet towards  the other 
planet. Focusing on the transfer from the inner planet to the 
horseshoe region  of the outer planet, we recall the expression for 
it given by the left hand side of Equation~(\ref{eq:transtohorseshoe}) 
in the form 
\begin{equation}
\label{eq:transtohorseshoe1}
\Gamma_{12hs}=|\Gamma_{L(1s)}| 2\lambda\sqrt{q_2/h}
\end{equation}
We can estimate this from the single planet run for which  the left 
panel of Figure~\ref{fig:amf-compare} illustrates the  angular momentum 
flux produced by the inner planet when isolated.  From this figure, we 
can see that the total angular momentum flux produced is 
$\sim 1.7\times 10^{-10}$ in code units and accordingly we estimate  
$\Gamma_{12hs}\sim 0.83\lambda \times 10^{-11}.$
The location of a  3:2 resonance with such a  outer planet is at 
$r=1.31r_1.$ From Figure~\ref{fig:amf-compare} we see that the angular 
momentum flux at this location is $\sim 2\times 10^{-11}.$ Referring 
back to Figure~\ref{fig:two-torque} which indicates the torque deficit
between the actual torque and the expected type~I torque acting on the 
outer planet in the two planet system, we see that this is also  
$\sim 2\times 10^{-11}.$
Thus we see there is consistency  with the picture presented here
of this deficit being supplied by the inner planet if the emitted 
flux that reaches the horseshoe region is mostly absorbed there and 
$\lambda \sim 2.4,$ being of order unity.


\section{The effect of the planets on the evolution of the surface 
density and the slowing of the inward migration of the outer planet}
\label{sec:surfacedensity}
In order to evaluate the effect of the planets in structuring the 
surface density profile of the disk, we illustrate the evolution of 
the  surface density profile, $\Sigma(r,t),$ by plotting a sequence 
of  surface density distributions obtained at different moments in 
time in our simulation with $q_1=1.3\times 10^{-5}$, 
$q_2=1.185\times 10^{-5}$ in the left panel of Figure~\ref{fig:disk_evo}.
The initial surface density profile was determined by 
Equation~(\ref{disk}) with $\alpha=0.5$ and $\Sigma_0=6\times 10^{-5}.$

For  comparison, we report the evolution of the surface density 
$\Sigma(r,t)$  without the planets at the same moments in time in the 
right panel of Figure~\ref{fig:disk_evo}. Considering the evolution 
of the disk without planets, which is driven by the disk viscosity, we  
see that the maximum value of the surface density moves outwards, 
the trajectory being indicated by the dashed-dotted line in the figure.
This has the consequence that the disk surface density profile becomes 
flatter as the evolution  time increases.

In the disk with the embedded planets, the formation of partial gaps
modifies the profile significantly. In particular, were the induced 
partial gaps to be filled in, the initial negative slope of the surface
density distribution at the outer planet position can be seen to be  
transformed into a positive one.
In addition the local maximum of the disk surface density that resides 
between the planets moves  slightly inwards along with the migration 
of two planets.

By analogy with the operation of the horseshoe drag at a planet trap, 
the transformation to a positive surface density  slope at the location 
of the outer planet might be expected to lead to a slowing down of its 
migration. Indeed it is notable that this occurs at a much earlier 
stage when compared to the evolution of the planets in a disk with the 
initial power law density profile which does not exhibit a positive 
slope on filling in the partial gaps
(compare Figures~\ref{fig:figure2} and  \ref{fig:disk_evo}).
But note that this effect is not ultimately responsible for the 
divergent migration which eventually occurs in both systems.

\begin{figure*}[htb!]
\plottwo{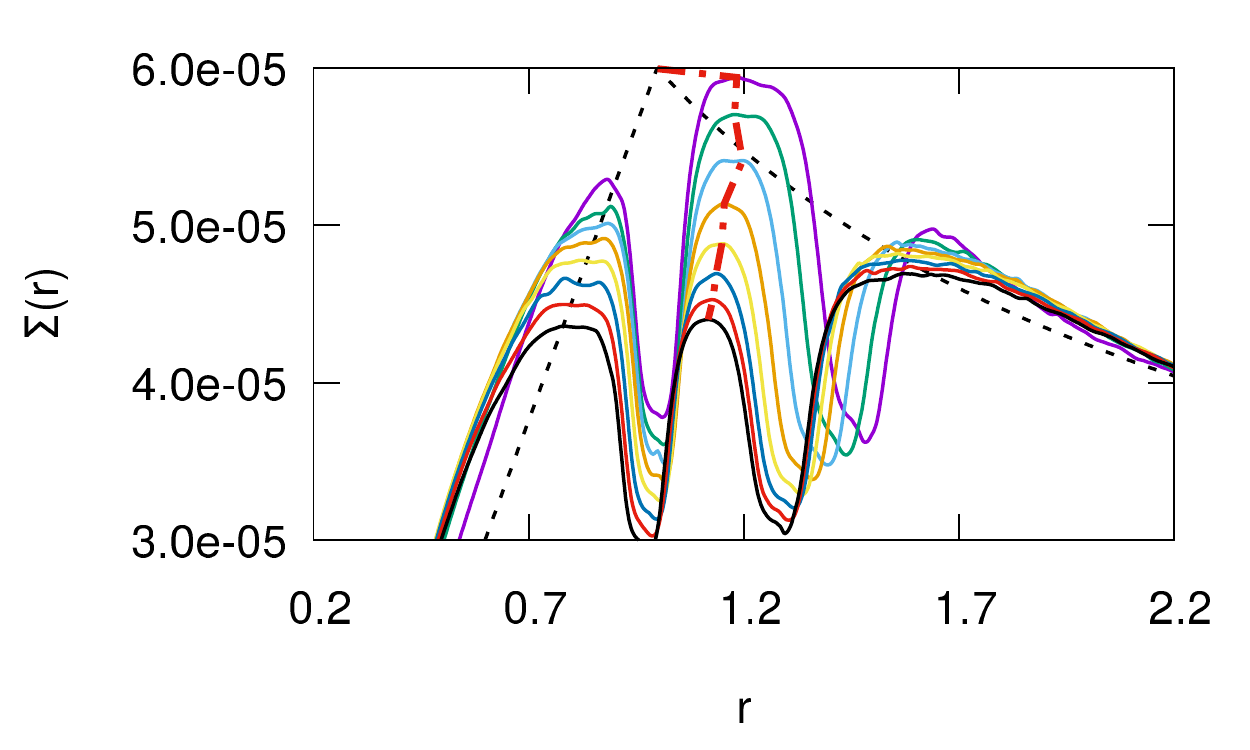}{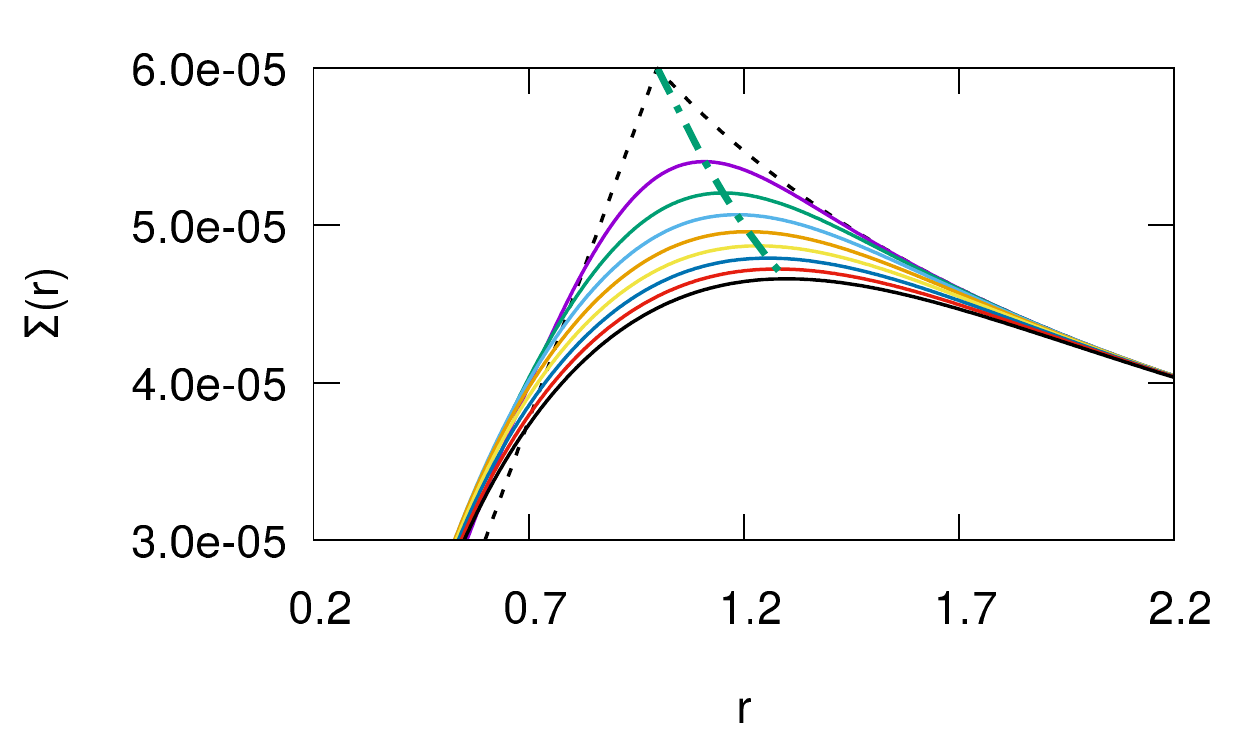}
\caption{The surface density profile of the disk with 
$\Sigma_0=6 \times 10^{-5}$  at each moment of time, with
embedded planets with $q_1=1.3\times 10^{-5}$ and 
$q_2=1.185\times 10^{-5}$ (left panel), and without planets 
(right panel).
The dashed black line indicates the initial surface density profile 
in the simulations. The solid lines show the surface density profiles 
at $t = 796, 1592, 2389, 3185, 3981, 4777, 5573$ and $6369$ orbits
respectively indicated with solid lines coloured in order from purple 
to black.
The red dashed-dotted line in the left panel indicates the position
of the local maximum of the surface density profile between the
locations of the planets while the green dashed-dotted line in the 
right panel shows the position of the maximum of the surface density 
profile of the disk.
}
\label{fig:disk_evo}
\end{figure*}

\textcolor{red}{
\section{The \textcolor{red}{dependence on} \textcolor{red}{ the assumed equation of state and the}  \textcolor{red} {effect of neglecting}
the self-gravity of the disk}\label{sec:selfg}
In this Section we investigate} \textcolor{red}{ the  dependence  on  two  
assumptions made  up to now} \textcolor{red}{ in our simulations.}
\textcolor{red} {These relate to the equation of state and the self-gravity of the disk
\subsection{The equation of state}\label{sec:adiab}
 The equation of state prescribed the disk to be  locally-isothermal.} \textcolor{red}{ 
 One aspect of this  that has been noted by \cite{Mira2019}} \textcolor{red}{
is that unlike when  an adiabatic 
or barotropic  equation of state applies,} \textcolor{red}{
 there is no strict conservation of wave action for small 
amplitude waves. There is accordingly the possibility of loss or gain 
from  the background. We  also expect coorbital torques to differ 
 on account of different behaviour of the state 
variables within the horseshoe region.
In order to check the influence of the equation of state on the results
of our simulations, we have performed a simulation adopting an adiabatic
equation of state.
The conclusion based on this 
 is that} \textcolor{red}{
 the generic
outcome of the simulations remains unchanged
independently of which 
equation of state is used. }\textcolor{red}{
This means that in both 
cases  repulsion between two planets caused by wave planet interaction 
is effective.}  \textcolor{red}{Additional details and further} \textcolor{red}{ discussion of this particular calculation 
are given in Appendix A.}
\textcolor{red}{\subsection{ Disk self-gravity}\label{sec:selfg1} }\textcolor{red}{
 In the simulations described} \textcolor{red}{up to now,} \textcolor{red}{
  the self-gravity of the disk has been neglected, as the 
surface densities of the gas used in our simulations are relatively low 
corresponding to a Toomre $Q$ value $\sim 20.$
One of the effects of including disk self-gravity  is to cause a
shift of the location of  Lindblad and corotation resonances, which 
leads to  changes in the torques acting on a planet thus affecting 
its migration \citep[see eg.][]{BarMas2008,Ataiee2020}.
In order to} \textcolor{red}{investigate}  \textcolor{red}{ the effect of self-gravity we have 
performed additional simulations described in detail in Appendix B.} 
\textcolor{red}{ From the results of these} \textcolor{red}{ 
 we can infer that it is unlikely that  
 self-gravity will have a significant  influence on the results 
of our simulations.}

\textcolor{red}{
\section{ The  effect of  a uniform surface density reduction on the Migration of two super-Earths}\label{sec:lowdensity}}
\textcolor{red}{
In Section~\ref{Sigmascale} we}  \textcolor{red}{explored the effect
of  modestly reducing the surface density scale on  super-Earth pair migration 
finding that separation from  resonance
continues to take place.
We investigated} \textcolor{red}{ the situation in which 
the outer planet is more massive than the inner one, and the 
relative  migration rate is not too fast for  3:2 resonance 
capture to take place. A reduction  factor of 1.33
was sufficient for our purpose. }

\textcolor{red}{An} \textcolor{red} {issue} \textcolor{red}{  arises as to whether the repulsion between
planets described in the previous Sections will also be present in 
a disk with} \textcolor{red}{a much smaller surface density, scaled down  4 or even 16 times
compared to that  considered up to now.  The motivation for this arises when we note that in the final stages
of the evolution and dispersal of the protoplanetary disk the surface density is expected to decrease 
to significantly smaller values than  expected for the minimum 
mass solar nebula.  }

\textcolor{red}{ We remark that there are good reasons
for expecting that the character of the orbital evolution does not change with such a scaling.
 Noting that if the planets are on fixed orbits,
the response induced in the disk will be proportional to  the surface density
scaling (the latter being everywhere scaled by the same constant factor). This is the case independently if self-gravity is not important
(increasingly the case when the surface density is reduced) and also when the effect of self-gravity is investigated following the approach
of Section \ref{sec:selfg1} and Appendix B.
This means that \textcolor{red}{as long as the system is not in resonance}  the rate of orbital evolution of the planet resulting from disk-planet torques
 should also follow this scaling. Provided the response adjustment rate is fast
 compared to the orbital evolution rate, the latter should also follow the same scaling.
 \textcolor{red}{Given that as shown by \citet{Baruteau2013} and Section \ref{sec:viscousdisk}  below the planet-planet repulsion
 mechanism works independently of the interaction between the planets we expect the procedure
 described here to be able to validate the orbital evolution rate scaling
 as well as confirm that planet planet interaction is not important in the later stages
 of the evolution as we find to be the case. } }

\textcolor{red}{\subsection{Numerical approach}
Because  following  the complete} \textcolor{red}{ planet evolution
in a disk with a very low surface density is computationally 
expensive, we have adopted} \textcolor{red}{ the following practical approach  to study this 
question.}\textcolor{red}{ 
The idea consists in starting  new calculations with a 
rescaled surface density profile taken from one of our already 
performed simulations.}\textcolor{red}{The form and direction of the orbital evolution
may then  be checked at different stages and then pieced togetether.}

\textcolor{red}{ For this purpose we} \textcolor{red}{select} \textcolor{red}{the surface
density profile obtained at $t = 10000$ orbits in the simulation 
shown in Figure~\ref{fig:fix_scaling}. Next, we 
reduce $\Sigma $ at each  grid point by a constant 
factor and then use such a density profile as the initial one} \textcolor{red}{ 
for a} \textcolor{red}{  new simulation. }

\textcolor{red}{Our choice of the time  $t = 10000$ 
orbits as the moment for  restarting  our calculation with a scaled down surface 
density was determined  by the need to start the evolution when the 
torque from the disk acting on the outer planet is already positive 
(see Figure~\ref{fig:two-torque}).
 We performed two new 
simulations, one with the surface density uniformly scaled down  by a
factor of 4 and another by a factor of 16. The initial surface 
density profiles in those two simulations are illustrated in
Figure~\ref{fig:sden-pratio-compare} together with the original
$\Sigma(r)$ from which they have been obtained.}

\begin{figure}[htbp]
\centerline{
\vbox{
\hbox{
\includegraphics[width=0.9\columnwidth]{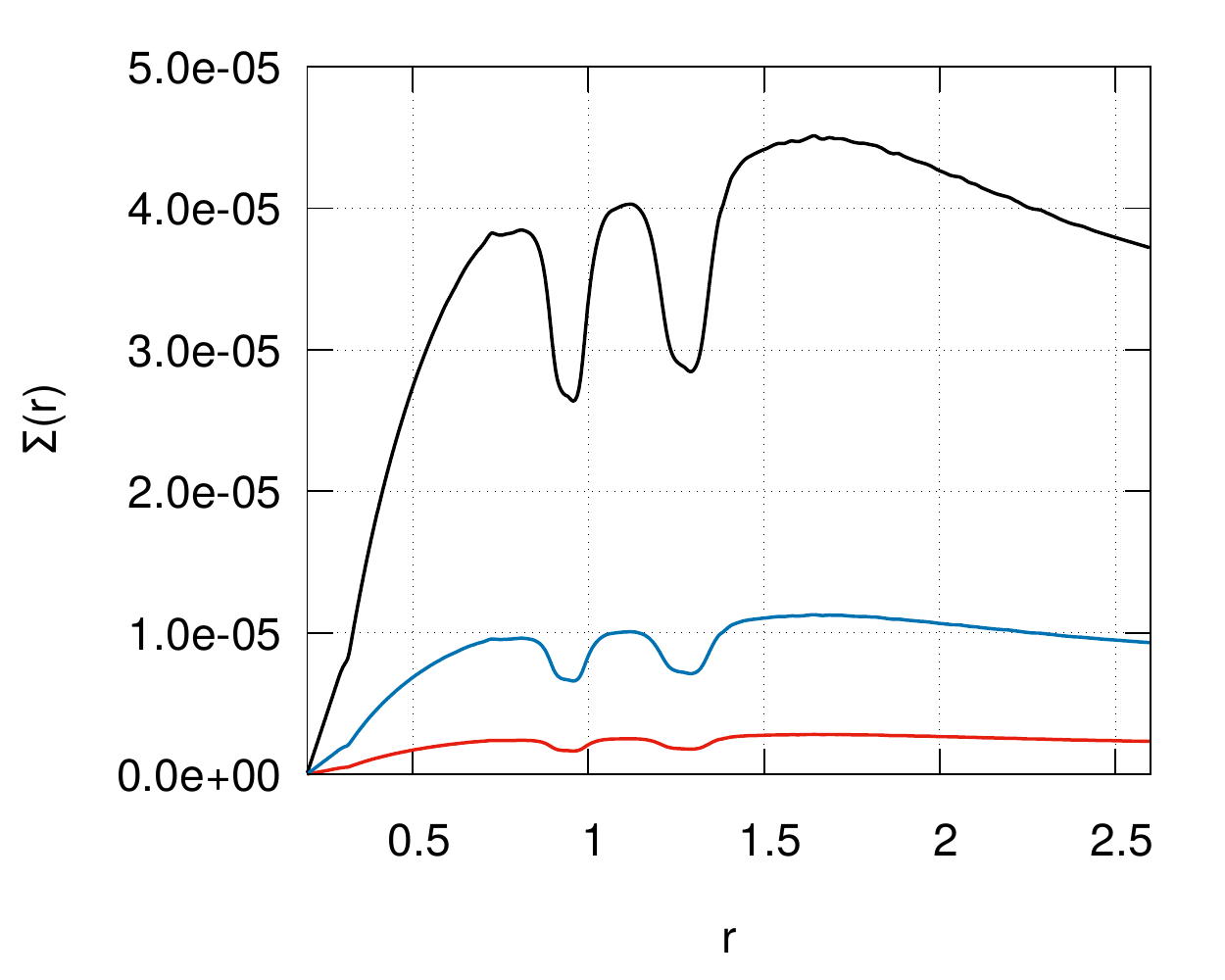}
}
}
}
\caption{The azimuthally averaged surface density at t = 10000 
orbits in the original simulation illustrated in Figure~\ref{fig:fix_scaling} (black line) and in the restarted 
ones with $\Sigma$ reduced by a factor of 4 (dark blue line) and by 
a factor of 16 (red line). 
}
\label{fig:sden-pratio-compare}
\end{figure}

\textcolor{red}{
The evolution of planets in the disk with the surface density
reduced by a factor of 4 and that with the surface density reduced 
by a factor of 16  is divergent}
\textcolor{red}{just as it was in the original one shown in Figure~\ref{fig:fix_scaling}.}
\textcolor{red}{We illustrate this in Figure~\ref{fig:ratio-rescale} showing
the period ratio of the two planets in the restarted runs (left panels)
together with the semi-major axes of the inner and outer planets 
(middle and right panels respectively).
The outer planet in both cases migrates outwards and the inner one
migrates inwards resulting in divergent migration.} \textcolor{red}{As expected the
orbital evolution  is  four times faster in the case for which the initial surface density
reduced by a factor of four.} 

\begin{figure*}[htbp]
\centerline{
\vbox{
\hbox{
\includegraphics[width=0.6\columnwidth]{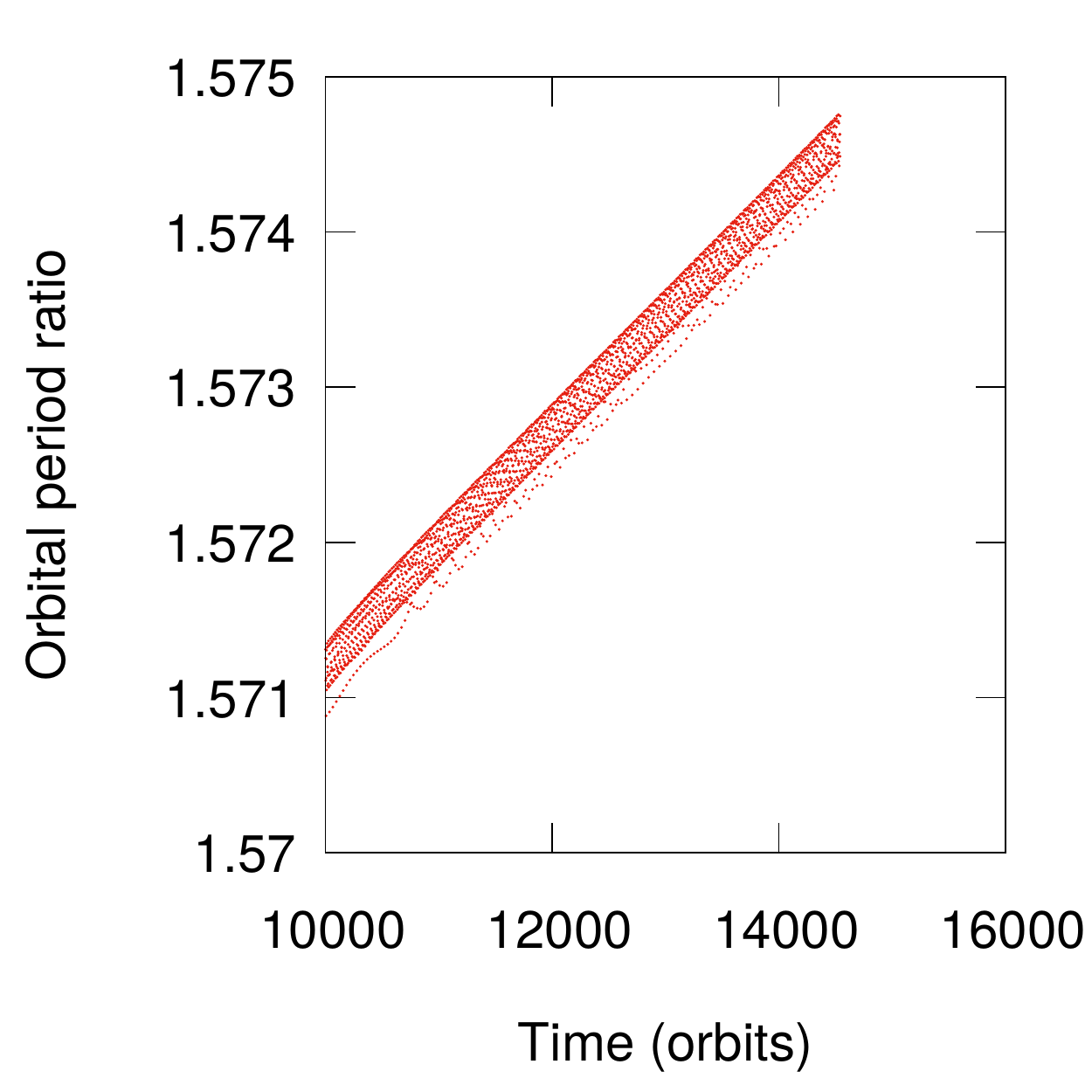}
\includegraphics[width=0.6\columnwidth]{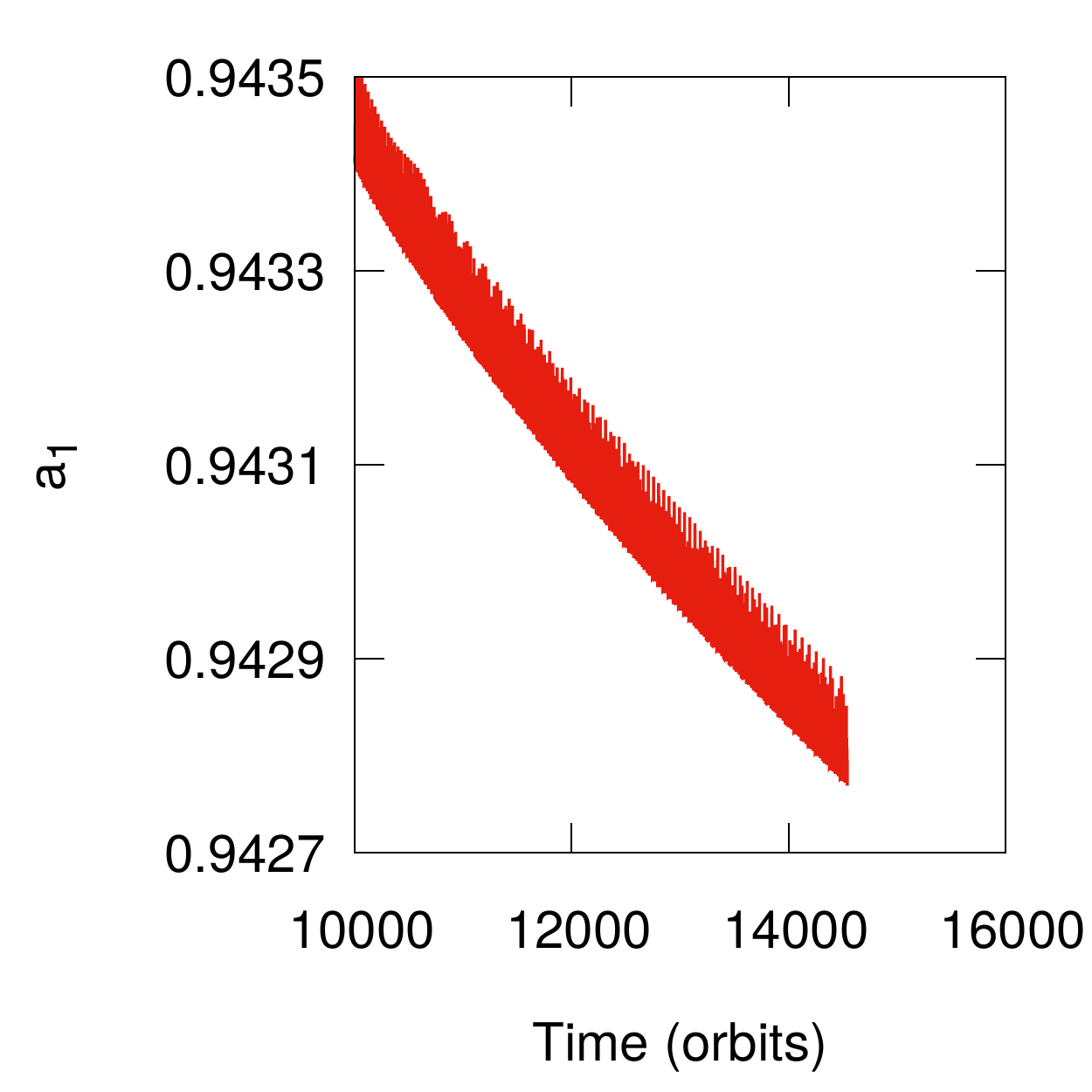}
\includegraphics[width=0.6\columnwidth]{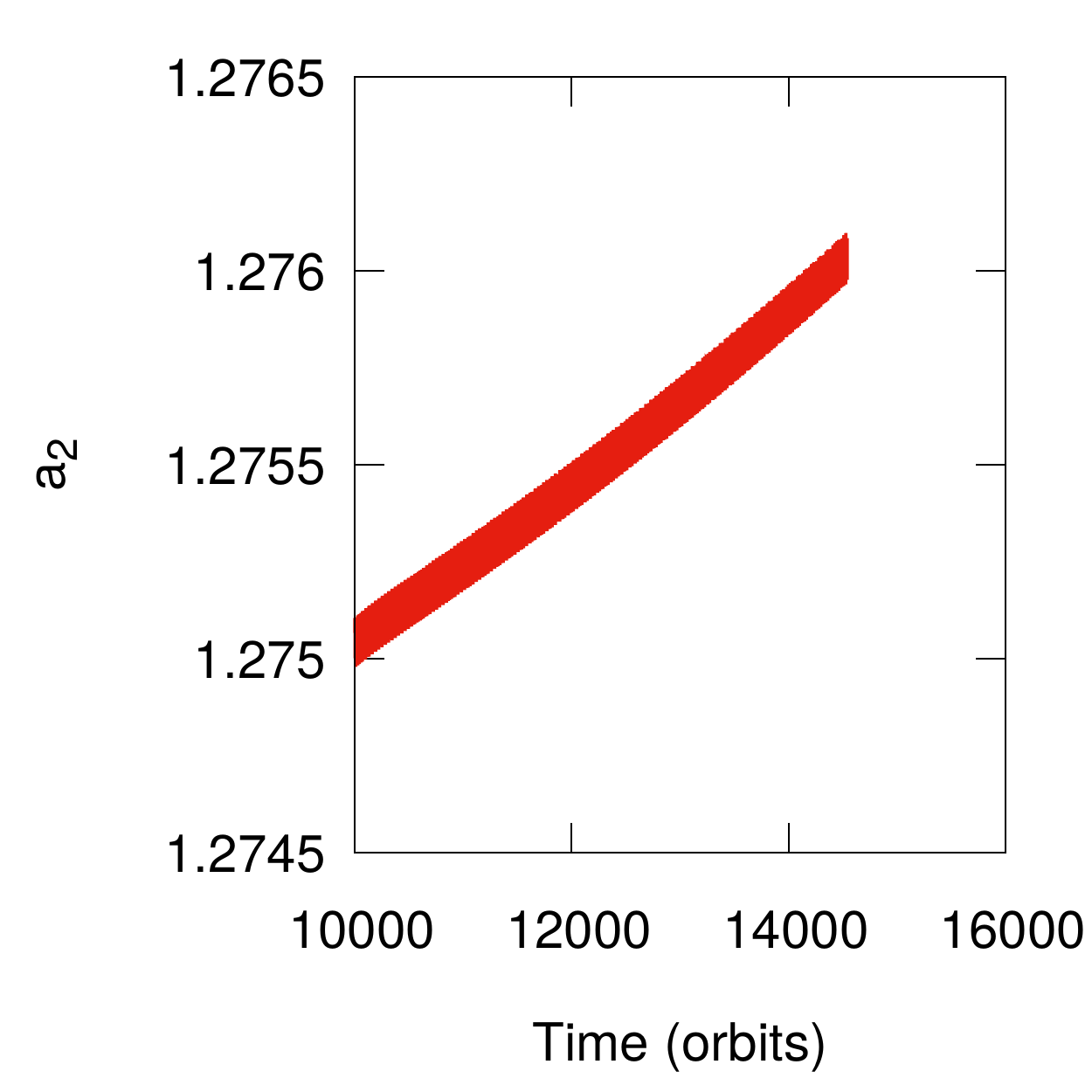}
}
\hbox{
\includegraphics[width=0.6\columnwidth]{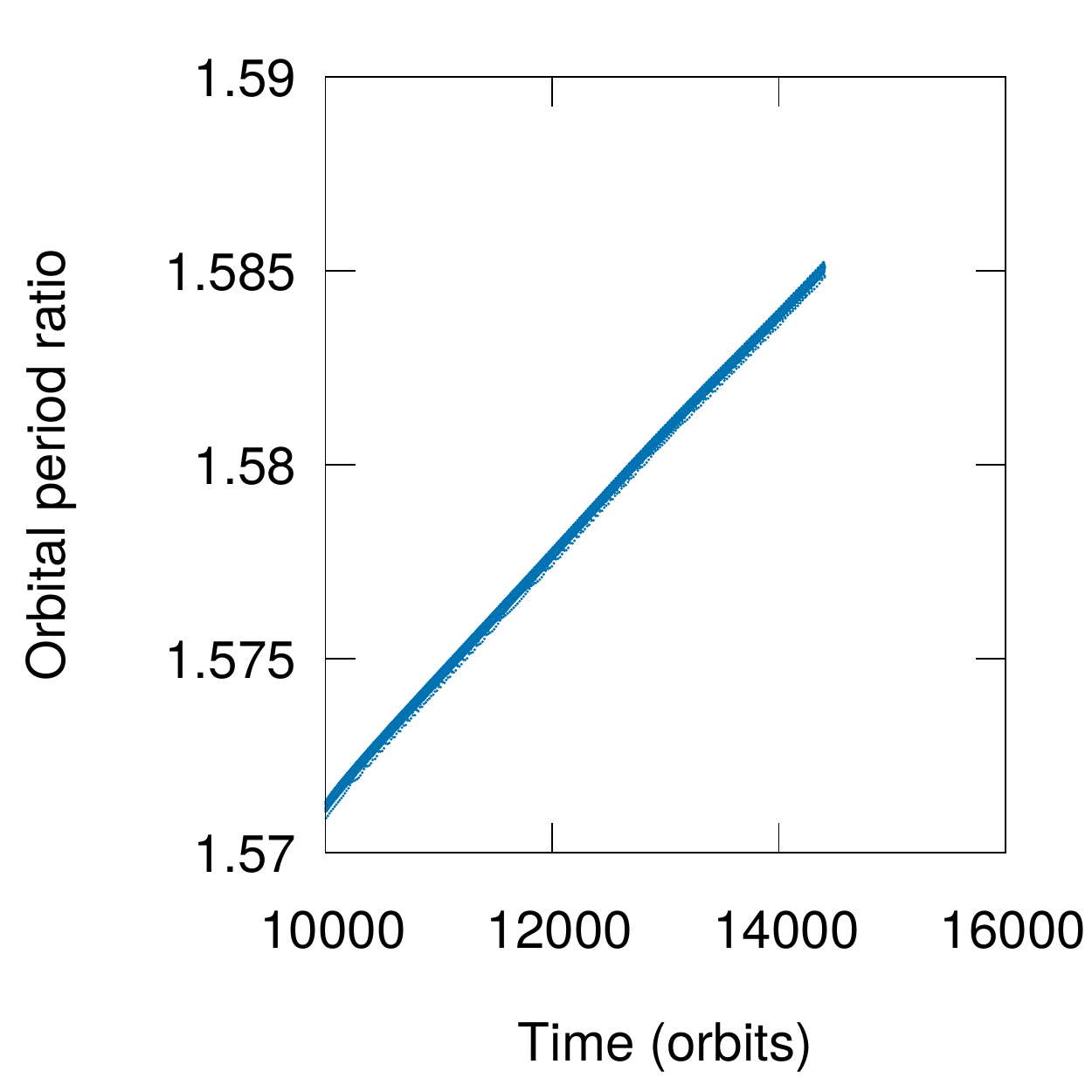}
\includegraphics[width=0.6\columnwidth]{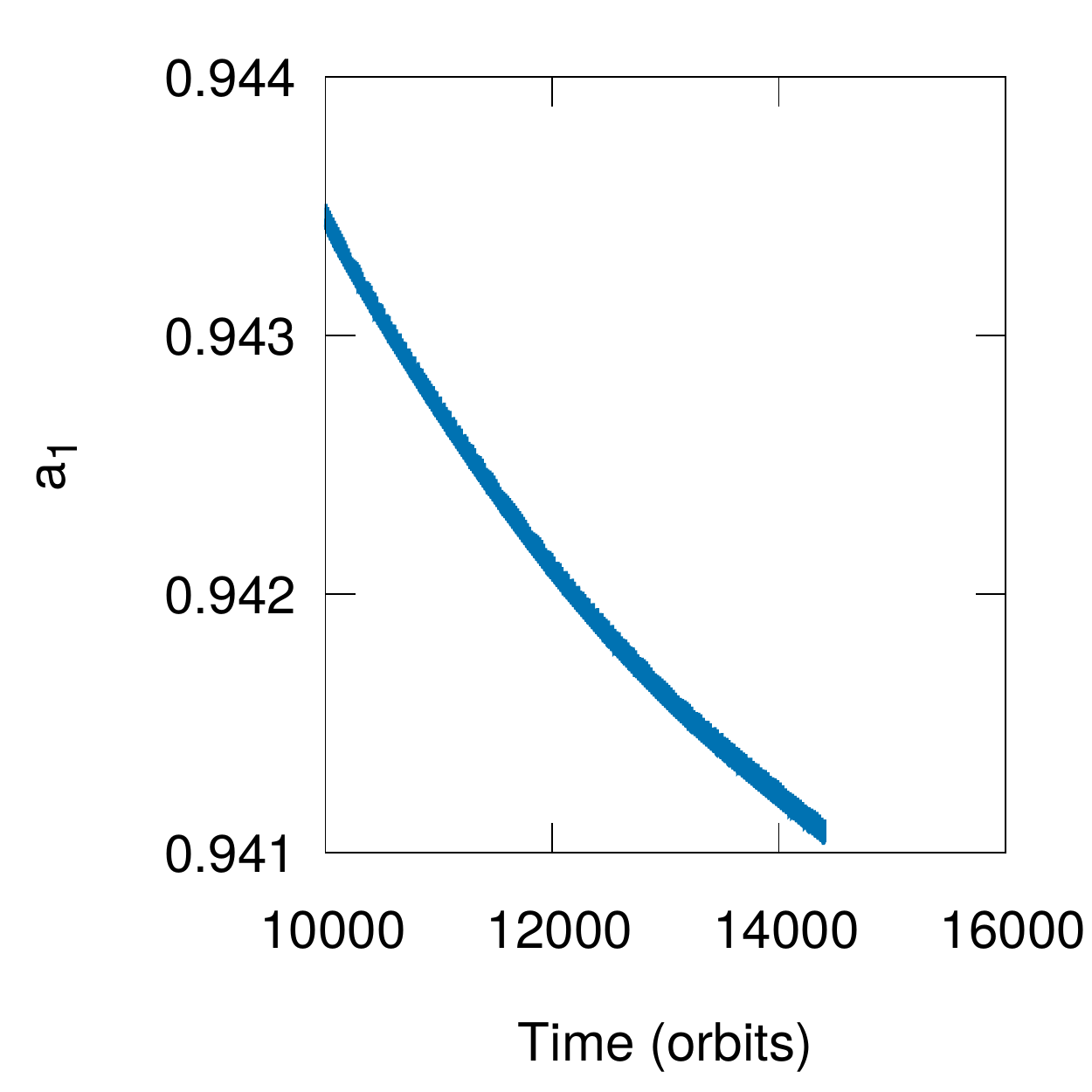}
\includegraphics[width=0.6\columnwidth]{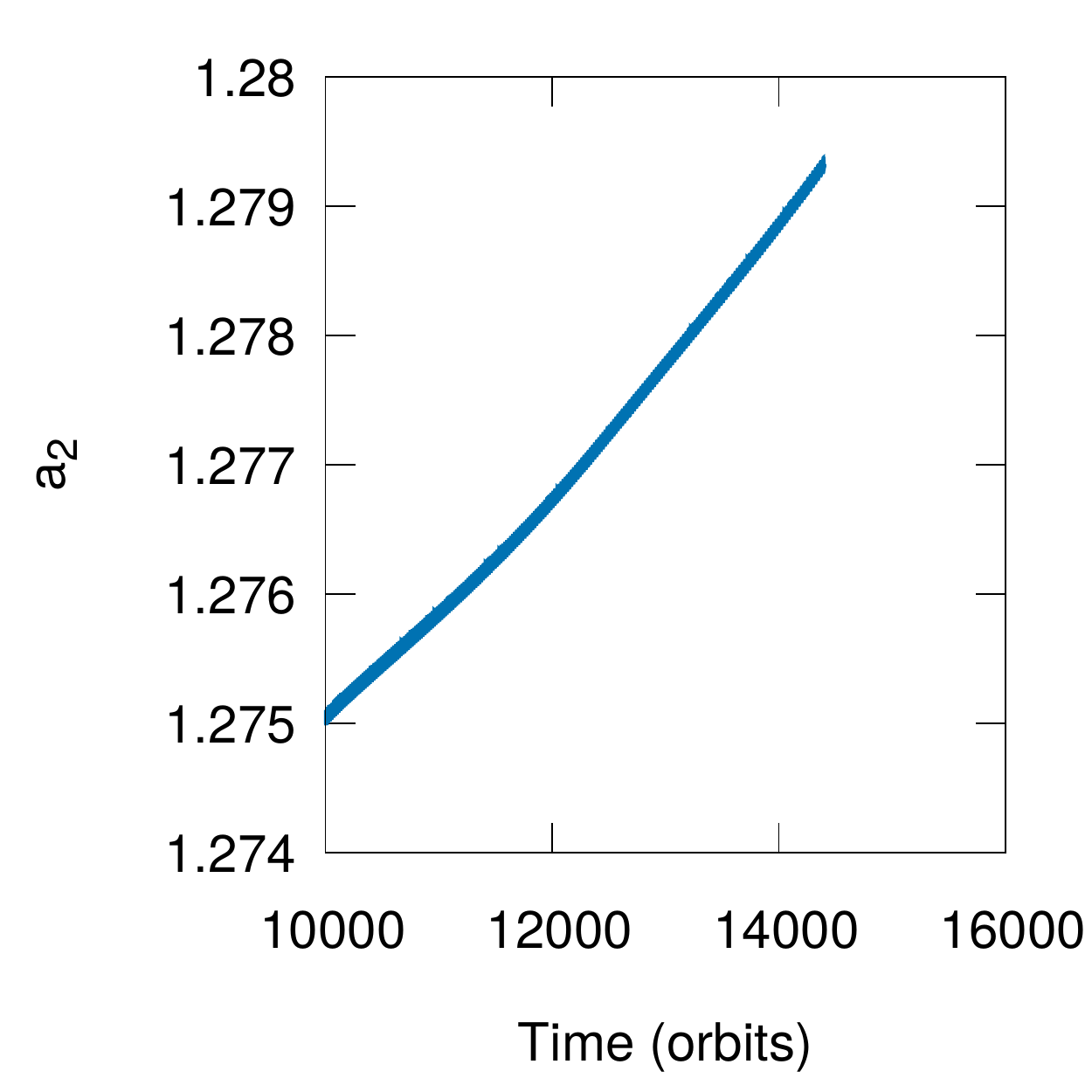}
}
}
}
\caption{Left: The evolution of the period ratios of two super-Earths 
in the restarted simulation with the  surface density reduced
by a factor of 4 (indicated by the dark blue line) and 16 (indicated
by the red line). Note that the evolution is four times slower in the former case
and sixteen times in the latter; 
Middle and Right: Evolution of the semi-major axis of the inner and 
outer planets in the restarted simulation with surface density reduced
by a factor of 4 (indicated by the dark blue line) and 16 (indicated
by the red line). 
}
\label{fig:ratio-rescale}
\end{figure*}

\textcolor{red}{
As a final} \textcolor{red}{demonstration} \textcolor{red}{of the efficiency of the planet
repulsion due to the wave planet interaction in the disk with
a substantially reduced surface density, we present a
comparison of the relevant torques in our simulations.
In the left panel of Figure~\ref{fig:torque-compare-rescale},
we compare the torque from the disk acting on the outer planet 
in the simulations with the  surface density} \textcolor{red}{reduced by a factor of 16}  
\textcolor{red}{ (red solid line) with the torque expected from a simple scaling 
procedure applied to the 
original calculation shown in Figure~\ref{fig:fix_scaling} \textcolor{red}{(denoted by the dashed yellow line).} 
The scaling procedure consists in decreasing the total
torque by a factor of 16 and increasing the time scale of the orbital
evolution by the same factor, as this time scale should scale 
inversely with the surface density.
In addition, in order to compare the same phases of the evolution
in both calculations the initial time for the lower density run  
is shifted to $t = 160000$ orbits.
 \textcolor{red}{The two torques are in very good agreement.} 
}

\begin{figure*}[htbp]
\centerline{
\hbox{
\includegraphics[width=0.9\columnwidth]{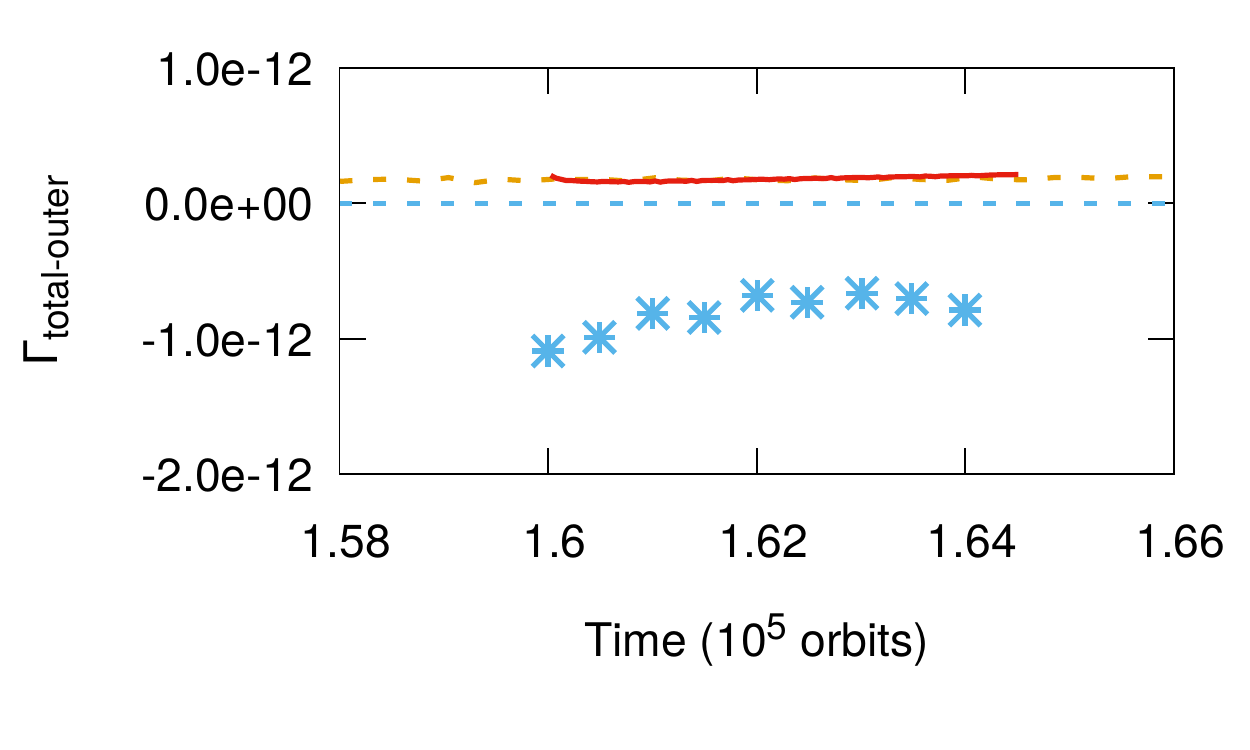}

\hspace{1.0cm}

\includegraphics[width=0.9\columnwidth]{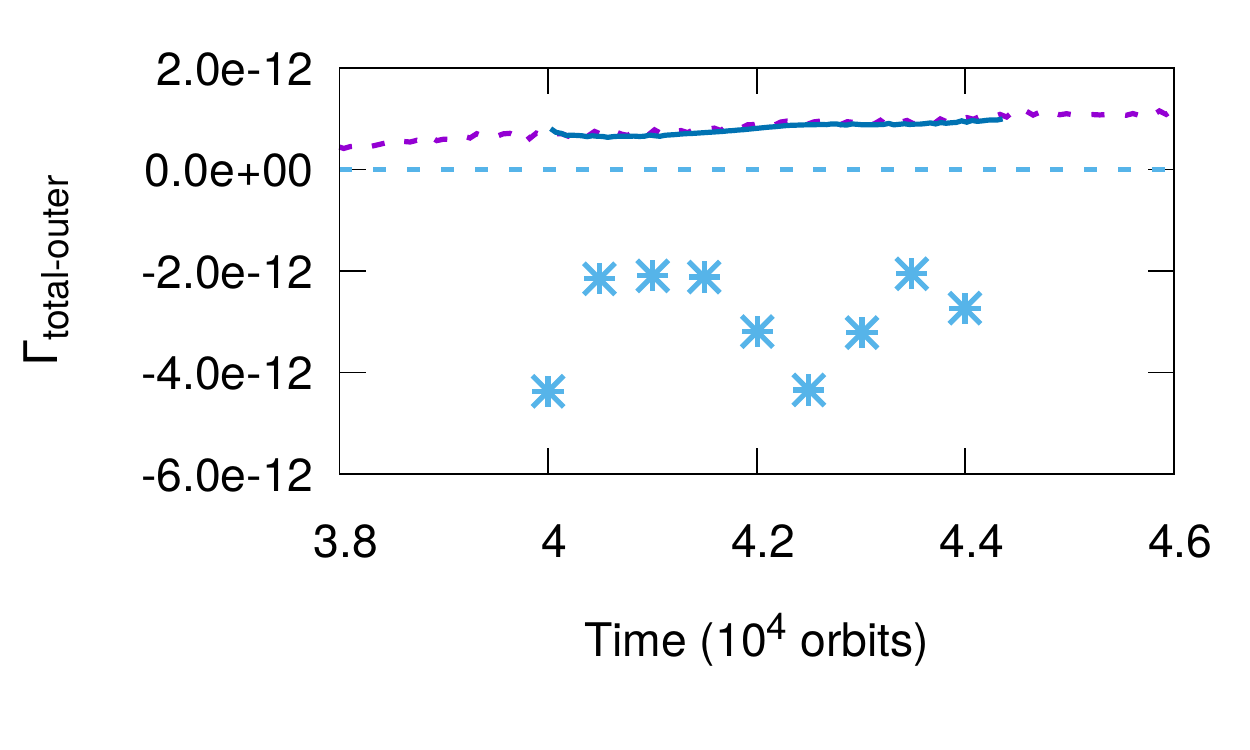}}
}
\caption{The locally time averaged total torque from the disk acting on the 
outer planet in the restarted simulations with the surface 
density respectively scaled down by a factor of 16 (left panel red solid line) and 4
 (right panel dark 
blue solid line).
For comparison the expected average total  torques obtained by the appropriate  scaling
applied to the original simulation 
are  denoted by the dashed yellow and  purple lines in the left and right panels respectively. 
The blue asterisks indicate  torques calculated from 
Equation~(\ref{eq:numerical-fit}).}
\label{fig:torque-compare-rescale}
\end{figure*}

\textcolor{red}{
In the right panel of Figure~\ref{fig:torque-compare-rescale}, we show 
the \textcolor{red}{average} torque in the run with the 4 times lower surface density (dark 
blue line) and the one expected from the scaling procedure described 
above (dashed purple line). Here the initial time for the low density 
calculation has been shifted to $t= 40000$ orbits.
\textcolor{red}{ These are in very good agreement.}
In Figure~\ref{fig:torque-compare-rescale} we also \textcolor{red}{show similar plots for the case
when the surface density was reduced by a factor of 16
and } the torques 
calculated from Equation~(\ref{eq:numerical-fit}), indicated by the 
blue asterisks.} 

\textcolor{red}{
One the basis of these results we can conclude that the repulsion
of the planets is  also present  in the disks with a 
reduced surface density relative to that used in  previous Sections.} 

\textcolor{red}{
\section{Planets in the super-Earth mass range \textcolor{red}{migrating}  in a
 protoplanetary disk \textcolor{red}{with local heating and cooling}}
\label{sec:viscousdisk} 
We have investigated the process of \textcolor{red}{ repulsion between planets arising from wave-planet interaction
 for a range of super-Earth masses 
and a range of values for the characteristic magnitude of  the disk surface density.
However, so far only a locally isothermal equation of state has been extensively considered.
We now extend consideration to an initial  disk model in which the balance  between local heating and cooling 
determines  the local disk temperature.
This is very different from a local isothermal prescription and could potentially
lead to substantially different dynamics in the vicinity of gaps induced by  orbiting planets.  
The particular disk model adopted is  viscous 
 while initially supporting a } 
 constant angular momentum flux. \textcolor{red}{As noted in Section \ref{sec:intro} such a situation may arise 
 when the inner disk interacts with a stellar magnetosphere  prior to
 dispersal \citep[eg.][]{Clarke1996}. We remark that the magnetosphere may be associated with a protoplanet trap
 but we do not model this here.}}

\textcolor{red}{
\subsection{A viscous protoplanetary disk \textcolor{red}{with local balance between heating and cooling}} 
Let us consider a model with constant angular momentum flux, 
$\cal{F}$, which reads 
\begin{equation}
{\cal F}= 3 \pi \nu \Sigma \sqrt{G M_{\star} r}
\label{flux}
\end{equation}
We adopt the general power law scalings with
$$\Sigma \propto r^{-\beta_1},\hspace{2mm}
h \propto r^{-\beta_2} \hspace{2mm} {\rm and}\hspace{2mm}
\nu \propto r^{-\beta_3}.$$}
\textcolor{red}{To provide scalings  for these quantities 
we set}
\textcolor{red}{
\begin{equation}
\Sigma R^2/M_{\star} = 6.0\times 10^{-5} f_{\Sigma} (r/R)^{-\beta_1}\label{FVS}
\end{equation}
\begin{equation}
h =0.02 f_{h}  (r/R)^{-\beta_2}\hspace{2mm}{\rm and}\label{FVh}
\end{equation}
\begin{equation}
\nu = 1.2 \times 10^{-6} f_{\nu}\sqrt{G M_{\star} R} (r/R)^{-\beta_3}\label{FVn}
\end{equation}
\textcolor{red}{where  we adopt the fixed parameters
$M_{\star}=1 M_{\odot}$ and  \textcolor{red}{the radius $R$  which determines that radial scale of the 
model to correspond to} 0.1 au with 
$f_h$, $f_{\Sigma}$ and $f_{\nu}$ being dimensionless scaling constants 
expected to be of order unity.
We remark that with the above specifications the characteristic disk mass within
radius, $R,$ is $6\times 10 ^{-5}\pi f_{\Sigma}M_{\star}.$ For
$R = 0.1$ au, $M_{\star}=1M_{\odot}$ and $f_{\Sigma}=1$  this is about 1.5  times smaller than extrapolated for the minimum mass solar nebula. 
In this context we recall
that this may be reduced by scaling down the surface density Then following the  discussion  in Section \ref{sec:lowdensity}
we might expect that under appropriate conditions and scaling that the form of the disk planets interaction would be preserved.
 }}
\textcolor{red}{
The constancy of  ${\cal F}$ requires that $\beta_1 + \beta_3 =1/2$.
Assuming a constant kinematic viscosity we have $\beta_3=0$
\textcolor{red}{and thus} $\beta_1 =1/2.$
Using \textcolor{red}{this and Equations (\ref{FVS}) - (\ref{FVn}), Equation~(\ref{flux}) leads to} 
\begin{equation}
\frac{\cal F}{\sqrt{GM_{\star}R}} = 
1.35 \times 10^{-7} f_{\nu} f_{\Sigma}M_{\odot}y^{-1} \label{barF} 
\end{equation}
 }
 
\textcolor{red}{
To complete the model we need to consider the energy balance. This
gives another relationship between the $\beta_i$. Thus together with 
the requirements specified above,  these indices \textcolor{red}{are completely specified.}
We start by considering 
the properties of the opacity.
The opacity is mostly due to grains as can be seen from 
\cite{Ferguson2005}.  From their Figures 10 and 11, \textcolor{red}{for the parameter ranges of interest}  the grain opacity 
is approximately constant and scales with metal abundance, $Z$, so that 
$\kappa \sim (Z/0.02)$. This scaling holds down to very small $Z$ as 
when $Z = 0$, $\kappa \sim 10^{-8}$, which is associated with Rayleigh 
scattering due to Hydrogen molecules. Thus if there is significant unknown  grain 
depletion as a result of the planet formation process, the opacity is highly uncertain translating to uncertainty 
in the disk models.}

\textcolor{red}{
The surface density evaluated \textcolor{red}{from Equation~(\ref{FVS})}  assuming $M_{\star}= M_{\odot}$ and $R=0.1$ 
au is given by 
\begin{equation}
\Sigma = 53333 f_{\Sigma}(r/R)^{1/2} 
\end{equation}
For standard metalicity  (no grain depletion) this means that the disk is very optically 
thick with $\tau = 5 \times 10^{4} f_{\Sigma}$ at $r=R$.}

\textcolor{red}{
In the optically thick case the rate of cooling per unit mass is 
given by
\begin{equation}
\label{eq:qminus}
Q^{-}=(2 \sigma T_e^4)/\Sigma
\end{equation}
where \textcolor{red}{ we assume that radial diffusion of heat is not important, this should be reasonable
for disturbances on a radial scale exceeding the scale height with smaller scales being   smoothed 
by physical and numerical viscosity,} $\sigma$ is the Stefan-Boltzmann constant and $T_e$ is the 
effective temperature
which has to be related to the mid-plane temperature. Let us 
consider the balance between viscous heating and cooling.
The rate of viscous heating per unit mass $Q^{+}$ is given by
\begin{equation}
\label{eq:qplus}
Q^{+}=\nu r^2 \left( \frac{d\Omega}{dr}\right)^{2} =
2.7 \times 10^{-6}f_{\nu}(GM_{\star})^{3/2} R^{-5/2} (r/R)^{-3}
\end{equation}
Equating this to $Q^{-}$ given above we can find $T_e$ which is 
expressed  by 
\begin{equation}
T_e^4= 
\frac{7.2}{\sigma}\times 10^{-2}f_{\nu} 
f_{\Sigma}(GM_{\star})^{3/2}R^{-5/2} (r/R)^{-7/2}
\end{equation}
or
\begin{equation}
T_e = 917.8 (f_{\nu} f_{\Sigma})^{1/4} (r/R)^{-7/8}\ \rm{K}.
\end{equation}
Next, we can consistently estimate $h$. Assuming that the mid-plane 
temperature $T=f_{T}T_e$, where $f_{T}$, 
\textcolor{red}{is a scaling parameter (see below),} 
 we have 
\begin{equation}
h= \frac{\sqrt{{\cal R}T/\mu}}{r\Omega}=
0.019 (f_{T})^{1/2} (f_{\nu} f_{\Sigma})^{1/8} (r/R)^{1/16}
\end{equation}
where ${\cal R}$ is the gas constant and $\mu$ is the mean 
molecular weight. 
Here we adopt $\mu =2.35$ (see eg. \cite{Flaig2012}) corresponding 
to \textcolor{red}{predominantly} molecular Hydrogen.}

\textcolor{red}
{The determination of $f_T$ depends  on how energy is transported from the disk midplane to
the surface. This is highly uncertain as the optical depth is affected by uncertain grain depletion
and turbulent transport due to the magnetorotational instability operates.
However, we remark that the simulations of \cite{Flaig2012} indicate only modest  mean vertical variation
of the temperature for models with optical depth  $< \sim 100.$
Accordingly in order to have a working model, we shall assume appropriate grain depletion and adopt $f_T=1.$     }

\textcolor{red}{
Summarizing, the complete specification of the initial model is 
then
\begin{equation}
\label{initialsigma}
\Sigma R^2/M_{\star} = 6.0\times 10^{-5} f_{\Sigma} (r/R)^{-1/2},
\end{equation}
\begin{equation}
\label{initialh}
h =0.019 f_{T}^{1/2} (f_{\nu} f_{\Sigma})^{1/8} (r/R)^{1/16} \hspace{3mm}{\rm and}
\end{equation}
\begin{equation}
\label{initialnu}
\nu = 1.2 \times 10^{-6} f_{\nu}\sqrt{GM_{\star}R} 
\end{equation}
\textcolor{red} {For practical purposes} we also set the scaling factors  $f_{\Sigma}$ and 
$f_{\nu}$ equal to unity.}

\textcolor{red}{
\subsection{Numerical setup}
We run  simulations  in which two planets are embedded in the 
disk with constant angular momentum flux described 
above (Equations~(\ref{initialsigma} - \ref{initialnu})). 
As \textcolor{red}{ mentioned above,  we adopt  the  unit of length 
$R$ = 0.1au and the unit of mas $M_{\star}  = M_\odot$}. 
The energy equation \textcolor{red} {solved}  in the simulations  is 
\begin{equation}
\frac{\partial e}{\partial t} + \nabla \cdot (e{\bf{v}}) = 
-P \nabla \cdot {\bf{v}} + Q^{+}\Sigma - Q^{-} \Sigma \label{NEOM}
\end{equation}
where $e$ is the internal energy per unit area, ${\bf{v}}$
is the gas velocity, $P$ is the integrated pressure,  
$Q^{-}$ and $Q^{+}$ are the heating and cooling functions 
defined in Equations~(\ref{eq:qminus}) and (\ref{eq:qplus}) 
respectively. 
An ideal equation of state is \textcolor{red} {adopted which closes the hydrodynamic 
equations.} Thus
\begin{equation}
P=\frac{{\cal R}}{\mu}\Sigma T.
\end{equation}
The internal \textcolor{red}{energy per unit area  is then}  linked to the temperature through
\begin{equation}
e=\frac{{\cal R}}{\mu}\frac{\Sigma T}{\gamma-1}
\end{equation}
where $\gamma$ is the adiabatic index
 is taken to be 9/7.}

\textcolor{red}{At this point we consider whether there is a surface density scaling down
transformation of the type discussed in Section \ref{sec:lowdensity} that will preserve the disk response to planets
in fixed orbits. In order for this to exist we see from Equation~(\ref{NEOM})
the effective  cooling rate must be similarly scaled down in order to preserve the form of the temperature.
This could come about through assumed additional heating via reprocessing of radiation from the central star
or an assumption that the disk transforms from being approximately of unit optical depth to becoming optically thin
or a combination of these effects.
\textcolor{red}{Alternatively one could  envisage that the disk  becomes optically thicker
through increasing the opacity with changing grain depletion as the surface density is scaled down.}
 }

\textcolor{red}{
We consider here \textcolor{red} {an initial}  model with constant angular
momentum flux, ${\cal F}$, \textcolor{red}{ for which }
 the inner and outer boundaries correspond 
to imperfectly slippery rigid boundaries 
\textcolor{red}{at which the constant flux is respectively transported away from and towards 
 by viscous stresses.}
Thus \textcolor{red}{in each case}  the boundary condition \textcolor{red} {takes  the}  form 
\begin{equation}
{\cal F} = - 2\pi \nu \Sigma r^3 \frac{d(v_{\phi}/r)}{dr},
\end{equation} 
where $v_{\phi}$ is the azimuthal  component \textcolor{red}{ of the velocity}.
\textcolor{red}{In our simulations we retain this boundary condition when the planets are introduced.
In doing so we remark that the boundary conditions may evolve on a long time scale
as physical conditions at the boundaries change. This requires greatly extended simulations that are
 a matter for future investigation.  }
}


\textcolor{red}{
\subsection{A case with the equal mass planets}
\textcolor{red} {In the  disk model}  described above, we locate two 
super-Earths with $q_1 = q_2 = 2.6 \times 10^{-5}$ at the initial 
positions of $r_1 = 1.0$ and $r_2 = 1.36$, respectively 
\textcolor{red}{These radii are the initial radii of the planets
expressed in units of $R$. From now on the general radius, $r,$ is also expressed in this unit}. The disk 
was relaxed for 800 orbits \textcolor{red}{ in order to ensure convergence
to the expected steady state}  before adding the planets to it.
The evolution of the period ratio, semi-major axes, eccentricities 
and the resonant angles for this pair of planets are presented in 
Figure~\ref{fig:evo-alpha-ss-26-26-136}. 
Both planets migrate inwards. After a short \textcolor{red}{ initial period} of divergent 
migration, the planets migrate convergently and at $t \sim$ 900 
orbits they arrive at the 3:2 MMR. The eccentricities are excited 
to $e_{1} \sim 0.012$ and $e_{2} \sim 0.013$. The 3:2 resonant 
angles  librate. When $t \sim 1300$ orbits, the planets start
to migrate divergently, and their distance from  exact \textcolor{red}{commensurability}
increases. \textcolor{red}{At later times,}  the eccentricities decrease and the resonant 
angles \textcolor{red}{cease}  librating.}

\begin{figure*}[htbp]
\vbox{
\hbox{
\includegraphics[width=0.5\columnwidth]{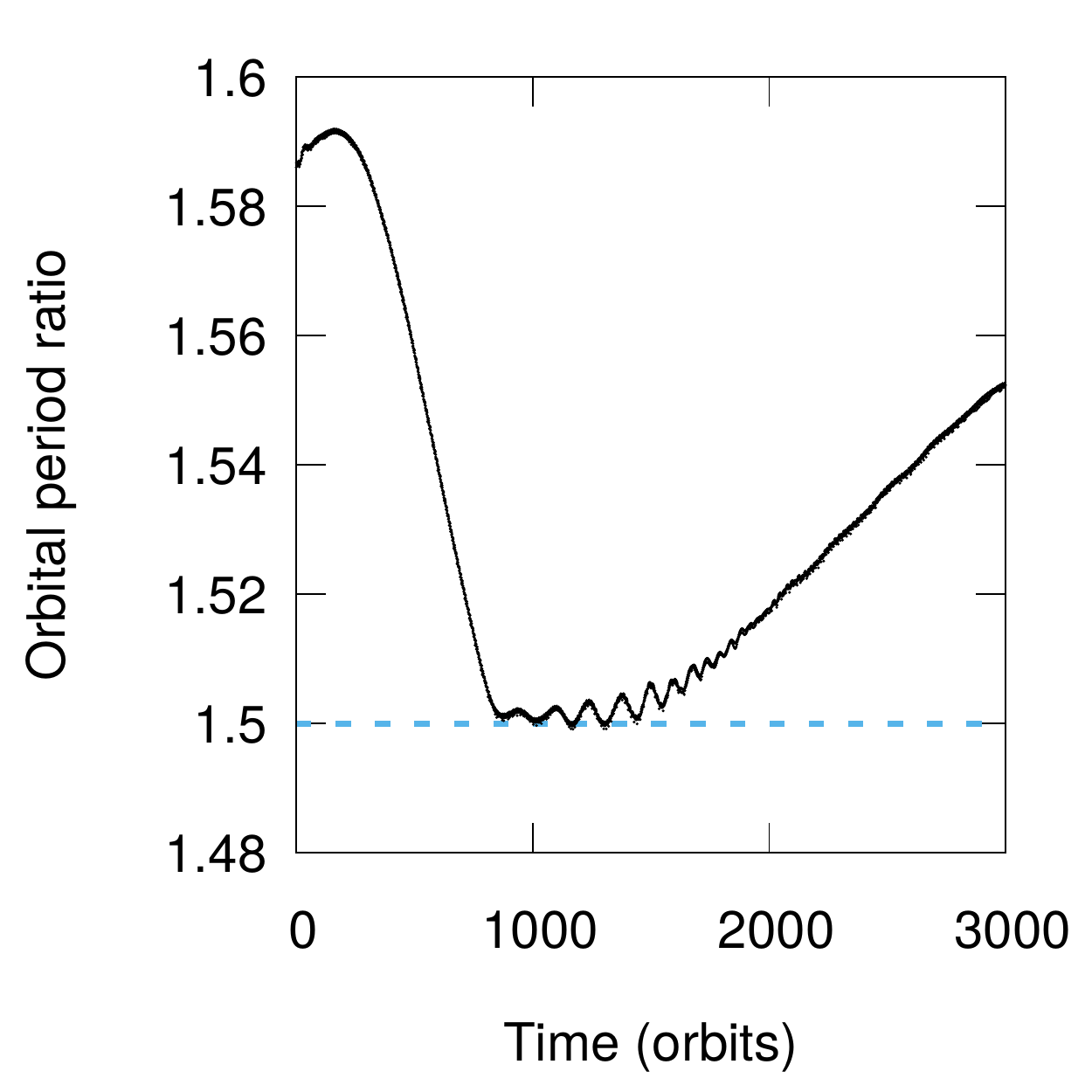}
\includegraphics[width=0.5\columnwidth]{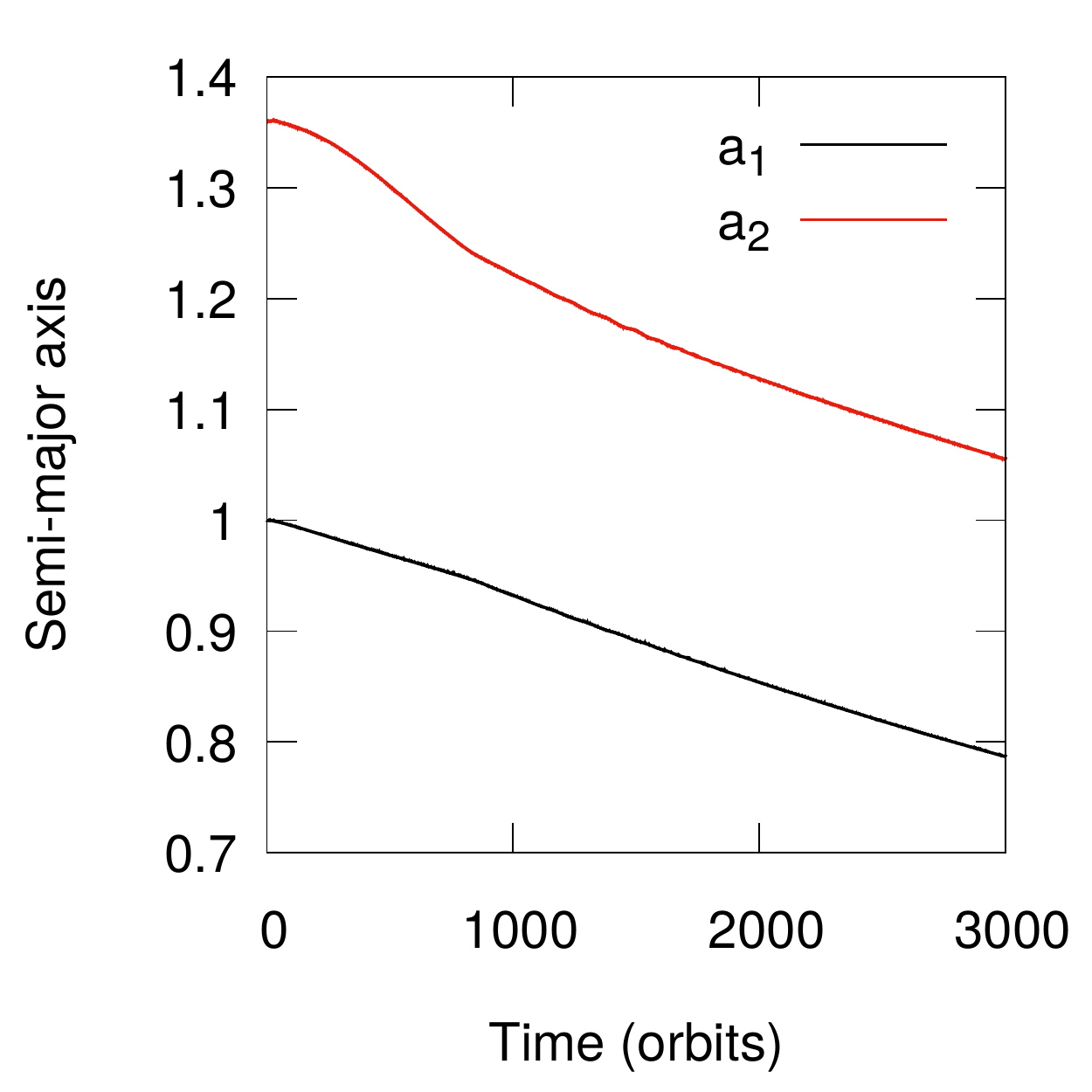}
\includegraphics[width=0.5\columnwidth]{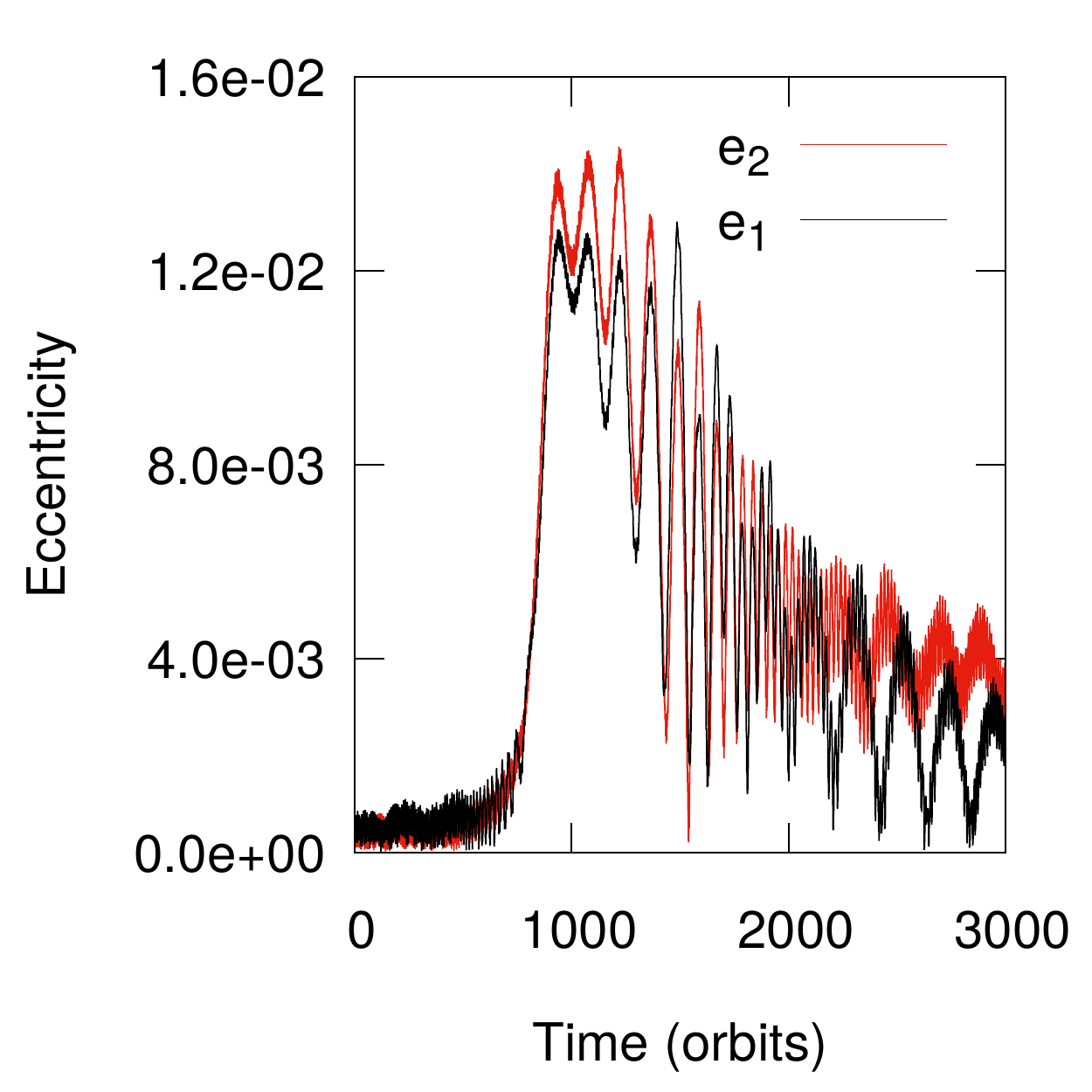}
\includegraphics[width=0.5\columnwidth]{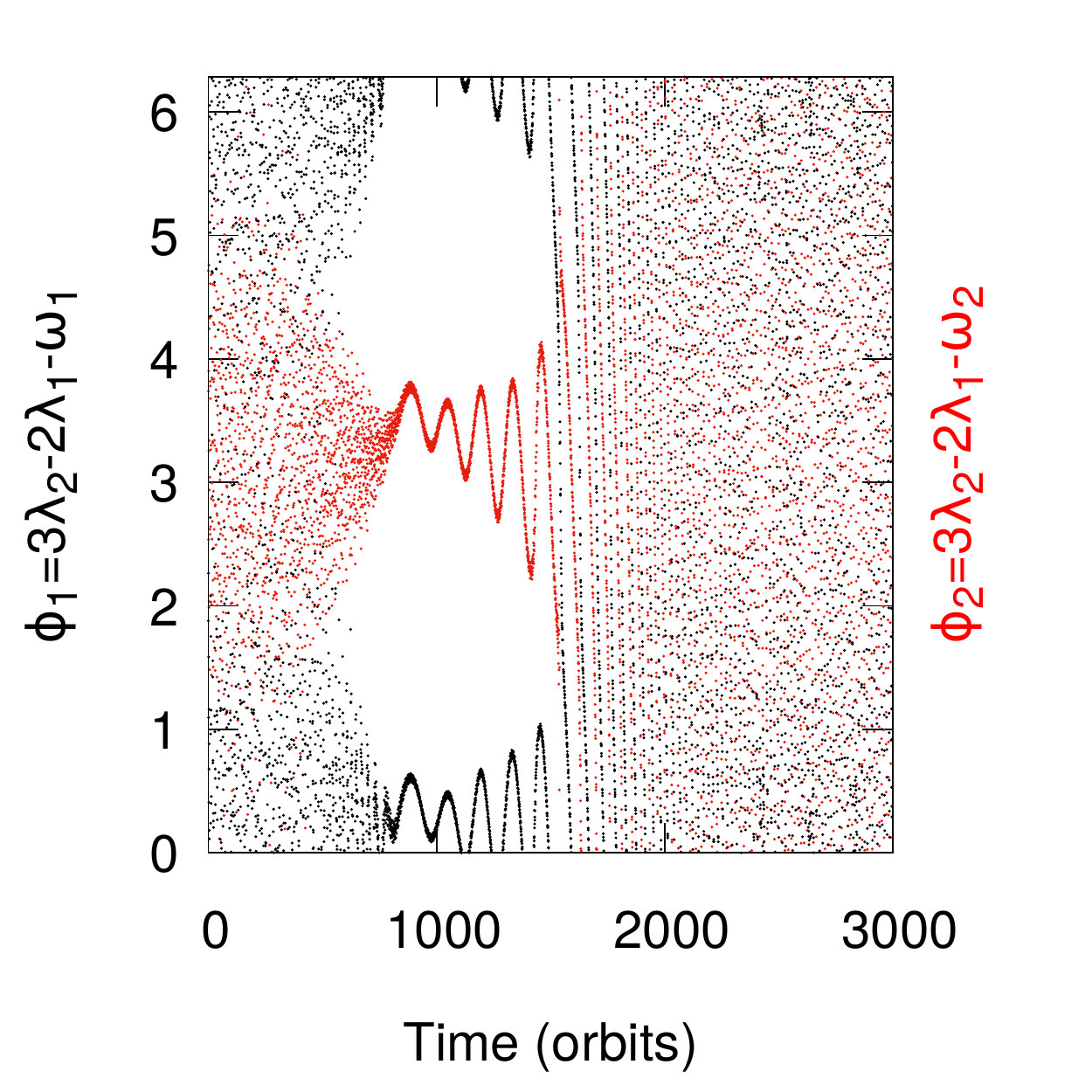}
}
}
\caption{Results of the hydrodynamical simulation of two 
super-Earths with  $q_1 = q_2 = 2.6 \times 10^{-5}$ migrating in the 
disk with local heating and cooling. }
\label{fig:evo-alpha-ss-26-26-136}
\end{figure*}

\textcolor{red}{
Next, we rerun the simulation  but switch off the planet-planet 
interaction in the calculation. In this way, the influence of 
MMR, \textcolor{red}{brought about through the gravitational influence of the planets on each other,}
 on the migration of the planets is excluded.  \textcolor{red}{Rather  we  observe
 effects} caused only by the interaction of planets with the gas 
in the disk \textcolor{red}{structured  by the wakes they produce.} 
The orbital period ratio and the semi-major axes of 
two planets in this simulation are shown in 
Figure~\ref{fig:compare-alpha-ss-26-26-136}. \textcolor{red}{These  are respectively}
illustrated 
 in the left and right panels.}

\textcolor{red}{
Finally, we run two simulations in which only a single planet with
$q = 2.6 \times 10^{-5}$ is put into the relaxed disk \textcolor{red}{at the 
initial radii} $r= 1.0$ and $r = 1.36$. The evolution of the 
semi-major axis of the planets  are shown in the 
right panel of Figure~\ref{fig:compare-alpha-ss-26-26-136}. 
\textcolor{red}{These}  are represented by the red solid and dashed lines.
\textcolor{red}{From  the orbital periods $P_{1}$ and $P_{2}$ obtained from  these two cases
we calculate the period ratio $P_{1}/P_{2}$ and plot it in the 
left panel of Figure~\ref{fig:compare-alpha-ss-26-26-136} where it  is  indicated }
by the red solid line.}

\begin{figure*}[htbp]
\centerline{
\vbox{
\hbox{
\includegraphics[width=0.8\columnwidth]{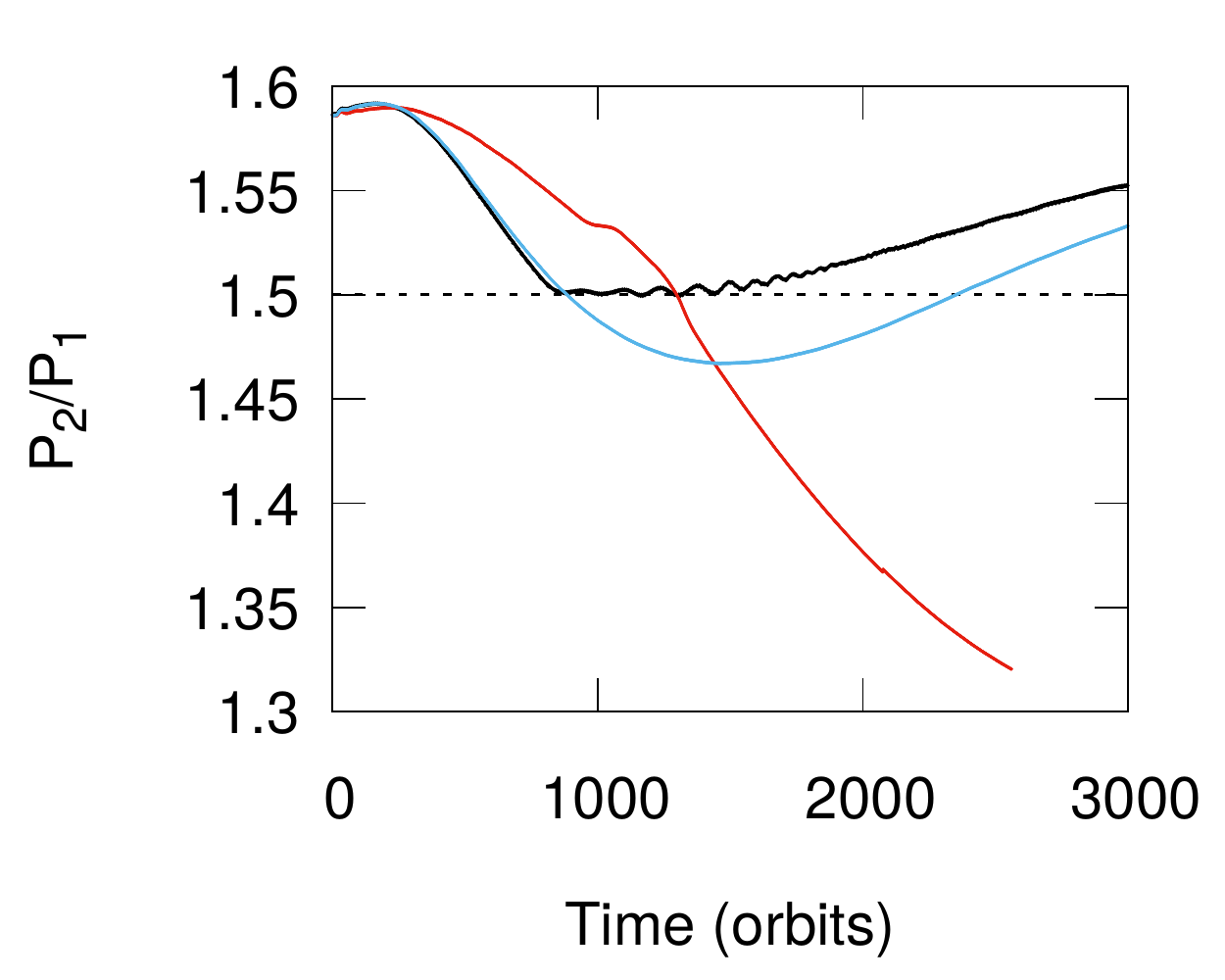}

\hspace{1.cm}

\includegraphics[width=0.8\columnwidth]{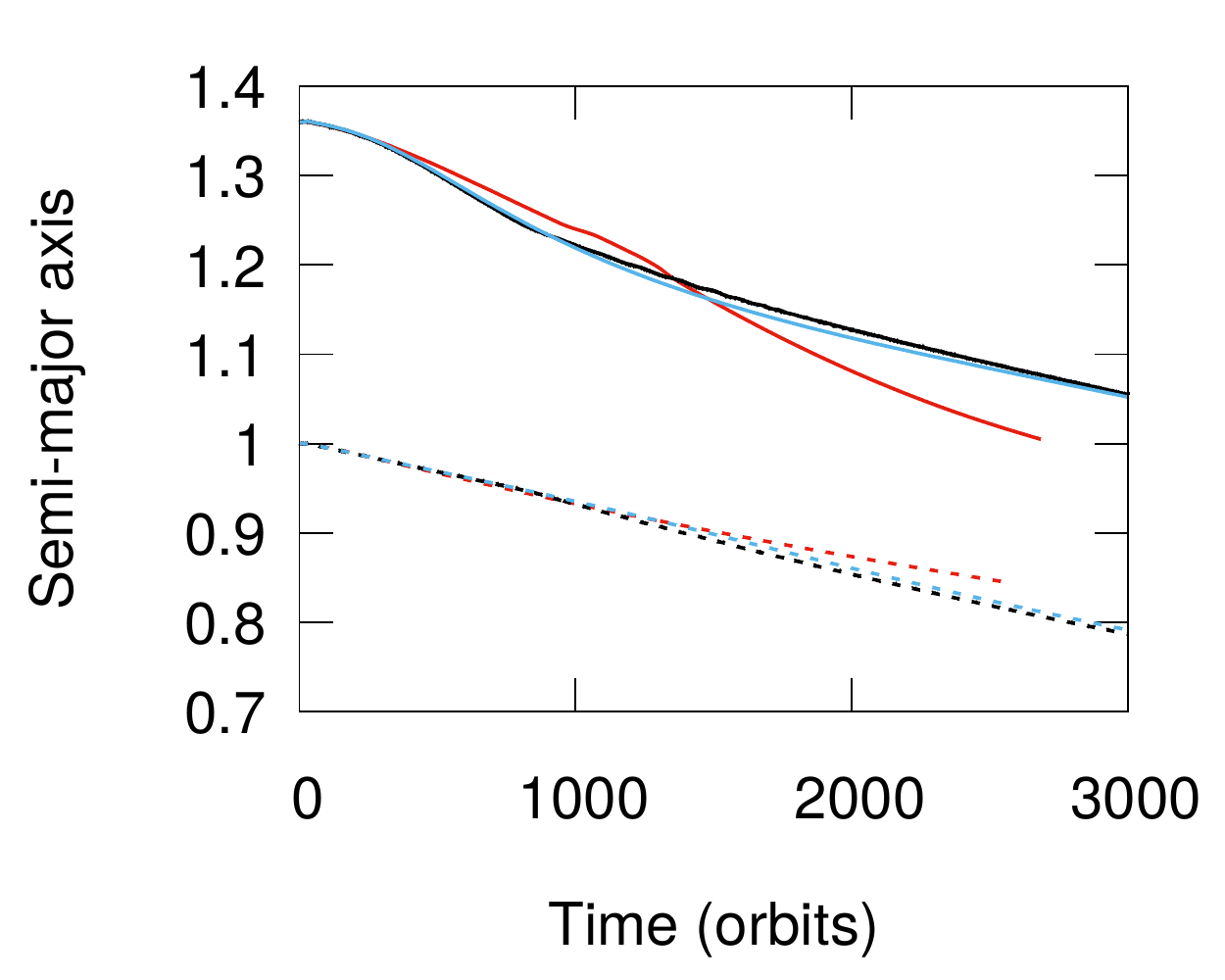}
}
}
}
\caption{Left: Comparison of the period ratio evolution of  the planets in the 
simulation of two super-Earths with  $q_1 = q_2 = 2.6 \times 10^{-5}$ 
 with (black solid line) and 
without (blue solid line) the planet planet interaction taken 
into account. The red solid line represents the 
period ratio calculated from the results of  single planet 
runs for $q = 2.6 \times 10^{-5}$ starting from  the initial positions 
 $r = 1.0$ and $r = 1.36$.
Right: Evolution of the semi-major axes of the two planets in the 
 simulations shown  in the left panel. The results of the 
simulations with and without the planet planet interaction are 
indicated by the black solid and dashed lines  and  the  blue  solid and dashed lines, respectively. 
The red dashed and solid lines represent the result of the single 
planet simulations initiated with $r= 1.0$ and $r = 1.36$, respectively.
}
\label{fig:compare-alpha-ss-26-26-136}
\end{figure*}

\textcolor{red}{
The single planet with $q=2.6\times 10^{-5}$, \textcolor{red}{initiated} at $r=1.36$
 migrates inwards. \textcolor{red}{ This is also the case when} 
 it is \textcolor{red}{initiated}  at $r=1.00$. Moreover, the planet \textcolor{red}{initiated}
at larger distance from the central star migrates faster than the 
one starting its evolution closer to the star. 
\textcolor{red}{Knowing this, one
 predicts  that the relative migration of a pair of such planets
should be convergent. \textcolor{red}{ In this context we remark that  temperature and surface density variations  in the disk
alone could produce divergent migration through their effect on the disk - planet interaction
\citep[see eg.][]{Marzari2020}. However, we do not find this here.
 Nonetheless convergent migration only persists  for 
roughly 1000 orbits,  at which point} they enter into the 3:2 MMR, 
when the interaction between the planets is retained, or }
for  1500 orbits if we switch off the gravitational interaction
between the  planets. The former situation could be considered as
an expected capture in the resonance, but the commensurability
is not mantained and the planets migrate divergently, till the
end of the simulation, \textcolor{red}{ as is also ultimately the case for which
 the interaction between the planets switched off. The ultimate divergent migration in
 the latter case must be produced by interaction with the perturbed disk
 with no resonant capture taking place. In this context we remark that the 
 eccentricities of both planets are always small in this case}}     

\textcolor{red}{
We have  also performed  a simulation for a case with 
$q_1 = q_2 = 1.95\times 10^{-5}$. The convergent relative migration 
rate of two planets is slower at the beginning of the simulation \textcolor{red}{ in this case,} 
which favours \textcolor{red}{ultimate}  repulsion between the planets. \textcolor{red}{Indeed, the 
planets are } found to  migrate divergently even  before arriving at the 3:2 MMR. 
We will return to this case in a summary at the end of this
Section.}
\textcolor{red}{
\subsection{A case with the unequal mass planets}
Here, we perform \textcolor{red}{a simulation} in which the inner planet is less
massive than the outer one. The inner planet has 
 $q_1 = 1.3 \times 10^{-5}$ and \textcolor{red} {is initiated} at 
$r_1 = 1.0$. The outer planet has the \textcolor{red}{same value}
 of $q_2 = 2.6 \times 10^{-5}$ and 
\textcolor{red}{ the  initial  location  $r_2 = 1.36$, as
in our previous simulation.}
The same disk model \textcolor{red} {as was adopted in the previous simulation}  
was  relaxed for 800 orbits before adding the planets .}

\textcolor{red}{
The evolution of the period ratio, semi-major axes, eccentricities 
and the 4:3 resonant angles are illustrated in 
Figure~\ref{fig:evo-alpha-ss-13-26-136}. In this case, 
the \textcolor{red}{early} relative
migration is convergent and much faster than in the equal planet
case. As a consequence, the planets passed through 
\textcolor{red}{the} 3:2 MMR at 
about $t \sim$ 500 orbits and at $t \sim$ 1200 orbits they arrived 
at the 4:3 MMR. The eccentricities of the  two planets get excited 
when the planets are passing through the 3:2 MMR and when arriving 
at the 4:3 MMR. After 3000 orbits, the period ratio is \textcolor{red}{slowly} 
increasing and the eccentricities \textcolor{red}{slowly}  
decreasing. Both 4:3 resonant 
angles \textcolor{red} { librate until}  the end of the simulation. 
During the \textcolor{red}{entire} calculation, the inner and outer planets migrate 
inwards.}

\begin{figure*}[htbp]
\vbox{
\hbox{
\includegraphics[width=0.5\columnwidth]{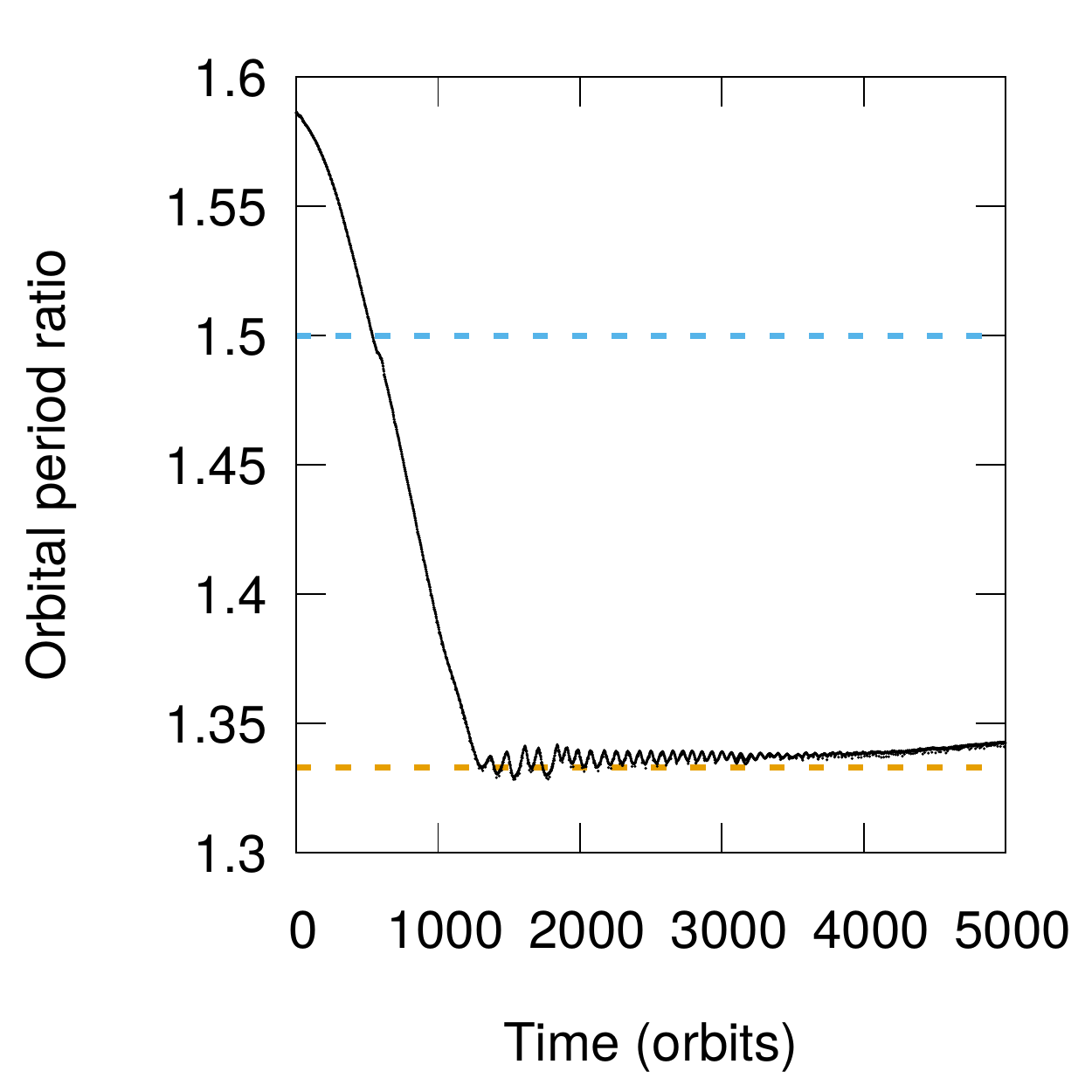}
\includegraphics[width=0.5\columnwidth]{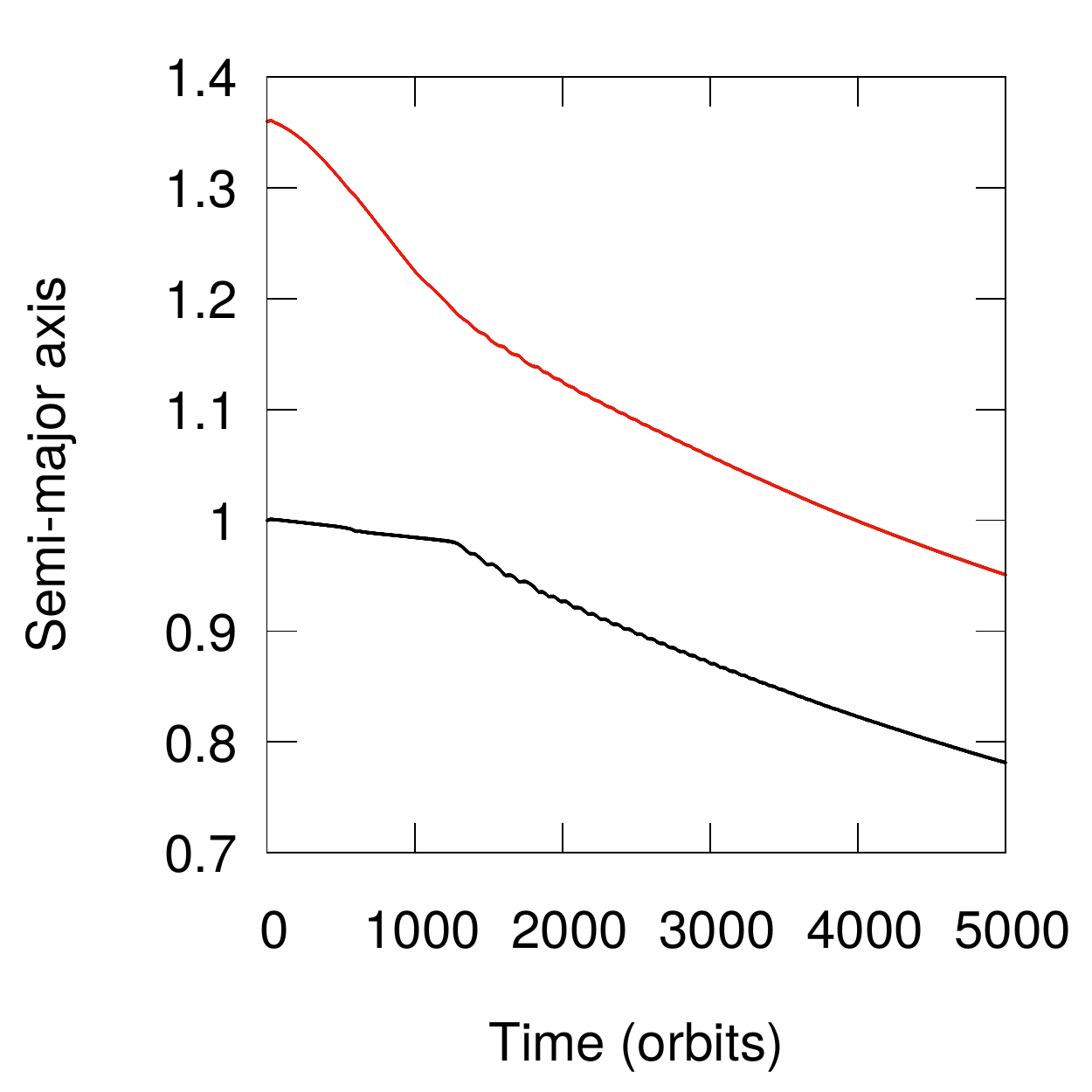}
\includegraphics[width=0.5\columnwidth]{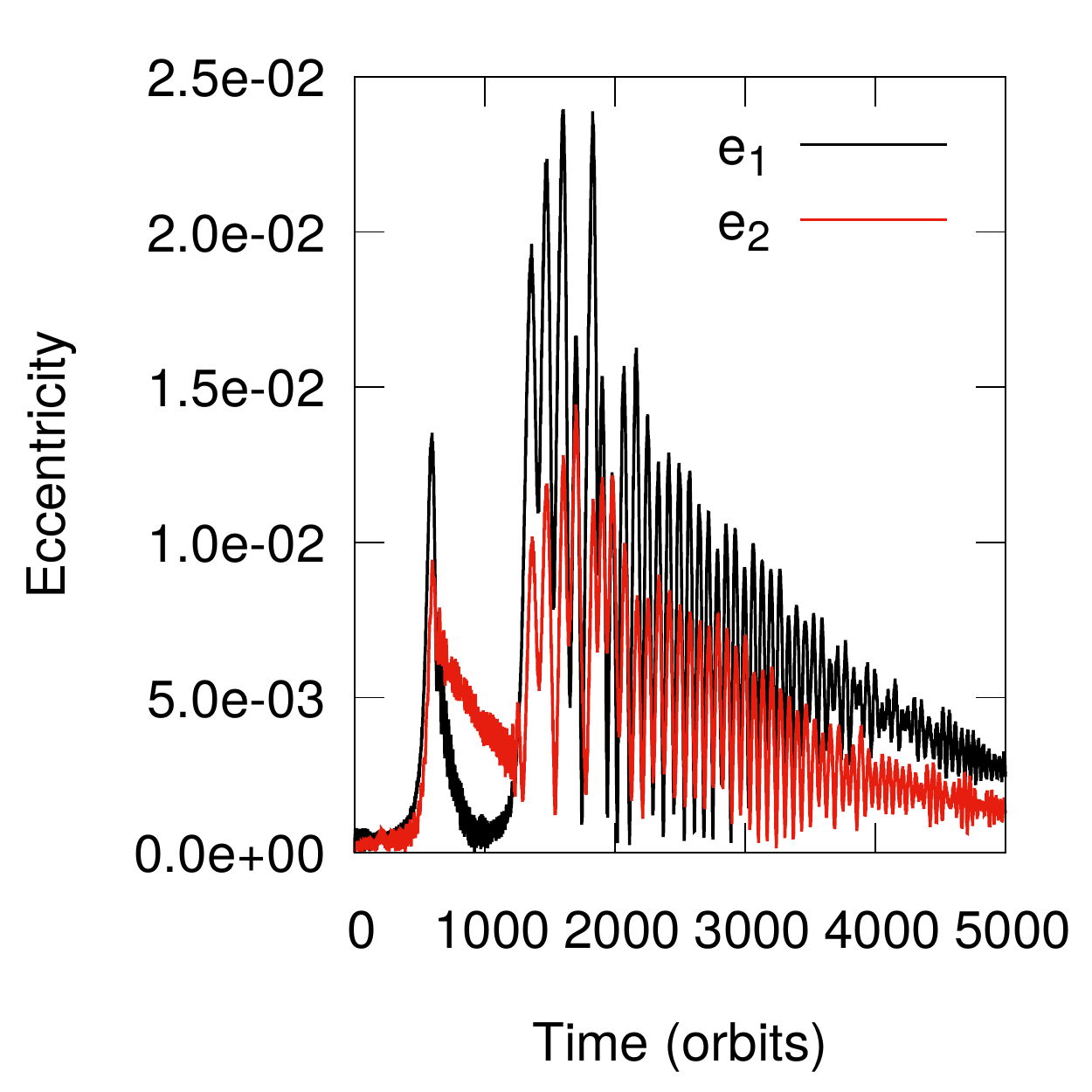}
\includegraphics[width=0.5\columnwidth]{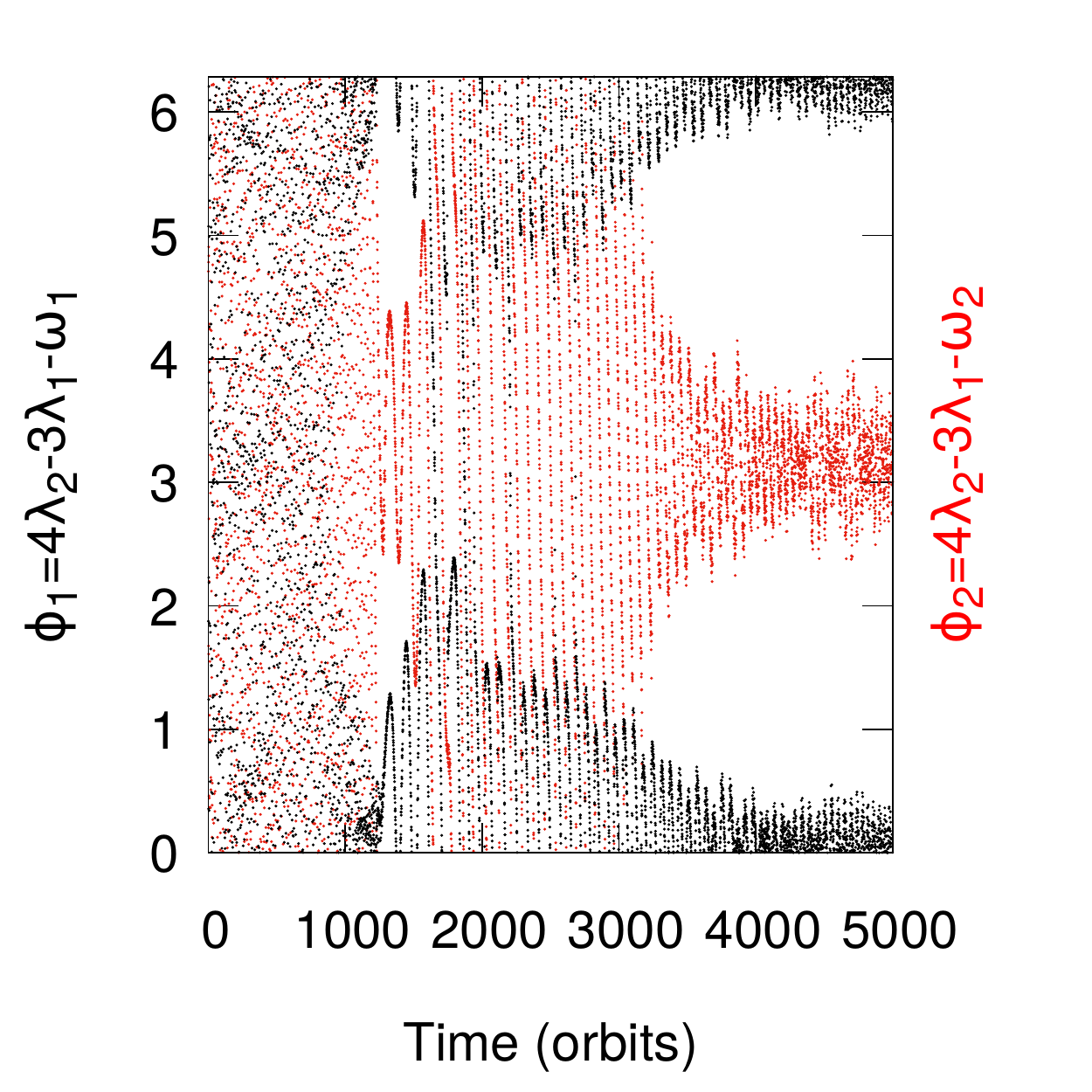}
}
}
\caption{Results of the hydrodynamical simulation of the evolution of 
two super-Earths with $q_1 = 1.3 \times 10^{-5}$ and $q_2 = 2.6 \times 10^{-5}$ 
migrating in the  disk with local heating and cooling.
}
\label{fig:evo-alpha-ss-13-26-136}
\end{figure*}

\textcolor{red}{
\textcolor{red}{As in the equal mass planet case, we ran}  a
simulation in which the planet planet interaction is switched 
off. The orbital period ratio and the semi-major axes of the two 
planets in this simulation are shown in 
Figure~\ref{fig:compare-alpha-ss-13-26-136}. They are \textcolor{red}{
respectively indicated by 
the blue solid  line in the left panel  and  the blue solid and dashed lines 
in the right panels.
We  also ran  simulations for which single planets with 
$q = 1.3 \times 10^{-5}$ and $q = 2.6 \times 10^{-5}$ were inserted in}
the relaxed disk with the initial positions at $r= 1.0$ and 
$r = 1.36$, respectively.}

\textcolor{red}{The evolution of the \textcolor{red}{ semi-major axes of 
the planets in these simulations} is shown in the right panel of 
Figure~\ref{fig:compare-alpha-ss-13-26-136},  and  are represented 
by the red dashed and solid lines. 
 Based on the orbital periods 
$P_{1}$ and $P_{2}$ respectively obtained from the single planet 
cases \textcolor{red}{initiated with} $r= 1.0$ and $r = 1.36$, we \textcolor{red}{determine}
 the period ratio 
$P_{1}/P_{2}$ and \textcolor{red}{ plot}  it in the left panel of 
Figure~\ref{fig:compare-alpha-ss-13-26-136} \textcolor{red}{ where it} is indicated by the red 
solid line.}

\textcolor{red}{ When the planet-planet interaction was removed
the early convergent migration  was rapid enough to enable the planets to pass through the 
4:3 and 5:4  commensurabilities. However, on approach to the 6:5 commensurability
the migration became divergent and the period ratio subsequently  increased slowly. This evolution is
of the same character as that in the equal mass case.}
\begin{figure*}[htbp]
\centerline{
\vbox{
\hbox{
\includegraphics[width=0.8\columnwidth]{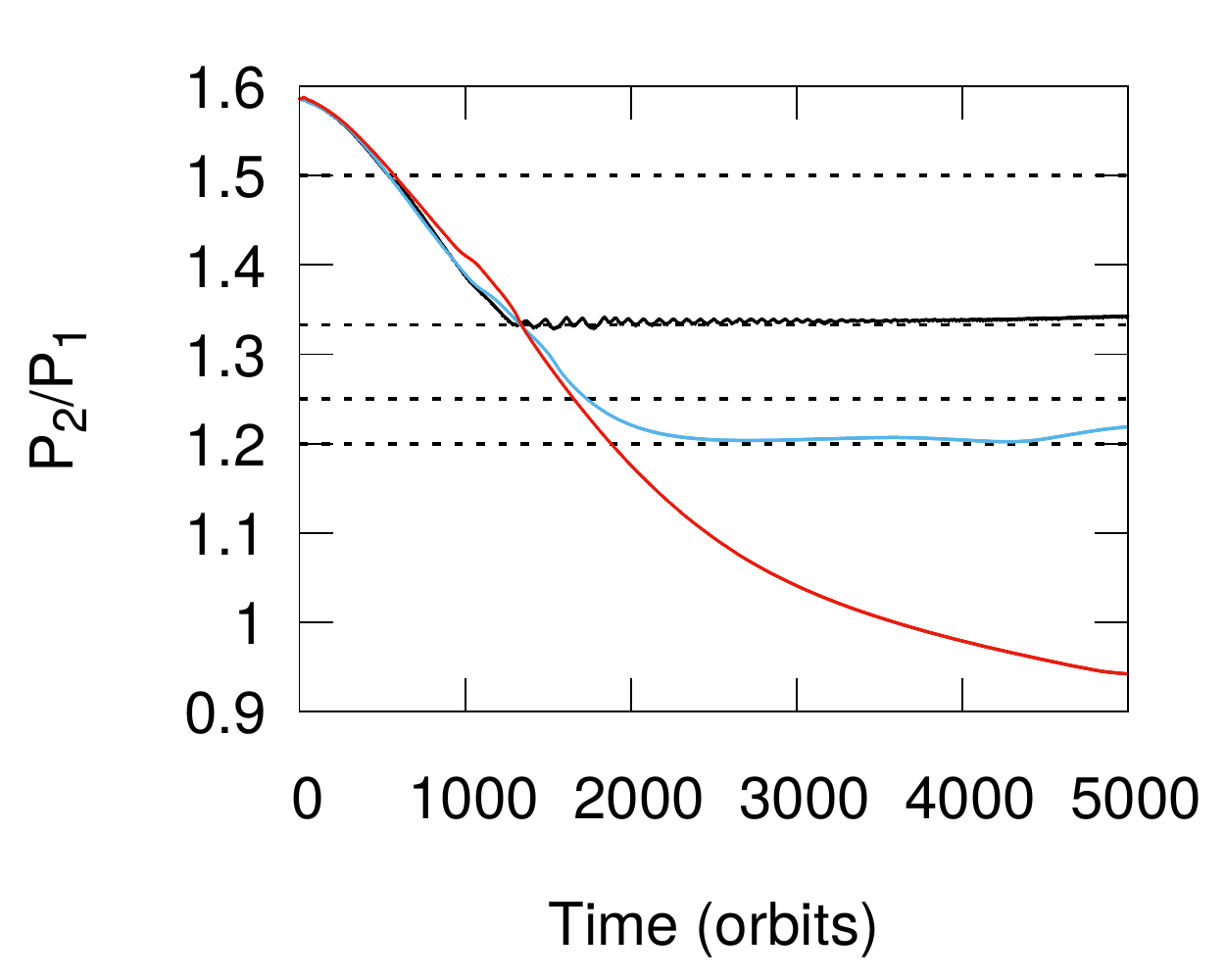}

\hspace{1.cm}

\includegraphics[width=0.8\columnwidth]{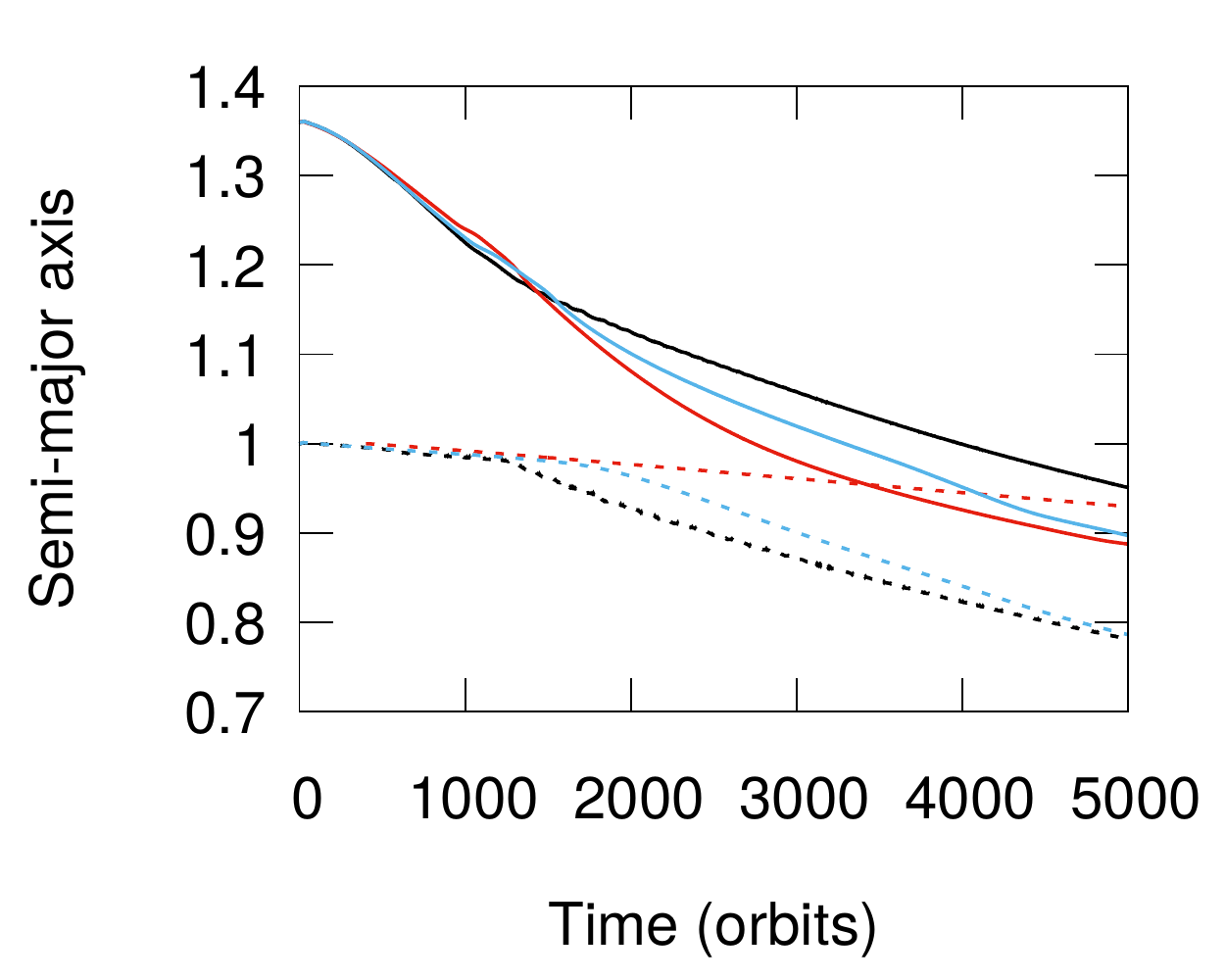}
}
}
}
\caption{Left: Comparison of the period ratio of planets in the 
simulations of the evolution of two super-Earths with 
$q_1 = 1.3 \times 10^{-5}$ and $q_2 = 2.6 \times 10^{-5}$ migrating 
in a disk with local heating and cooling with and 
without the planet-planet interaction.  These are respectively 
indicated by the black and blue solid lines while the red solid 
line represents the period ratio  based on results 
from  the single planet cases for $q = 1.3 \times 10^{-5}$ initiated at
 $r = 1.0$ and $q = 2.6 \times 10^{-5}$ initiated at 
$r= 1.36$.
Right: Evolution of the semi-major axes of two planets in the 
 simulations illustrated in the left panel. The results from the 
simulations with and without the planet planet interaction are respectively
denoted by the black and  blue dashed and solid lines. 
The red dashed and solid lines show  the results of the single 
planet simulations with $q = 1.3 \times 10^{-5}$ and 
$q = 2.6 \times 10^{-5}$ respectively.
}
\label{fig:compare-alpha-ss-13-26-136}
\end{figure*}

\textcolor{red}{
\subsection{The effectiveness of the repulsion between
planets in a disk with \textcolor{red}{ local heating and cooling } 
}
We have demonstrated  the repulsion between
planets with a mass in the super-Earth range in our simulations.
\textcolor{red} {This is due to the disturbance in the disk induced by one planet 
interacting with the other. As waves are launched by the planets in this interaction
we describe  this as 
 wave planet interaction.} The repulsion \textcolor{red}{ was found to be}
very strong between planets with  equal mass as can
be clearly noticed in Figure~\ref{fig:alpha-ss-crit1crit2}
(left panel). 
The initial relative migration in those cases is slow favouring the 
mechanism at work.  We have applied the criteria for the 
effectiveness of the repulsion derived in 
Section~\ref{Coorbgaps} 
 that were  
applied to our previous simulations,  and plotted in Figure~\ref{fig:crit1crit2} (see the discussion in 
Section~\ref{sec:innermass} for details),
to the three  \textcolor{red}{simulations discussed here.}
 The results are shown in Figure~\ref{fig:alpha-ss-crit1crit2}
(middle and right panels). \textcolor{red}{They are consistent with the criteria
we derived for the repulsion mechanism to work as they were for our previous simulations.}
\begin{figure*}[htbp]
\vbox{
\hbox{
\includegraphics[width=0.7\columnwidth]{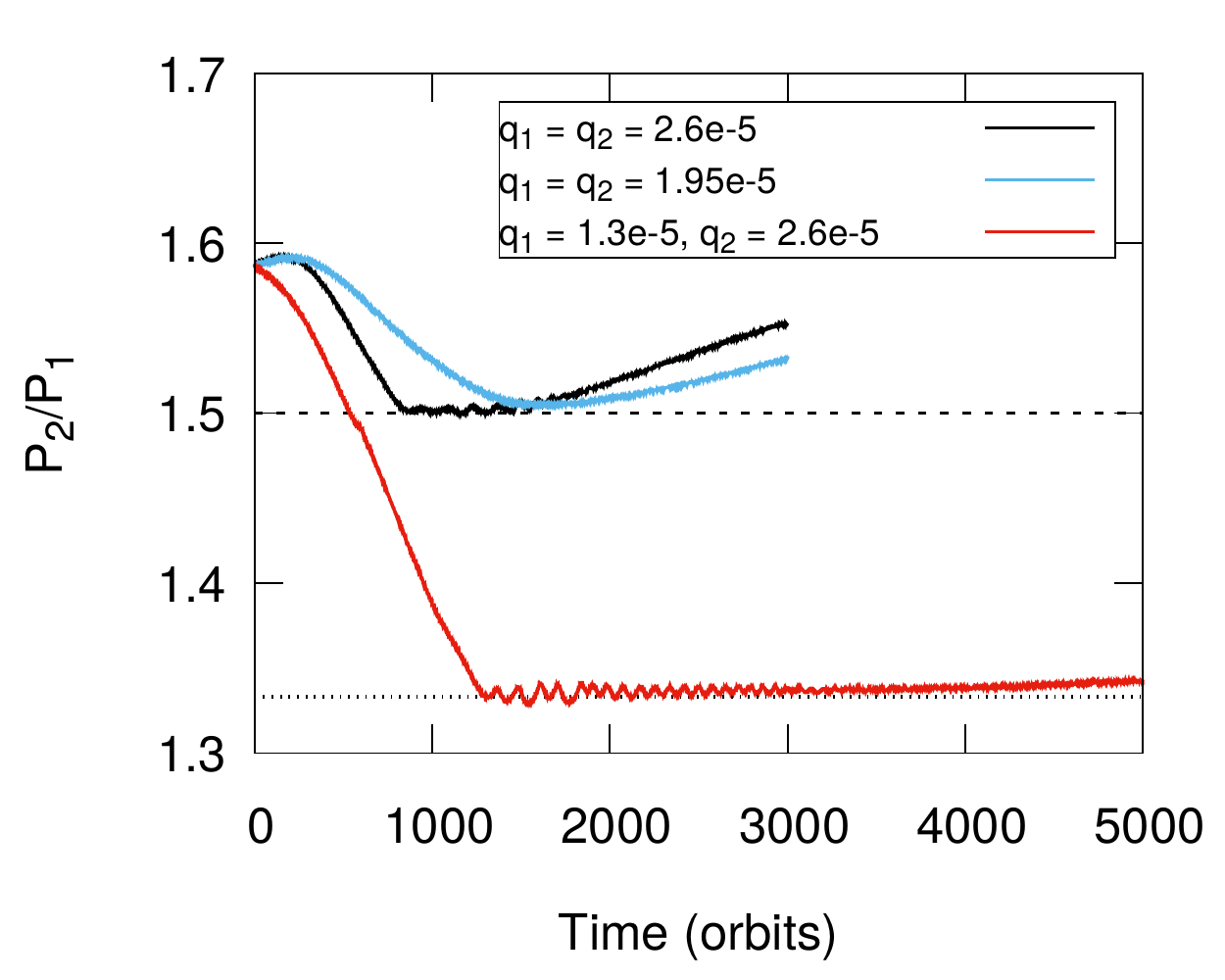}
\includegraphics[width=0.7\columnwidth]{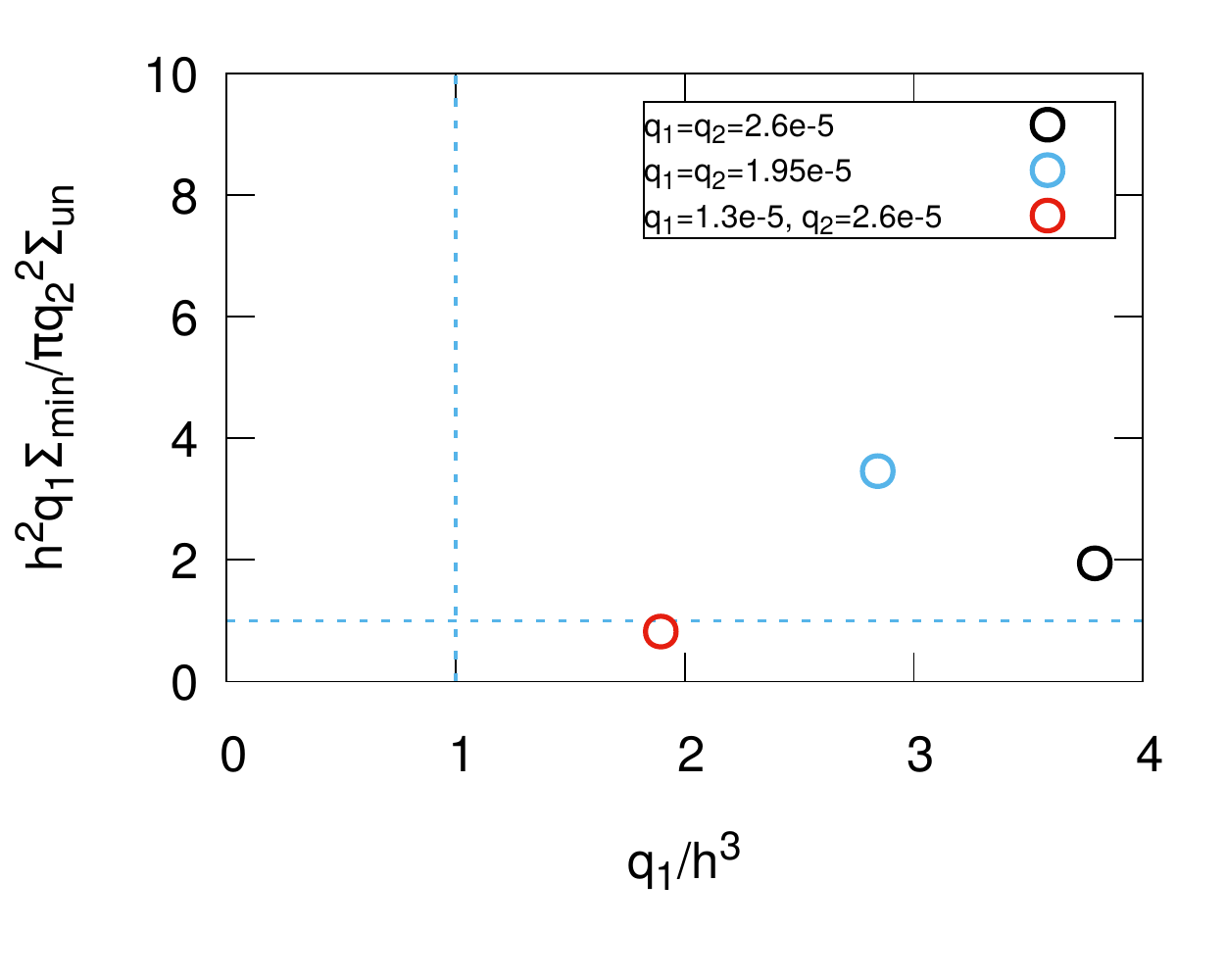}
\includegraphics[width=0.7\columnwidth]{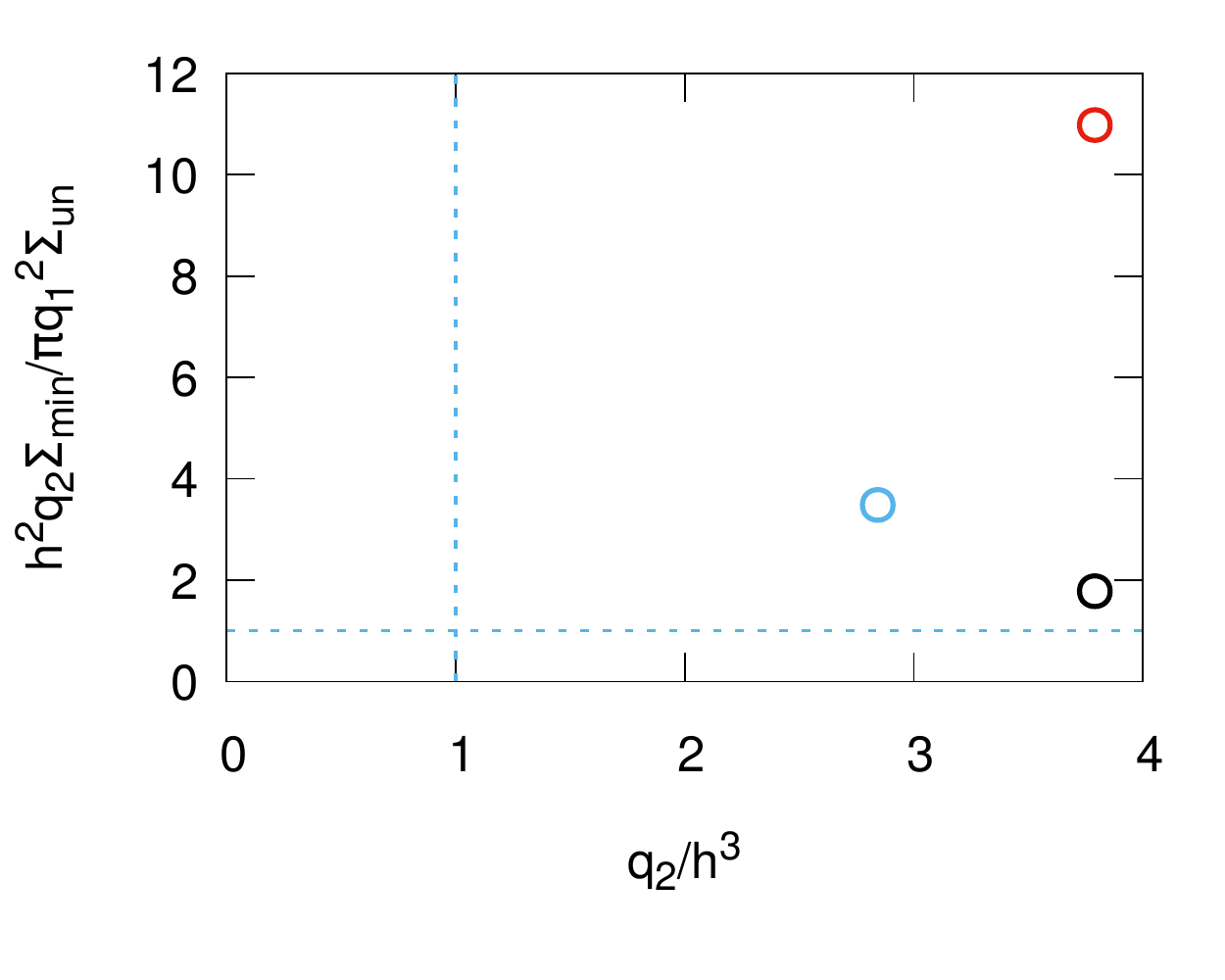}
}
}
\caption{Left:The period ratio of the two planets as a function of time 
for the simulations of two super-Earths with mass ratios of 
$q_1 = q_2  = 2.6 \times 10^{-5}$ (black line), 
$q_1 = q_2 = 1.95 \times 10^{-5}$ (blue line) and 
$q_1 = 1.3 \times 10^{-5}, q_2 = 2.6 \times 10^{-5}$ (red line) 
migrating in a disk with local heating and cooling. 
The horizontal dashed and dotted lines indicate the locations of the 3:2 
and 4:3 resonances, respectively. 
Middle and Right: The results of the simulations illustrated in the left panel
are plotted  in the planes used to indicate the effectiveness 
of the repulsion due to wave planet interaction as was  
done for our prior simulations in Figure~\ref{fig:crit1crit2} (see the discussion in 
Section~\ref{sec:innermass})
}
\label{fig:alpha-ss-crit1crit2}
\end{figure*}
}


\section{Discussion and Conclusions}
\label{sec:discussion}

In this paper we have demonstrated that wave planet repulsion can
be effective in converting convergent migration to divergent 
migration as two planets  in the super-Earth range, embedded in a 
thin low viscosity gaseous protoplanetary disk, approach a 
first-order commensurability.
In Section~\ref{sec:results} we found this for an inner planet with 
mass ratio $q_{1}=1.3\times 10^{-5}$ and an outer planet with mass 
ratio $q_{2}=1.185\times 10^{-5}$ in a disk with aspect ratio, 
$0.02,$ and kinematic viscosity, $1.2\times 10^{-6}.$ This occurred 
for  both a power law surface density profile and one allowing for  
an inner cavity.

In order to exhibit effects arising solely because of the presence 
of two planets we performed simulations of an isolated outer planet 
in Section~\ref{sec:single}.
Because in our simulations the planets always make partial gaps
we found that standard type I migration does not apply even when 
scaled with the surface density in the gap.
We developed  a new fit to the migration rate applicable to local
power law surface density profiles 
(Equation~(\ref{eq:numerical-fit})).
In Section~\ref{sec:twoplanetcase} we used this result to show the 
presence of an additional torque acting on the outer planet, which 
we identified as being due to wave planet interaction, and 
characterised its magnitude.

A simplified  description of wave planet interactions was developed 
in Section~\ref{Coorbgaps}. This was based on density waves emitted 
by one planet being absorbed in the horseshoe region of the other 
and then transmitted to it, thus being able to supply an additional 
torque needed to change from  convergent to divergent migration.
Approximate conditions for this to be effective were given 
(criteria~(\ref{eq:criterion1b}) and (\ref{eq:criterion2}) and their 
counterparts for the other planet).
These imply that the planets should be massive enough to make a 
partial gap but that this should not be so deep that there is 
inadequate material in  horseshoe regions to be able to transfer 
the angular momentum transported by waves to the associated planet.
They were verified for our simulations in Section~\ref{sec:innermass} 
for a range of mass ratios $(1.185 - 2.6)\times 10^{-5}$ adopted
for both  planets.

We investigated the dependence of the rate of initial convergent 
migration in Section~\ref{sec:repulsion}, showing that it increased 
on increasing the disk surface density or the mass ratio of the 
outer planet relative to the inner one. Increasing the initial 
rate of convergence resulted in the planets approaching each other
more closely before switching to divergent migration. 

In addition we checked that the angular momentum being transported 
between the planets was consistent with theoretical expectation in 
the case of density waves generated by the planets' interaction 
with the disk in Section~\ref{sec:horseshoe}.
It was also shown that this was consistent with the magnitude of 
the torque required to convert from convergent to divergent migration.

In Sections~\ref{sec:surfacedensity} - \ref{sec:lowdensity} we 
established the robustness of the mechanism  to  changes of the form 
of the surface density profile, changes to the equation of state and 
a correction to allow for the effect of self-gravity. \textcolor{red}{ 
\textcolor{red}{ In addition we investigated the effects of}
 significant  changes  to the magnitude of  the surface density in the disk in
which the planets are embedded by considering reductions in this by more than an order
of magnitude obtained through application of a uniform scaling factor.}
\textcolor{red}{ Finally in Section~\ref{sec:viscousdisk} we demonstrated 
the effectiveness of the
repulsion between planets in a viscous disk
 \textcolor{red}{ in which the  temperature was determined
by a local balance between heating and cooling.}}

Thus we have investigated the mechanism underlying the repulsion 
between two planets migrating in the disk that enables the 
transition between convergent and divergent migration in some detail. 
It has been found that density waves emitted by the planets and the 
horseshoe drag play an  important role  in the process.

\textcolor{red} { Significantly the phenomenon continues to occur when  the gravitational interaction between the planets is 
omitted which supports this view.}  A close approach to strict commensurability is not necessary to 
induce divergent migration. However, the mean-motion resonances 
may play a significant role in the process of planet repulsion.
Our findings provide one of the possible reasons, why there are 
not as many observed pairs of planets locked in \textcolor{red}{strict}  mean-motion 
resonances, as could be inferred from discussions taking into 
account only standard type I migration.
\textcolor{red}{ However, a full assessment requires  an extensive
study incorporating the entire final stages of the evolution of 
proptoplanetary disks that also incorporates the formation of the planets
which we hope can be undertaken in future.}

\acknowledgments
We  would like to thank Kazuhiro Kanagawa for the stimulating discussions 
in the early stages  of the preparation of this paper. We are indebted to
Franco Ferrari for his continuous support in the development of our 
computational techniques and computer facilities \textcolor{red}{ as well as
for his help in performing the calculations on GPU machine in his lab in 
Szczecin}. 
J.C.B.P. thanks the 
Faculty of Mathematics and Physics, University of Szczecin for hospitality. 
We would like to acknowledge the support by Polish National Science Center 
MAESTRO grantDEC-2012/06/A/ST9/00276.  \textcolor{red}{ Most of the  simulations}  were  
performed 
on HPC cluster HAL9000 of the Computing Center of the Faculty of Mathematics 
and Physics at the University of Szczecin. \textcolor{red}{ However, some calculations
would not have been completed without support from  DAMTP, University
of Cambridge, and E.S. is grateful for having been granted access to  
computational facilities. 
} 

\newpage
%

\vspace{5mm}


\software{FARGO3D \citep{BenitezM2016}} 



\newpage
\appendix
\textcolor{red}{
\section{The migration of two super-Earths in an adiabatic disk}
For the adiabatic case we adopt the equation of state in the form
\begin{eqnarray}
\label{eos-adia}
P=(\gamma-1)e\Sigma
\end{eqnarray}
}
\textcolor{red}{
\noindent where, $P$ is the vertically integrated pressure, $\gamma$ 
is the adiabatic index and $e$ is the specific internal energy.  \textcolor{red}{As an objective is to probe the 
effect of there being strict wave action conservation for linear waves, we follow \citet{Mira2019} and set}
the value of $\gamma$  to be 1.001.  \textcolor{red}{On account of being}  close to unity.
this has the effect that the equation of state approaches the limit  
in which the temperature is conserved on fluid elements. Physically 
in this limit the  temperature does not change on account of the very large 
number of degrees of freedom within the gas. \textcolor{red}{In this limit the temperature does not 
change in a Lagrangian sense in comparison with the locally-isothermal 
equation of state for which it does not change in an Eulerian sense.}}

\textcolor{red}{
The remaining details of the numerical setup are the same as in the 
flagship case for which
$q_{1}=1.3 \times 10^{-5}$ and $q_{2}=1.185 \times 10^{-5}$
and the initial surface density profile was
determined by Equation~(\ref{disk}) with $\alpha=0.5$ and
$\Sigma_0=6\times 10^{-5}$ \textcolor{red}{ (see Section \ref{innertrap}).}
However, we do not used the killing-zone treatment near the
boundaries of the disk  in this simulation.}

\begin{figure*}[htb!]
\centerline{
\vbox{
\hbox{
\includegraphics[width=0.30\columnwidth]{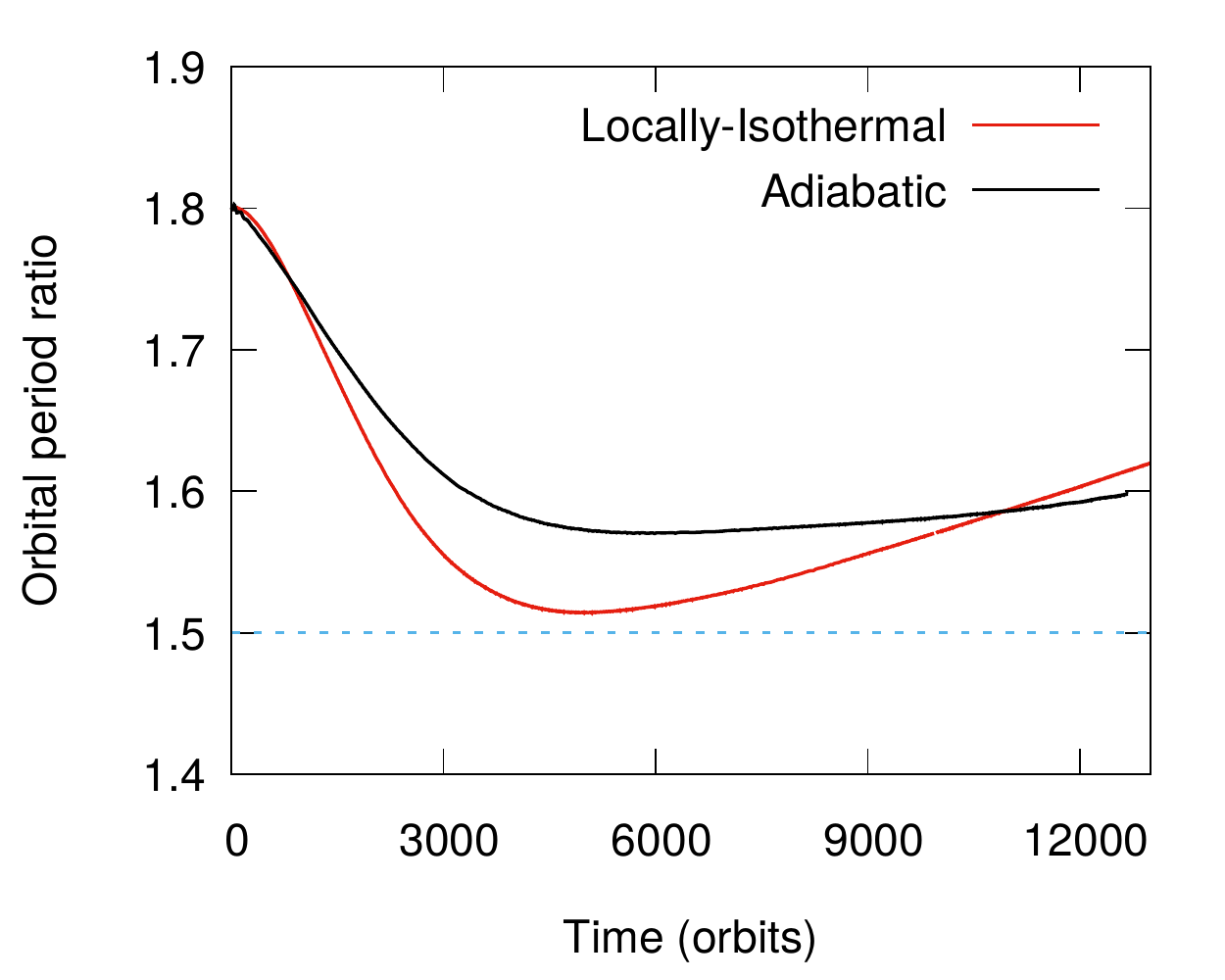}
\includegraphics[width=0.30\columnwidth]{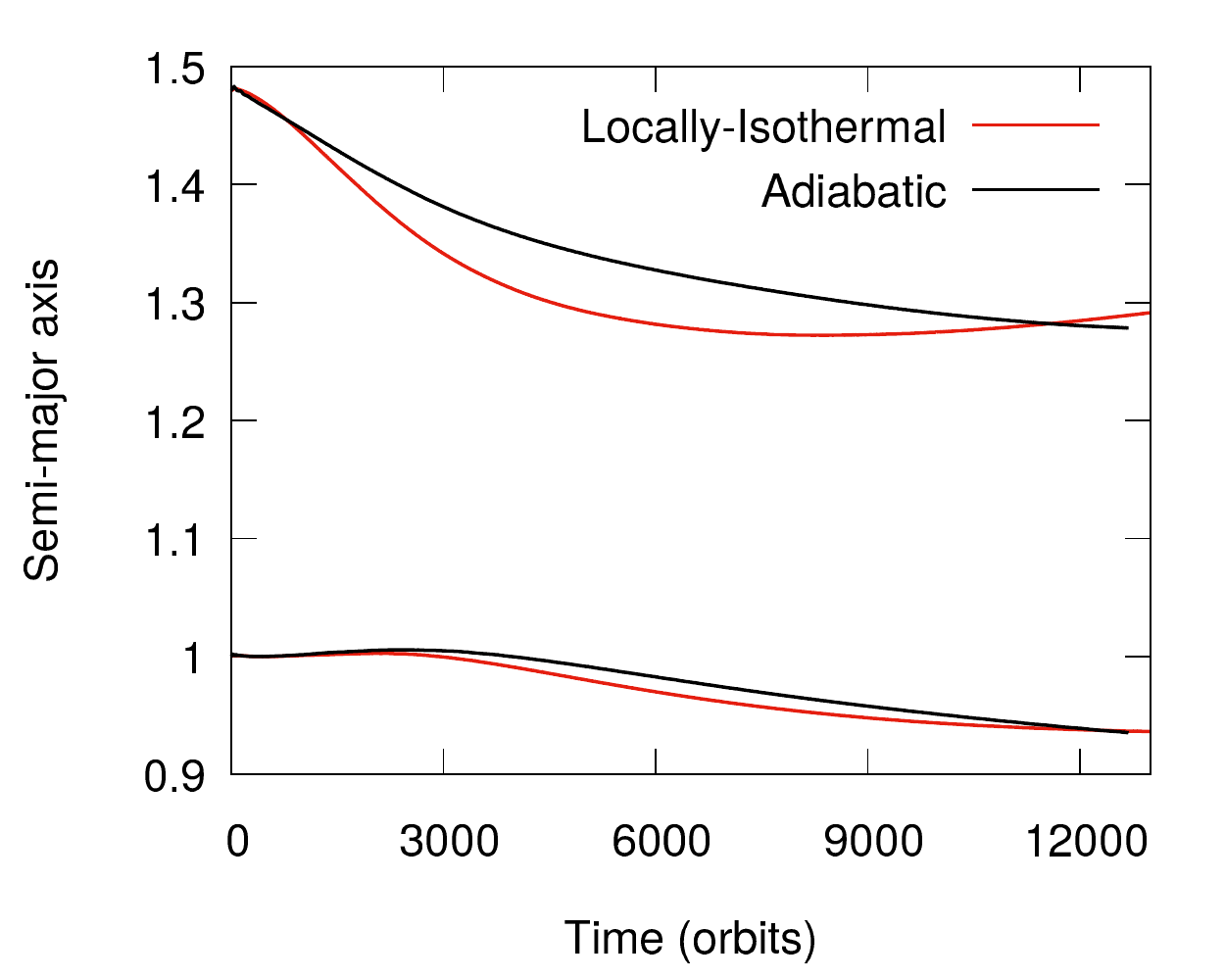}
\includegraphics[width=0.30\columnwidth]{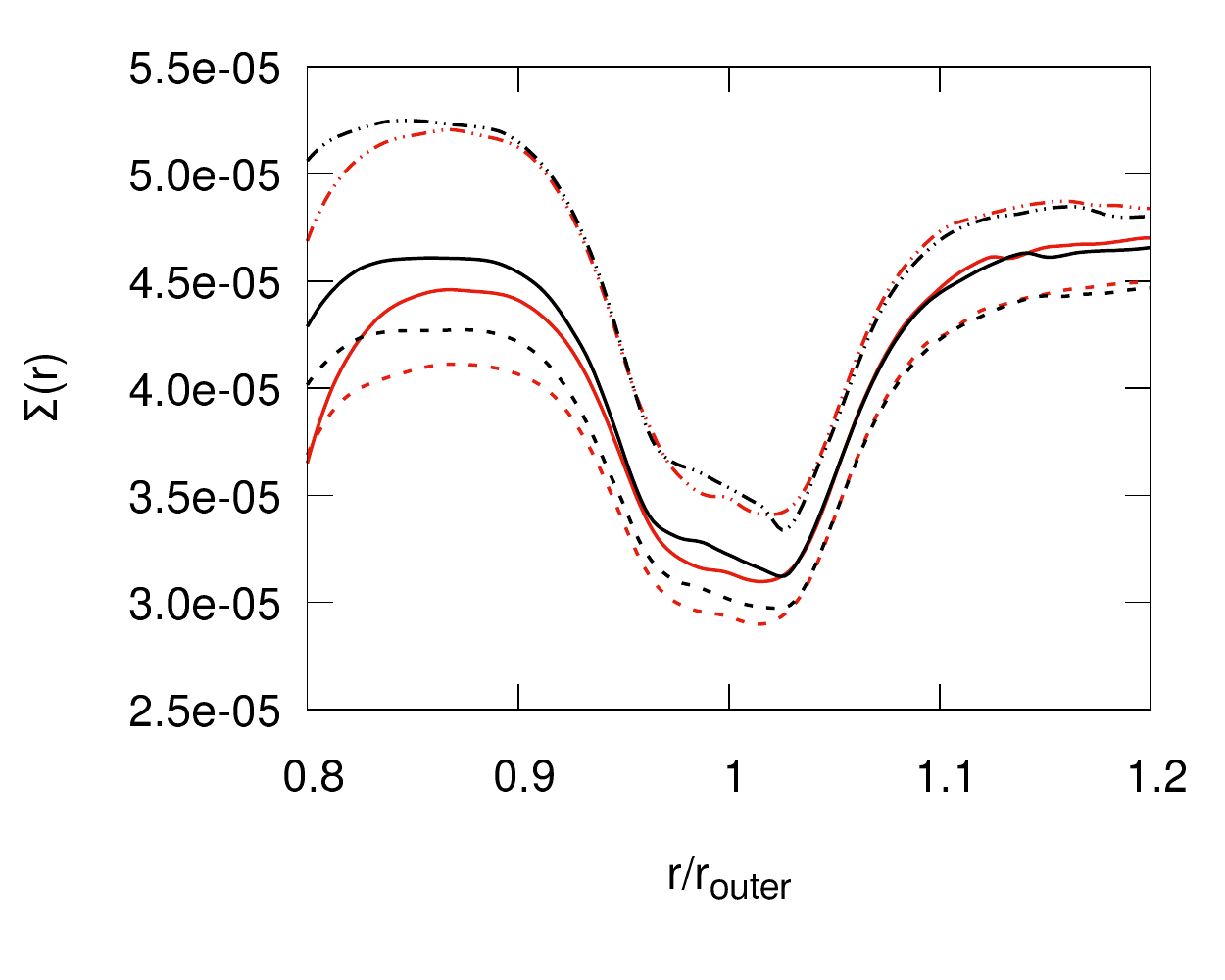}
}
}
}
\caption{Left and middle: The orbital period ratio and semi-major
axes of the  two planets in the simulations with a locally-isothermal
disk and an adiabatic equation of state. Right: The surface density 
profile in the vicinity of the outer planet in the simulations with 
a locally-isothermal disk (red lines) and the disk with an adiabatic 
equation of state (black lines) at t=3000, 6000 and 9000 orbits, 
these being respectively indicated by the dashed-dotted lines, solid 
lines and dashed lines.}
\label{fig:compare-adia-iso}
\end{figure*}

\textcolor{red}{
In Figure~\ref{fig:compare-adia-iso}, we show a comparison of 
the evolution of the orbital period ratio, semi-major axes of two 
planets and the surface density profile in the vicinity of
the outer planet in the simulation with the locally-isothermal 
disk with the corresponding quantities obtained for the adiabatic 
case. It is clear that the two planets start to undergo divergent 
migration  in the adiabatic case when the period ratio is around 
1.58 at $t \sim 6000$ orbits. From a comparison of the orbital 
period ratios presented in the left panel of 
Figure~\ref{fig:compare-adia-iso}, we can see
that in the adiabatic disk, the divergent migration occurs later
and when the distance between two planets is larger than in the
locally-isothermal disk. As shown in the middle panel of
Figure~\ref{fig:compare-adia-iso}, the migration
of the inner planet in both disks is similar while,
the migration rate of the outer planet is slower in the adiabatic
disk than in the locally-isothermal one. In the adiabatic case 
the outer planet migrates inwards till the end of the simulations, 
although at that stage it is very slow.
While in the locally-isothermal case, it migrates outwards at the
final stage of the calculation.}

\textcolor{red}{
In the right panel of Figure~\ref{fig:compare-adia-iso}, the
surface density profiles at t=3000, 6000 and 9000 orbits in two
cases are shown. It is clear that at t=3000 orbits, the
surface density profiles in the vicinity of the outer planet
in both simulations are similar, though their forms at the
bottom of the partial gap opened by the planet differ slightly.
The gap is slightly deeper in the locally isothermal case with 
difference between that and the adiabatic case being larger 
in the inner region of the gap.
This is most noticeable at t=6000 orbits, when the divergent
migration starts and at t=9000 orbits, in the adiabatic case, 
the maximum deviation of the  surface density in the partial 
gap from that found in the locally-isothermal case still occurs 
in the inner region.}

\textcolor{red}{
The above features indicate differing behaviour in the coorbital 
region associated with different temperature profiles 
(see the left panel of Figure~\ref{fig:temperature}). 
In  the case of a locally isothermal equation of state the 
temperature profile is simply determined by location.
On the other hand in the adiabatic case, in the limit 
$\gamma \rightarrow 1$, it will depend on the history of 
the material and possibly affected by phase mixing. 
The explicit comparison between those two cases at t=3000 orbits
is shown in the left panel of Figure~\ref{fig:temperature}. 
When the adiabatic equation of state is adopted the temparature 
in the gaps is reduced below the background. This is expected 
if the planet migrates inwards dragging its horseshoe region 
with it. Another feature of the temperature profile, present 
beyond the outer gap edge, namely the increase of the temperature 
above the background, is due to the inner material passage 
through the separatrix to the outer disk. 
In the right panel of Figure~\ref{fig:temperature} we present
a contour plot of the temperature in the horseshoe region of the
outer planet at t=3000 orbits.
The conditions in the gap are very smooth indicating that 
viscous and numerical diffusion have dealt well with phase 
mixing in the horseshoe region making the temperature quite 
uniform.
The features in the coorbital dynamics, described here, 
which are the consequences of the conservation of entropy 
lead to differences in the horseshoe drag that are not 
easy to determine and in turn to a different torque that 
we calculate.
Indeed such effects seem to slow down the convergent migration 
of the system at an early stage in the adiabatic case  leading 
to an onset of divergent migration at larger period ratios.}
\begin{figure*}[htb!]
\vbox{
\hbox{
\includegraphics[width=0.35\columnwidth]{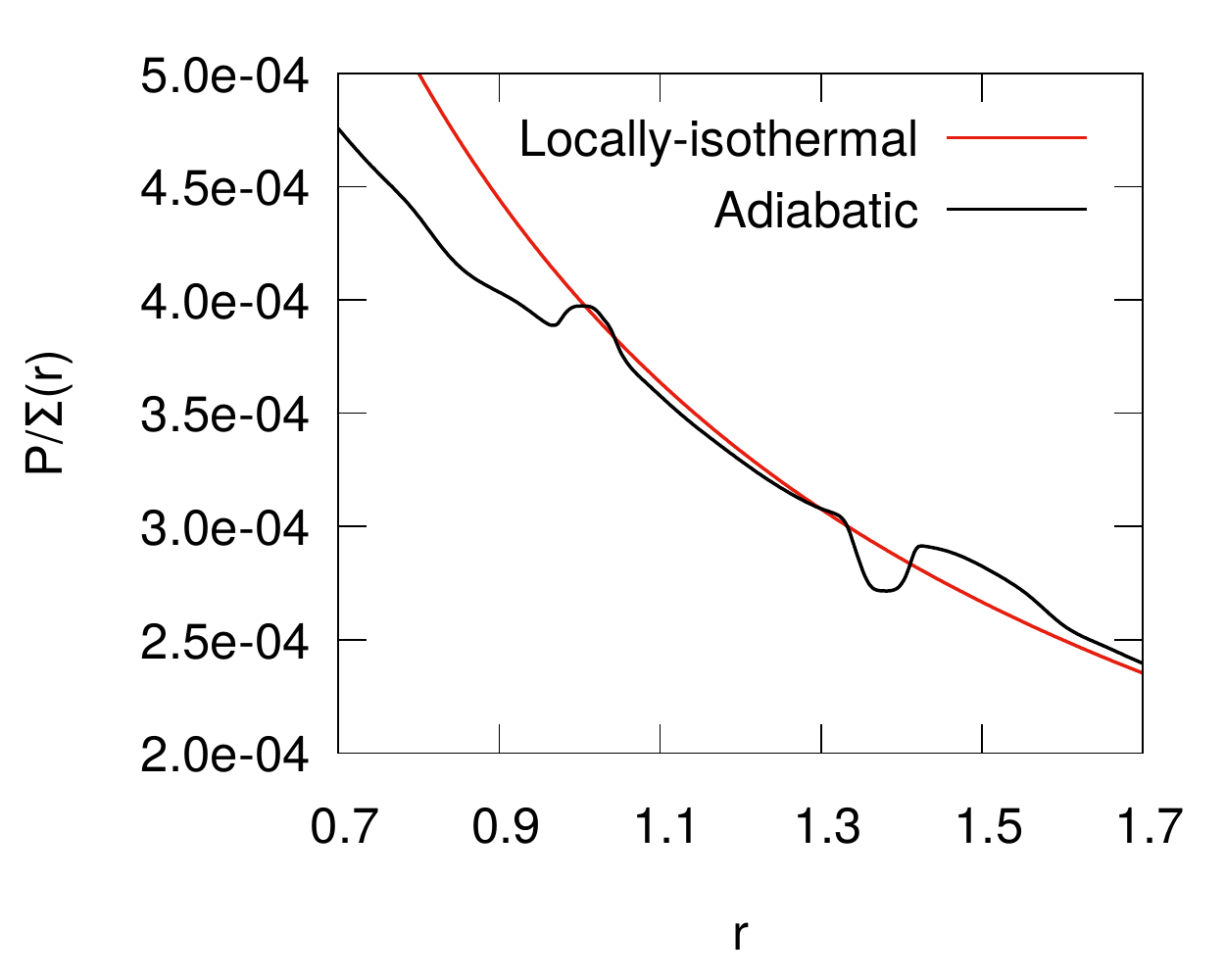}
\includegraphics[width=0.55\columnwidth]{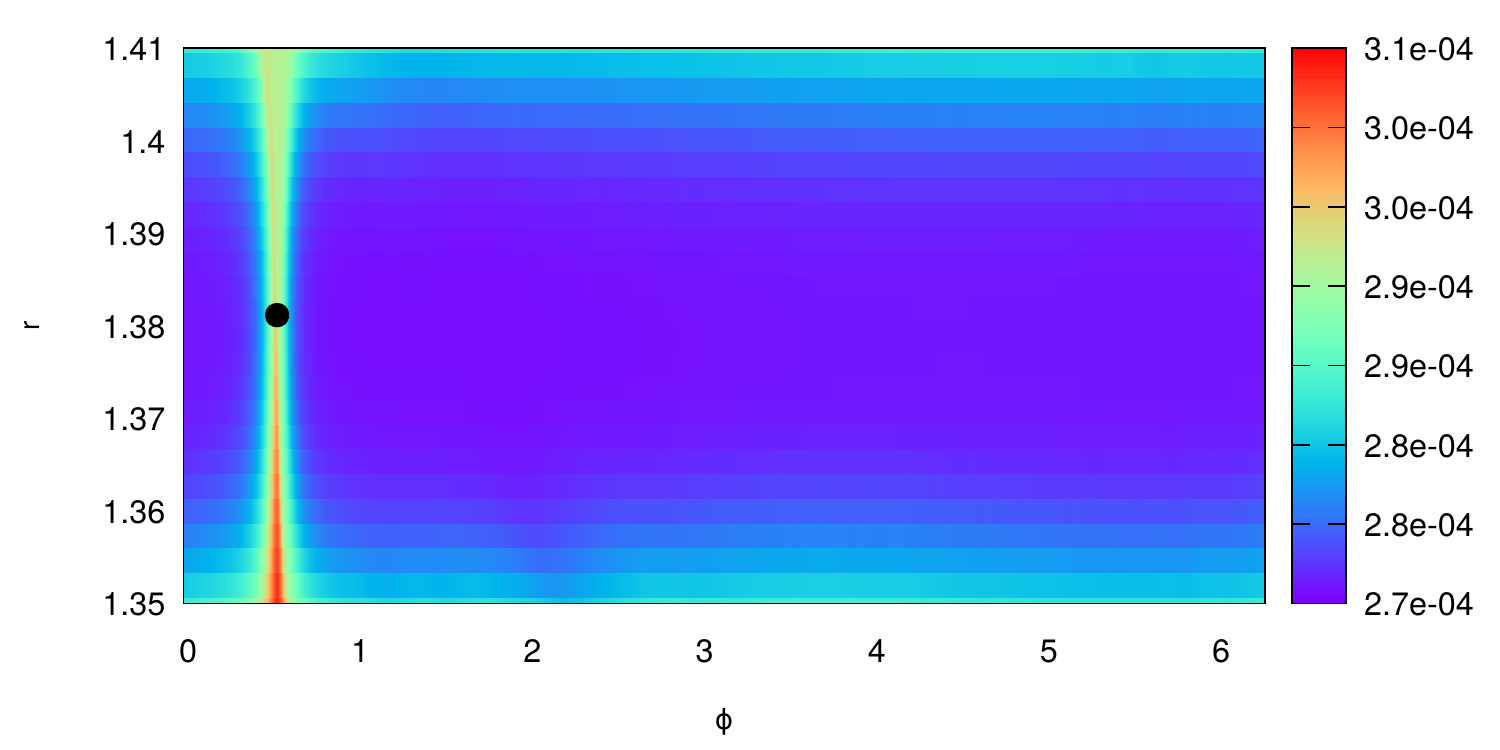}
}
}
\caption {Left: The temperature profile in the vicinity of two 
planets at t = 3000 orbits in the simulations with a 
locally-isothermal disk and an adiabatic equation of state. 
Right: A contour plot of the temperature in the horseshoe region 
of the outer planet at t = 3000 orbits in the simulation with an 
adiabatic equation of state. The position of the planet is 
indicated by the black solid circle. 
}
\label{fig:temperature}
\end{figure*}

\textcolor{red}{
\section{The effect of disk self-gravity on the migration of the 
two super-Earths}
In order to take account of the effect of self-gravity
we follow the procedure of \citet{Benitez2016} who found that 
this could be done by removing the azimuthally averaged surface 
density from  the density of each cell prior to the  calculation
of the force due to self-gravity acting on the planet.
This ensures that the background forces acting on the background 
rotating disk and  planets are applied in a consistent manner.}

\textcolor{red}{
Accordingly we run another simulation, in which all the numerical 
parameters are the same as adopted previously in the case
of a single planet with $q=1.185 \times 10^{-5}$ evolving in \textcolor{red}{a locally isothermal}
disk with $\alpha = 0.5$ starting with the surface density profile 
given by Equation~(\ref{disk}) with $\Sigma_0=6\times 10^{-5},$
$h=0.02$ and  $\nu=1.2\times 10^{-6}.$
Results for that case  have already been presented in 
Figure~\ref{fig:single-torque}.
In this case  the difference is that the forces acting on
the planet are evaluated according to the procedure  of
\citet{Benitez2016} outlined above.}

\textcolor{red}{
The evolution of the semi-major axis and the torque acting 
on the planet in this run are shown
in Figure~\ref{fig:self-gravity-check}. For comparison, 
the results  of the  simulation for the single planet
case with $\alpha = 0.5$ shown in Figure~\ref{fig:single-torque}
are also shown in this figure.}
\begin{figure*}[htb!]
\vbox{
\hbox{
\includegraphics[width=0.45\columnwidth]{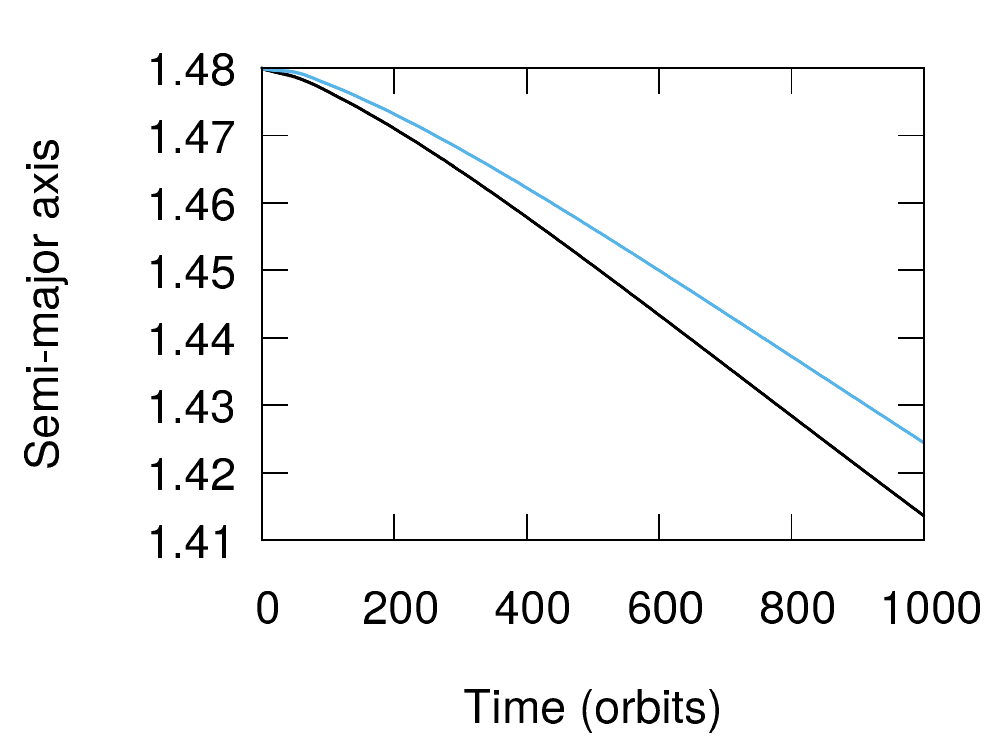}
\includegraphics[width=0.45\columnwidth]{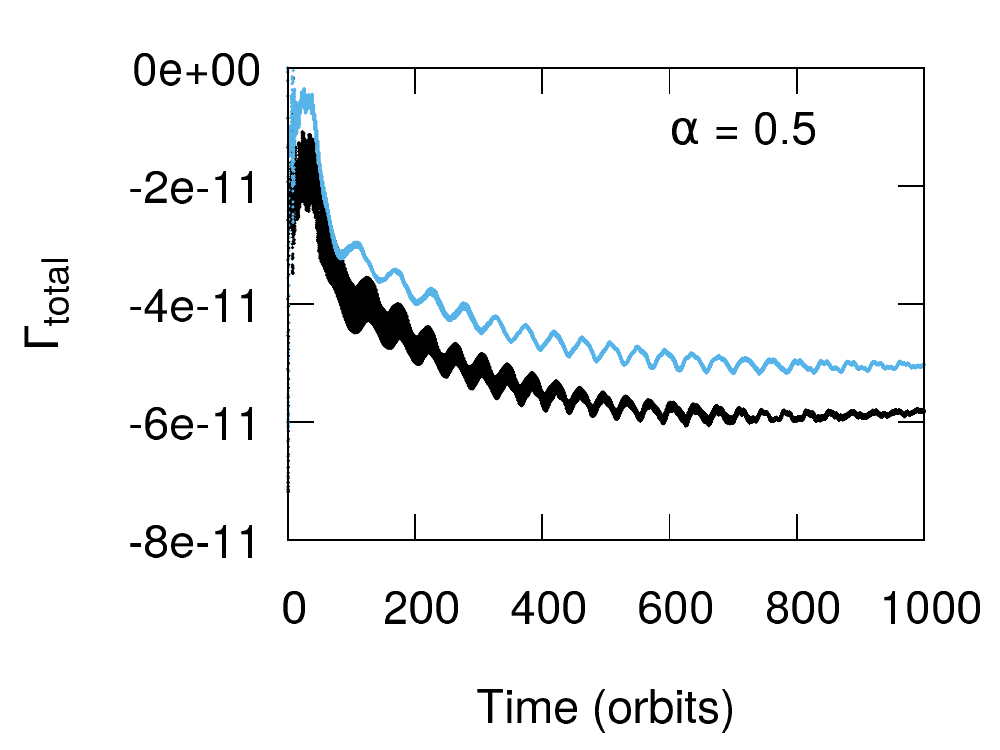}
}
}
\caption{Left: The evolution of the semi-major axis of the single planet 
in the non-self-gravitating disk (black line) and with the correction to 
allow for the effect of the disk self-gravity (blue line). 
Right: The torque from the disk acting on the planet in both cases.}
\label{fig:self-gravity-check}
\end{figure*}
\begin{figure}[htb!]
\centerline{
\vbox{
\hbox{
\includegraphics[width=0.45\columnwidth]{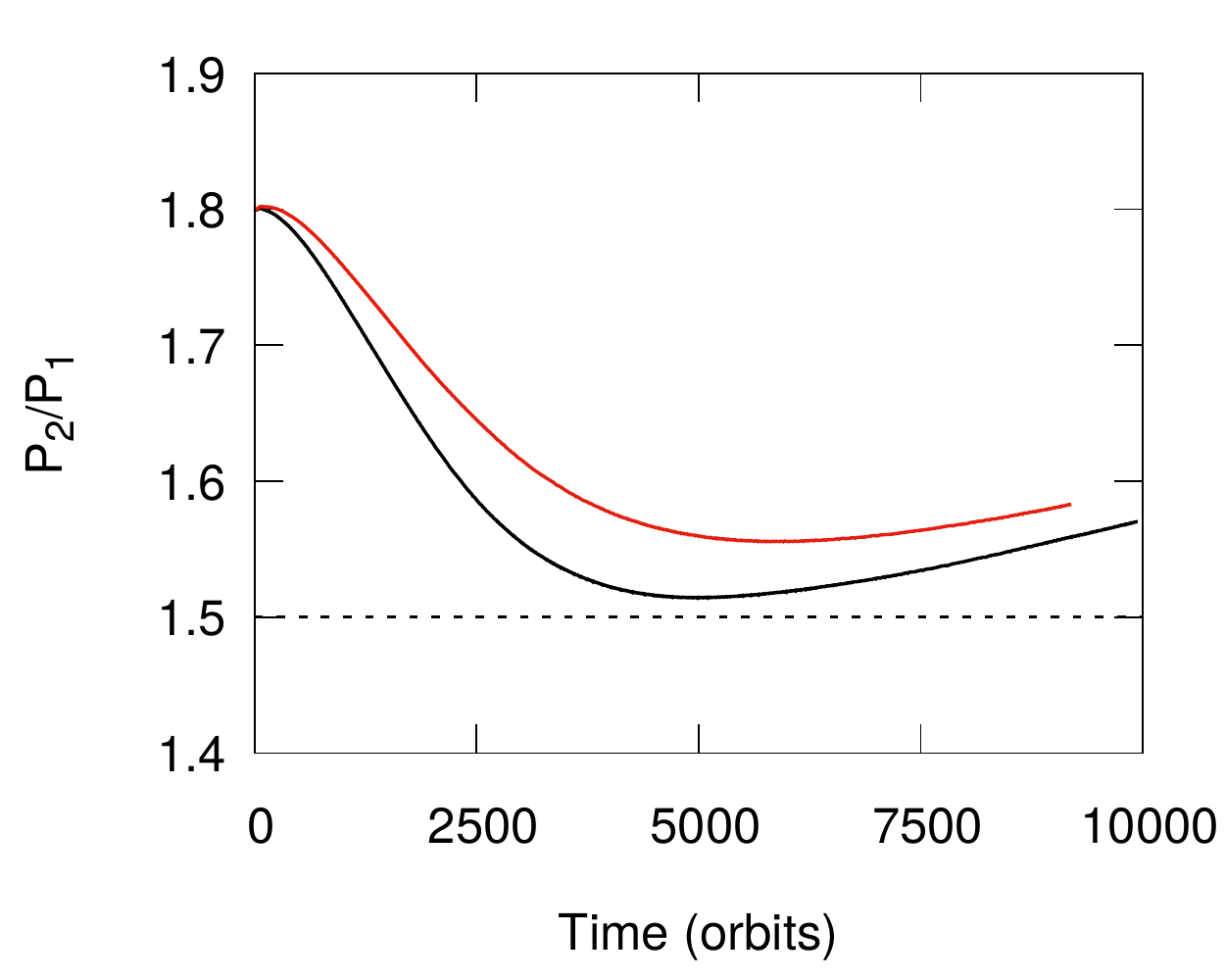}
}
}
}
\caption {The orbital period ratio of the planets in the 
non-self-gravitating disk  (black line) and with the 
correction to allow for the effect of the disk self-gravity (red line).}
\label{fig:two-self-gravity-check}
\end{figure}

\textcolor{red}{
From Figure~\ref{fig:self-gravity-check}, it is clear that after
applying the procedure to account for self-gravity, the migration 
of the planet in the disk is slightly slower than in the previous 
case.  The total torque from the disk acting on the planet after 
700 orbits is $\Gamma_{\rm total} \sim -5 \times 10^{-11}$ while 
in the case without the correction the total torque is around 
$-6 \times 10^{-11}$.
From this comparison, we see that the effect of introducing the new 
procedure is similar to what would be attained by reducing the 
background surface density by a modest amount and thus  could be 
compensated for by increasing it (see the discussion 
Section~\ref{Sigmascale}).}

\textcolor{red}{
After investigating the effect of disk self-gravity on the 
migration of a single planet we are quite confident that taking into
account this effect in the two-planet case will not change 
qualitatively the outcome of our calculations. To demonstrate this,
we perform the simulations for our flagship run \textcolor{red}{ (see Section \ref{innertrap} )} again, but this time
implementing the same procedure as described above for a single planet.
The results are shown in Figure~\ref{fig:two-self-gravity-check}
(red line) together with the original flagship case (black line).   
In this figure, we show a comparison of the evolution of
the orbital period ratios in both simulations.   
It is clear  that the divergent migration occurs later when the
self-gravity is taken into account. In addition, the divergent 
migration takes place when the distance between two planets is larger 
than in the disk without self-gravity included. The observed effect 
of disk self-gravity is similar to that obtained by reducing the 
background surface density, as illustrated in 
Figure~\ref{fig:fix_inmass2} and discussed in Section~\ref{Sigmascale}. }



\bibliography{sample63}{}

\begin{thebibliography}{100}

\bibitem[Ataiee \& Kley(2020)]{Ataiee2020} 
	Ataiee, S., Kley, W. 2020, \aap, 635, A204 
\bibitem[Baruteau \& Masset(2008)]{BarMas2008} 
	Baruteau, C., Masset, F. 2008, \apj, 678, 483 
\bibitem[Baruteau \& Papaloizou(2013)]{Baruteau2013} 
	Baruteau, C., Papaloizou, J.C.B. 2013, \apj, 778, 7
\bibitem[Batygin \& Morbidelli(2013)]{Batygin2013} 
	Batygin, K., Morbidelli, A. 2013, \aj, 145, 1
\bibitem[Benitez-Llambay \& Masset (2016)]{BenitezM2016} 
	Benitez-Llambay, P., Masset, F.S. 
	2016, \apjs, 223, 11
\bibitem[Benitez-Llambay et al.(2016)]{Benitez2016} 
	Benitez-Llambay, P., Ramos, X.S., Beauge, C., Masset, F.S. 
	2016, \apj, 826, 13
\bibitem[Carter et al.(2012)]{Carter2012} 
        Carter, J.A., Agol, E., \& Chaplin, W.J., et al. 
	2012, Science, 337, 556
\bibitem[Clarke \& Armitage (1996)]{Clarke1996} 	 
	 Clarke, C.J., Armitage, P.J., 1996, MNRAS, 280, 458
\bibitem[Crida et al.(2006)]{Crida2006} 
       Crida, A., Morbidelli, A., Masset, F. 2006, ICARUS, 181, 587
\bibitem[Delisle \& Laskar(2014)]{Delisle2014} 
	Delisle, J.B., Laskar, J. 2014, \aap, 570, L7
\bibitem[de Val-Borro et al.(2006)]{Val2006} 
	de Val-Borro, M., Edgar, R.G., \& Artymowicz, P., et al. 2006, 
	MNRAS, 370, 529
\bibitem[Dong et al.(2011)]{Dong2011} 
	Dong, R., Rafikov, R.R., Stone, J.M., Petrovich, C. 2011, \apj, 
	741, 56
\bibitem[Duffell et al.(2014)]{duffell2014} 
	Duffell, P.C., Haiman, Z., MacFadyen, A.I., D’Orazio, D.J., 
	Farris, B.D. 2014, ApJL, 792, L10
\bibitem[D\"urmann \& Kley(2015)]{durmann2015} 
	D\"urmann, C., Kley, W. 2015, \aap, 574, A52
\bibitem[D\"urmann \& Kley(2017)]{durmann2017} 
	D\"urmann, C., Kley, W. 2017, \aap, 598, A80
\bibitem[Fabrycky et al.(2014)]{Fabrycky2014} 
	Fabrycky, D.C., Lissauer, J.J., \& Ragozzine, D., et al. 
	2014, \apj, 790, 146
\bibitem[Ferguson et al. (2005)]{Ferguson2005}
	Ferguson, J. W., Alexander, D. R., Allard, F., Barman, T., 
	Bodnarik, J. G., Hauschildt, P.H., Heffner-Wong, A. \& Tamanai, A., 
	2005, \apj 623, 585
\bibitem[Goodman \& Rafikov(2001)]{Goodman} 
	Goodman, J., Rafikov, R.R. 2001, \apj, 552,793
\bibitem[Hadden \& Lithwick(2016)]{Hadden2016} 
	Hadden, S., Lithwick, Y. 2016, \apj, 828, 44
\bibitem[Hadden \& Lithwick(2017)]{Hadden2017} Hadden, S., 
	Lithwick, Y. 2017, \apj, 154, 5
\bibitem[Flaig et al. (2012)]{Flaig2012} 
	Flaig, M., Ruoff, P., Kley, W., Kissmann, R. 2012,
	MNRAS 420, 2419	
\bibitem[Jontof-Hutter et al.(2016)]{Jontof2016} Jontof-Hutter, D., 
	Ford, E.B., Rowe, J.F. 2016, \apj, 820, 39
\bibitem[Kanagawa et al.(2018)]{kanagawa2018} 
	Kanagawa, K., Tanaka, H., Szuszkiewicz, E. 2018, ApJ, 835, 140
\bibitem[Kanagawa \& Szuszkiewicz(2020)]{kanagawa2020} 
        Kanagawa, K., Szuszkiewicz, E. 2020, ApJ, 894, 59
\bibitem[Kley et al.(2004)]{Kley2004} 
	Kley, W., Peitz, J., Bryden, G. 2004, \aap, 414, 735
\bibitem[Kley \& Nelson (2012)]{KleyNelson2012} 
	Kley, W., Nelson, R.P. 2012, Annual Review of Astronomy and Astrophysics , 50, 211
\bibitem[Korycansky \& Papaloizou (1996)]{Kory1996} 
         Korycansky, D. G., Papaloizou, J.C.B. 1996, \apjsupp, 105, 181
\bibitem[Lee et al.(2013)]{Lee2013} 
	Lee, M.H., Fabrycky, D., Lin, D.N.C. 2013, \apj, 774, 52
\bibitem[Lin \& Papaloizou(1979)]{Lin1979} 
	Lin, D.N.C., Papaloizou, J.C.B. 1979, MNRAS, 186, 799 
\bibitem[Lin \& Papaloizou(1993)]{Lin1993} 
	Lin, D.N.C., \& Papaloizou, J.C.B. 1993, Protostars and Planets III
	(Univ. of Arizona Press, Tucson, AZ)
\bibitem[Lissauer et al.(2011)]{Lissauer2011} 
	Lissauer, J.J., Ragozzine, D., \& Fabrycky, D.C., et al. 2011, 
	\apj, 197, 8
\bibitem[Lithwick \& Wu(2012)]{Lithwick2012} 
	Lithwick, Y., Wu, Y. 2012, \apj, 756, L11
\bibitem[Marzari \& D' Angelo(2020)]{Marzari2020} 
Marzari, F., D' Angelo, G., 2020, \aap, 641, 125
\bibitem[Masset et al.(2006)]{Masset2006b} 
	Masset, F.S., Morbidelli, A., Crida, A., Ferreira, J. 2006, 
	\apj, 642, 478
\bibitem[Miranda \& Rafikov(2019)]{Mira2019} 
	Miranda, R., Rafikov, R.R. 2019, \apjl, 878, L9
\bibitem[Nelson \& Papaloizou(2002)]{Nelson2002} 
	Nelson, R.P., Papaloizou, J.C.B. 2002, MNRAS, 333, 26
\bibitem[Paardekooper \& Papaloizou(2009)]{Paardekooper2009} 
	Paardekooper, S.J., Papaloizou, J.C.B. 2009, MNRAS, 394, 2283
\bibitem[Paardekooper et al.(2010)]{Paardekooper2010} 
	Paardekooper, S.J., Baruteau, C., Crida, A., Kley, W.
	2010, MNRAS, 401, 1950
\bibitem[Papaloizou \& Lin(1984)]{Papa1984} 
	Papaloizou, J.C.B., Lin, D.N.C. 1984, \apj, 285, 818 
\bibitem[Papaloizou \& Szuszkiewicz(2005)]{Papa2005} 
	Papaloizou, J.C.B., Szuszkiewicz, E. 2005, MNRAS, 363, 153
\bibitem[Papaloizou et al.(2007)]{Papa2007} 
        Papaloizou, J.C.B., Nelson, R.P., Kley, W., Masset, F.S.,
        \& Artymowicz, P. 2007, Protostars and Planets V, 655 
        (University of Arizona Press, Tucson)
\bibitem[Papaloizou \& Terquem(2010)]{JohnCaroline2010} 
	Papaloizou, J.C.B., Terquem, C. 2010, MNRAS, 405, 573
\bibitem[Papaloizou(2011)]{John2011} 
	Papaloizou, J.C.B. 2011, Celest. Mech. Dyn. Astron., 111, 83
\bibitem[Podlewska-Gaca et al.(2012)]{Podlewska2012} 
	Podlewska-Gaca, E., Papaloizou, J.C.B., Szuszkiewicz, E. 2012, 
	MNRAS, 421, 1736
\bibitem[Saad-Olivera et al.(2020)]{Olivera2020} Saad-Olivera, X., 
	Martinez, C.F., Costa de Souza, A., Roig, F., Nesvoruy, D. 2020, MNRAS, 491, 5238
\bibitem[\protect\citeauthoryear{Shakura \& Sunyaev}{1973}]{Sha1973} 
       Shakura, N.I., Sunyaev, R. A. 1973, \aap, 24, 337
\bibitem[Steffen \& Hwang(2015)]{Steffen2015} 
	Steffen, J.H., Hwang, J. A. 2015, MNRAS, 448, 1956
\bibitem[Vissapragada et al.(2020)]{Vissapragada2020} 
	Vissapragada, S., Jontof-Hutter, D., \& Shporer, A., et al. 2020, \aj, 159, 108

\end{thebibliography}
\bibliographystyle{aasjournal}



\end{document}